\documentclass[11pt]{article}
\newcounter{ourcount}
\usepackage{amsmath,amsfonts,amssymb,graphicx,epsfig,pdflscape,multirow,theorem,gensymb}
\usepackage[matrix,arrow,curve]{xy} 
\numberwithin{equation}{section}
\graphicspath{{./eps/}}
\usepackage[nosort]{cite}
\usepackage[usenames]{color}
\usepackage{pstricks}
\usepackage[backref=false]{hyperref}
\usepackage[capitalise,noabbrev]{cleveref}
\hypersetup{
colorlinks=true,
citecolor=red,
linkcolor=darkblue,
urlcolor=darkblue
}
\definecolor{darkblue}{rgb}{0,0,.8}
\definecolor{red}{rgb}{1,0,0}

\theorembodyfont{\itshape} 
\theoremheaderfont{\scshape}
\theoremstyle{plain}

\numberwithin{equation}{section}

\newcommand{\nc}{\newcommand}
\def\arxiv#1#2{\href{http://arxiv.org/abs/#1}{\textsf{arXiv:#1\,#2}}}

\nc{\ir}{\mathrm{i}}
\nc{\dd}{\mathrm{d}}   
\nc{\eE}{\mathsf{e}}
\renewcommand{\ge}{\geqslant}
\renewcommand{\le}{\leqslant}
\nc{\ceff}{c_\text{eff}}
\nc{\Deltamin}{\Delta_\text{Min}}
\nc{\ddo}{\frac\dd{\dd\omega}}

\nc{\bib}{\bibitem}
\nc{\be}{\begin{equation}}
\nc{\ee}{\end{equation}}
\nc{\chit}{\raisebox{0.25ex}{$\chi$}}
\nc{\Dbh}{\mbox{\boldmath $\hat D$}}
\nc{\Dh}{\mbox{$\hat D$}}
\nc{\Dbb}{\mbox{\boldmath $\bar D$}}
\nc{\Dbm}{\mbox{\boldmath $\mathcal Q$}}
\nc{\Dm}{\mbox{$\mathcal Q$}}
\nc{\Dbmh}{\mbox{\boldmath $\mathcal {\hat  Q}$}}
\nc{\Dmh}{\mbox{$\mathcal {\hat  Q}$}}
\nc{\Dbt}{\mbox{\boldmath $\tilde{D}$}}
\nc{\Tbt}{\mbox{\boldmath $\tilde{T}$}}

\nc{\tl}{\mathsf{TL}}
\nc{\eptl}{\mathsf{\mathcal EPTL}}
\nc{\lambdabr}{\textrm{\raisebox{-0.06cm}{$\bar \lambda$}}}

\nc{\db}{\mbox{\boldmath $d$}}
\nc{\Ab}{\mbox{\boldmath $A$}}
\nc{\Bb}{\mbox{\boldmath $B$}}
\nc{\Cb}{\mbox{\boldmath $C$}}
\nc{\Db}{\mbox{\boldmath $D$}}
\nc{\eb}{\mbox{\boldmath $e$}}
\nc{\Fb}{\mbox{\boldmath $F$}}
\nc{\Fbt}{\mbox{\boldmath $\tilde{F}$}}
\nc{\fb}{\mbox{\boldmath $f$}}
\nc{\fbt}{\mbox{\boldmath $\tilde{f}$}}
\nc{\Gb}{\mbox{\boldmath $G$}}
\nc{\Hb}{\mbox{\boldmath $H$}}
\nc{\Ib}{\mbox{\boldmath $I$}}
\nc{\Jb}{\mbox{\boldmath $J$}}
\nc{\Kb}{\mbox{\boldmath $K$}}
\nc{\Lb}{\mbox{\boldmath $L$}}
\nc{\Mb}{\mbox{\boldmath $M$}}
\nc{\Ob}{\mbox{\boldmath $0$}}
\nc{\Pb}{\mbox{\boldmath $P$}}
\nc{\Qb}{\mbox{\boldmath $Q$}}
\nc{\Rb}{\mbox{\boldmath $R$}}
\nc{\Tb}{\mbox{\boldmath $T$}}
\nc{\Tbb}{\mbox{\boldmath $\bar T$}}
\nc{\Tbm}{\mbox{\boldmath $\mathcal T$}}
\nc{\tb}{\mbox{\boldmath $t$}}
\nc{\tbb}{\mbox{\boldmath $\bar t$}}
\nc{\Ub}{\mbox{\boldmath $U$}}
\nc{\Vb}{\mbox{\boldmath $V$}}
\nc{\Wb}{\mbox{\boldmath $W$}}
\nc{\Xb}{\mbox{\boldmath $X$}}
\nc{\xb}{\mbox{\boldmath $x$}}
\nc{\xbb}{\mbox{\boldmath $\bar x$}}
\nc{\yb}{\mbox{\boldmath $y$}}
\nc{\ybb}{\mbox{\boldmath $\bar y$}}
\nc{\zb}{\mbox{\boldmath $z$}}
\nc{\zbb}{\mbox{\boldmath $\bar z$}}
\nc{\Zb}{\mbox{\boldmath $Z$}}
\nc{\Lambdab}{\boldsymbol{\Lambda}}
\nc{\afb}{\mbox{\boldmath $\mathfrak a$}}
\nc{\bfb}{\mbox{\boldmath $\mathfrak b$}}
\nc{\cfb}{\mbox{\boldmath $\mathfrak c$}}
\nc{\dfb}{\mbox{\boldmath $\mathfrak d$}}
\nc{\Afb}{\mbox{\boldmath $\mathfrak A$}}
\nc{\Bfb}{\mbox{\boldmath $\mathfrak B$}}
\nc{\Cfb}{\mbox{\boldmath $\mathfrak C$}}
\nc{\Dfb}{\mbox{\boldmath $\mathfrak D$}}
\nc{\amf}{\mbox{$\mathfrak a$}}
\nc{\bmf}{\mbox{$\mathfrak b$}}
\nc{\cmf}{\mbox{$\mathfrak c$}}
\nc{\dmf}{\mbox{$\mathfrak d$}}
\nc{\fmf}{\mbox{$\mathfrak f$}}
\nc{\gmf}{\mbox{$\mathfrak g$}}
\nc{\Amf}{\mbox{$\mathfrak A$}}
\nc{\Bmf}{\mbox{$\mathfrak B$}}
\nc{\Cmf}{\mbox{$\mathfrak C$}}
\nc{\Dmf}{\mbox{$\mathfrak D$}}
\nc{\asf}{\mbox{$\mathsf a$}}
\nc{\bsf}{\mbox{$\mathsf b$}}
\nc{\csf}{\mbox{$\mathsf c$}}
\nc{\dsf}{\mbox{$\mathsf d$}}
\nc{\fsf}{\mbox{$\mathsf f$}}
\nc{\gsf}{\mbox{$\mathsf g$}}
\nc{\Asf}{\mbox{$\mathsf A$}}
\nc{\Bsf}{\mbox{$\mathsf B$}}
\nc{\Csf}{\mbox{$\mathsf C$}}
\nc{\Dsf}{\mbox{$\mathsf D$}}
\nc{\stanV}{\mathsf{V}}
\nc{\stanW}{\mathsf{W}}
\nc{\sddots}{
\begin{pspicture}[shift=0.0cm](0,0)(0,0)
\rput(0,0.1){\psdots[dotsize=0.9pt](0,0)(0.09,-0.05)(-0.09,0.05)}
\end{pspicture}
}
\nc{\sddotss}{
\begin{pspicture}[shift=0.0cm](0,0)(0,0)
\rput(0,0.1){\psdots[dotsize=0.9pt](0,0)(0.09,-0.07)(-0.09,0.07)}
\end{pspicture}
}
\def\svdots{
\begin{pspicture}[shift=0.0cm](-0.05,-0.15)(0.05,0.15)
\psdots[dotsize=0.9pt](0,0)(0,-0.09)(0,0.09)
\end{pspicture}
}
\newrgbcolor{darkgreen}{0., 0.733333, 0.0621042}

\definecolor{lightblue}{rgb}{.7,.7,1}
\definecolor{lightestblue}{rgb}{.95,.95,1}
\definecolor{lightlightblue}{rgb}{.85,.85,1}
\definecolor{midblue}{rgb}{.7,.7,1}
\definecolor{purple}{rgb}{1,0,1}

\def\facegrid#1#2{
\psframe[fillstyle=solid,fillcolor=lightlightblue,linewidth=0pt]#1#2
\psgrid[gridlabels=0pt,subgriddiv=1]#1#2}

\def\loopa{
\psframe[linewidth=.25pt](0,0)(1,1)
}
\def\loopb{
\psframe[linewidth=.25pt](0,0)(1,1)
\psarc[linewidth=1.5pt,linecolor=blue](0,1){.5}{-90}{0}
}
\def\loopc{
\psframe[linewidth=.25pt](0,0)(1,1)
\psarc[linewidth=1.5pt,linecolor=blue](1,0){.5}{90}{180}
}
\def\loopd{
\psframe[linewidth=.25pt](0,0)(1,1)
\psarc[linewidth=1.5pt,linecolor=blue](0,0){.5}{0}{90}
}
\def\loope{
\psframe[linewidth=.25pt](0,0)(1,1)
\psarc[linewidth=1.5pt,linecolor=blue](1,1){.5}{180}{270}
}
\def\loopf{
\psframe[linewidth=.25pt](0,0)(1,1)
\psline[linewidth=1.5pt,linecolor=blue](0.5,0)(0.5,1)
}
\def\loopg{
\psframe[linewidth=.25pt](0,0)(1,1)
\psline[linewidth=1.5pt,linecolor=blue](0,0.5)(1,0.5)
}
\def\looph{
\psframe[linewidth=.25pt](0,0)(1,1)
\psarc[linewidth=1.5pt,linecolor=blue](1,0){.5}{90}{180}
\psarc[linewidth=1.5pt,linecolor=blue](0,1){.5}{-90}{0}
}
\def\loopi{
\psframe[linewidth=.25pt](0,0)(1,1)
\psarc[linewidth=1.5pt,linecolor=blue](0,0){.5}{0}{90}
\psarc[linewidth=1.5pt,linecolor=blue](1,1){.5}{180}{270}
}

\begin{document}

\topmargin -5mm
\oddsidemargin 5mm

\vspace*{-2cm}

\setcounter{page}{1}

\vspace{22mm}
\begin{center}
{\huge {\bf Groundstate finite-size corrections and \\[0.1cm]  dilogarithm identities for the twisted  \\[0.1cm] $\boldsymbol{A_1^{(1)}}$, $\boldsymbol{A_2^{(1)}}$ and $\boldsymbol{A_2^{(2)}}$ models}}
\end{center}

\vspace{10mm}
\begin{center}
{\vspace{-5mm}\Large Alexi Morin-Duchesne$^{a,b}$, Andreas Kl\"umper$^c$, Paul A.~Pearce$^{d,e}$}
\\[.5cm]
{\em {}$^a$Institut de Recherche en Math\'ematique et Physique\\ Universit\'e catholique de Louvain, Louvain-la-Neuve, B-1348, Belgium}
\\[.4cm]
{\em {}$^b$Max Planck Institut f\"ur Mathematik, 53111 Bonn, Germany}
\\[.4cm]
  {\em {}$^c$Fakult\"at f\"ur Mathematik und Naturwissenschaften \\ Bergische
 Universit\"at Wuppertal, 42097 Wuppertal, Germany}
\\[.4cm]
{\em {}$^d$School of Mathematics and Statistics, University of Melbourne\\
Parkville, Victoria 3010, Australia}
\\[.4cm]
{\em {}$^e$School of Mathematics and Physics, University of Queensland}\\
{\em St Lucia, Brisbane, Queensland 4072, Australia}
\\[.4cm] 
{\tt alexi.morin.duchesne\,@\,gmail.com}
\qquad
{\tt kluemper\,@\,uni-wuppertal.de}
\qquad
{\tt papearce\,@unimelb.edu.au}

\end{center}


\vspace{8mm}
\centerline{{\bf{Abstract}}}
\vskip.4cm
\noindent 
We consider the $Y$-systems satisfied by the $A_1^{(1)}$, $A_2^{(1)}$, $A_2^{(2)}$ vertex and loop models at roots of unity with twisted boundary conditions on the cylinder. The vertex models are the 6-, 15- and Izergin-Korepin 19-vertex models respectively. The corresponding loop models are the dense, fully packed and dilute Temperley-Lieb loop models respectively. For all three models, our focus is on roots of unity values of $\eE^{\ir\lambda}$ with the crossing parameter $\lambda$ corresponding to the principal and dual series of these models. Converting the known functional equations to nonlinear integral equations in the form of Thermodynamic Bethe Ansatz (TBA) equations, we solve the $Y$-systems for the finite-size $\frac 1N$ corrections to the groundstate eigenvalue following the methods of Kl\"umper and Pearce. The resulting expressions for $c-24\Delta$, where $c$ is the central charge and $\Delta$ is the conformal weight associated with the groundstate, are simplified using various dilogarithm identities. Our analytic results are in agreement with previous results obtained by different methods.


\newpage
\tableofcontents
\newpage
\hyphenpenalty=30000

\setcounter{footnote}{0}

%
\section{Introduction}
%

Two-dimensional lattice models on the square lattice are Yang-Baxter integrable~\cite{BaxterBook1982} if their Boltzmann face weights, dependent on the spectral parameter $u$ and crossing parameter $\lambda$,  satisfy the Yang-Baxter equation. Solutions to the Yang-Baxter equation are typically classified~\cite{Bazh85,Jimbo86} by Lie algebras. Perhaps the simplest families of Yang-Baxter integrable lattice models are classified by the trilogy of Lie algebras $A_1^{(1)}$, $A_2^{(1)}$ and $A_2^{(2)}$. The representations of these algebras admit at least three types, namely, vertex type, loop type and height or Restricted Solid-On-Solid (RSOS) type.
Specifically, the vertex models are the 
\mbox{6-vertex} model~\cite{Lieb1967,Baxter1972,Baxter1973,ZP95,BazhMang2007,FrahmMDP2019}, 
15-vertex model~\cite{KR82,BabelonEtAl1982,dV89,AlcarazMartins1990,dVGR94,KNS94,ZinnJustin1998} and 
Izergin-Korepin 19-vertex model~\cite{IK1981,WBN1992,AMN1995,ZB97,VJS2014} respectively. The corresponding loop models are the 
dense Temperley-Lieb loop model~\cite{Nienhuis82,BloteNienhuis89,YB95,PRZ2006,SAPR2009,MDPR2014,MDKP2017}, 
fully packed Temperley-Lieb loop model~\cite{Resh91,DEI2016,MDPR2019} and 
dilute Temperley-Lieb loop model~\cite{DJS2010,SAPR2012,G12,FeherNien2015,GarbNien2017,MDP19} respectively. 
Algebraically, these loop models are described by the dense and dilute Temperley-Lieb algebras~\cite{TempLieb1971,GP93,P94,Grimm96,planarJones1999}.
In this paper, we are interested in representations of vertex and loop type models at roots of unity, where the crossing parameter $\frac{\lambda}{\pi}\in\Bbb Q$ is parameterised as
\be
\lambda=\left\{\begin{array}{cl}
\displaystyle\frac{\pi(p'-p)}{p'},\quad&\mbox{$A_1^{(1)}$, $A_2^{(1)}$\!,}\\[8pt]
\displaystyle\frac{\pi(b-a)}{2b},\quad
&A_2^{(2)}\!,
\end{array}
\right.
\qquad p,p',a,b \in \mathbb Z, \qquad \mathrm{gcd}(p,p') = \mathrm{gcd}(a,b) = 1.
\label{RootsUnity}
\ee
Our investigation will in fact focus on a subset of these values, namely on two one-parameter series of values, the {\it principal series} and the {\it dual series}:
\be
A_1^{(1)}\!, A_2^{(1)}: \quad
\left\{\begin{array}{ll}
p = p'-1, & \textrm{Principal series,}
\\[0.1cm]
p=1, & \textrm{Dual series,} 
\end{array}\right.
\qquad
A_2^{(2)}: \quad
\left\{\begin{array}{ll}
a = b-1, & \textrm{Principal series,}
\\[0.1cm]
a=1-b, & \textrm{Dual series.} 
\end{array}\right.
\ee

We consider periodic boundary conditions on a cylinder with a twist $\omega = \eE^{\ir \gamma}$, where the vertex and loop models are described by single-row transfer matrices $\Tb(u)$. The Yang-Baxter equation ensures the commutativity of the transfer matrices at different values of the spectral parameter. The groundstate eigenvalue $T(u)$ of the transfer matrix admits the $\frac1N$ expansion
\be
\label{eq:fsc.general}
\log T(u) = -N f_{\rm bulk}(u) + \frac{\pi \sin \vartheta(u)}{6N}(c - 24 \Delta) + o(\tfrac1N).
\ee
The first term involves the bulk free energy $f_{\rm bulk}(u)$, a non-universal quantity that can be computed using the inversion identities satisfied by the transfer matrices \cite{YS1979,B82}. In contrast, the $\frac1N$ finite-size correction term is universal. Its prefactor is written in terms of the conformal data of the associated Conformal Field Theory (CFT) realized in the continuum scaling limit, namely the central charge $c$ and the dimensions $\Delta = \bar \Delta$ of the conformal field corresponding to the groundstate. It also involves the anisotropy angle $\vartheta(u)$, which is model dependent. 

The integrability of the $A_1^{(1)}$, $A_2^{(1)}$ and $A_2^{(2)}$ models is manifest in the $T$- and $Y$-systems of functional equations~\cite{Zam1991a,Zam1991b,KP92,KunibaNS9310,KNS2011} satisfied by the transfer matrices. Closed finite $T$- and $Y$-systems for the 
$A_1^{(1)}$, $A_2^{(1)}$, $A_2^{(2)}$ vertex and loop models at roots of unity have been obtained in \cite{MDPR2014,MDPR2019,MDP19} respectively. 
The $Y$-systems are {\em universal}~\cite{CMP2001} in the sense that they do not depend on the boundary conditions or the topology. Following the methods of Kl\"umper and Pearce~\cite{PK91,KP91,KP92}, each $Y$-system can be solved for the finite-size $\frac 1N$ correction term in \eqref{eq:fsc.general}, allowing one to compute $c - 24 \Delta$ and $\vartheta (u)$ exactly.

For the vertex models, a twist $\omega = \eE^{\ir \gamma}$ is introduced between the first and last columns in the transfer matrix $\Tb(u)$ via a twist matrix $\mathbf{\Theta}$ that acts on the auxiliary space. The transfer matrix is then given by a trace over that auxiliary space:
\be
\label{eq:twists}
\mathop{\mbox{Tr}}  \big(\mathbf{\Theta}\,\Tb(u)\big),
\qquad 
\mathbf{\Theta}=\left\{
\begin{array}{cc}
    \begin{pmatrix}
      \omega & 0 \\ 0 & \omega^{-1}\end{pmatrix},&\quad A_1^{(1)}\!,\\[0.6cm]
    \begin{pmatrix}
      \omega & 0&0 \\ 0 &1&0\\ 0&0&\omega^{-1}\end{pmatrix},&\quad A_2^{(1)}\!, A_2^{(2)}\!.
     \end{array}\right.
\ee
Because of the $s\ell(3)$ symmetry, the $A_2^{(1)} $ models allow for a two-parameter twist matrix but we only consider the one-parameter twist matrix.

For the loop models, the relevant representations of the Temperley-Lieb algebras are the so-called {\it standard modules}. Their construction in terms of link states is well-known and here we will use the convention of \cite{MDKP2017,MDPR2019,MDP19} for these modules for the $A_1^{(1)}$, $A_2^{(1)}$ and $A_2^{(2)}$ models, respectively. The twist $\omega = \eE^{\ir \gamma}$, with $\gamma\in(-\pi,\pi)$, is a parameter associated with these representations. It measures the winding of the defects around the cylinder. 
Each defect that crosses the vertical line where the cylinder is cut contributes a factor $\omega$ if it travels towards the left, whereas it contributes $\omega^{-1}$ if it travels towards the right as it progresses down the cylinder. Clearly, for loop models, powers of $\omega$ only appear in standard modules with a positive number of defects. In the zero defect module, the representation instead depends on the fugacity $\alpha$ of the non-contractible loops. This fugacity  is parametrised as $\alpha=\omega+\omega^{-1} =2\cos\gamma$ allowing for the simultaneous treatment of vertex and loop models in our calculations.

Some special values of the twist parameter are distinguished. The untwisted vertex model is obtained by specialising the twist parameter to $\omega = 1$. In the loop model, this corresponds to setting the fugacity of
the non-conctractible loops to $\alpha = \omega+\omega^{-1} =2$. For this value, the vertex and loop models share a common modular invariant partition function. Another natural choice for the loop models is to assign equal fugacities to the contractible and non-contractible loops: $\alpha = \beta$. This is achieved by specialising the twist parameter to
\be
\label{TwistChoices}
\gamma=
\left\{\begin{array}{ll}
\lambda,&A_1^{(1)}\!, A_2^{(1)}\!,\\[0.1cm]
4\lambda-\pi,&A_2^{(2)}\!, \quad 0 < \lambda < \frac \pi 2,\\[0.1cm]
3\pi-4\lambda,&A_2^{(2)}\!, \quad \frac \pi 2 < \lambda < \pi.
\end{array}\right.
\ee
For the $A_2^{(1)}$ case, the choice $\gamma=\lambda$ coincides with the Fully Packed Loop (FPL) model~\cite{BatchBNY,KondevGN}. Another interesting value is $\gamma=2\lambda$ for which the continuum scaling limit of this model is described~\cite{DEI2016} by a non-rational $W_3$ conformal field theory, which we believe can be obtained as the logarithmic limit ($m,m'\to\infty, m/m'\to p/p'$) of the rational minimal $W_3(m,m')$ models~\cite{Zamo1985,BouwkMP,IlesWatts}.

From the analytic calculation of finite-size corrections, our results for the conformal data of the twisted models are
\begingroup
\allowdisplaybreaks
\begin{subequations}
\label{eq:central.charges}
\begin{alignat}{4}
&A_1^{(1)}:\qquad&&
\displaystyle c - 24 \Delta = \left\{\begin{array}{ll}
\displaystyle 1-\frac{6\gamma^2 p'}{\pi^2(p'-1)},\\[0.3cm]
\displaystyle 1-\frac{6\gamma^2 p'}{\pi^2},
\end{array}\right.
\quad&&
\begin{array}{ll}
\textrm{\raisebox{-0.05cm}{Principal series,}}\\[0.2cm]
\textrm{\raisebox{-0.37cm}{Dual series,}}
\end{array}
\\
&A_2^{(1)}: \qquad&&
\displaystyle c - 24 \Delta = \left\{\begin{array}{ll}
\displaystyle 2-\frac{6\gamma^2 p'}{\pi^2(p'-1)},\\[8pt]
\displaystyle 2-\frac{6\gamma^2 p'}{\pi^2},
\end{array}\right.
\quad&&
\begin{array}{ll}
\textrm{\raisebox{-0.05cm}{Principal series,}}\\[0.2cm]
\textrm{\raisebox{-0.37cm}{Dual series,}}
\end{array}\\
\label{A22dualresult}&A_2^{(2)}:\qquad&&
\displaystyle c - 24 \Delta = \left\{\begin{array}{ll}
\displaystyle 1-\frac{3\gamma^2 b}{\pi^2(b-1)},\\[8pt]
\displaystyle \frac32-\frac{3\gamma^2 b}{\pi^2},
\end{array}\right.
\quad&&
\begin{array}{ll}
\textrm{\raisebox{-0.05cm}{Principal series,}}\\[0.2cm]
\textrm{\raisebox{-0.37cm}{Dual series.}}
\end{array}
\end{alignat}
\end{subequations}
\endgroup
These are to be compared with the analytic results from other methods, such as Non-Linear Integral Equations (NLIEs) and Bethe Ansatz techniques:
\begin{subequations}
\label{eq:central.charges.2}
\begin{alignat}{4}
&A_1^{(1)}:\qquad&
\displaystyle c - 24 \Delta &= 
\displaystyle 1-\frac{6\gamma^2}{\pi(\pi-\lambda)},\hspace{1.1cm}\mbox{$0<\lambda<\pi$ (dense phase~\cite{HQB87,KBP91}),}\\[8pt]
&A_2^{(1)}: \qquad&
\displaystyle c - 24 \Delta &= 
\displaystyle 2-\frac{6\gamma^2}{\pi(\pi-\lambda)}, \hspace{1.1cm}
\mbox{$0<\lambda<\pi$ (Regimes I \& II~\cite{AlcarazMartins1990,ZinnJustin1998,DEI2016})},
\\
&A_2^{(2)}:\qquad&
\displaystyle c - 24 \Delta &= \left\{\begin{array}{ll}
\displaystyle 1-\frac{3\gamma^2}{\pi(\pi-2\lambda)},&\mbox{$0<\lambda<\frac{\pi}{2}$ (Regime I: dilute/dense phase~\cite{WBN1992,ZB97}),}\\[8pt]
\displaystyle \frac32-\frac{3\gamma^2}{2\pi(\pi-\lambda)},&\mbox{$\frac{2\pi}{3}<\lambda<\pi$ (Regime II~\cite{WBN1992,ZB97}).}
\end{array}\right.
\end{alignat}
\end{subequations}
The $A_2^{(1)}$ intervals in Regimes I \& II are $0<\lambda<\frac{\pi}{2}$ and $\frac{\pi}{2}<\lambda<\pi$ respectively.
The $A_2^{(2)}$ dilute/dense intervals in Regime I are $0<\lambda<\frac{\pi}{4}$ and $\frac{\pi}{4}<\lambda<\frac{\pi}{2}$ respectively. Note also that we do not consider the $A_2^{(2)}$ model in the non-compact regime $\frac \pi 2<\lambda<\frac{2\pi}3$~\cite{VJS2014}. 

The outline of the paper is as follows. We compute the groundstate finite-size corrections for the $A_1^{(1)}$\!, $A_2^{(2)}$\! and $A_2^{(1)}$\! models in \cref{sec:A11,sec:A22,sec:A21} respectively. While the $A_2^{(1)}$\! model may appear simpler as a lattice model, the analysis of its $Y$-system is more complicated than that of the $A_2^{(2)}$\! model, prompting us to present our analysis for the $A_2^{(2)}$ model first for pedagogical reasons. Each of these three sections is subdivided in the same way. Firstly, we review the definition of the model. Secondly, we review the functional relations and $Y$-systems that were obtained in previous papers. Thirdly and fourthly, we solve these equations for the $\frac 1N$ finite-size correction for the principal and dual series respectively. The final step of the calculations uses certain dilogarithm identities stated and proven in \cref{app:dilogs}. We make some concluding remarks in \cref{sec:conclusion}.

Throughout the paper, our algebraic and analytic results are checked by a computer implementation using Mathematica~\cite{Wolfram} on a Mac Pro with 8 parallel kernels and 256GB of RAM. The fused transfer matrices are coded symbolically out to system sizes $N=6$.  Fixing $\lambda$ and $\gamma$ numerically to 40 digits precision, the entries of fused transfer matrices are obtained to high precision as Laurent polynomials in $\eE^{\ir u}$ with constant coefficients for system sizes out to $N=9$. Using 18 digit machine precision, the fused transfer matrices were obtained out to system sizes $N=11$ or $N=12$. Since the common transfer matrix eigenvectors are independent of $u$, the eigenvalues are also obtained as Laurent polynomials in $\eE^{\ir u}$ with numerical coefficients. Importantly, this allows us to solve numerically for the locations of eigenvalue zeros in the complex $u$ plane and to confirm the qualitative patterns of zeros and analyticity information we use to solve the $Y$-systems. Specializing $u$, $\lambda$ and $\gamma$, direct confirmation of the finite-size corrections were carried out using the Arnoldi method~\cite{Arnoldi} to find the dominant eigenvalues of numerical unfused transfer matrices out to system sizes $N=12$ and applying Vanden Broeck-Schwartz~\cite{vBS} sequence extrapolation.

%
\section{Finite-size corrections for the $\boldsymbol{A_1^{(1)}}\!$ models}\label{sec:A11}

\subsection[Definition of the $A_1^{(1)}$ models]{Definition of the $\boldsymbol{A_1^{(1)}}\!$ models}

The loop and vertex models in the $A_1^{(1)}$ family are the dense loop model and the 6-vertex model. The dense loop model is a face model on the square lattice, where each face takes on one of two possible local configurations designated by tiles. 
The elementary face operator for the loop model is defined by the linear combination
\be
\begin{pspicture}[shift=-.40](0,0)(1,1)
\facegrid{(0,0)}{(1,1)}
\psarc[linewidth=0.025]{-}(0,0){0.16}{0}{90}
\rput(.5,.5){$u$}
\end{pspicture}
\ \ =s_1(-u)\ \
\begin{pspicture}[shift=-.40](0,0)(1,1)
\facegrid{(0,0)}{(1,1)}
\rput[bl](0,0){\looph}
\end{pspicture}
\ + s_0(u) \ \
\begin{pspicture}[shift=-.40](0,0)(1,1)
\facegrid{(0,0)}{(1,1)}
\rput[bl](0,0){\loopi}
\end{pspicture}\ \, , \qquad
s_k(u) = \frac{\sin(k \lambda + u)}{\sin \lambda},
\label{eq:faceop.A11}
\ee
where $u$ is the spectral parameter and $\lambda$ is the crossing parameter. On the cylinder, the fugacities of the contractible and non-contractible loops are
\be
\beta = 2 \cos \lambda, \qquad \alpha = \omega + \omega^{-1} = 2 \cos \gamma,
\label{betaq}
\ee
where $\omega = \eE^{\ir \gamma}$ is a free parameter.
The $\check R$-matrix of the six-vertex model is
\be
\check R(u) = 
\begin{pmatrix}
      s_1(-u)&0&0&0 \\[0.1cm]
      0& \eE^{\ir u} &s_0(u) &0\\[0.1cm]
      0& s_0(u) & \eE^{-\ir u}&0\\[0.1cm]
      0&0&0& s_1(-u)
\end{pmatrix}=s_1(-u)I+s_0(u)\check R(\lambda).
\ee
The twist matrix is \eqref{eq:twists} with $\omega = \eE^{\ir \gamma}$. Both the vertex and loop $A_1^{(1)}$ models are described by the Temperley-Lieb algebra \cite{TempLieb1971}, with its parameter fixed to $\beta = 2 \cos \lambda$. On the cylinder, the single-row transfer matrices are elements of the enlarged periodic Temperley-Lieb algebra \cite{L91,MS93,GL98,PRVcyl2010}. For the vertex model, the Temperley-Lieb generators are defined on $({\Bbb C}^2)^{\otimes N}$ by
\be
e_j=I\otimes I \otimes \cdots \otimes\check R(\lambda)\otimes\cdots\otimes I,\qquad j=1,2,\ldots,N,
\ee
where $\check R(\lambda)$ occurs in slots $j$ and $j+1$.

The fundamental regime of the $A_1^{(1)}\!$ models is
\be
0<\lambda<\pi, \qquad 0<u<\lambda.
\ee
The roots of unity values of $\lambda$ are those for which $\frac{\lambda}{\pi}\in{\Bbb Q}$. We parameterise them in terms of two integers $p,p'$ as
\be
\lambda = \lambda_{p,p'} =  \frac{\pi(p'-p)}{p'},\qquad  1\le p<p',\qquad\mbox{gcd}(p,p')=1.
\ee
Our calculation of the finite-size corrections below focuses on two series:
\be
\label{eq:series.A11}
\begin{array}{lll}
\textrm{Principal series:}\quad & (p,p') = (p'-1,p'),\quad & 0<u<\lambda,\\[0.15cm]
\textrm{Dual series:}\quad & (p,p') = (1,p'), & 0<u<\lambda.
\end{array}
\ee
We fix $N$ to be an even number. For the vertex model, the groundstate lies in the zero magnetisation sector. For the loop model, it lies in the standard module with zero defects, $\stanW_{N,0}$. (We follow the convention used in \cite{MDKP2017} for these modules.) 
In these sectors, the spectrum of the transfer matrix $\Tb(u)$ of the $A_1^{(1)}$ models is invariant (up to an irrelevant overall sign) under the involution
\be
\lambda \leftrightarrow \pi-\lambda,\quad u \leftrightarrow -u.
\ee
The dual series can therefore be alternatively specified by 
\be
\label{eq:dualseries.A11}
\begin{array}{ll}
\textrm{Dual series:}\quad & (p,p') = (p'-1,p'),\qquad \lambda-\pi<u<0.
\end{array}
\ee

In this section, we study the groundstate of the transfer matrix for even $N$. We focus on values of
$u$ in the neighborhood of $u = \frac \lambda 2$ and restrict to values
of the twist parameter $\omega = \eE^{\ir \gamma}$  with $\gamma$ in the 
interval $(0,\frac{\pi}{p'})$. We will compute the $\frac1N$ finite-size correction 
term for the groundstate eigenvalue $T(u)$ to confirm the conformal 
prediction \eqref{eq:fsc.general} with 
\be
\label{eq:fsc.A11}
c-24 \Delta = \left\{\begin{array}{ll}
\displaystyle 1-\frac{6\gamma^2 p'}{\pi^2(p'-1)}, \quad &\textrm{Principal series,}\\[0.35cm]
\displaystyle 1-\frac{6\gamma^2 p'}{\pi^2},\quad &\textrm{Dual series,}
\end{array}\right.
\qquad
\vartheta(u) = \frac {\pi u}{\lambda}.
\ee

\subsection{Functional relations}

The fused transfer matrices $\Tb^{n}(u)$ for the $A_1^{(1)}$ models are defined recursively \cite{MDPR2014,MDKP2017} from the fusion hierarchy relations, as functions of the fundamental transfer matrix $\Tb^{1}(u) = \Tb(u)$. The $T$-system equations are
\be
\label{eq:Trelations.A11}
\Tb^n_0 \Tb^n_1 = f_{-1}f_{n}\Ib + \Tb^{n+1}_0 \Tb^{n-1}_1, \qquad n \ge 0,
\ee
where we use the notations
\be
\Tb^1(u) = \Tb(u),\qquad\Tb^n_k = \Tb^{n} (u+k \lambda), \qquad  \Tb^0_k = f_{k-1} \Ib, \qquad \Tb^{-1}_k = \Ob,\qquad f_k = \bigg(\frac{\sin(k\lambda + u)}{\sin \lambda}\bigg)^N.
\ee
For $\lambda = \lambda_{p,p'}$, the infinite $Y$-system closes into a finite system involving $p'-1$ functions:
\be
\tb^{n}_0 = \frac{\Tb^{n+1}_0\Tb^{n-1}_1}{f_{-1}f_n}, \qquad \xb_0 = (-1)^{Np/2}\frac{\Tb^{p'-2}_1}{f_{-1}},
\ee
where $n = 1, \dots, p'-2$.
The $Y$-system equations are
\begin{subequations}
\begin{alignat}{2}
\tb^n_0 \tb^n_1 &= (\Ib + \tb^{n-1}_1)(\Ib + \tb^{n+1}_0), \hspace{1.5cm} n = 1, \dots, p'-3,
\\[0.15cm]
\tb^{p'-2}_0 \tb^{p'-2}_1 &= (\Ib + \tb^{p'-3}_1)(\Ib+\eE^{\ir \Lambdab}\xb_0)(\Ib+\eE^{-\ir \Lambdab}\xb_0), 
\\[0.15cm]
\xb_0 \xb_1 &= (\Ib + \tb^{p'-2}_1).
\end{alignat}
\end{subequations}
In the loop model, the operator $\eE^{\ir \Lambdab}$ is diagonal on the standard module $\stanW_{N,0}$, with its unique eigenvalue given by $\omega^{p'}$. Likewise in the vertex model, in the zero-magnetisation sector, the matrix $\eE^{\ir \Lambdab}$ is diagonal with the unique eigenvalue $\omega^{p'}$.

\subsection{The principal series}\label{sec:A11.principal}

In this subsection, we fix $p = p'-1$ with $p' \in \mathbb N_{\ge 2}$, so that $\lambda = \frac{\pi}{p'}$. We compute the $\frac1N$ term in \eqref{eq:fsc.general} explicitly, for $N$ even, $\gamma \in (0,\frac\pi{p'})$ and $u$ in the neighborhood of $\frac \lambda 2$.

\subsubsection[Analyticity properties and symmetric $Y$-system]{Analyticity properties and symmetric $\boldsymbol Y$-system}\label{sec:analyticity.A11}

\begin{figure}
\centering
\begin{tabular}{ccc}
\begin{tabular}{c}
$T^{1}(u)$ \\[0.1cm]
\includegraphics[width=.25\textwidth]{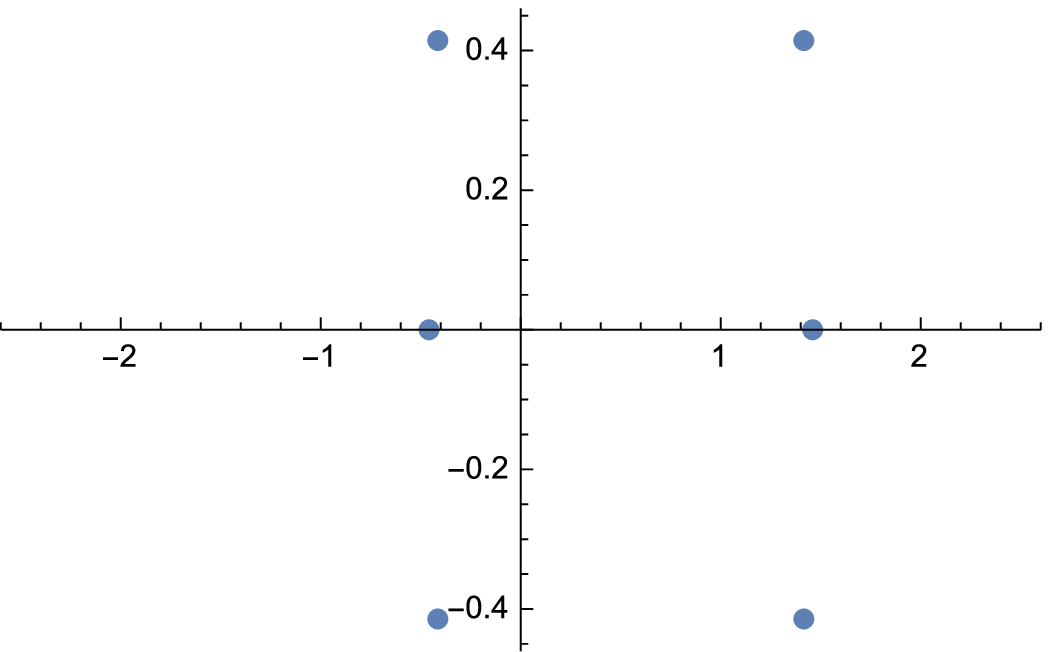} \\[0.3cm]
$T^{3}(u)$ \\[0.1cm]
\includegraphics[width=.25\textwidth]{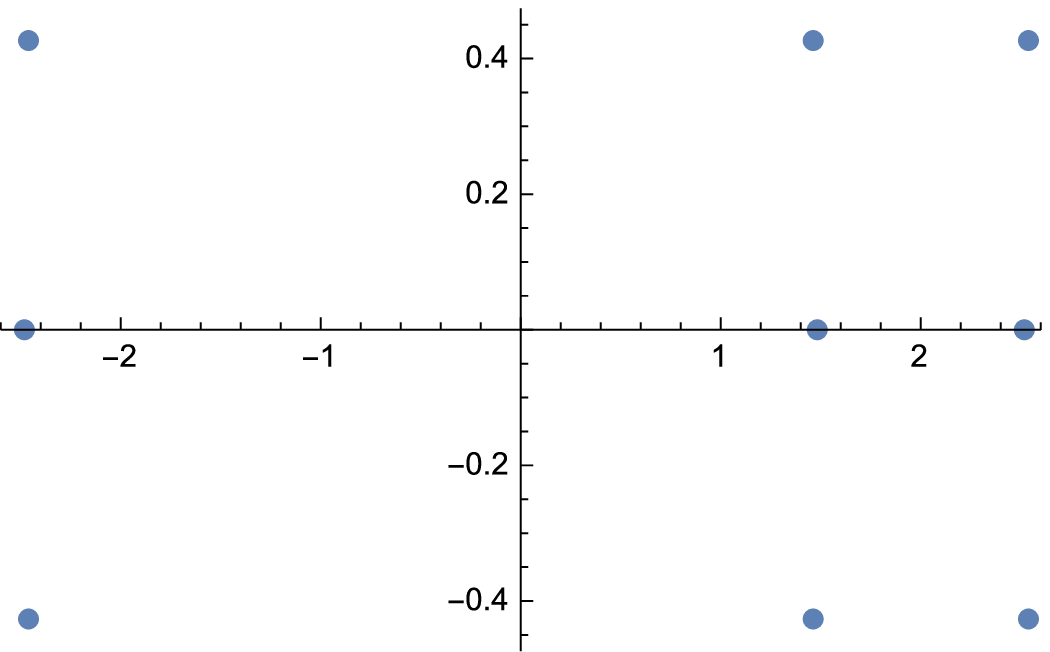}
\end{tabular}
&\qquad&
\begin{tabular}{c}
$T^{2}(u)$ \\[0.1cm]
\includegraphics[width=.25\textwidth]{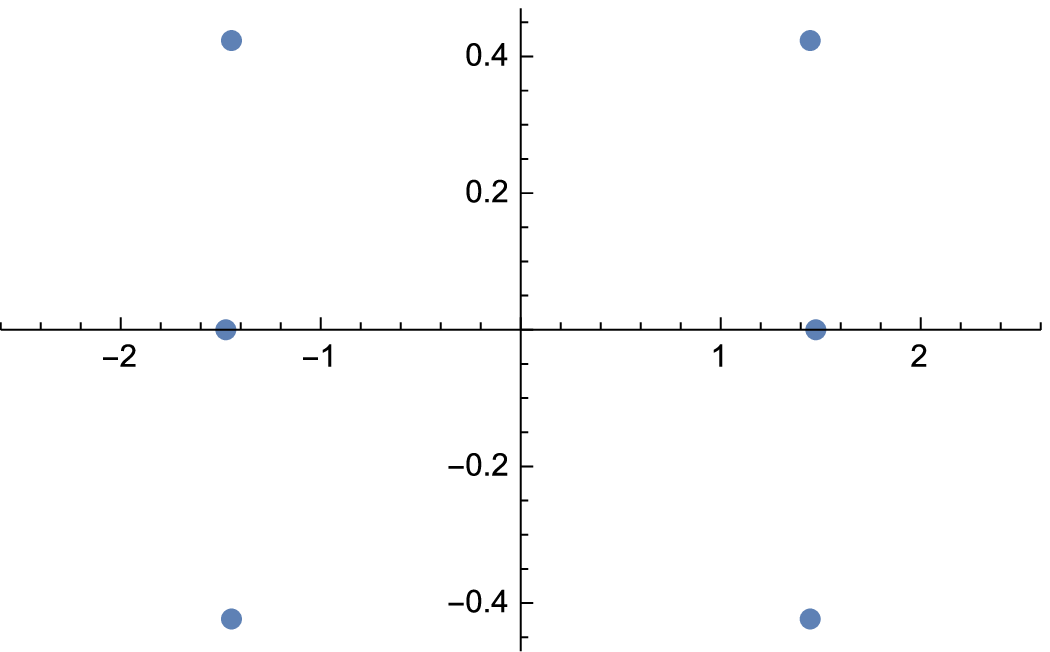}
\\[0.3cm]
$T^{4}(u)$ \\[0.1cm]
\includegraphics[width=.25\textwidth]{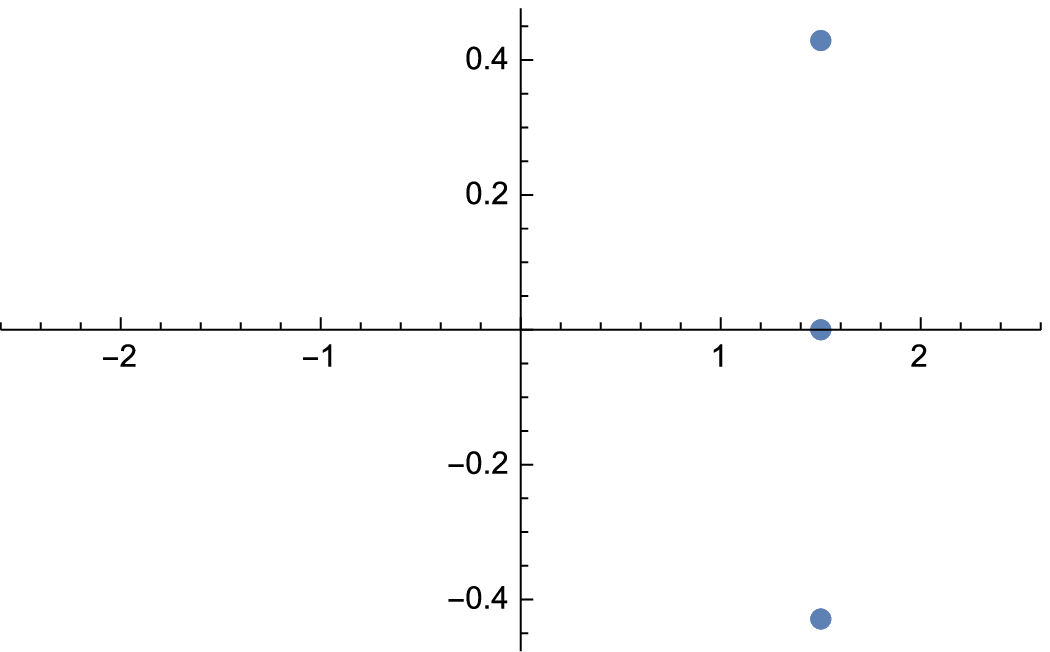}
\end{tabular}
\end{tabular}
\caption{The patterns of zeros for the groundstate of $\Tb(u)$ for $N=6$, $(p,p') = (4,5)$ and $\omega = 1$, in the complex $u$-plane. The horizontal axis is divided in units of $\lambda = \frac\pi5$. Each zero of $T^{4}(u)$ is doubly degenerate. This degeneracy is lifted for $\omega\neq1$, with the zeros remaining on the same vertical lines for $\omega$ on the unit circle.}
\label{fig:patterns.A11.principal}
\end{figure}

Our computer implementation of the transfer matrices reveals a number of properties for the groundstate for $\lambda = \frac\pi{p'}$. In the complex $u$-plane, the zeros of $T^1(u)$ for the groundstate approximately lie on the vertical lines Re$(u) = -\frac \lambda 2,\frac{3\lambda}2$. Because of the periodicity property $T^1(u+\pi) = T^1(u)$, these patterns are repeated in each vertical strip of width $\pi$.  Likewise, the zeros of $T^2(u)$ lie on the vertical lines Re$(u) = -\frac{3\lambda}2, \frac{3\lambda}2$. More generally, for $T^n(u)$, the zeros lie on two vertical lines: Re$(u) = (\frac12-n)\lambda, \frac{3\lambda}2$. The center of the analyticity strip is therefore situated at Re$(u) = (2-n)\frac{\lambda}2$. This holds for $n = 1, \dots, p'-1$. To illustrate, the patterns of zeros for $N=6$, $(p,p') = (4,5)$ and $\omega=1$ are given in \cref{fig:patterns.A11.principal}.

With this information, we deduce the positions of the zeros and poles of the functions $t^n(u)$, $1+t^n(u)$, $x(u)$ and $(1+\omega^{p'} x_0)(1+\omega^{-p'} x_0)$. This is achieved using the relations
\be
\label{eq:t.props.A11}
t^n_0 = \frac{T^{n+1}_0T^{n-1}_1}{f_{-1}f_n}, \qquad 1+ t^n_0 = \frac{T^{n}_0T^{n}_1}{f_{-1}f_n}, \qquad 1+t^{p'-1}_0 = (1+\omega^{p'} x_0)(1+\omega^{-p'} x_0).
\ee
In particular, we note that $t^1(u)$ has a zero of order $N$ at $u=0$ and poles at $u = -\lambda, \lambda$, also of order~$N$. The functions $t^n(u)$ with $n>1$ have poles of order $N$ at $u = \lambda$ and $u = -n\lambda$, but no zeros of order~$N$. Finally, $x(u)$ has a pole of order $N$ at $u =\lambda$. The same deductions are repeated for $1+t^n(u)$ and $\big(1+\omega^{p'} x(u)\big)\big(1+\omega^{-p'} x(u)\big)$. The calculation of the finite-size correction of this subsection uses certain assumptions for the analyticity strips of these functions. These are given in \cref{tab:analyticity.A11}. Crucially, these analyticity strips are free of zeros and poles, except for the order-$N$ zero of $t^1(u)$.

\begin{table}
\begin{center}
\begin{tabular}{c|c|c}
& width is larger than & centered at 
\\[0.1cm]\hline
&&\\[-0.3cm]
$t^n(u)$ & $\lambda$ & $(1-n)\frac{\lambda}2$
\\[0.1cm]
$1+t^n(u)$ & 0 & $(1-n)\frac{\lambda}2$
\\[0.1cm]
$x(u)$ & $\lambda$ & $(2-p')\frac{\lambda}2$
\\[0.1cm]
$\big(1+\omega^{p'}x(u)\big)\big(1+\omega^{-p'}x(u)\big)$ & 0 & $(2-p')\frac{\lambda}2$
\end{tabular}
\caption{The analyticity strips for the various functions.}
\label{tab:analyticity.A11}
\end{center}
\end{table}

The properties of these analyticity strips lead us to define new $Y$-functions where the parameterisation of the arguments is different. This is done in such a way that, for the new functions, the central line of the strip coincides with the real axis:
\begin{subequations}
\begin{alignat}{2}
t^{n}(u) &= \amf^n\Big(\!-\!\tfrac {\ir \pi}{\lambda}\big(u+\tfrac{(n-1)\lambda}2\big)\Big), 
\qquad &&\Amf^n(z) = 1+\amf^n(z),
\qquad n = 1, \dots, p'-2,
\\[0.15cm]
x(u) &= \amf^{p'-1}\Big(\!-\!\tfrac {\ir \pi}{\lambda}\big(u+\tfrac {(p'-2)\lambda} 2\big)\Big), 
\qquad &&\Amf^{p'-1}(z) = \big(1 + \omega^{p'} \amf^{p'-1}(z)\big)\big(1 + \omega^{-p'} \amf^{p'-1}(z)\big).\label{eq:A11b}
\end{alignat}
\end{subequations}
With these new parameterisations, the $Y$-system takes a symmetric form:
\be
\amf^{n}(z - \tfrac{\ir \pi}2) \amf^{n}(z + \tfrac{\ir \pi}2) = \Amf^{n-1}(z) \Amf^{n+1}(z), \qquad n = 1, \dots, p'-1,
\ee
where we use the convention $\Amf^0(z) = \Amf^{p'}(z) = 1$. The $D$-type $Y$-system has thus been rewritten as a $Y$-system of $A$-type.

In terms of the variable $z$, the analyticity strips are horizontal and centered on the real line. For $\amf^n(z)$, these strips have a width larger than $\ir \pi$. Our computer implementation also reveals that, in the $z$-plane, the zeros of all the eigenvalues are symmetrically distributed between the upper and lower half-planes, implying that the eigenvalues are real for $z \in \mathbb R$. The patterns of zeros are also identical in the left and right half-planes. This implies that 
\be
\amf^n(z + \ir \xi) =  \amf^n(z - \ir \xi) = \amf^n(-z + \ir \xi),\qquad z,\xi \in \mathbb R.
\ee
This is not true in general, for other eigenstates of the transfer matrix.

\subsubsection{Bulk and finite contributions}

The eigenvalue of the elementary transfer matrix is related to the first $Y$-system function by
\be
T^1_0T^1_1 = f_{-1}f_1 (1 + t^{1}_0).
\ee
The eigenvalue $T(u) = T^1(u)$ is written as the product of its bulk and finite contributions: 
\be
T(u) = \kappa(u)^N T_{\rm f}(u).
\ee
These satisfy the functional equations
\be
\label{eq:func.rel.A11}
\kappa(u)\kappa(u+\lambda)= \frac{\sin(\lambda +u)\sin(\lambda - u)}{\sin^2\lambda},
\qquad
T_{\rm f}(u)T_{\rm f}(u+\lambda) = 1+t^1(u).
\ee
The solution for the bulk contribution is known \cite{B82}:
\be
 \log\kappa(u)=\int_{-\infty}^\infty \frac{\dd t}t \frac{\cosh(\pi-2\lambda)t\sinh ut\sinh(\lambda-u)t}{\sinh\pi t\cosh\lambda t},\qquad 0<u<\lambda<\pi.
\ee
For the finite term, we define
\be
T_{\rm f}(u) = \bmf\big(\!-\!\tfrac {\ir \pi}{\lambda}(u-\tfrac \lambda 2)\big),
\ee
and rewrite the second relation in \eqref{eq:func.rel.A11} as
\be
\label{eq:bbb.A11}
\bmf(z - \tfrac{\ir \pi}2) \bmf(z + \tfrac{\ir \pi}2) = \Amf^1(z).
\ee

\subsubsection{Non-linear integral equations}

The order-$N$ zero of $t^1(u)$ lies inside its analyticity strip, whereas its order-$N$ poles lie outside of this strip. The corresponding zero of $\amf^1(z)$ lies at $z = 0$. We then define the functions $\ell^n(z)$ as
\be
\label{eq:ell.a.A11}
\ell^1(z) = \frac{\amf^1(z)}{\eta(z)^N},\qquad \ell^n(z) = \amf^n(z), \qquad n = 2, \dots, p'-1,
\ee
where
\be
\eta(z) = \tanh (\tfrac z2), \qquad
\eta(z-\tfrac{\ir \pi}2)\eta(z+\tfrac{\ir \pi}2) = 1.
\ee
As a result, we have
\be
\label{eq:sym.Y.L.A11}
\ell^n(z - \tfrac{\ir \pi}2) \ell^n(z + \tfrac{\ir \pi}2) = \Amf^{n+1}(z)\Amf^{n-1}(z), \qquad n = 1, \dots, p'-1.
\ee

The functions $\ell^n(z)$ are analytic and non-zero inside their respective analyticity strips. As discussed in \cref{sec:braid.and.bulk.A11}, for generic values of $\omega$, their asymptotic values for $z \to \pm \infty$ are finite and nonzero. This allows us to define the Fourier transform of their logarithmic derivative:
\be
\label{eq:Fourier.transforms.A11}
L^n(k) = \frac1{2\pi} \int_{-\infty}^\infty \dd z\, \eE^{-\ir k z}\big(\log \ell^n(z)\big)', \qquad 
A^n(k) = \frac1{2\pi} \int_{-\infty}^\infty \dd z\, \eE^{-\ir k z}\big(\log \Amf^n(z)\big)'.
\ee
The non-linear integral equations are obtained by first taking the Fourier
transform of the logarithmic derivative of \eqref{eq:sym.Y.L.A11} yielding
\be
L^n(k) = \frac 1{2\cosh(\frac{\pi k}2)} \big(A^{n-1}(k)+A^{n+1}(k)\big), \qquad n = 1, \dots, p'-1,
\ee
where we use the conventions $A^0(k) = A^{p'}(k) = 0$.
Applying the inverse transform, we find
\be
\label{eq:NLIE.ell.A11}
\big(\log \ell^{n}(z)\big)' = K * \big(\log \Amf^{n-1}\big)' + K * \big(\log \Amf^{n+1}\big)', \qquad n = 1, \dots, p'-1,
\ee
where the convolution of two functions is
\be
\label{eq:convolution}
(f * g)(z) = \int_{-\infty}^\infty \dd y f(z-y) g(y) =  \int_{-\infty}^\infty \dd y f(y) g(z-y).
\ee
The kernel is given by 
\be
K(z) = \frac1{2\pi}\int_{-\infty}^\infty \dd k\, \frac{\eE^{\ir k z}}{2\cosh\frac{\pi k}{2}}= \frac{1}{2\pi\cosh z}. 
\ee
Integrating \eqref{eq:NLIE.ell.A11} over $z$ removes the derivatives and introduces overall additive constants. Using \eqref{eq:ell.a.A11}, we obtain the non-linear integral equations for $\amf^n(z)$:
\be
\label{eq:NLIEs.A11}
\log \amf^n(z) - \phi_n = \fmf^n(z) + K * \log \Amf^{n-1} + K * \log \Amf^{n+1}, \qquad n = 1, \dots, p'-1,
\ee
where $\phi_1$, \dots, $\phi_{p'-1}$ are the integration constants and the driving terms are
\be
\fmf^n(z) = \left\{
\begin{array}{cl}
N \log \eta(z) & n = 1,\\[0.15cm]
0 & n = 2, \dots, p'-1.
\end{array}\right.
\ee

\subsubsection{Scaling functions and scaling non-linear integral equations}
In \eqref{eq:NLIEs.A11}, the dependence on $N$ appears only in the driving term $N \log \eta(z)$. For $z$ of order $\pm \log N$ with $N$ large, this function converges to an exponential:
\be
\lim_{N\to \infty}  N \log \eta \big(\!\pm\!(z + \log N)\big) = -2 \, \eE^{-z},
\ee
where we recall that $N$ is assumed to be even. To compute the finite-size correction at order $\frac 1N$, we assume that the $Y$-system functions appearing in \eqref{eq:NLIEs.A11} are well-defined in this limit. The patterns of zeros are all symmetric with respect to the imaginary $z$-axis, so the scaling functions behave identically in the left and right half-planes:
\be
\mathsf a^n(z) = \lim_{N\to \infty} \amf^n\big(\!\pm\!(z + \log N)\big), \qquad \mathsf A^n(z) = \lim_{N\to \infty} \Amf^n\big(\!\pm\!(z + \log N)\big).
\ee
These satisfy the following integral equations:
\be
\label{eq:scalingNLIEs.A11}
\log \asf^n(z) - \phi_n = \fsf^n(z) +  K * \log \Asf^{n-1} + K * \log \Asf^{n+1}, \qquad n = 1, \dots, p'-1,
\ee
where
\be
\fsf^n(z) = 
\left\{
\begin{array}{cl}
\!-2\, \eE^{-z}, & n = 1,\\[0.15cm]
0, & n = 2, \dots, p'-1.
\end{array}
\right.
\ee

\subsubsection{Braid and bulk behavior}\label{sec:braid.and.bulk.A11}

The scaling functions have finite asymptotics for $z \to \pm\infty$. For $z \to \infty$, these are obtained from the braid limits of the transfer matrix eigenvalues:
\be
\asf^n_\infty = \frac{(\omega^n-\omega^{-n})(\omega^{n+2}-\omega^{-n-2})}{(\omega-\omega^{-1})^2}, \qquad
\asf^{p'-1}_\infty = \frac{\omega^{p'-1}-\omega^{-(p'-1)}}{\omega-\omega^{-1}},
\label{eq:braid.an.A11}
\ee
where $n = 1, \dots, p'-2$. These values satisfy the following $Y$-system:
\begin{subequations}
\label{eq:braid.Y.system.A11}
\begin{alignat}{2}
(\asf^n_{\infty})^2 &= (1 + \asf_{\infty}^{n-1})(1 + \asf_{\infty}^{n+1}), \qquad \quad n = 1, \dots, p'-3,
\\[0.15cm]
(\asf^{p'-2}_{\infty})^2 &= (1 + \asf_{\infty}^{p'-3})(1+\omega^{p'} \asf^{p'-1}_{\infty})(1+\omega^{-p'} \asf^{p'-1}_{\infty}),
\\[0.15cm](\asf^{p'-1}_{\infty})^2 &= (1+\asf^{p'-2}_{\infty}).
\end{alignat}
\end{subequations}
These asymptotic values allow us to compute the constants $\phi_1, \dots, \phi_{p'-1}$, by studying the $z\to \infty$ asymptotics of \eqref{eq:scalingNLIEs.A11}. On the interval $\gamma \in (0,\frac \pi {p'})$, all the functions in \eqref{eq:braid.an.A11} are positive and finite. We use 
\begin{subequations}
\begin{alignat}{2}
\lim_{z\to \infty} K* \mathsf X &=  \mathsf X_\infty \int_{-\infty}^\infty \dd y \, K(y) = \tfrac12 \mathsf X_\infty
\end{alignat}
\end{subequations}
and find that the constants are all zero:
\be
\phi_n = 0, \qquad n = 1, \dots, p'-1, \qquad \textrm{for } \gamma \in (0,\tfrac \pi {p'}).
\ee

For the bulk behavior at $z \to -\infty$, we recall that the function $\amf^1(z)$ has a zero of order $N$ near the origin, so that $\asf^1_{-\infty} = 0$. The bulk asymptotic values $\asf^n_{-\infty}$ satisfy the following $Y$-system equations:
\begin{subequations}
\label{eq:bulk.Y.system.A11}
\begin{alignat}{2}
(\asf^n_{-\infty})^2 &= (1 + \asf_{-\infty}^{n-1})(1 + \asf_{-\infty}^{n+1}), \qquad \quad n = 2, \dots, p'-3,
\\[0.15cm]
(\asf^{p'-2}_{-\infty})^2 &= (1 + \asf_{-\infty}^{p'-3})(1+\omega^{p'} \asf^{p'-1}_{-\infty})(1+\omega^{-p'} \asf^{p'-1}_{-\infty}),
\\[0.15cm](\asf^{p'-1}_{-\infty})^2 &= (1+\asf^{p'-2}_{-\infty}).
\end{alignat}
\end{subequations}
This system has more than one solution, and we wish to retain the unique one that has all positive values in the range $\gamma \in (0,\frac \pi {p'})$. To obtain the solution, we note that the system \eqref{eq:bulk.Y.system.A11} is obtained from \eqref{eq:braid.Y.system.A11} by simultaneously shifting the indices by one unit, changing $p'$ for $p'-1$ and changing $\omega$ for $\omega^{p'/(p'-1)}$. The resulting solution is
\be
\asf^n_{-\infty} = \frac{(\bar\omega^{n-1}-\bar\omega^{-(n-1)})(\bar\omega^{n+1}-\bar\omega^{-(n+1)})}{(\bar\omega - \bar\omega^{-1})^2}, 
\qquad 
\asf^{p'-1}_{-\infty} = \frac{\bar\omega^{p'-2}-\bar\omega^{-(p'-2)}}{\bar\omega-\bar\omega^{-1}}, 
\qquad 
\bar\omega = \omega^{p'/(p'-1)}.\label{eq:lowlim.A11}
\ee

\subsubsection{Finite-size correction and the dilogarithm technique}\label{sec:dilog.A11}

Applying the Fourier transform and subsequently the inverse transform of the logarithmic derivative to \eqref{eq:bbb.A11}, we find
\be
\log \bmf(z) - \phi_0 = K*\log\Amf^1 = \int_{-\infty}^\infty \dd y\,K(y-z)\log \Amf^1(y)
\ee
where $\phi_0$ is an integration constant. We express this in terms of integrals involving the scaling function $\Asf^1(y)$:
\begin{alignat}{2}
\log \bmf(z) - \phi_0 &= \int_{-\log N}^\infty \dd y \Big(K(y+\log N - z) \log \Amf^1(y+\log N) + K(-y-\log N - z) \log \Amf^1(-y-\log N)\Big)
\nonumber\\[0.15cm]
&\simeq \frac{1}{\pi N}(\eE^{z}+\eE^{-z})\int_{-\infty}^\infty \dd y\, \eE^{-y} \log \Asf^1(y),
\label{eq:log.b.int.A11}
\end{alignat}
where we used
\be
K(z + \log N) \simeq \frac{\eE^{-z}}{\pi N}.
\ee
Here, $\simeq$ indicates that higher-order terms in $\frac 1N$ have been omitted.

To apply the dilogarithm technique, we define the integral
\be
\mathcal J = \int_{-\infty}^\infty \dd y \bigg[\sum_{n=1}^{p'-1}(\log \asf^n)' \log \Asf^n  -\log \asf^n (\log \Asf^n)'\bigg],
\ee
where the dependence of the functions on the argument $y$ is dropped for ease of notation. This integral is evaluated in two ways. The first consists of replacing $\log \asf^n$ and its derivative by its expression~\eqref{eq:scalingNLIEs.A11}. Many terms cancel out because of the symmetry property $K(-z) = K(z)$ of the kernel. The only surviving contributions come from the driving terms, and the result reads
\be
\mathcal J = 4 \int_{-\infty}^\infty \dd y \, \eE^{-y} \log \Asf^1(y).
\ee
Up to an overall prefactor, this is precisely the integral that we wish to compute in \eqref{eq:log.b.int.A11}. The second way of computing the integral is to apply the derivatives explicitly, which yields
\be
\mathcal J = \int_{-\infty}^\infty \dd y \,  \bigg[\sum_{n=1}^{p'-2}\frac{\dd \asf^n}{\dd y}\bigg(\frac{\log \Asf^n}{\asf^n} -\frac{\log \asf^n}{\Asf^n}\bigg) + \frac{\dd \asf^{p'-1}}{\dd y}\bigg(\frac{\log \Asf^{p'-1}}{\asf^{p'-1}}  - \frac{\dd \Asf^{p'-1}}{\dd \asf^{p'-1}}\frac{\log \asf^{p'-1}}{\Asf^{p'-1}}\bigg)\bigg].
\ee
Dividing the integral into two parts and changing the integration variables from $y$ to $\asf^n$, we find
\be
\mathcal J = \sum_{n=1}^{p'-2}\int_{\asf^n_{-\infty}}^{\asf^n_\infty} \dd \asf^n \bigg(\frac{\log \Asf^n}{\asf^n}  -\frac{\log \asf^n}{\Asf^n}\bigg) + \int_{\asf^{p'-1}_{-\infty}}^{\asf^{p'-1}_\infty} \dd \asf^{p'-1}  \bigg(\frac{\log \Asf^{p'-1}}{\asf^{p'-1}} - \frac{\dd \Asf^{p'-1}}{\dd \asf^{p'-1}}\frac{\log \asf^{p'-1}}{\Asf^{p'-1}}\bigg)
\ee
where 
\begin{subequations}
\begin{alignat}{2}
\Asf^n &= 1 + \asf^n, \qquad n = 1, \dots, p'-2, \\[0.15cm]
 \Asf^{p'-1} &= \big(1 + \omega^{p'} \asf^{p'-1}\big)\big(1 + \omega^{-p'} \asf^{p'-1}\big).
\end{alignat}
\end{subequations}
The result is therefore a combination of regular integrals. Setting $\omega = \eE^{\ir \gamma}$, the integral evaluates to
\be
\mathcal J = \frac{\pi^2}3\bigg(1-\frac{6\gamma^2 p'}{\pi^2 (p'-1)}\bigg) = \frac{\pi^2}3\bigg(1-\frac{6 \gamma^2}{\pi(\pi-\lambda)}\bigg), \qquad \gamma \in (0, \tfrac \pi {p'}).
\ee
The proof of this result is given in \cref{app:A11.principal}. The final result is
\be
\log \bmf(z) \simeq \frac{\pi \cosh z}{6 N} \bigg(1-\frac{6\gamma^2 p'}{\pi^2 (p'-1)}\bigg), \qquad 
\log T_{\rm f}(u) \simeq \frac{\pi \sin \frac{\pi u}{\lambda}}{6 N} \bigg(1-\frac{6\gamma^2 p'}{\pi^2 (p'-1)}\bigg),
\ee
where the constant $\phi_0$ was found to equal zero using $T_{\rm f}(u=0)=1$. This result is precisely \eqref{eq:fsc.general} with $c-24 \Delta$ and $\vartheta(u)$ given in \eqref{eq:fsc.A11}.

\subsection{The dual series}\label{sec:A11.dual}

In this subsection, we fix $p = 1$, so that $\lambda = \frac{\pi(p'-1)}{p'}$, and consider $p'\in \mathbb N_{\ge 2}$. We also define $\bar\lambda = \pi - \lambda = \frac{\pi}{p'}$. We compute the $\frac1N$ term in \eqref{eq:fsc.general} explicitly, for $N$ even, $\gamma \in (0,\frac\pi{p'})$ and $u$ in the neighborhood of $\frac \lambda 2$.

\subsubsection[Analyticity properties and symmetric $Y$-system]{Analyticity properties and symmetric $\boldsymbol Y$-system}\label{sec:analyticity.A11.dual}

\begin{figure}
\centering
\begin{tabular}{ccc}
\begin{tabular}{c}
$T^{1}(u)$ \\[0.1cm]
\includegraphics[width=.25\textwidth]{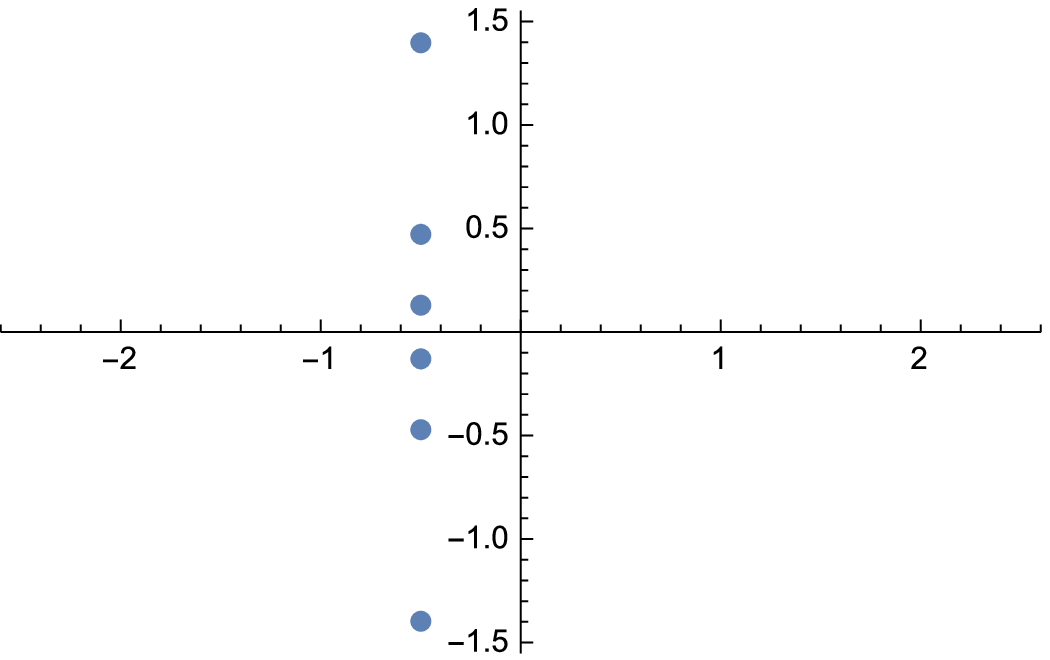} \\[0.3cm]
$T^{3}(u)$ \\[0.1cm]
\includegraphics[width=.25\textwidth]{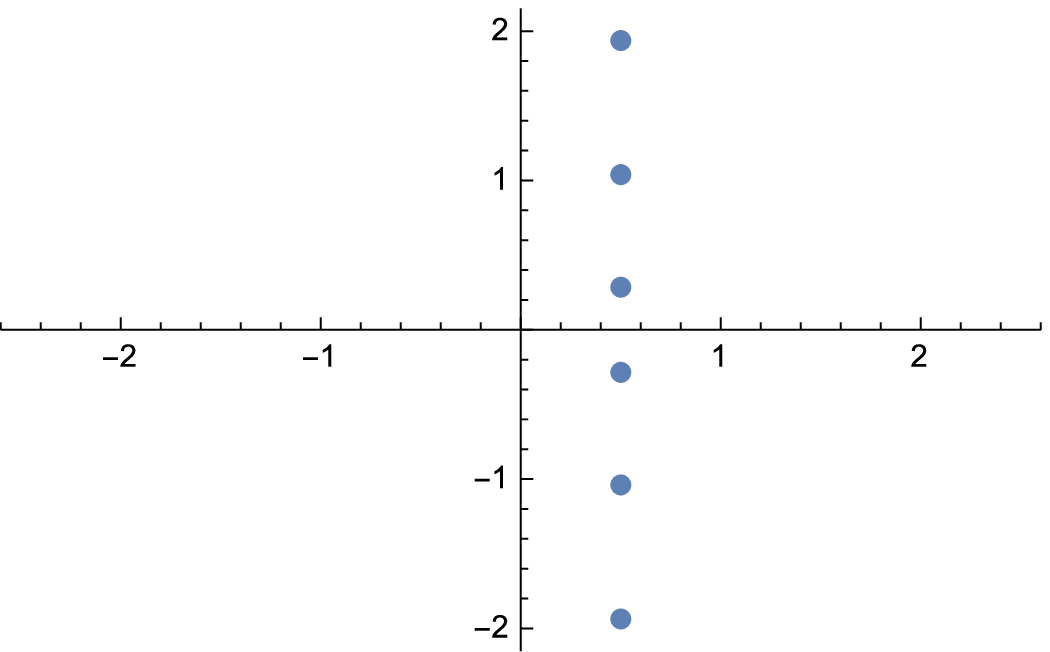}
\end{tabular}
&\qquad&
\begin{tabular}{c}
$T^{2}(u)$ \\[0.1cm]
\includegraphics[width=.25\textwidth]{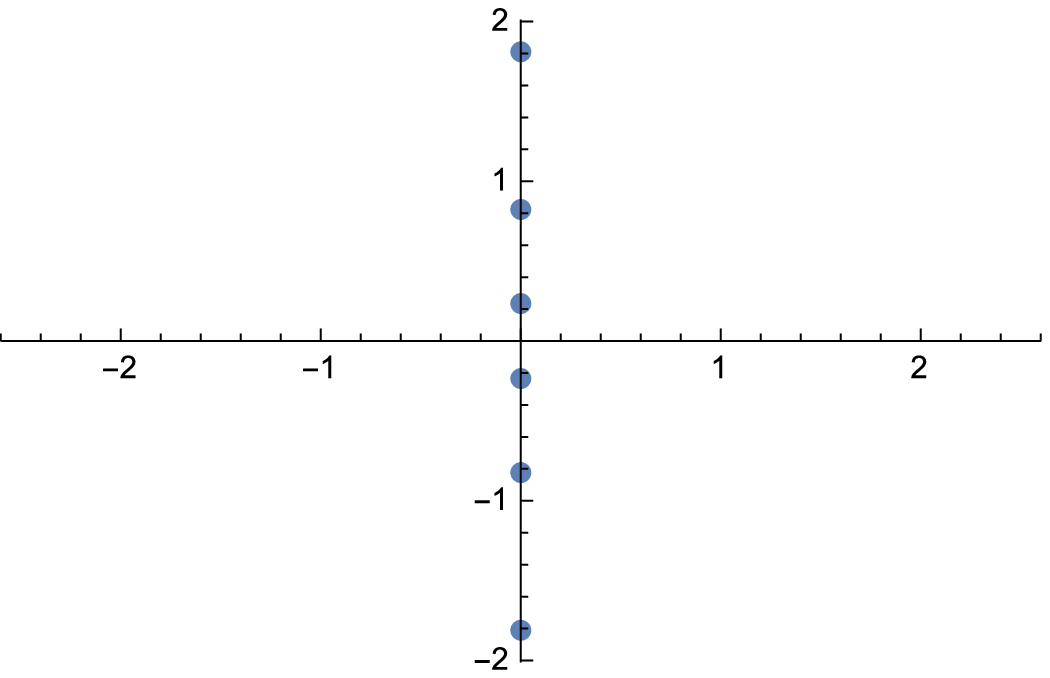}
\\[0.3cm]
$T^{4}(u)$ \\[0.1cm]
\includegraphics[width=.25\textwidth]{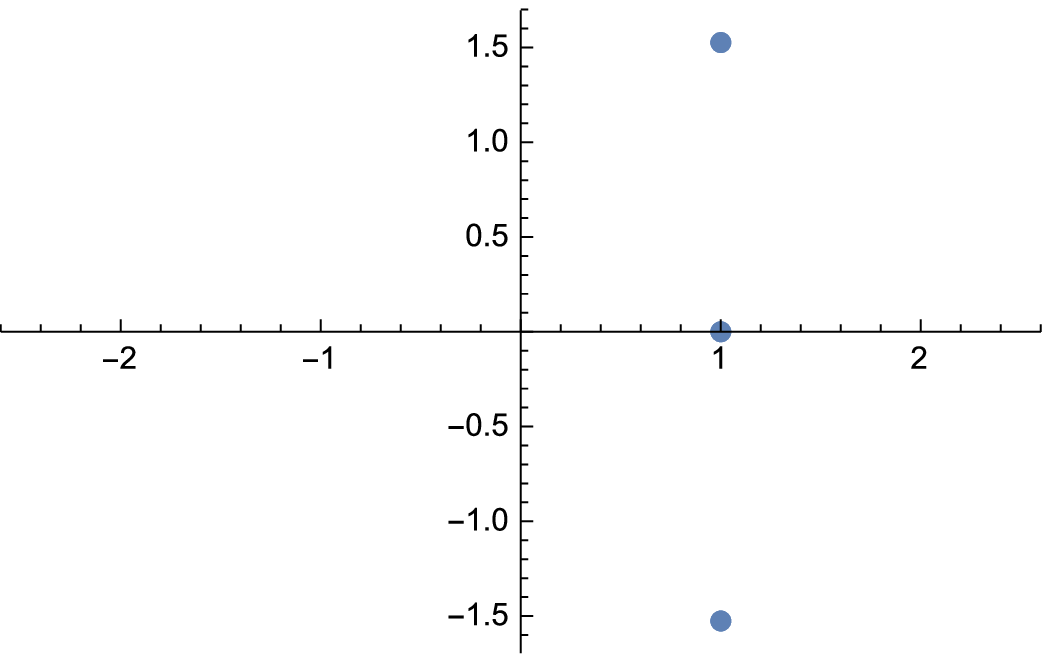}
\end{tabular}
\end{tabular}
\caption{The patterns of zeros for the groundstate of $\Tb(u)$ for $N=6$, $(p,p') = (1,5)$ and $\omega = 1$, in the complex $u$-plane. The horizontal axis is divided in units of $\bar \lambda = \frac\pi5$. Each zero of $T^{4}(u)$ is doubly degenerate. This degeneracy is lifted for $\omega\neq1$, with the zeros remaining on the same vertical lines for $\omega$ on the unit circle.}
\label{fig:patterns.A11.dual}
\end{figure}

Our computer implementation of the transfer matrices reveals a number of properties for the groundstate for $\lambda = \frac{\pi(p'-1)}{p'}$. In the complex $u$-plane, the zeros of $T^1(u)$ for the groundstate approximately lie on the vertical line  Re$(u) = -\frac {\bar \lambda} 2$. Because of the periodicity property $T^1(u+\pi) = T^1(u)$, these patterns are repeated in each vertical strip of width $\pi$.  Likewise, the zeros of $T^2(u)$ lie on the vertical line Re$(u) =0$. More generally, for $T^n(u)$, the zeros lie on the vertical line Re$(u) = \frac{(n-2)\bar\lambda}2$. This holds for $n = 1, \dots, p'-1$. We consider analyticity strips for the functions $T^n(u)$ that are centered at Re$(u) =\frac{\lambda}2+\frac{(n-1)\bar \lambda}2$. To illustrate, the patterns of zeros for $N=6$, $(p,p') = (1,5)$ and $\omega=1$ are given in \cref{fig:patterns.A11.dual}.

With this information, we deduce the positions of the zeros and poles of the functions $t^n(u)$, $1+t^n(u)$, $x(u)$ and $\big(1+\omega^{p'} x(u)\big)\big(1+\omega^{-p'} x(u)\big)$ using the relations \eqref{eq:t.props.A11}. In particular, $t^1(u)$ has a zero of order $N$ at $u=0$. In the derivation below, this zero will not play any role, as it will sit outside the analyticity strip of $t^1(u)$. This function also has poles of order $N$ at $u = \bar\lambda,(p'-1)\bar\lambda$. Likewise, the functions $t^n(u)$ with $n=1, \dots, p'-2$ have poles of order $N$ at $u = n \bar\lambda$ and $u = (p'-1)\bar\lambda$. 
Finally, $x(u)$ has a pole of order $N$ at $u =\lambda \equiv \frac{\lambda}2+\frac{(p'-1)\bar \lambda}2$, where $\equiv$ indicates an equality modulo $\pi$. 
The poles of these functions will play a crucial role in the following. To work with functions that remain finite inside their analyticity strips, we make a change of variable and work with the reciprocals of these functions:
\be 
\tilde t^n(u) = t^n(u)^{-1}, \qquad \tilde x(u) = x(u)^{-1}.
\ee 
Inside their analyticity strips, these functions have order-$N$ zeros instead of order-$N$ poles. The assumptions that we use for the properties of the analyticity strips of these functions are given in \cref{tab:analyticity.A11.dual}. Crucially, except for the order-$N$ zeros of $\tilde t^{p'-2}(u)$ and $\tilde x(u)$, these analyticity strips are free of zeros and poles.

We proceed with a second change of variables that modifies the argument in such a way that the central line of the strip coincides with the real axis:
\begin{subequations}
\begin{alignat}{2}
\tilde t^n(u) &= \amf^n\Big(\!-\!\tfrac {\ir \pi}{\textrm{\raisebox{-0.06cm}{$\bar \lambda$}}}\big(u-\tfrac{\lambda + n \bar\lambda}2\big)\Big), 
\qquad &&\Amf^n(z) = 1+\amf^n(z),
\qquad n = 1, \dots, p'-2,
\\[0.15cm]
\tilde x(u) &= \amf^{p'-1}\Big(\!-\!\tfrac {\ir \pi}{\textrm{\raisebox{-0.06cm}{$\bar \lambda$}}}\big(u-\tfrac {\lambda + (p'-1)\bar\lambda} 2\big)\Big),
\qquad &&\Amf^{p'-1}(z) = \big(1 + \omega^{p'} \amf^{p'-1}(z)\big)\big(1 + \omega^{-p'} \amf^{p'-1}(z)\big).\label{eq:A11b.dual}
\end{alignat}
\end{subequations}
In terms of these functions, the $Y$-system reads:
\begin{subequations}
\label{eq:Ysystem.a.A11.dual}
\begin{alignat}{2}
\frac{\amf^{2}(z)}{\amf^{1}(z - \tfrac{\ir \pi}2) \amf^{1}(z + \tfrac{\ir \pi}2)} &= \Amf^{2}(z), 
\\[0.15cm]
\frac{\amf^{n-1}(z)\amf^{n+1}(z)}{\amf^{n}(z - \tfrac{\ir \pi}2) \amf^{n}(z + \tfrac{\ir \pi}2)} &= \Amf^{n-1}(z) \Amf^{n+1}(z), \qquad n = 2, \dots, p'-3,
\\[0.15cm]
\frac{\amf^{p'-3}(z)\big(\amf^{p'-1}(z)\big)^2}{\amf^{p'-2}(z - \tfrac{\ir \pi}2) \amf^{p'-2}(z + \tfrac{\ir \pi}2)} &= \Amf^{p'-3}(z) \Amf^{p'-1}(z),
\\[0.15cm]
\frac{\amf^{p'-2}(z)}{\amf^{p'-1}(z - \tfrac{\ir \pi}2) \amf^{p'-1}(z + \tfrac{\ir \pi}2)} &= \Amf^{p'-2}(z).
\end{alignat}
\end{subequations}

\begin{table}
\begin{center}
\begin{tabular}{c|c|c}
& width is larger than & centered at 
\\[0.1cm]\hline
&&\\[-0.3cm]
$\tilde t^n(u)$ & $\bar\lambda$ & $\frac{\lambda + n\bar \lambda}2$
\\[0.1cm]
$1+\tilde t^n(u)$ & 0 & $\frac{\lambda + n\bar \lambda}2$
\\[0.1cm]
$\tilde x(u)$ & $\bar\lambda$ & $\frac{\lambda + (p'-1)\bar \lambda}2$
\\[0.1cm]
$\big(1+\omega^{p'} \tilde x(u)\big)\big(1+\omega^{-p'}\tilde x(u)\big)$ & 0 & $\frac{\lambda+(p'-1)\bar \lambda}2$
\end{tabular}
\caption{The analyticity strips for the various functions.}
\label{tab:analyticity.A11.dual}
\end{center}
\end{table}

The analyticity strips of $\amf^n(z)$ are horizontal strips centered at the origin and with a width larger than $\ir \pi$. The two order-$N$ zeros of $\amf^n(z)$ sit at $\pm\frac{\ir \pi}2(p'-n-1)$ for $n = 1, \dots, p'-2$. For $\amf^{p'-1}(z)$, there is a single order-$N$ zero sitting at the origin. Our computer implementation also reveals that, in the $z$-plane, the zeros of all the eigenvalues are symmetrically distributed between the upper and lower half-planes, implying that these eigenvalues are real for $z \in \mathbb R$. The patterns of zeros for the groundstate are also identical in the left and right half-planes. This implies that
\be
\amf^n(z + \ir \xi) =  \amf^n(z - \ir \xi) = \amf^n(-z + \ir \xi),\qquad z,\xi \in \mathbb R.
\ee
This is specific to the groundstate and does not hold for all the other eigenstates of the transfer matrix. 

\subsubsection{Bulk and finite contributions}

The eigenvalues $T^n(u)$ of the fused transfer matrices are related to the $Y$-system functions by
\be
\frac{T^n_0T^n_1}{T^{n+1}_0T^{n-1}_1} = 1 +\tilde t^{n}_0, \qquad \frac{T^{p'-1}_0 T^{p'-1}_1}{T^{p'-2}_2} = (1+\omega^{p'}\tilde x_0)(1+\omega^{-p'} \tilde x_0).
\ee
They are written as the product of bulk and finite contributions: 
\be
T^n(u) = \big(\kappa^n(u)\big)^N T^n_{\rm f}(u).
\ee
The bulk and finite contributions satisfy the functional equations
\begin{subequations}
\begin{alignat}{3}
\frac{\kappa^n(u)\kappa^n(u-\bar\lambda)}{\kappa^{n+1}(u)\kappa^{n-1}(u-\bar\lambda)}&= 1,
&\frac{\kappa^{p'-1}(u)\kappa^{p'-1}(u-\bar\lambda)}{\big(\kappa^{p'-2}(u-\bar\lambda)\big)^2}&= 1,
\\[0.2cm]
\label{eq:func.rel.A11.dual}
\frac{T_{\rm f}^n(u)T_{\rm f}^n(u-\bar\lambda)}{T_{\rm f}^{n+1}(u)T_{\rm f}^{n-1}(u-\bar\lambda)}&= 1 + \tilde t^{n}(u),
\qquad
&\frac{T_{\rm f}^{p'-1}(u)T_{\rm f}^{p'-1}(u-\bar\lambda)}{\big(T_{\rm f}^{p'-2}(u-\bar\lambda)\big)^2}&= \big(1+\omega^{p'}\tilde x(u)\big)\big(1+\omega^{-p'} \tilde x(u)\big),
\end{alignat}
\end{subequations}
where $n = 1, \dots, p'-2$.
The initial condition is $\kappa^0(u) = \frac{\sin(\lambda-u)}{\sin \lambda}$ and $T_{\rm f}^0(u) = 1$. For the finite contribution, we define
\be
T^n_{\rm f}(u) = \bmf\Big(\!-\!\tfrac {\ir \pi}{\textrm{\raisebox{-0.06cm}{$\bar \lambda$}}}\big(u-\tfrac {\lambda + (n-1)\bar \lambda} 2\big)\Big)
\ee
and rewrite the relations in \eqref{eq:func.rel.A11.dual} in a symmetric way as
\be
\label{eq:bbb.A11.dual}
\frac{\bmf^n(z - \tfrac{\ir \pi}2) \bmf^n(z + \tfrac{\ir \pi}2)}{\bmf^{n-1}(z)\bmf^{n+1}(z)} = \Amf^n(z), \qquad
\frac{\bmf^{p'-1}(z - \tfrac{\ir \pi}2) \bmf^{p'-1}(z + \tfrac{\ir \pi}2)}{(\bmf^{p'-2}(z))^2} = \Amf^{p'-1}(z), 
\ee
where $n = 1, \dots, p'-2$.

\subsubsection{Non-linear integral equations}

The $Y$-system functions $\amf^n(z)$ have order-$N$ zeros on the imaginary axis. We define new functions $\ell^n(z)$ where those zeros are removed:
\be
\label{eq:ell.a.A11.dual}
\ell^n(z) = \frac{\amf^n(z)}{\Big[\eta\big(z-\frac{\ir \pi}2(p'\!-\!1\!-\!n)\big)\eta\big(z+\frac{\ir \pi}2(p'\!-\!1\!-\!n)\big)\Big]^N},\qquad  \ell^{p'-1}(z) = \frac{\amf^{p'-1}(z)}{\eta(z)^N},
\ee
where $n = 1, \dots, p'-2$ and
\be
\eta(z) = \tanh \tfrac z{2(p'-1)}, \qquad \eta\big(z-\tfrac{\ir \pi}2(p'\!-\!1)\big)\eta\big(z+\tfrac{\ir \pi}2(p'\!-\!1)\big) = 1.
\ee
Rewriting the $Y$-system in terms of the functions $\ell^n(z)$, we find that it has exactly the same form as \eqref{eq:Ysystem.a.A11.dual}, but with each $\amf^n(z)$ on the left-hand sides replaced by the corresponding $\ell^n(z)$:
\begin{subequations}
\label{eq:Ysystem.ell.A11.dual}
\begin{alignat}{2}
\frac{\ell^{2}(z)}{\ell^{1}(z - \tfrac{\ir \pi}2) \ell^{1}(z + \tfrac{\ir \pi}2)} &= \Amf^{2}(z), 
\\[0.15cm]
\frac{\ell^{n-1}(z)\ell^{n+1}(z)}{\ell^{n}(z - \tfrac{\ir \pi}2) \ell^{n}(z + \tfrac{\ir \pi}2)} &= \Amf^{n-1}(z) \Amf^{n+1}(z), \qquad n = 2, \dots, p'-3,
\\[0.15cm]
\frac{\ell^{p'-3}(z)\big(\ell^{p'-1}(z)\big)^2}{\ell^{p'-2}(z - \tfrac{\ir \pi}2) \ell^{p'-2}(z + \tfrac{\ir \pi}2)} &= \Amf^{p'-3}(z) \Amf^{p'-1}(z),
\\[0.15cm]
\frac{\ell^{p'-2}(z)}{\ell^{p'-1}(z - \tfrac{\ir \pi}2) \ell^{p'-1}(z + \tfrac{\ir \pi}2)} &= \Amf^{p'-2}(z).
\end{alignat}
\end{subequations}

The functions $\ell^n(z)$ are analytic and non-zero inside their respective analyticity strips. As discussed in \cref{sec:braid.and.bulk.A11.dual}, for generic values of $\omega$, their asymptotic values for $z \to \pm \infty$ are finite and nonzero. This allows us to define the Fourier transforms $L^n(k)$ and $A^n(k)$ of their logarithmic derivative, as in \eqref{eq:Fourier.transforms.A11}. We take the Fourier transform of the logarithmic derivative of \eqref{eq:Ysystem.ell.A11.dual} and find 
\be
\left(\begin{smallmatrix}
-2 \cosh \frac {\pi k}2 & 1 & 0 &  &  \\
1 & -2 \cosh \frac {\pi k}2 & 1 & \sddots &  \\
0 & 1 & -2 \cosh \frac {\pi k}2 & 1 & 0 \\
 & \sddots & 1 & -2 \cosh \frac {\pi k}2 & 2 \\
 &  & 0 & 1 & -2 \cosh \frac {\pi k}2
\end{smallmatrix}\right)
\left(\begin{smallmatrix}
L^1 \\[0.1cm] L^2 \\[0.1cm] \svdots \\[0.1cm] L^{p'-2} \\[0.1cm] L^{p'-1}
\end{smallmatrix}\right)
 = 
 \left(\begin{smallmatrix}
0 &&& 1 &&& 0 &&&  &&&  \\[0.15cm]
1 &&& 0 &&& 1 &&& \sddotss &  \\[0.15cm]
0 &&& 1 &&& 0 &&& 1 &&& 0 \\[0.15cm]
 &&& \sddotss &&& 1 &&& 0 &&& 1 \\[0.15cm]
 &&&  &&& 0 &&& 1 &&&0 
\end{smallmatrix}\right)
\left(\begin{smallmatrix}
A^1 \\[0.1cm] A^2 \\[0.1cm] \svdots \\[0.1cm] A^{p'-2} \\[0.1cm] A^{p'-1}
\end{smallmatrix}\right).
\ee
We compute the inverse of the matrix in the left side, apply it to the right side and find
\be
\vec L = \hat K \cdot \vec A.
\ee
Here $\hat K$ is a symmetric matrix that can be computed explicitly.
Applying the inverse transform, we find
\be
\label{eq:NLIE.ell.A11.dual}
\big(\log \ell^{n}(z)\big)' = \sum_{m=1}^{p'-1} K_{nm} * \big(\log \Amf^{m}\big)', \qquad n = 1, \dots, p'\!-\!1,
\ee
where the kernel functions are given by
\be
\label{eq:Kij.A11.dual}
K_{nm}(z) = \frac1{2\pi}\int_{-\infty}^\infty \dd k\, \eE^{\ir k z}\hat K_{nm}.
\ee
Integrating \eqref{eq:NLIE.ell.A11.dual} over $z$ removes the derivatives and introduces overall additive constants. Using \eqref{eq:ell.a.A11.dual}, we obtain the non-linear integral equations for $\amf^n(z)$:
\be
\label{eq:NLIEs.A11.dual}
\log \amf^n(z) - \phi_n = \fmf^n(z) + \sum_{m=1}^{p'-1} K_{nm} * \log \Amf^{m}, \qquad n = 1, \dots, p'\!-\!1,
\ee
where $\phi_1$, \dots, $\phi_{p'-1}$ are the integration constants and the driving terms are
\be
\fmf^n(z) = \left\{
\begin{array}{cl}
N \log \Big[\eta\big(z-\frac{\ir \pi}2(p'\!-\!1\!-\!n)\big)\eta\big(z+\frac{\ir \pi}2(p'\!-\!1\!-\!n)\big) \Big]\quad& n = 1,\dots, p'\!-\!2,\\[0.3cm]
N \log \eta(z) & n = p'\!-\!1.
\end{array}\right.
\ee

\subsubsection{Scaling functions and scaling non-linear integral equations}
In \eqref{eq:NLIEs.A11.dual}, the dependence on $N$ appears only in the driving terms. For $z$ of order $(p'\!-\!1) \log N$ with $N$ large, these functions converge to exponentials:
\be
\fsf^n(z) = \lim_{N\to \infty} \fmf^n\big(z+(p'\!-\!1)\log N\big) = 
\left\{\begin{array}{cl}
-4 \sin\big(\frac{\pi n}{2(p'-1)}\big)\,\eE^{-z/(p'-1)}\quad& n = 1,\dots, p'\!-\!2,\\[0.3cm]
-2\, \eE^{-z/(p'-1)} & n = p'\!-\!1.
\end{array}\right.
\ee
To compute the finite-size correction at order $\frac 1N$, we assume that the $Y$-system functions appearing in \eqref{eq:NLIEs.A11.dual} are well-defined in this limit. The patterns of zeros are all symmetric with respect to the imaginary $z$-axis, so the scaling functions behave identically in the left and right half-planes:
\begin{subequations}
\be
\mathsf a^n(z) = \lim_{N\to \infty} \amf^n\big(\!\pm\!(z + (p'\!-\!1) \log N)\big), \qquad \mathsf A^n(z) = \lim_{N\to \infty} \Amf^n\big(\!\pm\!(z +(p'\!-\!1) \log N)\big).
\ee
\end{subequations}
These satisfy the following set of integral equations:
\be
\label{eq:scalingNLIEs.A11.dual}
\log \asf^n(z) - \phi_n =  \fsf^n(z) + \sum_{m=1}^{p'\!-\!1} K_{nm} * \log \Asf^{m}.
\ee

\subsubsection{Braid and bulk behavior}\label{sec:braid.and.bulk.A11.dual}

The scaling functions have finite asymptotics for $z \to \pm \infty$. For $z \to \infty$, these are obtained from the braid limits of the transfer matrix eigenvalues:
\be
\asf^n_\infty = \frac{(\omega-\omega^{-1})^2}{(\omega^n-\omega^{-n})(\omega^{n+2}-\omega^{-n-2})}, \qquad
\asf^{p'-1}_\infty = \frac{\omega-\omega^{-1}}{\omega^{p'-1}-\omega^{-(p'-1)}},
\label{eq:braid.an.A11.dual}
\ee
where $n = 1, \dots, p'-2$. These values are constant solutions to the $Y$-system \eqref{eq:Ysystem.a.A11.dual}. They are positive and finite on the interval $\gamma \in (0,\frac \pi {p'})$. By studying the $z\to \infty$ asymptotics of \eqref{eq:scalingNLIEs.A11.dual}, we are able to compute the constants $\phi_1, \dots, \phi_{p'-1}$:
\be
\phi_n = 0, \qquad n = 1, \dots, p'-1, \qquad \gamma\in (0,\tfrac\pi{p'}).
\ee

The behavior of the functions $\asf^n(z)$ for $z \to - \infty$ is dictated by the driving terms in the non-linear integral equations. These originated from the order-$N$ zeros of these functions that lie on the imaginary axis in the $z$-plane. As a result, the bulk limits of these functions all vanish:
\be
\asf^n_{-\infty} = 0, \qquad n = 1, \dots, p'\!-\!1.
\label{eq:bulk.an.A11.dual}
\ee 

\subsubsection{Finite-size correction and the dilogarithm technique}\label{sec:dilog.A11.dual}

We define the Fourier transform of the logarithmic derivative of the functions $\bmf^n(z)$: 
\be
B^n(k) = \frac1{2\pi} \int_{-\infty}^\infty \dd z\, \eE^{-\ir k z}\big(\log \bmf^n(z)\big)'.
\ee
Applying the Fourier transform and subsequently the inverse transform to \eqref{eq:bbb.A11.dual}, we find
\be
\left(\begin{smallmatrix}
2 \cosh \frac {\pi k}2 & -1 & 0 &  &  \\
-1 & 2 \cosh \frac {\pi k}2 & -1 & \sddots &  \\
0 & -1 & 2 \cosh \frac {\pi k}2 & -1 & 0 \\
 & \sddots & -1 & 2 \cosh \frac {\pi k}2 & -1 \\
 &  & 0 & -2 & 2 \cosh \frac {\pi k}2
\end{smallmatrix}\right)
\left(\begin{smallmatrix}
B^1 \\[0.1cm] B^2 \\[0.1cm] \svdots \\[0.1cm] B^{p'-2} \\[0.05cm] B^{p'-1}
\end{smallmatrix}\right)
 = 
\left(\begin{smallmatrix}
A^1 \\[0.1cm] A^2 \\[0.1cm] \svdots \\[0.1cm] A^{p'-2} \\[0.05cm] A^{p'-1}
\end{smallmatrix}\right).
\ee
Denoting by $M$ the matrix on the left-hand side, we invert $M$ and apply it to both sides of the equation. The matrix elements of the first row of $M^{-1}$ are
\be
(M^{-1})_{1n} = \frac{\cosh\big(\frac{\pi k}2(p'\!-\!1\!-\!n)\big)}{\cosh\big(\frac{\pi k}2(p'\!-\!1)\big)}\times 
\left\{\begin{array}{cl}
1\quad& n = 1, \dots, p'\!-\!2,\\[0.15cm]
\frac12\quad & n = p'\!-\!1.
\end{array}\right.
\ee
As a result, we find
\be
\log \bmf^1(z) - \phi_0 = \sum_{n=1}^{p'-2} \tilde K_n * \log \Amf^n + \tfrac12 \tilde K_{p'-1} \log \Amf^{p'-1}, 
\ee
where
\be
\tilde K_n (z) = \frac1{2\pi} \int_{-\infty}^\infty \dd k\, \eE^{\ir k z} \frac{\cosh\big(\frac{\pi k}2(p'\!-\!1\!-\!n)\big)}{\cosh\big(\frac{\pi k}2(p'\!-\!1)\big)}
= \frac{1}{\pi(p'-1)} \frac{\sin\big(\frac{\pi n}{2(p'-1)}\big) \cosh\big(\frac{z}{p'-1}\big)}{\sinh\Big(\frac{z-\frac{\ir \pi n}2}{p'-1}\Big)\sinh\Big(\frac{z+\frac{\ir \pi n}2}{p'-1}\Big)}.
\ee
In the scaling limit, we have
\be
\tilde K_n\big(z + (p'\!-\!1)\log N\big) \simeq \frac{2\sin\big(\frac{\pi n}{2(p'-1)}\big)\eE^{-z/(p'-1)}}{\pi N (p'\!-\!1)}
\ee
where $\simeq$ indicates that higher-order terms in $\frac1N$ have been omitted.
Using this result, we find
\begin{alignat}{2}
\bmf^1(z) - \phi_0 &= \int_{-(p'-1)\log N}^\infty \dd y \bigg[\sum_{n=1}^{p'-2} \tilde K_n\big(y+(p'\!-\!1)\log N - z\big) \log \Amf^n\big(y+(p'\!-\!1)\log N\big) 
\nonumber\\[0.15cm]
&\hspace{3cm}+ \tfrac12 \tilde K_{p'-1}\big(y+(p'\!-\!1)\log N - z\big) \log \Amf^{p'-1}\big(y+(p'\!-\!1)\log N\big) 
\nonumber\\[0.15cm]
&\hspace{3cm}+ \sum_{n=1}^{p'-2} \tilde K_n\big(\!-y-(p'\!-\!1)\log N - z\big) \log \Amf^n\big(\!-y-(p'\!-\!1)\log N\big) 
\nonumber\\[0.15cm]
&\hspace{3cm} +\tfrac12 \tilde K_{p'-1}\big(\!-y-(p'\!-\!1)\log N - z\big) \log \Amf^{p'-1}\big(\!-y-(p'\!-\!1)\log N\big) \bigg]
\nonumber\\
& \simeq \frac {2\big(\eE^{z/(p'-1)}+\eE^{-z/(p'-1)}\big)}{\pi N (p'\!-\!1)} \int_{-\infty}^\infty \dd y\, \eE^{-y/(p'-1)} \bigg[\sum_{n=1}^{p'-2} \sin\Big(\frac{\pi n}{2(p'\!-\!1)}\Big)\log \Asf^n + \tfrac12 \Asf^{p'-1}\bigg].
\label{eq:log.b.int.A11.dual}
\end{alignat}

To apply the dilogarithm technique, we define the integral
\be
\mathcal J = \int_{-\infty}^\infty \dd y \bigg[\sum_{n=1}^{p'-1}(\log \asf^n)' \log \Asf^n  -\log \asf^n (\log \Asf^n)'\bigg].
\ee
This integral is evaluated in two ways. The first consists of replacing $\log \asf^n$ and its derivative by its expression \eqref{eq:scalingNLIEs.A11.dual}. Many terms cancel out because of the symmetries $K_{nm}(z) = K_{nm}(-z) = K_{mn}(z)$ of the kernel functions \eqref{eq:Kij.A11.dual}. The only surviving contributions come from the driving terms, and the result reads
\be
\mathcal J = \frac8{p'\!-\!1} \int_{-\infty}^\infty \dd y \, \eE^{-y/(p'-1)} \bigg[\sum_{n=1}^{p'-2} \sin\frac {\pi n}{2(p'\!-\!1)}\, \log \Asf^n + \tfrac12 \Asf^{p'-1}\bigg].
\ee
Up to an overall prefactor, this is precisely the integral we wish to compute in \eqref{eq:log.b.int.A11.dual}. The second way of computing the integral is to apply the derivatives explicitly, which yields
\be
\mathcal J = \int_{-\infty}^\infty \dd y \,  \bigg[\sum_{n=1}^{p'-2}\frac{\dd \asf^n}{\dd y}\bigg(\frac{\log \Asf^n}{\asf^n} -\frac{\log \asf^n}{\Asf^n}\bigg) + \frac{\dd \asf^{p'-1}}{\dd y}\bigg(\frac{\log \Asf^{p'-1}}{\asf^{p'-1}}  - \frac{\dd \Asf^{p'-1}}{\dd \asf^{p'-1}}\frac{\log \asf^{p'-1}}{\Asf^{p'-1}}\bigg)\bigg].
\ee
Dividing the integral into two parts and changing the integration variables from $y$ to $\asf^n$, we find
\be
\mathcal J = \sum_{n=1}^{p'-2}\int_{\asf^n_{-\infty}}^{\asf^n_\infty} \dd \asf^n \bigg(\frac{\log \Asf^n}{\asf^n}  -\frac{\log \asf^n}{\Asf^n}\bigg) + \int_{\asf^{p'-1}_{-\infty}}^{\asf^{p'-1}_\infty} \dd \asf^{p'-1}  \bigg(\frac{\log \Asf^{p'-1}}{\asf^{p'-1}} - \frac{\dd \Asf^{p'-1}}{\dd \asf^{p'-1}}\frac{\log \asf^{p'-1}}{\Asf^{p'-1}}\bigg)
\ee
where 
\begin{subequations}
\begin{alignat}{2}
\Asf^n &= 1 + \asf^n, \qquad n = 1, \dots, p'-2, \\[0.15cm]
 \Asf^{p'-1} &= \big(1 + \omega^{p'} \asf^{p'-1}\big)\big(1 + \omega^{-p'} \asf^{p'-1}\big).
\end{alignat}
\end{subequations}
The result is therefore a combination of regular integrals, with the upper and lower bounds given by \eqref{eq:braid.an.A11.dual} and \eqref{eq:bulk.an.A11.dual} respectively. Setting $\omega = \eE^{\ir \gamma}$, the integral evaluates to
\be
\mathcal J = \frac{\pi^2}3\bigg(1-\frac{6\gamma^2 p'}{\pi^2}\bigg) = \frac{\pi^2}3\bigg(1-\frac{6 \gamma^2}{\pi(\pi-\lambda)}\bigg), \qquad \gamma \in (0, \tfrac \pi {p'}).
\ee
The proof of this result is given in \cref{app:A11.dual}. The final result is
\be
\log \bmf(z) \simeq \frac{\pi \cosh \frac{z}{p'-1}}{6 N} \bigg(1-\frac{6\gamma^2 p'}{\pi^2}\bigg), \qquad 
\log T_{\rm f}(u) \simeq \frac{\pi \sin \frac{\pi u}{\lambda}}{6 N} \bigg(1-\frac{6\gamma^2 p'}{\pi^2}\bigg),
\ee
where the constant $\phi_0$ was found to vanish using $T_{\rm f}(u=0)=1$.
This result is precisely \eqref{eq:fsc.general} with $c-24 \Delta$ and $\vartheta(u)$ given in \eqref{eq:fsc.A11}.

%
\section{Finite-size corrections for the $\boldsymbol{A_2^{(2)}}$ models}\label{sec:A22}
%

\subsection[Definition of the $A_2^{(2)}$ models]{Definition of the $\boldsymbol{A_2^{(2)}}$ models}\label{sec:A22def}

The loop and vertex models in the $A_2^{(2)}$ family are the dilute Temperley-Lieb loop model and the Izergin-Korepin 19-vertex model. The $A_2^{(2)}$ loop model is a face model on the square lattice, where each face takes on one of nine possible local configurations. The elementary face operator is defined by the linear combination\\[-10pt]
\begin{alignat}{2}
\label{eq:face.op}
\begin{pspicture}[shift=-.40](0,0)(1,1)
\facegrid{(0,0)}{(1,1)}
\psarc[linewidth=0.025]{-}(0,0){0.16}{0}{90}
\rput(.5,.5){$u$}
\end{pspicture}
\ \ = \ &\rho_1\ \
\begin{pspicture}[shift=-.40](0,0)(1,1)
\facegrid{(0,0)}{(1,1)}
\rput[bl](0,0){\loopa}
\end{pspicture}
\ \ + \rho_2\ \
\begin{pspicture}[shift=-.40](0,0)(1,1)
\facegrid{(0,0)}{(1,1)}
\rput[bl](0,0){\loopb}
\end{pspicture}
\ \ + \rho_3\ \
\begin{pspicture}[shift=-.40](0,0)(1,1)
\facegrid{(0,0)}{(1,1)}
\rput[bl](0,0){\loopc}
\end{pspicture}
\ \ + \rho_4\ \
\begin{pspicture}[shift=-.40](0,0)(1,1)
\facegrid{(0,0)}{(1,1)}
\rput[bl](0,0){\loopd}
\end{pspicture}
\ \ + \rho_5\ \
\begin{pspicture}[shift=-.40](0,0)(1,1)
\facegrid{(0,0)}{(1,1)}
\rput[bl](0,0){\loope}
\end{pspicture}
\nonumber\\[0.2cm] \ &\hspace{0.2cm}+ \rho_6\ \
\begin{pspicture}[shift=-.40](0,0)(1,1)
\facegrid{(0,0)}{(1,1)}
\rput[bl](0,0){\loopf}
\end{pspicture}
\ \ + \rho_7\ \
\begin{pspicture}[shift=-.40](0,0)(1,1)
\facegrid{(0,0)}{(1,1)}
\rput[bl](0,0){\loopg}
\end{pspicture}
\ \ + \rho_8\ \
\begin{pspicture}[shift=-.40](0,0)(1,1)
\facegrid{(0,0)}{(1,1)}
\rput[bl](0,0){\looph}
\end{pspicture}
\ \ + \rho_9\ \
\begin{pspicture}[shift=-.40](0,0)(1,1)
\facegrid{(0,0)}{(1,1)}
\rput[bl](0,0){\loopi}
\end{pspicture}\ \, ,
\end{alignat}
where the local Boltzmann weights are
\begin{subequations}
\label{eq:weights}
\begin{alignat}{4}
&\rho_1=1+\frac{\sin u\sin(3\lambda-u)}{\sin 2\lambda \sin 3\lambda}, \qquad
&&\rho_2 = \rho_3= \frac{\sin(3\lambda-u)}{\sin 3\lambda}, \qquad
&&\rho_4 = \rho_5=\frac{\sin u}{\sin3\lambda},\\[0.15cm]
&\rho_6 = \rho_7 = \frac{\sin u\sin(3\lambda-u)}{\sin2\lambda\sin3\lambda}, \qquad
&&\rho_8= \frac{\sin(2\lambda-u)\sin(3\lambda-u)}{\sin2\lambda\sin3\lambda}, \qquad
&&\rho_9= -\frac{\sin u\sin(\lambda-u)}{\sin2\lambda\sin3\lambda}.
\end{alignat}
\end{subequations}
The fugacities of the contractible and non-contractible loops are
\be
\beta = -2 \cos 4 \lambda, \qquad \alpha = \omega + \omega^{-1} = 2 \cos \gamma,
\ee
where $\omega = \eE^{\ir\gamma}$ is a free parameter. The $\check R$-matrix of the 19-vertex model is
\be
\label{eq:A22.Rcheck}
\check R(u) = \left(
\begin{array}{ccccccccc}
 \rho _8 & 0 & 0 & 0 & 0 & 0 & 0 & 0 & 0 \\
 0 & \rho _2 & 0 & \rho _7 & 0 & 0 & 0 & 0 & 0 \\
 0 & 0 & \rho _8-\rho _9\,\eE^{4\ir \lambda}  & 0 & \ir \rho _4\,\eE^{2\ir \lambda} & 0 & \rho _9 & 0 & 0 \\
 0 & \rho _6 & 0 & \rho _3 & 0 & 0 & 0 & 0 & 0 \\
 0 & 0 & \ir\rho _5\,\eE^{2\ir \lambda} & 0 & \rho _1 & 0 & -\ir\rho _5\,\eE^{-2\ir \lambda} & 0 & 0 \\
 0 & 0 & 0 & 0 & 0 & \rho _3 & 0 & \rho _6 & 0 \\
 0 & 0 & \rho _9 & 0 & -\ir\rho _4\,\eE^{-2\ir \lambda} & 0 & \rho _8-\rho _9\,\eE^{-4\ir \lambda} & 0 & 0
   \\
 0 & 0 & 0 & 0 & 0 & \rho _7 & 0 & \rho _2 & 0 \\
 0 & 0 & 0 & 0 & 0 & 0 & 0 & 0 & \rho _8 \\
\end{array}
\right).
\ee
The twist matrix is given in \eqref{eq:twists} with $\omega = \eE^{\ir \gamma}$. Both the vertex and loop $A_2^{(2)}$ models are described by the dilute Temperley-Lieb algebra \cite{GP93,P94,BSA14}, with its parameter $\beta$ fixed to $\beta = -2 \cos 4\lambda$. On the cylinder, the single-row transfer matrices are elements of the periodic dilute Temperley-Lieb algebra \cite{MDPR2019}. The $\check R(u)$ can be written as $\check R(u)=\sum_{\nu=1}^9 \rho_\nu g^{(\nu)}$ where the matrices $g^{(\nu)}$ with $\nu = 1, \dots, 9$ are the matrix representatives of the nine tiles in \eqref{eq:face.op} in the vertex representation of the dilute Temperley-Lieb algebra. The explicit form of these matrices can be directly read off from \eqref{eq:A22.Rcheck}.

The relevant regimes of the $A_2^{(2)}$ models are
\begin{eqnarray}
\begin{array}{lll}
\mbox{I (Dilute; Dense):} &0<u<3\lambda, \qquad &0<\lambda<\tfrac{\pi}{4};\ \ \tfrac{\pi}{4}<\lambda<\tfrac{\pi}{2},\\[4pt]
\mbox{II (Dual):}  &0<u<3\lambda-2\pi, \qquad &\tfrac{2\pi}{3}<\lambda<\pi.
\end{array}
\end{eqnarray}
Regime III with $\tfrac{\pi}{2}<\lambda<\tfrac{2\pi}{3}$ is the non-compact regime \cite{VJS2014} which will not be studied in this paper. 
For the vertex model in Regimes I and II, the groundstate appears in the zero magnetisation sector. For the loop model in Regimes I and II, it lies in the standard module with zero defects, $\stanW_{N,0}$. (We follow the convention used in \cite{MDP19} for these modules.) 
In these sectors, up to an irrelevant overall sign, the transfer matrix $\Tb(u)$ of the $A_2^{(2)}$ models is invariant under the involution
\be
\lambda \leftrightarrow \pi-\lambda,\qquad u \leftrightarrow \pi-u.\label{A22duality}
\ee
It follows that the dual Regime II is equivalent to the regime $3\lambda\!-\!\pi<u<0$ with $0<\lambda<\tfrac{\pi}{2}$ considered in other papers.

The roots of unity values of $\lambda$ are those for which $\frac{\lambda}{\pi}\in{\Bbb Q}$. We parameterise them in terms of two integers $a,b$ as
\be
\lambda = \lambda_{a,b} = \frac{\pi(b-a)}{2b},\qquad  \mbox{gcd}(a,b)=1.
\ee
An alternative parameterisation for $\lambda$ of the $A_2^{(2)}$ models, used in earlier works such as \cite{WNS92,WPSN94,ZPG95,BatchS98,JunjiS04and05,SAPR2012}, is $\lambda = \frac{\pi(2p'-p)}{4p'}$. The precise relation between $(a,b)$ and $(p,p')$ relating the two parameterisations is 
\be
(p,p')=\begin{cases}
(2a,b),&\mbox{$b$ odd},\\
(a,b/2),&\mbox{$b$ even},
\end{cases}\qquad 
(a,b)=\begin{cases}
(p,2p'),&\mbox{$p$ odd},\\
(p/2,p'),&\mbox{$p$ even},
\end{cases}\qquad 1\le p<p',\ \ 1\le |a|<b.
\ee
Our calculation of the finite-size corrections below focuses on two series:
\be
\label{eq:series.A22}
\begin{array}{ll}
\textrm{Principal series:}\quad & (a,b) = (b-1,b),\qquad u>0, \qquad   \mbox{I (Dilute)},\\[0.15cm]
\textrm{Dual series:}\quad & (a,b) = (1-b,b),\qquad u>0, \qquad \mbox{II}.
\end{array}
\ee
From \eqref{A22duality}, the dual series can be alternatively specified by 
\be
\label{eq:dualseries.A22}
\begin{array}{ll}
\textrm{Dual series:}\quad & (a,b) = (b-1,b),\qquad 3\lambda-\pi<u<0.
\end{array}
\ee

In this section, we study the groundstate of the transfer matrix. 
We will focus on values of $u$ in the neighborhood of $u = \frac {3\lambda} 2$. We consider arbitrary integer $N$ for the principal series and $N\equiv0\textrm{ mod }2$ for the dual series. Furthermore, we will restrict to values of the twist parameter $\omega = \eE^{\ir \gamma}$ in the intervals 
\be
\gamma \in 
\left\{
\begin{array}{cl}
\big(0,\frac{2\pi(b-1)}{b^2}\big) & \textrm{Principal series},
\\[0.2cm]
\big(0,\frac{2\pi}{b+1}\big) & \textrm{Dual series}.
\end{array}
\right.
\ee
We will compute the $\frac1N$ finite-size correction term for the groundstate eigenvalue $T(u)$ and will confirm the conformal prediction \eqref{eq:fsc.general} with 
\be
\label{eq:fsc.A22}
c-24 \Delta =
\left\{\begin{array}{ll}
\displaystyle 1- \frac{3\gamma^2 b}{\pi^2 (b-1)} &\quad \textrm{Principal series,}\\[0.35cm]
\displaystyle \frac32-\frac{3\gamma^2 b}{\pi^2} &\quad \textrm{Dual series,}
\end{array}\right.
\quad
\vartheta(u) = \left\{\begin{array}{cl}
\displaystyle\frac {\pi u}{3\lambda} \quad &\textrm{Principal series,}\\[0.35cm]
\displaystyle\frac {\pi (u-\pi)}{3\lambda-2\pi}\quad &\textrm{Dual series.}
\end{array}\right.
\ee

\subsection{Functional relations}

The fused transfer matrices $\Tb^{m,n}(u)$ for the $A_2^{(2)}$ models are defined recursively in \cite{MDP19} from the fusion hierarchy relations, as functions of the fundamental transfer matrix $\Tb^{1,0}(u) = \Tb(u) = \Tb^{0,1}(u-\lambda)$. There, it was found that the $T$-system equations involve only the transfer matrices $\Tb^{m,0}(u)$ where the second index is zero:
\be
\label{eq:Trelations.A22}
\Tb^{m,0}_0 \Tb^{m,0}_2 = \sigma^m f_{-3}f_{2m}\Tb^{m,0}_1 + \Tb^{m+1,0}_0 \Tb^{m-1,0}_2, \qquad m \ge 0.
\ee
We use the compact notations
\begin{subequations}
\begin{alignat}{3}
&\Tb^{m,n}_k = \Tb^{m,n} (u+k \lambda), 
\qquad 
&&\Tb^{1,0}(u) = \Tb(u),
\qquad
&& \Tb^{-1,0}_k = \boldsymbol 0,
\qquad
\\[0.1cm]
&f_k = \bigg(\frac{\sin(u+k\lambda)}{(\sin 2 \lambda \sin 3\lambda)^{1/2}}\bigg)^N, 
\qquad 
&&
\Tb^{0,0}_k = f_{k-3}f_{k-2}\Ib,
\qquad
&&
\sigma = (-1)^{N}.
\end{alignat}
\end{subequations}

For $\lambda = \lambda_{a,b}$, the $Y$-system is finite and is defined in terms of a set of $b$ functions:
\be
\tb^n_0 = \frac{\Tb^{n+1,0}_0 \Tb^{n-1,0}_2}{\sigma^n f_{-3}f_{2n}\Tb^{n,0}_1}, \quad n = 1, \dots, b-2, \qquad \xb_0 = \sigma \frac{\Tb^{b-2,0}_2}{\Tb^{b-1,0}_1}, 
\qquad \yb_0 = \xb_{-1}\xb_0.
\ee
The functional equations are 
\begin{subequations}
\begin{alignat}{2}
\frac{\tb^n_0\tb^n_2}{\tb^n_1} &= \frac{(\Ib+\tb^{n-1}_2)(\Ib+\tb^{n+1}_0)}{\Ib+\tb^{n}_1}, \qquad\quad n = 1, \dots, b-3,
\\[0.15cm]
\frac{\tb^{b-2}_0 \tb^{b-2}_2}{\tb^{b-2}_1} &= \frac{(\Ib+\tb^{b-3}_2)(\Ib+\eE^{\ir \Lambdab}\xb_0)(\Ib+\xb_0)(\Ib+\eE^{-\ir \Lambdab}\xb_0)}{(\Ib+\tb^{b-2}_1)(\Ib-\yb_0)(\Ib-\yb_1)},
\\[0.15cm]
\frac{\xb_0 \xb_2}{\xb_1} &= \frac{(\Ib + \tb^{b-2}_2)(\Ib-\yb_1)(\Ib-\yb_2)}{(\Ib+\eE^{\ir \Lambdab}\xb_1)(\Ib+\xb_1)(\Ib+\eE^{-\ir \Lambdab}\xb_1)},
\\[0.15cm]
\frac{\yb_0 \yb_2}{\yb_1} &=  \frac{(\Ib + \tb^{b-2}_1)(\Ib + \tb^{b-2}_2)(\Ib-\yb_0)(\Ib-\yb_1)^2(\Ib-\yb_2)}{(\Ib+\eE^{\ir \Lambdab}\xb_0)(\Ib+\xb_0)(\Ib+\eE^{-\ir \Lambdab}\xb_0)(\Ib+\eE^{\ir \Lambdab}\xb_1)(\Ib+\xb_1)(\Ib+\eE^{-\ir \Lambdab}\xb_1)}.
\end{alignat}
\end{subequations}
Another relation that will play an important role in the derivations below is the fusion hierarchy relation for $\Tb^{b-1,b-1}(u)$ obtained in \cite{MDP19}, which can be conveniently written as
\be
\Ib-\yb_0 = \frac{\sigma^{b-a}f_{-3} \Tb^{b-1,b-1}_1}{\Tb^{b-1,0}_0\Tb^{b-1,0}_1}.
\ee

In the loop model, the matrix $\eE^{\ir \Lambdab}$ is diagonal on the standard module $\stanW_{N,0}$, with its unique eigenvalue given by $\omega^{b}$. Likewise in the vertex model, in the sector of zero magnetisation, the matrix $\eE^{\ir \Lambdab}$ is also diagonal with the unique eigenvalue $\omega^{b}$.

\subsection{The principal series}\label{sec:A22.principal}

In this subsection, we fix $a = b-1$ with $b \in \mathbb N_{\ge 4}$, so that $\lambda = \frac{\pi}{2b}$. We compute the $\frac1N$ term in \eqref{eq:fsc.general} explicitly for $\gamma \in \big(0,\frac{2\pi(b-1)}{b^2}\big)$ and $u$ in the neighborhood of $\frac {3\lambda}2$.

\subsubsection[Analyticity properties and symmetric $Y$-system]{Analyticity properties and symmetric $\boldsymbol Y$-system}\label{sec:analyticity.A22.principal}

\begin{figure}
\centering
\begin{tabular}{ccccc}
\begin{tabular}{c}
$T^{1,0}(u)$ \\[0.1cm]
\includegraphics[width=.25\textwidth]{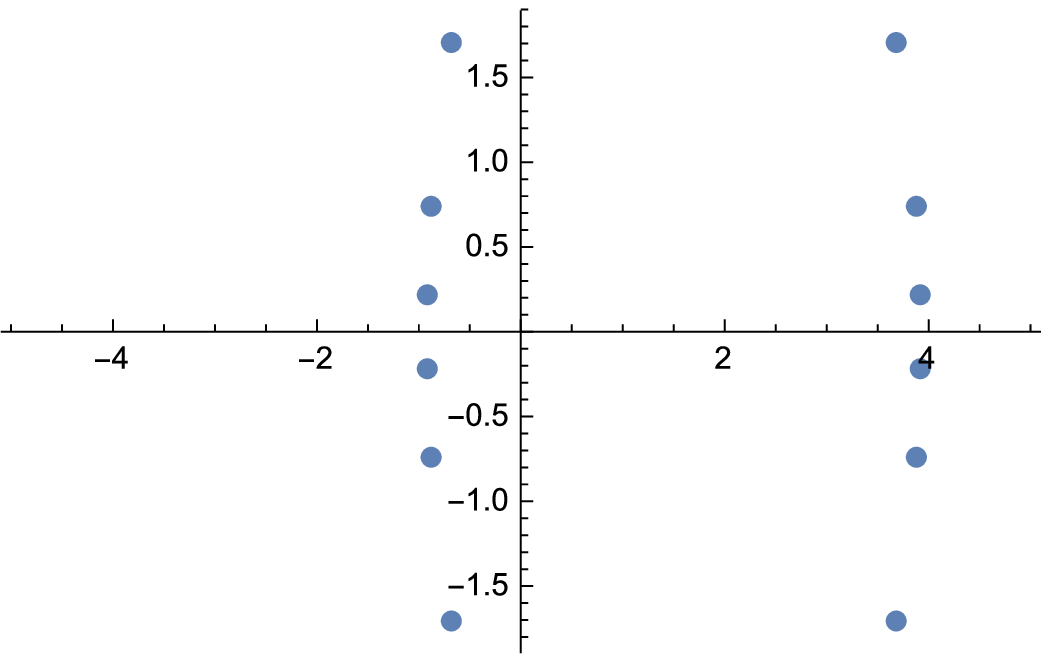} \\[0.3cm]
$T^{3,0}(u)$ \\[0.1cm]
\includegraphics[width=.25\textwidth]{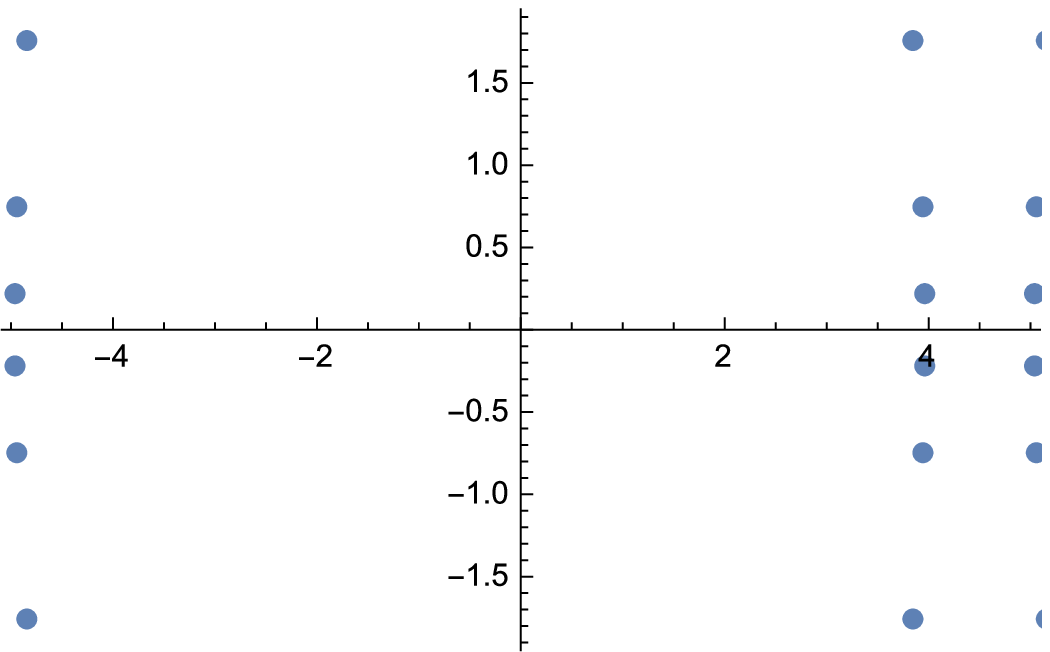}
\end{tabular}
&\quad&
\begin{tabular}{c}
$T^{2,0}(u)$ \\[0.1cm]
\includegraphics[width=.25\textwidth]{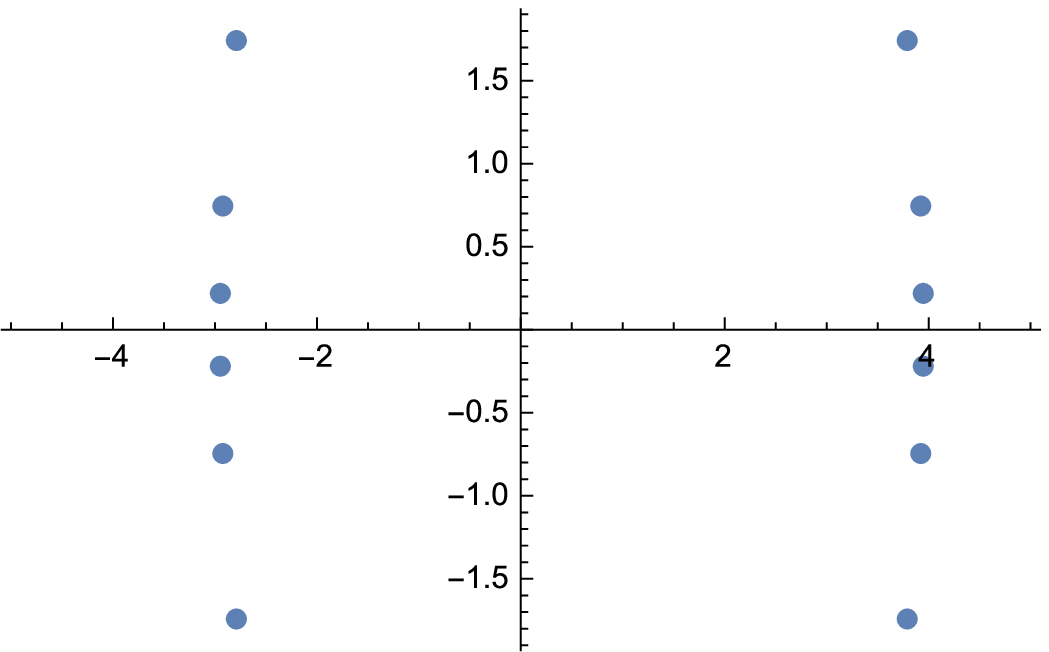}
\\[0.3cm]
$T^{4,0}(u)$ \\[0.1cm]
\includegraphics[width=.25\textwidth]{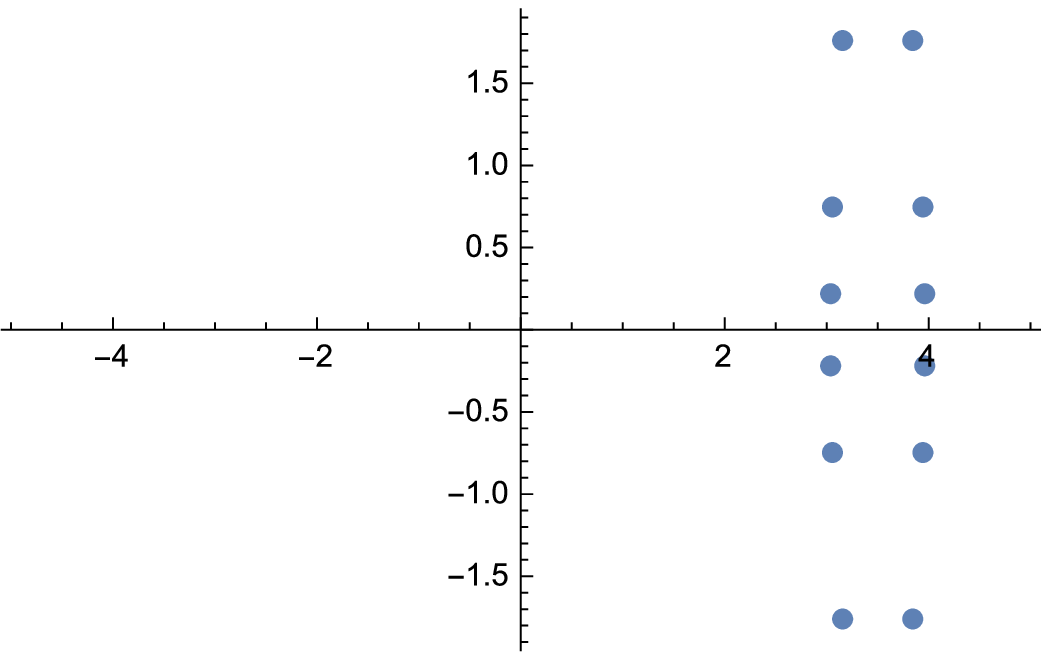}
\end{tabular}
&\quad&
\begin{tabular}{c}
$T^{4,4}(u)$ \\[0.1cm]
\includegraphics[width=.25\textwidth]{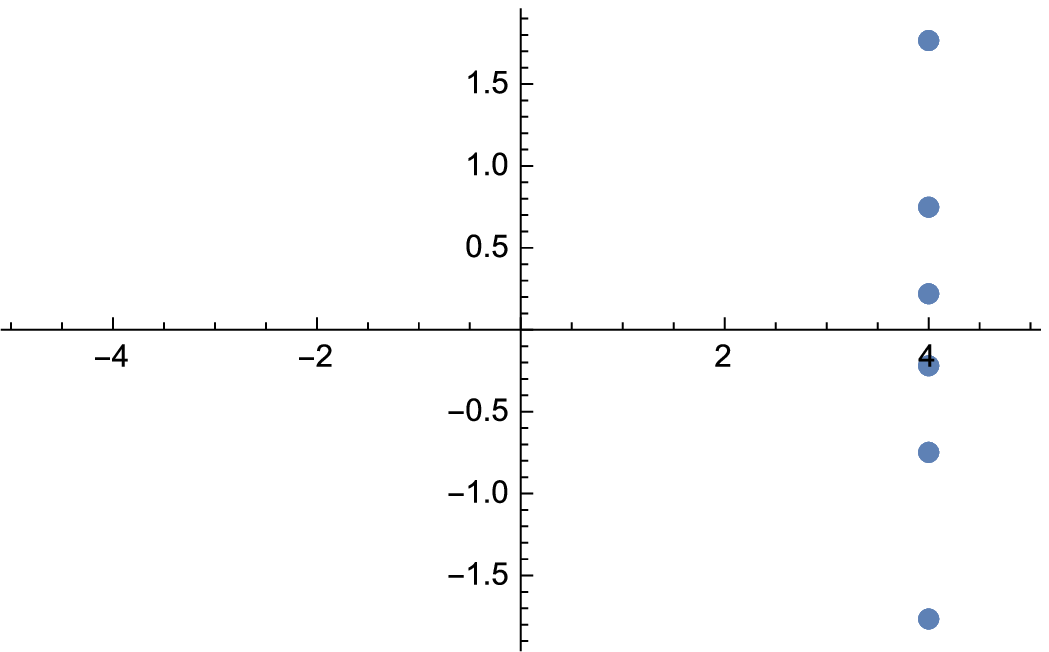} 
\end{tabular}
\end{tabular}
\caption{The patterns of zeros for the groundstate of $\Tb(u)$ for $N=6$, $(a,b) = (4,5)$ and $\omega = 1$, in the complex $u$-plane. The horizontal axis is divided in units of $\lambda = \frac\pi{10}$. Each zero of $T^{4,4}(u)$ is triply degenerate. This degeneracy is lifted for $\omega\neq1$, with the zeros remaining on the same vertical lines for $\omega$ on the unit circle.}
\label{fig:patterns.A22.principal}
\end{figure}

Our computer implementation of the transfer matrices reveals that, in the
complex $u$-plane, the zeros of $T^{1,0}(u)$ for the groundstate approximately
lie on the vertical lines with Re$(u) = -\lambda, 4\lambda$. Its analyticity
strip is therefore centered at $\frac{3\lambda}2$. Likewise, the zeros of
$T^{2,0}(u)$ lie on the vertical lines Re$(u) = -3 \lambda, 4\lambda$. These
patterns are repeated in each vertical strip of width $\pi$. In general, the
zeros of $T^n(u)$ lie on the vertical lines Re$(u) = -(2n-1) \lambda,
4\lambda$. This holds for $n = 1, \dots, b-1$.
Finally, the zeros of $T^{b-1,b-1}(u)$ lie on the vertical line Re$(u) = 4\lambda$.
To illustrate, the patterns of zeros for $N=6$, $(a,b) = (4,5)$ and $\omega=1$ are 
given in \cref{fig:patterns.A22.principal}.

This allows us to find the positions of the zeros and poles of the functions of the $Y$-system. In particular, $t^1(u)$ has zeros of order $N$ at $u=0, \lambda$ and poles at $u = -2\lambda,3\lambda$. Those zeros of order $N$ are inside its analyticity strip and will play an important role in the following. In contrast, its poles lie outside of this strip and will not play an important role. The functions $t^n(u)$ with $n = 2, \dots, b-2$ have no zeros of order $N$, but have poles of order $N$ on the real axis that lie outside of their analyticity strip. The same investigation with our computer program is repeated for the analyticity strips of $1+t^n(u)$, $\big(1+\omega^b x(u)\big)\big(1+x(u)\big)\big(1+\omega^{-b} x(u)\big)$ and $1-y(u)$. This is done using the relations
\be
\label{eq:extra.y.relations.A22}
1+t^n_0 = \frac{T^{n,0}_0T^{n,0}_2}{\sigma^n f_{-3}f_{2n} T^{n,0}_1}, \qquad 1-y_0 = \frac{\sigma^{b-a}f_{-3} T^{b-1,b-1}_1}{T^{b-1,0}_0T^{b-1,0}_1}, \qquad 1+t^{b-1}_0 = \frac{(1+\omega^{b}x_0)(1+x_0)(1+\omega^{-b}x_0)}{(1-y_0)(1-y_1)}.
\ee
Our derivation below uses certain assumptions for the analyticity strips of these functions. These are given in \cref{tab:analyticity.A22}. Crucially, except for the order-$N$ zeros of $t^1(u)$, these analyticity strips are free of zeros and poles.
\begin{table}
\begin{center}
\begin{tabular}{c|c|c}
& width is larger than & centered at 
\\[0.1cm]\hline
&&\\[-0.3cm]
$t^n(u)$ & $2 \lambda$ & $(\frac32-n)\lambda$
\\[0.1cm]
$1+t^n(u)$ & $0$ & $(\frac32-n)\lambda$
\\[0.1cm]
$x(u)$ & $2\lambda$ & $(\frac52-b)\lambda$
\\[0.1cm]
$\big(1+\omega^b x(u)\big)\big(1+x(u)\big)\big(1+\omega^{-b} x(u)\big)$ & $0$ & $(\frac52-b)\lambda$
\\[0.1cm]
$y(u)$ & $2\lambda$ & $(3-b)\lambda$
\\[0.1cm]
$1-y(u)$ & $2\lambda$ & $(3-b)\lambda$
\end{tabular}
\caption{The analyticity strips for the various functions.}
\label{tab:analyticity.A22}
\end{center}
\end{table}

We make a change of variables for the $Y$-system functions in such a way that the central lines of the analyticity strips coincide with the real axis:
\begin{subequations}
\begin{alignat}{2}
&t^n(u) = \amf^n\Big(\!-\!\tfrac {\ir \pi}{3 \lambda}\big(u+(n-\tfrac32) \lambda\big)\Big),
\qquad 
\hspace{0.25cm}\Amf^n(z) = 1+\amf^n(z),
\qquad n = 1, \dots, b-2,
\\[0.15cm]
&x(u) = \amf^{b-1}\Big(\!-\!\tfrac {\ir \pi}{3 \lambda}\big(u+(b-\tfrac52)\lambda\big)\Big), 
\\[0.15cm]
&\Amf^{b-1}(z) = \big(1 + \omega^b \amf^{b-1}(z)\big)\big(1 + \amf^{b-1}(z)\big)\big(1 + \omega^{-b} \amf^{b-1}(z)\big),\label{eq:A22c}
\\[0.15cm]\label{eq:y(u).A22}
&y(u) = \amf^{b}\Big(\!-\!\tfrac {\ir \pi}{3 \lambda}\big(u+(b-3)\lambda\big)\Big), 
\hspace{1.2cm} \Amf^{b}(z) = 1-\amf^{b}(z).
\end{alignat}
\end{subequations}
In terms of these functions, the $Y$-system takes a more symmetric form:
\begin{subequations}
\label{eq:sym.Y.A22}
\begin{alignat}{2}
\frac{\amf^{n}(z - \frac{\ir \pi}3) \amf^{n}(z + \frac{\ir \pi}3)}{\amf^{n}(z)} &= \frac{\Amf^{n-1}(z)\Amf^{n+1}(z)}{\Amf^{n}(z)},
\label{eq:sym.Ya.A22}\\[0.15cm]
\frac{\amf^{b-2}(z - \frac{\ir \pi}3) \amf^{b-2}(z + \frac{\ir \pi}3)}{\amf^{b-2}(z)} &= \frac{\Amf^{b-3}(z)\Amf^{b-1}(z)}{\Amf^{b-2}(z)\Amf^{b}(z - \frac{\ir \pi}6)\Amf^{b}(z + \frac{\ir \pi}6)}, 
\label{eq:sym.Yb.A22}\\[0.15cm]
\frac{\amf^{b-1}(z - \frac{\ir \pi}3) \amf^{b-1}(z + \frac{\ir \pi}3)}{\amf^{b-1}(z)} &= \frac{\Amf^{b-2}(z)\Amf^{b}(z - \frac{\ir \pi}6)\Amf^{b}(z + \frac{\ir \pi}6)}{\Amf^{b-1}(z)}, 
\label{eq:sym.Yc.A22}\\[0.15cm]
\frac{\amf^{b}(z - \frac{\ir \pi}3) \amf^{b}(z + \frac{\ir \pi}3)}{\amf^{b}(z)} &= \frac{\Amf^{b-2}(z-\frac{\ir \pi}6)\Amf^{b-2}(z+\frac{\ir \pi}6)\Amf^{b}(z - \frac{\ir \pi}3){\Amf^{b}(z)}^2\Amf^{b}(z + \frac{\ir \pi}3)}{\Amf^{b-1}(z-\frac{\ir \pi}6)\Amf^{b-1}(z+\frac{\ir \pi}6)}.\label{eq:sym.Yd.A22}
\end{alignat}
\end{subequations}

In terms of the variable $z$, the analyticity strips are horizontal and centered on the real line. For $\amf^n(z)$, the width of the strips is $\frac {2\ir \pi} 3$. Our computer implementation also reveals that, in the $z$-plane, the zeros of all the eigenvalues are symmetrically distributed between the upper and lower half-planes, and likewise between the right and left half-planes. This implies that the eigenvalues are real for $z \in \mathbb R$. The patterns of zeros for the groundstate are also identical in the left and right half-planes. This implies that
\be
\amf^n(z + \ir \xi) =  \amf^n(z - \ir \xi) = \amf^n(-z + \ir \xi),\qquad z,\xi \in \mathbb R.
\ee
This is not true for all eigenstates of the transfer matrix.

\subsubsection{Bulk and finite contributions}

The eigenvalue of the elementary transfer matrix is related to the $Y$-system functions by
\be
\frac{T^{1,0}_0 T^{1,0}_2}{T^{1,0}_1} = \sigma f_{-3}f_2  (1 + t^{1}_0).
\ee
The function $T(u) = T^{1,0}(u)$ is written as the product of its bulk and finite contributions: 
\be
T(u) = \kappa(u)^N T_{\rm f}(u).
\ee
These satisfy the functional equations
\be
\label{eq:func.rel.A22}
\frac{\kappa(u)\kappa(u+2\lambda)}{\kappa(u+\lambda)} = \frac{\sin(2\lambda +u)\sin(3\lambda - u)}{\sin 2\lambda\sin 3\lambda},
\qquad
\frac{T_{\rm f}(u)T_{\rm f}(u+2\lambda)}{T_{\rm f}(u+\lambda)} = 1+t^1(u).
\ee
The known solution for the bulk contribution is \cite{WBN1992}:
\be
\log\kappa(u)=2\!\int_0^\infty\!\! \dd t\;\frac{\sinh \lambda t \sinh\frac{(\pi-2\lambda)t}{2}\cosh \frac{(2u-3\lambda)t}{2}}{t \sinh\frac{\pi t}{2}\cosh\frac{3\lambda t}{2}}+
\log\Big|\frac{\sin u\sin (3\lambda-u)}{\sin 2\lambda \sin 3\lambda}\Big|,\quad \mbox{Regime I}.
\ee
For the finite term, we define
\be
T_{\rm f}(u) = \bmf\big(\!-\!\tfrac {\ir \pi}{3 \lambda}(u-\tfrac {3\lambda} 2)\big)
\ee
and rewrite the second relation in \eqref{eq:func.rel.A22} as
\be
\label{eq:bbb.A22}
\frac{\bmf(z - \frac{\ir \pi}3) \bmf(z + \frac{\ir \pi}3)}{\bmf(z)} = \Amf^1(z).
\ee

\subsubsection{Non-linear integral equations}

The order-$N$ zeros of $t^1(u)$ lie inside its analyticity strip, whereas its order-$N$ poles lie outside of this strip. The corresponding zeros of $\amf^1(z)$ lie at $z = \pm \frac{\ir \pi}6$. We then define the functions $\ell^n(z)$ as
\be
\label{eq:ell.a.A22}
\ell^1(z) = \frac{\amf^1(z)}{\eta(z)^N}, \qquad \ell^n(z) = \amf^n(z), \qquad n = 2, \dots, b,
\ee
where
\be
\eta(z) = \tanh\tfrac12(z-\tfrac{\ir \pi}6)\tanh\tfrac12(z+\tfrac{\ir \pi}6),
\qquad
\frac{\eta(z-\tfrac{\ir \pi}3)\eta(z+\tfrac{\ir \pi}3)}{\eta(z)} = 1.
\ee
The function $\ell^1(z)$ has no zeros inside its analyticity strip. Moreover we have
\be
\label{eq:sym.Y.L.A22}
\frac{\ell^1(z - \frac{\ir \pi}3) \ell^1(z + \frac{\ir \pi}3)}{\ell^1(z)} = \frac{\Amf^2(z)}{\Amf^1(z)}.
\ee
The other equations in the $Y$-system are identical to those in \eqref{eq:sym.Y.A22}, with the functions $\amf^n(z)$ on the left-hand side replaced by the corresponding functions $\ell^n(z)$.

The functions $\ell^n(z)$ are all analytic and non-zero inside their respective analyticity strips. As discussed in \cref{sec:braid.and.bulk.A22}, for generic values of $\omega$, these functions have asymptotic values for $z \to \pm \infty$ that are finite and nonzero. This allows us to define the Fourier transform of their logarithmic derivative:
\be
L^n(k) = \frac1{2\pi} \int_{-\infty}^\infty \dd z\, \eE^{-\ir k z}\big(\log \ell^n(z)\big)', \qquad 
A^n(k) = \frac1{2\pi} \int_{-\infty}^\infty \dd z\, \eE^{-\ir k z}\big(\log \Amf^n(z)\big)',
\ee
where $n = 1, \dots, b$.
The non-linear integral equations for the eigenvalues are obtained by first taking the Fourier transform of the logarithmic derivative of the $Y$-system equations yielding
\begin{subequations}
\begin{alignat}{2}
L^n(k) &= \frac 1{2 \cosh\frac{\pi k}3 - 1} \Big[A^{n-1}(k) -  A^n(k) + A^{n+1}(k)\Big], \qquad n = 1, \dots, b-3,
\\[0.15cm]L^{b-2}(k) &= \frac 1{2 \cosh\frac{\pi k}3 - 1} \Big[ A^{b-3}(k)- A^{b-2}(k) +  A^{b-1}(k) - 2\cosh(\tfrac{\pi k}6)A^{b}(k)\Big],
\\[0.15cm]L^{b-1}(k) &= \frac 1{2 \cosh\frac{\pi k}3 - 1} \Big[ A^{b-2}(k)- A^{b-1}(k) +  2\cosh(\tfrac{\pi k}6)A^{b}(k)\Big],
\\[0.15cm]
L^{b}(k) &=  \frac 1{2 \cosh\frac{\pi k}3 - 1} \Big[2\cosh\tfrac{\pi k}6\,\big(A^{b-2}(k) -A^{b-1}(k)\big) + 3 A^{b}(k) \Big] + A^{b}(k).
\end{alignat}
\end{subequations}
Applying the inverse Fourier transform, we obtain the non-linear integral equations for the functions $\ell^n(z)$. We use \eqref{eq:ell.a.A22} to express these in terms of the functions $\amf^n(z)$, integrate once with respect to $z$ and find
\begin{subequations}
\label{eq:NLIEs.A22}
\begin{alignat}{2}
\log \amf^{1}(z) - \phi_{1} &= N \log \eta(z) - K * \log \Amf^{1} + K * \log \Amf^{2}, 
\\[0.15cm]
\log \amf^{n}(z) - \phi_{n} &= K * \log \Amf^{n-1} - K * \log \Amf^{n} + K * \log \Amf^{n+1}, \qquad n = 2, \dots, b-3,
\\[0.15cm]
\log \amf^{b-2}(z) - \phi_{b-2} &= K * \log \Amf^{b-3} - K * \log \Amf^{b-2} + K * \log \Amf^{b-1} - \tilde K * \log \Amf^{b},
\\[0.15cm]
\log \amf^{b-1}(z) - \phi_{b-1} &= K * \log \Amf^{b-2} - K * \log \Amf^{b-1} + \tilde K * \log \Amf^{b},
\\[0.15cm]
\log \amf^{b}(z) - \phi_b &= \tilde K * \log \Amf^{b-2} - \tilde K * \log \Amf^{b-1} + 3 K * \log \Amf^{b} + \log \Amf^{b}(z),\label{eq:scalingNLIEs.A22e}
\end{alignat}
\end{subequations}
where $\phi_1$, \dots, $\phi_b$ are the integration constants. The convolution is defined in \eqref{eq:convolution} and the kernels are given by
\begin{subequations}
\begin{alignat}{2}
K(z) &= \frac1{2\pi}\int_{-\infty}^\infty \dd k \, \eE^{\ir k z} \frac{1}{2\cosh(\tfrac{\pi k}3) - 1}= \frac{\sqrt 3}{\pi} \frac{\sinh(2z)}{\sinh(3z)}, 
\\[0.15cm]
\tilde K(z) &=\frac1{2\pi} \int_{-\infty}^\infty \dd k \, \eE^{\ir k z}\frac{2\cosh(\tfrac{\pi k}6)}{2\cosh(\tfrac{\pi k}3) - 1}= \frac{3}{\pi} \frac{\cosh(2z)}{\cosh(3z)}.
\end{alignat}
\end{subequations}

\subsubsection{Scaling functions and scaling non-linear integral equations}
In \eqref{eq:NLIEs.A22}, the dependence on $N$ appears only in the driving term $N \log \eta(z)$. For $z$ of order $\pm \log N$ with $N$ large, this function behaves as an exponential:
\be
\lim_{N\to \infty} N \log \eta\big(\!\pm\!(z + \log N)\big)  = -2 \sqrt 3\, \eE^{-z}.
\ee
To compute the finite-size correction at order $\frac 1N$, we assume that the unknown functions appearing in \eqref{eq:NLIEs.A22} are well-defined in this limit. For the groundstate, the patterns of zeros are all symmetric with respect to the imaginary $z$-axis, so the scaling functions behave identically in both limits:
\be
\mathsf a^n(z) = \lim_{N\to \infty} \amf^n\big(\!\pm\!(z + \log N)\big), \qquad \mathsf A^n(z) = \lim_{N\to \infty} \Amf^n\big(\!\pm\!(z + \log N)\big),\qquad n = 1, \dots, b.
\ee
These satisfy the following set of integral equations:
\begin{subequations}
\label{eq:scalingNLIEs.A22}
\begin{alignat}{2}
\log \asf^{1}(z) - \phi_{1} &=-2 \sqrt 3\, \eE^{-z} - K * \log \Asf^{1} + K * \log \Asf^{2}, 
\\[0.15cm]
\log \asf^{n}(z) - \phi_{n} &= K * \log \Asf^{n-1} - K * \log \Asf^{n} + K * \log \Asf^{n+1}, \qquad n = 2, \dots, b-3,
\\[0.15cm]
\log \asf^{b-2}(z) - \phi_{b-2} &= K * \log \Asf^{b-3} - K * \log \Asf^{b-2} + K * \log \Asf^{b-1} - \tilde K * \log \Asf^{b},
\\[0.15cm]
\log \asf^{b-1}(z) - \phi_{b-1} &= K * \log \Asf^{b-2} - K * \log \Asf^{b-1} + \tilde K * \log \Asf^{b},
\\[0.15cm]
\log \asf^{b}(z) - \phi_b &= \tilde K * \log \Asf^{b-2} - \tilde K * \log \Asf^{b-1} + 3 K * \log \Asf^{b} + \log \Asf^{b}(z).
\end{alignat}
\end{subequations}
Multiplying the last equation by an overall minus sign, we see that the kernel terms on the right sides become symmetric.

\subsubsection{Braid and bulk behavior}\label{sec:braid.and.bulk.A22}

The scaling functions have finite asymptotics for $z \to \pm \infty$. For $z \to \infty$, these are obtained directly from the braid limits of the transfer matrix eigenvalues:
\be
\asf^n_\infty = \frac{(\omega^{n/2}-\omega^{-n/2})(\omega^{(n+3)/2}-\omega^{-(n+3)/2})}{(\omega^{1/2}-\omega^{-1/2})(\omega-\omega^{-1})}, 
\qquad 
\asf^{b-1}_\infty =\frac{\omega^{(b-1)/2}-\omega^{-(b-1)/2}}{\omega^{(b+1)/2}-\omega^{-(b+1)/2}}, 
\qquad \asf^{b}_\infty = (\asf^{b-1}_\infty)^2,\label{eq:braid.an.A22}
\ee
where $n = 1, \dots, b-2$.
For $\gamma \in (0, \frac {2\pi}{b+1})$, these constants are all positive and finite.
These values allow us to compute the constants $\phi_n$ by studying the $z\to \infty$ asymptotics of \eqref{eq:scalingNLIEs.A22}. We use 
\begin{subequations}
\begin{alignat}{2}
\lim_{z\to \infty} K* \mathsf X &=  \mathsf X_\infty \int_{-\infty}^\infty \dd y \, K(y) =  \mathsf X_\infty, \qquad
\lim_{z\to \infty} \tilde K* \mathsf X = \mathsf X_\infty \int_{-\infty}^\infty \dd y \, \tilde K(y) = 2\, \mathsf X_\infty,
\end{alignat}
\end{subequations}
and find that the constants vanish:
\be
\phi_n = 0, \qquad n = 1, \dots, b, \qquad \gamma \in (0,\tfrac {2\pi}{b+1}).
\ee

For the bulk behavior at $z \to -\infty$, we recall that the function $\amf^1(z)$ has a zero of order $N$ near the origin, so that $\asf^1_{-\infty} = 0$. From the $Y$-system equation, the asymptotic values $\asf^n_{-\infty}$ satisfy the relations
\begin{subequations}
\begin{alignat}{2}
\asf^n_{-\infty} &= \frac{(1+\asf^{n-1}_{-\infty})(1+\asf^{n+1}_{-\infty})}{(1+\asf^{n}_{-\infty})}, \qquad n = 2, \dots, b-3,
\\[0.15cm]
\asf^{b-2}_{-\infty} &=  \frac{(1+\asf_{-\infty}^{b-3})(1+\omega^b\asf^{b-1}_{-\infty})(1+\asf^{b-1}_{-\infty})(1+\omega^{-b}\asf^{b-1}_{-\infty})}{(1+\asf^{b-2}_{-\infty})\big(1-(\asf^{b-1}_{-\infty})^2\big)^2},
\\[0.15cm]
\asf^{b-1}_{-\infty} &= \frac{(1+\asf^{b-2}_{-\infty})\big(1-(\asf^{b-1}_{-\infty})^2\big)^2}{(1+\omega^b\asf^{b-1}_{-\infty})(1+\asf^{b-1}_{-\infty})(1+\omega^{-b}\asf^{b-1}_{-\infty})}.
\end{alignat}
\end{subequations}
Of the many solutions to this system, we select the only one for which all the functions are positive for $\omega = 1$. Indeed, in a small interval around this point, these functions are real for $z \in \mathbb R$, have positive asymptotics at $z \to \infty$ and have no zeros inside the analyticity strip, which implies that they also have positive bulk asymptotics. The solution is
\begin{subequations}
\label{eq:lowlims.A22}
\begin{alignat}{2}
\asf^n_{-\infty} &= \frac{(\bar\omega^{(n-1)/2}-\bar\omega^{-(n-1)/2})(\bar\omega^{(n+2)/2}-\bar\omega^{-(n+2)/2})}{(\bar\omega^{1/2}-\bar\omega^{-1/2})(\bar\omega-\bar\omega^{-1})}, \qquad n = 1, \dots, b-2,\label{eq:lowlim.A22_1}
\\[0.15cm]
\qquad \asf^{b-1}_{-\infty} &=\frac{\bar\omega^{(b-2)/2}-\bar\omega^{-(b-2)/2}}{\bar\omega^{b/2}-\bar\omega^{-b/2}}, 
\qquad \asf^{b}_{-\infty} = (\asf^{b-1}_{-\infty})^2,
\qquad \bar\omega = \omega^{b/(b-1)}.\label{eq:lowlim.A22_2}
\end{alignat}
\end{subequations}
These values are finite on the range $\gamma \in (0,\frac{2\pi(b-1)}{b^2})$.

\subsubsection{Finite-size correction and the dilogarithm technique}\label{sec:dilog.A22}

Applying the Fourier transform and subsequently the inverse transform of the logarithmic derivative of \eqref{eq:bbb.A22}, we find
\be
\log \bmf(z) - \phi_0 =\int_{-\infty}^\infty \dd y\,K(y-z)\log \Amf^1(y),
\ee
where $\phi_0$ is an integration constant. For large $N$, we express this in terms of integrals involving the scaling function $\Asf^1(z)$:
\begin{alignat}{2}
\log \bmf(z) - \phi_0 &= \int_{-\log N}^\infty \dd y \Big(K(y\!+\!\log N \!-\! z) \log \Amf^1(y\!+\!\log N) + K(-y\!-\!\log N \!-\! z) \log \Amf^1(-y\!-\!\log N)\Big)
\nonumber\\[0.15cm]
&\simeq \frac{\sqrt 3}{\pi N}(\eE^{z}+\eE^{-z})\int_{-\infty}^\infty \dd y\, \eE^{-y} \log \Asf^1(y),
\label{eq:log.b.int.A22}
\end{alignat}
where we used
\be
K(z + \log N) \simeq \frac{\sqrt 3\, \eE^{-z}}{\pi N}.
\ee
Here, $\simeq$ indicates that higher-order terms in $\frac1N$ are omitted.

To apply the dilogarithm technique, we define the integral
\begin{alignat}{2}
\mathcal J = \int_{-\infty}^\infty \dd y &\bigg[\sum_{n=1}^{b-1}\Big((\log \asf^n)' \log \Asf^n  -\log \asf^n (\log \Asf^n)'\Big) - \Big(\big(\log \asf^b\big)' \log \Asf^b -\log \asf^b\, (\log \Asf^b)' \Big) \bigg],
\end{alignat}
where the dependence on the argument $y$ is dropped for ease of notation. This integral is evaluated in two ways. The first consists of replacing $\log \asf^n$ by its expression in \eqref{eq:scalingNLIEs.A22}. Many terms cancel out because of the symmetry properties of the kernels
\be
K(-z) = K(z), \qquad \tilde K(-z) = \tilde K(z).
\ee
 The only surviving contributions come from the driving terms, and the result reads
\be
\mathcal J = 4 \sqrt 3 \int_{-\infty}^\infty \dd y \, \eE^{-y} \log \Asf^1(y).
\ee
Up to an overall prefactor, this is precisely the integral we wish to compute in \eqref{eq:log.b.int.A22}. The second way of computing the integral is to apply the derivatives explicitly, which yields
\be
\mathcal J = \int_{-\infty}^\infty \dd y \,  \bigg[\sum_{n=1}^{b-2}\frac{\dd \asf^n}{\dd y}\bigg(\frac{\log \Asf^n}{\asf^n} -\frac{\log \asf^n}{\Asf^n}\bigg) + \frac{\dd \asf^{b-1}}{\dd y}\bigg(\frac{\log \Asf^{b-1}}{\asf^{b-1}}  - \frac{\dd \Asf^{b-1}}{\dd \asf^{b-1}}\frac{\log \asf^{b-1}}{\Asf^{b-1}}\bigg) - \frac{\dd \asf^b}{\dd y} \bigg(\frac{\log \Asf^b}{\asf^b} + \frac{\log \asf^b}{\Asf^b}\bigg)\bigg].
\ee
Dividing the integral in three parts and changing the integration variables from $y$ to $\asf^n$, we find
\begin{alignat}{2}
\mathcal J &= \sum_{n=1}^{b-2}\int_{\asf^n_{-\infty}}^{\asf^n_\infty} \dd \asf^n \bigg(\frac{\log \Asf^n}{\asf^n}  -\frac{\log \asf^n}{\Asf^n}\bigg) + \int_{\asf^{b-1}_{-\infty}}^{\asf^{b-1}_\infty} \dd \asf^{b-1} \bigg(\frac{\log \Asf^{b-1}}{\asf^{b-1}} - \frac{\dd \Asf^{b-1}}{\dd \asf^{b-1}}\frac{\log \asf^{b-1}}{\Asf^{b-1}}\bigg) 
\nonumber\\[0.15cm]
&- \int_{\asf^b_{-\infty}}^{\asf^b_\infty} \dd \asf^b  \bigg(\frac{\log \Asf^b}{\asf^b} + \frac{\log \asf^b}{\Asf^b}\bigg)
\label{eq:J1.A22}
\end{alignat}
where
\begin{subequations}
\begin{alignat}{2}
\Asf^n &= 1 + \asf^n, \qquad n = 1, \dots, b-2,
\\[0.15cm] 
\Asf^{b-1} &= (1 + \omega^b \asf^{b-1})(1 + \asf^{b-1})(1 + \omega^{-b} \asf^{b-1}),
\\[0.15cm]  
\Asf^b &= (1 - \asf^b).
\end{alignat}
\end{subequations}
The resulting expression for $\mathcal J$ is therefore a combination of regular integrals. Setting $\omega = \eE^{\ir \gamma}$, the integral evaluates to
\be
\label{eq:J2.A22}
\mathcal J = \frac{\pi^2}3\bigg(1- \frac{3\gamma^2 b}{\pi^2 (b-1)}\bigg) = \frac{\pi^2}3\bigg(1-\frac{3\gamma^2}{\pi(\pi-2\lambda)}\bigg), \qquad \gamma \in \big(0,\tfrac{2\pi(b-1)}{b^2}\big).
\ee
The proof of this result is given in \cref{app:A22.principal}. The final result is
\be
\log \bmf(z) \simeq \frac{\pi \cosh z}{6 N} \bigg(1- \frac{3\gamma^2 b}{\pi^2 (b-1)}\bigg), \qquad 
\log T_{\rm f}(u) \simeq \frac{\pi \sin \frac{\pi u}{3\lambda}}{6 N} \bigg(1- \frac{3\gamma^2 b}{\pi^2 (b-1)}\bigg),
\ee
where the constant $\phi_0$ was found to equal zero using $T_{\rm f}(u=0)=1$.
This result is precisely \eqref{eq:fsc.general} with $c-24 \Delta$ and $\vartheta(u)$ given in \eqref{eq:fsc.A22}.

\subsection{The dual series}\label{sec:A22.dual}

In this subsection, we fix $a = 1-b$, so that $\lambda = \frac{\pi(2b-1)}{2b}$, and consider $b\in \mathbb N_{\ge 4}$. We also define $\bar\lambda = \pi - \lambda = \frac{\pi}{2b}$. We compute the $\frac1N$ term in \eqref{eq:fsc.general} explicitly for $N$ even, $\gamma \in \big(0,\frac{2\pi}{b+1}\big)$ and $u$ in the neighborhood of $\frac{3\lambda}2$.

\subsubsection[Analyticity properties and symmetric $Y$-system]{Analyticity properties and symmetric $\boldsymbol Y$-system}\label{sec:analyticity.A22.dual}

\begin{figure}
\centering
\begin{tabular}{ccccc}
\begin{tabular}{c}
$T^{1,0}(u)$ \\[0.1cm]
\includegraphics[width=.25\textwidth]{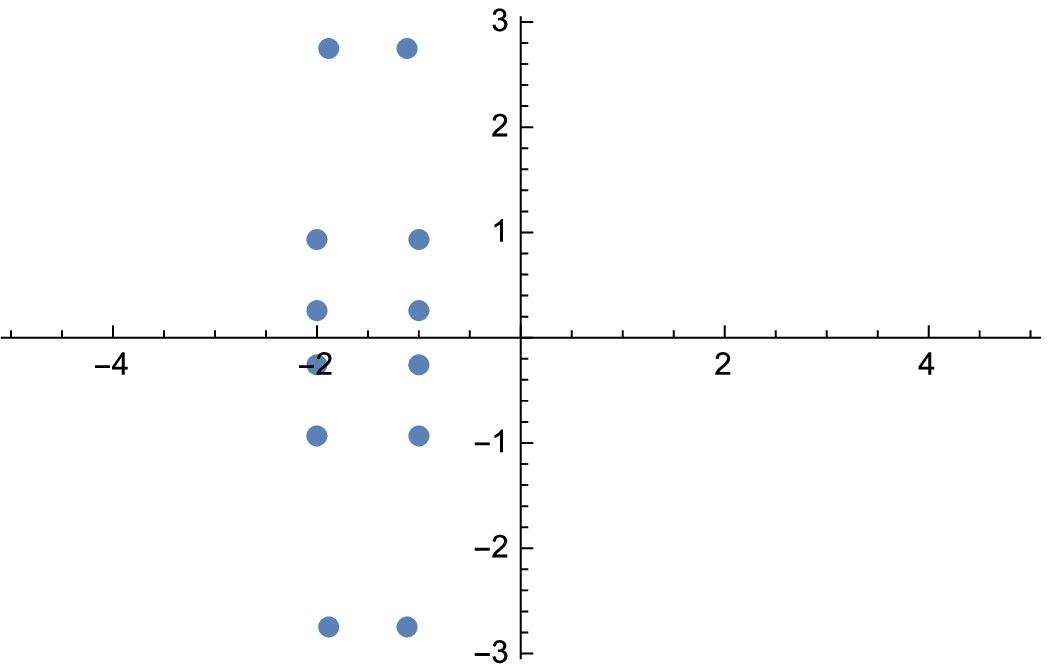} \\[0.3cm]
$T^{3,0}(u)$ \\[0.1cm]
\includegraphics[width=.25\textwidth]{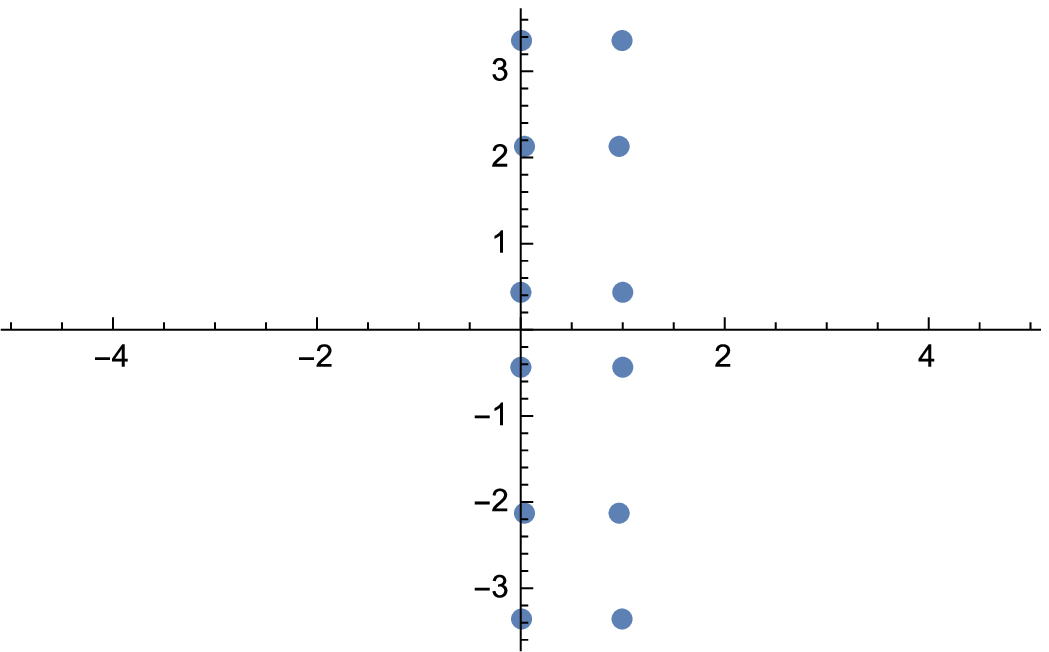}
\end{tabular}
&\quad&
\begin{tabular}{c}
$T^{2,0}(u)$ \\[0.1cm]
\includegraphics[width=.25\textwidth]{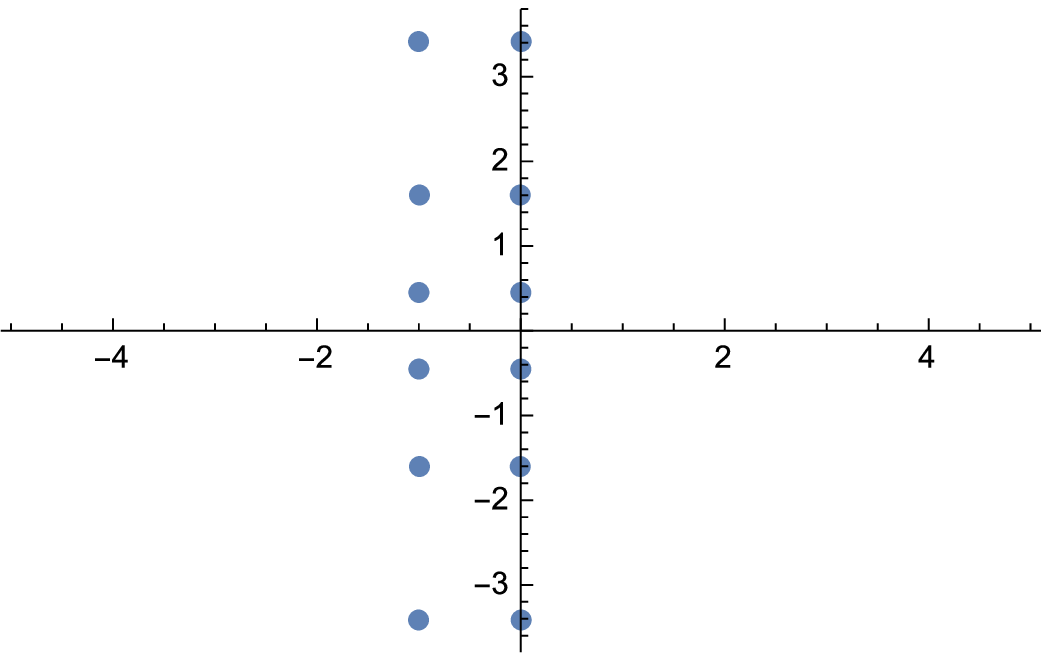}
\\[0.3cm]
$T^{4,0}(u)$ \\[0.1cm]
\includegraphics[width=.25\textwidth]{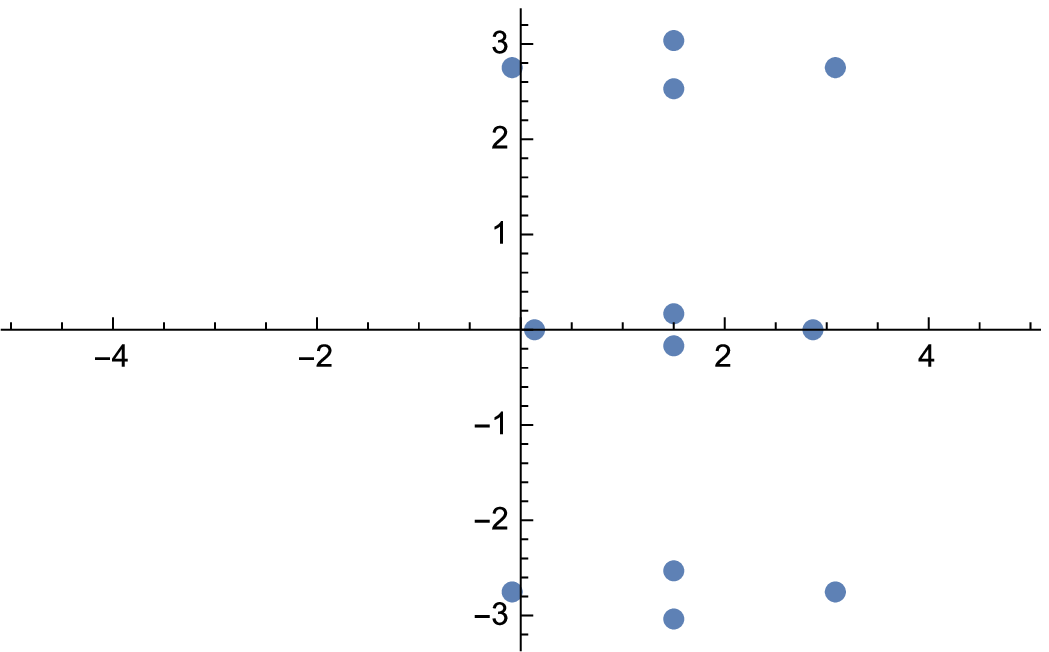}
\end{tabular}
&\quad&
\begin{tabular}{c}
$T^{4,4}(u)$ \\[0.1cm]
\includegraphics[width=.25\textwidth]{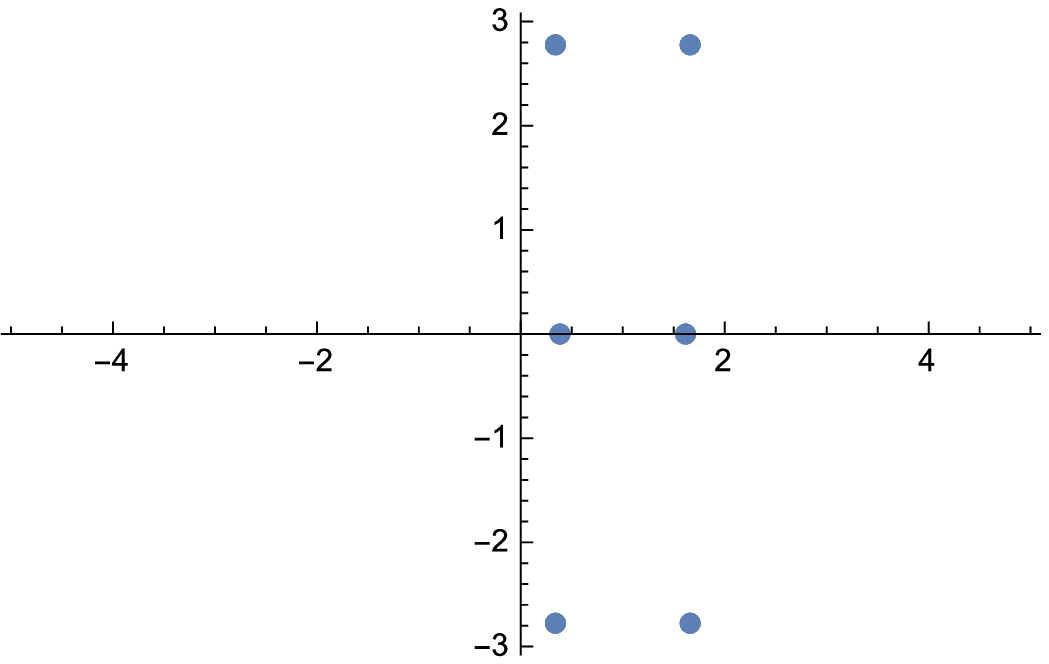} 
\end{tabular}
\end{tabular}
\caption{The patterns of zeros for the groundstate of $\Tb(u)$ for $N=6$, $(a,b) = (-4,5)$ and $\omega = 1$, in the complex $u$-plane. The horizontal axis is divided in units of $2\bar \lambda = \frac\pi5$. Each zero of $T^{4,4}(u)$ is triply degenerate. This degeneracy is lifted for $\omega\neq1$, with the zeros remaining on the same vertical lines for $\omega$ on the unit circle.}
\label{fig:patterns.A22.dual}
\end{figure}

Our computer implementation of the transfer matrices reveals that, in the complex $u$-plane, the zeros of $T^{1,0}(u)$ for the groundstate approximately lie on the vertical lines with Re$(u) = -\bar \lambda, -2 \bar \lambda$. Its analyticity strip is centered at $\frac{3\lambda}2$. Likewise, the zeros of $T^{2,0}(u)$ lie on the vertical lines Re$(u) = -\bar \lambda, 0$. These patterns are repeated in each vertical strip of width $\pi$. For $n=1, \dots, b-2$, the zeros of $T^{n,0}(u)$ lie on the vertical lines Re$(u) = -\bar \lambda, 2(n-2)\bar\lambda$. The zeros of $T^{b-1,0}(u)$ lie on the three vertical lines Re$(u) = -\bar \lambda, (b-\frac72)\bar \lambda, 2(b-3)\bar \lambda$. The center of the analyticity strip for $T^{n,0}(u)$ is at Re$(u) = \frac{3\lambda}2 + (n-1)\bar \lambda$. Finally, the zeros of $T^{b-1,b-1}(u)$ lie on the vertical lines Re$(u) = (b-\frac72)\bar\lambda, (b-\frac92)\bar\lambda$. To illustrate, the patterns of zeros for $N=6$, $(a,b) = (-4,5)$ and $\omega=1$ are given in \cref{fig:patterns.A22.dual}.

This allows us to find the positions of the zeros and poles of the functions of the $Y$-system. In particular, $t^1(u)$ has zeros of order $N$ at $u=0, \lambda$ and poles at $u = 2\bar\lambda,(2b-3)\bar\lambda$. Those zeros of order~$N$ will lie outside its analyticity strip and will play no role in the following. In contrast, its poles will play an important role. The functions $t^n(u)$ with $n = 2, \dots, b-2$ have no zeros of order $N$, but have two poles of order $N$ at $u = 2n\bar\lambda, (2b-3)\bar \lambda$. The functions $x(u)$ and $y(u)$ have no order-$N$ zeros and no order-$N$ poles.

The same investigation is repeated for the functions $1+t^n(u)$, $\big(1+\omega^b x(u)\big)\big(1+x(u)\big)\big(1+\omega^{-b} x(u)\big)$ and $1-y(u)$, using the relations \eqref{eq:extra.y.relations.A22}. In particular, we find that the function $1-y(u)$
has an order-$N$ zero at $u=3\lambda \equiv -3 \bar \lambda$, where $\equiv$ denotes an equality modulo $\pi$. In the following, we choose the last function entering the $Y$-system to be $w(u) = \big(1-y(u)\big)/y(u)$ instead of $y(u)$, as this function has an order-$N$ zero which will play an important role for the calculation. Moreover, to work with functions that remain finite inside their analyticity strips, we make a change of variables and work with the reciprocals of the functions $t^n(u)$ and $x(u)$: 
\be 
\tilde t^n(u) = t^n(u)^{-1}, \qquad \tilde x(u) = x(u)^{-1}.
\ee 
Inside their analyticity strips, the functions $\tilde t^n(u)$ have order-$N$
zeros instead of order-$N$ poles. Our derivation below uses certain assumptions for the analyticity strips of these functions. These are given in \cref{tab:analyticity.A22.dual}. Crucially, except for the order-$N$ zeros of $\tilde t^{b-2}(u)$ and $w(u)$, these analyticity strips are free of zeros and poles.
\begin{table}
\begin{center}
\begin{tabular}{c|c|c}
& width is larger than & centered at 
\\[0.1cm]\hline
&&\\[-0.3cm]
$\tilde t^n(u)$ & $2 \bar\lambda$ & $\frac{3\lambda}2 + n\bar \lambda$
\\[0.1cm]
$1+\tilde t^n(u)$ & $0$ & $\frac{3\lambda}2 + n\bar \lambda$
\\[0.1cm]
$\tilde x(u)$ & $2\bar\lambda$ & $\frac{3\lambda}2 + (b-1)\bar \lambda$
\\[0.1cm]
$\big(1+\omega^b \tilde x(u)\big)\big(1+\tilde x(u)\big)\big(1+\omega^{-b} \tilde x(u)\big)$ & $0$ & $\frac{3\lambda}2 + (b-1)\bar \lambda$
\\[0.1cm]
$w(u)$ & $\bar\lambda$ & $\frac{3\lambda}2 + (b-\frac32)\bar \lambda$
\\[0.1cm]
$1+w(u)$ & $0$ & $\frac{3\lambda}2 + (b-\frac32)\bar \lambda$
\end{tabular}
\caption{The analyticity strips for the various functions.}
\label{tab:analyticity.A22.dual}
\end{center}
\end{table}

We make a change of variables for the $Y$-system functions in such a way that the central lines of the analyticity strips coincide with the real axis:
\begin{subequations}
\begin{alignat}{2}
&\tilde t^n(u) = \amf^n\Big(\!-\!\tfrac {\ir \pi}{\textrm{\raisebox{-0.06cm}{$\bar \lambda$}}}\big(u-\tfrac{3\lambda}2-n \bar\lambda\big)\Big),
\qquad 
\hspace{1.25cm}\Amf^n(z) = 1+\amf^n(z),
\qquad n = 1, \dots, b-2,
\\[0.15cm]
&\tilde x(u) = \amf^{b-1}\Big(\!-\!\tfrac {\ir \pi}{\textrm{\raisebox{-0.06cm}{$\bar \lambda$}}}\big(u-\tfrac{3\lambda}2-(b-1) \bar\lambda\big)\Big), 
\\[0.15cm]
&\Amf^{b-1}(z) = \big(1 + \omega^b \amf^{b-1}(z)\big)\big(1 + \amf^{b-1}(z)\big)\big(1 + \omega^{-b} \amf^{b-1}(z)\big),\label{eq:A22c.dual}
\\[0.15cm]\label{eq:y(u).A22.dual}
&w(u) = \amf^{b}\Big(\!-\!\tfrac {\ir \pi}{\textrm{\raisebox{-0.06cm}{$\bar \lambda$}}}\big(u-\tfrac{3\lambda}2-(b-\tfrac32)\lambda\big)\Big), 
\hspace{1.2cm} \Amf^{b}(z) = 1+\amf^{b}(z).
\end{alignat}
\end{subequations}
In terms of these functions, the $Y$-system reads
\begin{subequations}
\label{eq:sym.Y.A22.dual}
\begin{alignat}{2}
\frac{\amf^{2}(z)}{\amf^{1}(z - \ir \pi) \amf^{1}(z + \ir \pi)} &= \frac{\Amf^{2}(z)}{\Amf^{1}(z)},
\\[0.15cm]
\frac{\amf^{n-1}(z)\amf^{n+1}(z)}{\amf^{n}(z - \ir \pi) \amf^{n}(z + \ir \pi)} &= \frac{\Amf^{n-1}(z)\Amf^{n+1}(z)}{\Amf^{n}(z)}, \qquad n = 2, \dots, b-3,
\\[0.15cm]
\frac{\amf^{b-3}(z)\amf^{b-1}(z)\amf^b(z-\frac{\ir \pi}2)\amf^b(z+\frac{\ir \pi}2)}{\amf^{b-2}(z - \ir \pi) \amf^{b-2}(z + \ir \pi)\amf^{b-1}(z-\ir \pi)\amf^{b-1}(z+\ir \pi)}
&= \frac{\Amf^{b-3}(z)\Amf^{b-1}(z)}{\Amf^{b-2}(z)}, 
\\[0.15cm]
\frac{\amf^{b-2}(z)}{\amf^{b}(z-\tfrac{\ir \pi}2) \amf^{b}(z +\tfrac{\ir \pi}2)} &= \frac{\Amf^{b-2}(z)}{\Amf^{b-1}(z)}, 
\\[0.15cm]
\amf^{b-1}(z-\tfrac{\ir \pi}2)\amf^{b-1}(z+\tfrac{\ir \pi}2) &= \Amf^b(z).\end{alignat}
\end{subequations}

In terms of the variable $z$, the analyticity strips are horizontal and centered on the real line. For $\amf^n(z)$ with $n = 1, \dots, b-2$, the width of the strips is $2\ir \pi$. Our computer implementation also reveals that, in the $z$-plane, the zeros of all the eigenvalues are symmetrically distributed between the upper and lower half-planes, and likewise between the right and left half-planes. This implies that the eigenvalues are real for $z \in \mathbb R$. The patterns of zeros for the groundstate are also identical in the left and right half-planes. This implies that
\be
\amf^n(z + \ir \xi) =  \amf^n(z - \ir \xi) = \amf^n(-z + \ir \xi),\qquad z,\xi \in \mathbb R.
\ee
This is not true for all eigenstates of the transfer matrix.

\subsubsection{Bulk and finite contributions}

The eigenvalues of the transfer matrices are related to the $Y$-system functions by
\begin{subequations}
\begin{alignat}{2}
\frac{T^{n,0}_0 T^{n,0}_2}{T^{n+1,0}_0 T^{n-1,0}_2} &= 1 + \tilde t^{n}_0, \qquad n = 1, \dots, b-2,
\\
\frac{w_0 w_1}{f_{-3}f_{-2}}\frac{T^{b-2,0}_1 T^{b-2,0}_3}{T^{b-2,0}_2} &= (1+ \omega^b\tilde x_0)(1+\tilde x_0)(1+\omega^{-b}\tilde x_0),
\\
\frac{T^{b-1,0}_0T^{b-1,0}_1}{T^{b-2,0}_1T^{b-2,0}_2} &= 1 + w_0.
\end{alignat}
\end{subequations}
The eigenvalues $T^{n,0}(u)$ and $w(u)$ are written as the product of their bulk and finite contributions: 
\be
T^{n,0}(u) = \big(\kappa^{n}(u)\big)^N T^{n}_{\rm f}(u), \qquad w(u) =  \big(\kappa_w(u)\big)^N w_{\rm f}(u).
\ee
These satisfy the functional equations
\begin{subequations}
\begin{alignat}{2}
\frac{\kappa^{n}(u) \kappa^{n}(u-2\bar\lambda)}{\kappa^{n+1}(u) \kappa^{n-1}(u-2\bar\lambda)} &= 1,\qquad n = 1, \dots, b-2,
\\[0.15cm]
\kappa_{\omega}(u)\kappa_{\omega}(u-\bar \lambda)\frac{\kappa^{b-2}(u-\bar\lambda)\kappa^{b-2}(u-3\bar\lambda)}{\kappa^{b-2}(u-2\bar\lambda)} &= \frac{\sin(3\lambda -u)\sin(2\lambda-u)}{\sin 3\lambda \sin 2\lambda},
\\[0.15cm]\frac{\kappa^{b-1}(u)\kappa^{b-1}(u-\bar\lambda)}{\kappa^{b-2}(u-\bar\lambda)\kappa^{b-2}(u-2\bar\lambda)} &= 1,
\end{alignat}
\end{subequations}
and
\begin{subequations}
\label{eq:func.rel.A22.dual}
\begin{alignat}{2}
\frac{T_{\rm f}^{n}(u) T_{\rm f}^{n}(u-2\bar\lambda)}{T_{\rm f}^{n+1}(u) T_{\rm f}^{n-1}(u-2\bar\lambda)} &= 1 + \tilde t^{n}(u),\qquad n = 1, \dots, b-2,
\\[0.15cm]
w_{\rm f}(u)w_{\rm f}(u-\bar \lambda)\frac{T_{\rm f}^{b-2}(u-\bar\lambda)T_{\rm f}^{b-2}(u-3\bar\lambda)}{T_{\rm f}^{b-2}(u-2\bar\lambda)} &= \big(1+\omega^b \tilde x(u)\big)\big(1+\tilde x(u)\big)\big(1+\omega^{-b} \tilde x(u)\big),
\\[0.15cm]
\frac{T_{\rm f}^{b-1}(u)T_{\rm f}^{b-1}(u-\bar\lambda)}{T_{\rm f}^{b-2}(u-\bar\lambda)T_{\rm f}^{b-2}(u-2\bar\lambda)} &= 1 + w(u).
\end{alignat}
\end{subequations}
The initial conditions are $\kappa^{0}(u) = \frac{\sin(3\lambda - u)\sin(2\lambda - u)}{\sin 3\lambda\sin 2\lambda}$ and $T^{0}_{\rm f}(u) = 1$. 
The known solution for the bulk $n=1$ contribution is \cite{WBN1992}:
\be
\log\kappa(u)=2\!\int_0^\infty\!\! \dd t\;\frac{\sinh (\pi -\lambda) t \sinh\frac{(2\lambda-\pi)t}{2}\cosh \frac{(2u-3\lambda+2\pi)t}{2}}{t \sinh\frac{\pi t}{2}\cosh\frac{(3\lambda-2\pi) t}{2}}+
\log\Big|\frac{\sin u\sin (3\lambda-u)}{\sin 2\lambda \sin 3\lambda}\Big|,\quad \mbox{Regime II}.
\ee
For the finite terms, we define
\be
T^{n}_{\rm f}(u) = \bmf^n\Big(\!-\!\tfrac {\ir \pi}{\textrm{\raisebox{-0.06cm}{$\bar \lambda$}}}\big(u-\tfrac{3\lambda}2-(n-1) \bar\lambda\big)\Big), \quad n = 1, \dots, b-1, 
\qquad 
w_{\rm f}(u) = \bmf^b\Big(\!-\!\tfrac {\ir \pi}{\textrm{\raisebox{-0.06cm}{$\bar \lambda$}}}\big(u-\tfrac{3\lambda}2-(b-\tfrac32) \bar\lambda\big)\Big),
\ee
and rewrite the relations \eqref{eq:func.rel.A22.dual} as
\begin{subequations}
\label{eq:bbb.A22.dual}
\begin{alignat}{2}
\frac{\bmf^n(z - \ir \pi) \bmf^n(z + \ir \pi)}{\bmf^{n-1}(z)\bmf^{n+1}(z)} &= \Amf^n(z), \qquad n = 1, \dots, b-2,
\\[0.15cm]
\frac{\bmf^b(z - \frac{\ir \pi}2) \bmf^b(z + \frac{\ir \pi}2)\bmf^{b-2}(z-\ir \pi)\bmf^{b-2}(z+\ir \pi)}{\bmf^{b-2}(z)} &= \Amf^{b-1}(z),
\\[0.1cm]\
\frac{\bmf^{b-1}(z-\frac{\ir \pi}2)\bmf^{b-1}(z+\frac{\ir \pi}2)}{\bmf^{b-2}(z-\frac{\ir \pi}2)\bmf^{b-2}(z+\frac{\ir \pi}2)} &= \Amf^{b}(z).
\end{alignat}
\end{subequations}

\subsubsection{Non-linear integral equations}

The functions $\amf^n(z)$ have order-$N$ zeros on the imaginary axis. We define new functions $\ell^n(z)$ where those zeros are removed:
\begin{subequations}
\label{eq:ell.a.A22.dual}
\begin{alignat}{2}
\ell^n(z) &= \frac{\amf^n(z)}{\Big[\eta\big(z-\ir \pi(b-n-\tfrac32)\big)\eta\big(z+\ir \pi(b-n-\tfrac32)\big)\Big]^N},  \qquad n = 1, \dots, b-2,
\\[0.15cm]
\ell^{b-1}(z) &= \amf^{b-1}(z), \qquad \ell^b(z) =  \frac{\amf^b(z)}{\eta(z)^N},
\end{alignat}
\end{subequations}
where
\be
\eta(z) = \tanh\tfrac z{2(2b-3)},
\qquad
\eta\big(z-\tfrac{\ir \pi}2(2b-3)\big)\eta\big(z+\tfrac{\ir \pi}2(2b-3)\big) = 1.
\ee
In terms of these functions, the $Y$-system reads
\begin{subequations}
\begin{alignat}{2}
\frac{\ell^{2}(z)}{\ell^{1}(z - \ir \pi) \ell^{1}(z + \ir \pi)} &= \frac{\Amf^{2}(z)}{\Amf^{1}(z)},
\\[0.15cm]
\frac{\ell^{n-1}(z)\ell^{n+1}(z)}{\ell^{n}(z - \ir \pi) \ell^{n}(z + \ir \pi)} &= \frac{\Amf^{n-1}(z)\Amf^{n+1}(z)}{\Amf^{n}(z)}, \qquad n = 2, \dots, b-3,
\\[0.15cm]
\frac{\ell^{b-3}(z)\ell^{b-1}(z)\ell^b(z-\frac{\ir \pi}2)\ell^b(z+\frac{\ir \pi}2)}{\ell^{b-2}(z - \ir \pi) \ell^{b-2}(z + \ir \pi)\ell^{b-1}(z-\ir \pi)\ell^{b-1}(z+\ir \pi)}
&= \frac{\Amf^{b-3}(z)\Amf^{b-1}(z)}{\Amf^{b-2}(z)}, 
\\[0.15cm]
\frac{\ell^{b-2}(z)}{\ell^{b}(z-\tfrac{\ir \pi}2) \ell^{b}(z +\tfrac{\ir \pi}2)} &= \frac{\Amf^{b-2}(z)}{\Amf^{b-1}(z)}, 
\\[0.15cm]
\ell^{b-1}(z-\tfrac{\ir \pi}2)\ell^{b-1}(z+\tfrac{\ir \pi}2) &= \Amf^b(z).\end{alignat}
\end{subequations}

The functions $\ell^n(z)$ are all analytic and non-zero inside their respective analyticity strips. As discussed in \cref{sec:braid.and.bulk.A22.dual}, for $\omega = \eE^{\ir \gamma}$ with $\gamma \in (0,\frac{2\pi}{b+1})$, these functions have asymptotic values for $z \to \pm \infty$ that are finite and nonzero. This allows us to define the Fourier transform of their logarithmic derivative:
\be
L^n(k) = \frac1{2\pi} \int_{-\infty}^\infty \dd z\, \eE^{-\ir k z}\big(\log \ell^n(z)\big)', \qquad 
A^n(k) = \frac1{2\pi} \int_{-\infty}^\infty \dd z\, \eE^{-\ir k z}\big(\log \Amf^n(z)\big)',
\ee
where $n = 1, \dots, b$. The non-linear integral equations for the eigenvalues are obtained by first taking the Fourier transform of the logarithmic derivative of the $Y$-system equations yielding
\begin{alignat}{2}
&\left(\begin{smallmatrix}
-2 \cos (\pi k) & 1 & 0 &  &  \\
1 & -2 \cos (\pi k) & 1 & \sddots &  \\
0 & 1 & -2 \cos (\pi k) & 1 & 0 & 0 \\
 & \sddots & 1 & -2 \cos (\pi k) & 1-2\cos(\pi k) & 2 \cos (\frac {\pi k}2)\\
 &  & 0 & 1 & 0 & -2 \cos (\frac {\pi k}2)\\
 & & & 0 & 2 \cos (\frac {\pi k}2) & 0
\end{smallmatrix}\right)
\left(\begin{smallmatrix}
L^1 \\[0.1cm] L^2 \\[0.1cm] \svdots \\[0.1cm] L^{b-2} \\[0.1cm] L^{b-1} \\[0.1cm] L^{b}
\end{smallmatrix}\right)
\nonumber\\[0.3cm]
\label{eq:LKA.A22.dual}
 &\hspace{7cm}= 
 \left(\begin{smallmatrix}
-1 &&& 1 &&& 0 &&&  &&&  \\[0.15cm]
1 &&& -1 &&& 1 &&& \sddotss &  \\[0.15cm]
0 &&& 1 &&& -1 &&& 1 &&& 0 \\[0.15cm]
 &&& \sddotss &&& 1 &&& -1 &&& 1 &&& 0 \\[0.15cm]
 &&&  &&& 0 &&& 1 &&&-1 &&& 0 \\[0.15cm]
 &&&  &&&  &&& 0 &&&0 &&& 1
\end{smallmatrix}\right)
\left(\begin{smallmatrix}
A^1 \\[0.1cm] A^2 \\[0.1cm] \svdots \\[0.1cm] A^{b-2} \\[0.1cm] A^{b-1}\\[0.1cm] A^{b}
\end{smallmatrix}\right).
\end{alignat}
We compute the inverse of the matrix in the left side, apply it to the right side and find
\be
\vec L = \hat K \cdot \vec A.
\ee
The matrix $\hat K$ is symmetric and its entries can be computed explicitly. Applying the inverse transform, we find
\be
\label{eq:NLIE.ell.A22.dual}
\big(\log \ell^{n}(z)\big)' = \sum_{m=1}^{b} K_{nm} * \big(\log \Amf^{m}\big)', \qquad n = 1, \dots, b,
\ee
where the kernel functions are given by
\be
\label{eq:Kij.A22.dual}
K_{nm}(z) = \frac1{2\pi}\int_{-\infty}^\infty \dd k\, \eE^{\ir k z}\hat K_{nm}.
\ee
Using \eqref{eq:ell.a.A22.dual}, we obtain the non-linear integral equations for $\amf^n(z)$:
\be
\label{eq:NLIEs.A22.dual}
\log \amf^n(z) - \phi_n = \fmf^n(z) + \sum_{m=1}^{b} K_{nm} * \log \Amf^{m}, \qquad n = 1, \dots, b,
\ee
where $\phi_1$, \dots, $\phi_{b}$ are the integration constants and the driving terms are
\be
\fmf^n(z) = \left\{
\begin{array}{cl}
N \log \Big[\eta\big(z-\ir \pi(b-n-\tfrac32)\big)\eta\big(z+\ir \pi(b-n-\tfrac32)\big)]\quad& n = 1,\dots, b-2,\\[0.2cm]
0 & n = b-1,\\[0.2cm]
N \log \eta(z) & n = b.
\end{array}\right.
\ee

\subsubsection{Scaling functions and scaling non-linear integral equations}
In \eqref{eq:NLIEs.A22.dual}, the dependence on $N$ appears only in the driving terms. For $z$ of order $\pm (2b-3)\log N$ with $N$ large, these terms behave as exponentials:
\be
\fsf^n(z) = \lim_{N\to \infty} \fmf^n\big(\!\pm\!(z + (2b-3)\log N)\big) = \left\{
\begin{array}{cl}
-4 \sin\big(\tfrac{\pi n}{2b-3}\big) \,\eE^{-z/(2b-3)} \quad& n = 1,\dots, b-2,\\[0.2cm]
0 & n = b-1,\\[0.2cm]
-2\, \eE^{-z/(2b-3)} & n = b.
\end{array}\right.
\ee
To compute the finite-size correction at order $\frac 1N$, we assume that the unknown functions appearing in \eqref{eq:NLIEs.A22.dual} are well-defined in this limit. For the groundstate, the patterns of zeros are all symmetric with respect to the imaginary $z$-axis, so the scaling functions behave identically in both limits:
\be
\mathsf a^n(z) = \lim_{N\to \infty} \amf^n\big(\!\pm\!(z +(2b-3) \log N)\big), \qquad \mathsf A^n(z) = \lim_{N\to \infty} \Amf^n\big(\!\pm\!(z +(2b-3) \log N)\big),
\ee
where $n = 1, \dots, b$.
These satisfy the following set of integral equations:
\be
\label{eq:scalingNLIEs.A22.dual}
\log \asf^n(z) - \phi_n = \fsf^n(z) + \sum_{m=1}^{b} K_{nm} * \log \Asf^{m}, \qquad n = 1, \dots, b.
\ee
We note that the kernel terms in these non-linear integral equations are symmetric.

\subsubsection{Braid and bulk behavior}\label{sec:braid.and.bulk.A22.dual}

The scaling functions have finite asymptotics for $z \to \pm\infty$. For $z \to \infty$, these are obtained directly from the braid limits of the transfer matrix eigenvalues:
\begin{subequations}
\label{eq:uplims.A22}
\begin{alignat}{2}
\asf^n_\infty &= \frac{(\omega^{1/2}-\omega^{-1/2})(\omega-\omega^{-1})}{(\omega^{n/2}-\omega^{-n/2})(\omega^{(n+3)/2}-\omega^{-(n+3)/2})}, \qquad n = 1, \dots, b-2,\label{eq:uplim.A22_1}
\\[0.15cm]
\asf^{b-1}_\infty &=\frac{\omega^{(b+1)/2}-\omega^{-(b+1)/2}}{\omega^{(b-1)/2}-\omega^{-(b-1)/2}}, 
\qquad \asf^{b}_\infty = \frac{(\omega^b-\omega^{-b})(\omega-\omega^{-1})}{(\omega^{(b-1)/2}-\omega^{-(b-1)/2})^2}.\label{eq:uplim.A22_2}
\end{alignat}
\end{subequations}
For $\gamma \in (0, \frac {2\pi}{b+1})$, these functions are positive and finite. These allow us to compute the constants $\phi_n$ by studying the $z\to \infty$ asymptotics of \eqref{eq:scalingNLIEs.A22.dual}. We find that the constants all vanish:
\be
\phi_n = 0.
\ee

The behavior of the functions $\asf^n(z)$ for $z \to - \infty$ is dictated by the driving terms in the non-linear integral equations \eqref{eq:scalingNLIEs.A22.dual}. These originated from the order-$N$ zeros of these functions that lie on the imaginary axis in the $z$-plane. As a result, we have
\be
\label{eq:bulk.an.A22.dual}
\asf^n_{-\infty} = 
\left\{\begin{array}{ll}
0 &\quad n = 1, 2, \dots, b-2, b,\\[0.1cm]
1 &\quad n = b-1.
\end{array}\right.
\ee 
The value for $\asf^{b-1}_{-\infty}$ was obtained as the positive solution to the $Y$-system relation $(\asf^{b-1}_{-\infty})^2 = 1+\asf^b_{-\infty}$.

\subsubsection{Finite-size correction and the dilogarithm technique}\label{sec:dilog.A22.dual}

We define the Fourier transform of the logarithmic derivative of the functions $\bmf^n(z)$: 
\be
B^n(k) = \frac1{2\pi} \int_{-\infty}^\infty \dd z\, \eE^{-\ir k z}\big(\log \bmf^n(z)\big)', \qquad n = 1, \dots, b.
\ee
Applying the Fourier transform to \eqref{eq:bbb.A22.dual}, we find
\be
\left(\begin{smallmatrix}
2 \cosh (\pi k) & -1 & 0 &  &  \\
-1 & 2 \cosh (\pi k) & -1 & \sddots &  \\
0 & -1 & 2 \cosh (\pi k) & -1 & 0 \\
 & \sddots & -1 & 2 \cosh (\pi k) & -1 & 0 \\
 &  & 0 & 2\cos(\pi k)-1 & 0 & 2 \cosh (\frac {\pi k}2)\\
  &  &  & -2 \cosh (\frac {\pi k}2) & 2 \cosh (\frac {\pi k}2) & 0
\end{smallmatrix}\right)
\left(\begin{smallmatrix}
B^1 \\[0.05cm] B^2 \\[0.05cm] \svdots \\[0.05cm] B^{b-2} \\[0.05cm] B^{b-1} \\[0.05cm] B^{b}
\end{smallmatrix}\right)
 = 
\left(\begin{smallmatrix}
A^1 \\[0.05cm] A^2 \\[0.05cm] \svdots \\[0.05cm] A^{b-2} \\[0.05cm] A^{b-1} \\[0.05cm] A^{b}
\end{smallmatrix}\right).
\ee
Denoting by $M$ the matrix on the left-hand side, we invert $M$ and apply it to both sides of the equation. The matrix elements of the first row of $M^{-1}$ are
\be
(M^{-1})_{1n} =
\left\{\begin{array}{cl}
\displaystyle\frac{\cosh\big(\frac{\pi k}2(2b-3-2n)\big)}{\cosh\big(\frac{\pi k}2(2b-3)\big)}\quad& n = 1, \dots, b-2,\\[0.5cm]
0\quad& n = b-1,\\[0.3cm]
\displaystyle\frac{1}{2\cosh\big(\frac{\pi k}2(2b-3)\big)}\quad & n = b.
\end{array}\right.
\ee
As a result, we find
\be
\log \bmf^1(z) - \phi_0 = \sum_{n=1}^{b-2} \tilde K_n * \log \Amf^n + \tfrac12 \tilde K_{(b-3)/2} * \log \Amf^b
\ee
where
\be
\tilde K_n (z) = \frac1{2\pi} \int_{-\infty}^\infty \dd k\, \eE^{\ir  k z} \frac{\cosh\big(\frac{\pi k}2(2b-3-2n)\big)}{\cosh\big(\frac{\pi k}2(2b-3)\big)} = \frac{1}{\pi(2b-3)} \frac{\sin\big(\frac{\pi n}{2(2b-3)}\big) \cosh\big(\frac{z}{2b-3}\big)}{\sinh\Big(\frac{z-\frac{\ir \pi n}2}{2b-3}\Big)\sinh\Big(\frac{z+\frac{\ir \pi n}2}{2b-3}\Big)}.
\ee
This leads to
\begin{alignat}{2}
\bmf^1(z) - \phi_0 &= \int_{-(2b\!-\!3)\log N}^\infty \dd y \bigg[\sum_{n=1}^{b-2} \tilde K_n\big(y+(2b\!-\!3)\log N - z\big) \log \Amf^n\big(y+(2b-3)\log N\big) 
\nonumber\\[0.15cm]
&\hspace{3cm}+ \tfrac12 \tilde K_{(b-3)/2}\big(y+(2b\!-\!3)\log N - z\big) \log \Amf^{b}\big(y+(2b\!-\!3)\log N\big) 
\nonumber\\[0.15cm]
&\hspace{3cm}+ \sum_{n=1}^{b-2} \tilde K_n\big(\!-y-(2b\!-\!3)\log N - z\big) \log \Amf^n\big(\!-y-(2b\!-\!3)\log N\big) 
\nonumber\\[0.15cm]
&\hspace{3cm} +\tfrac12 \tilde K_{(b-3)/2}\big(\!-y-(2b\!-\!3)\log N - z\big) \log \Amf^{b}\big(\!-y-(2b\!-\!3)\log N\big) \bigg]
\nonumber\\
& \simeq \frac {2\big(\eE^{z/(2b-3)}+\eE^{-z/(2b-3)}\big)}{\pi N (2b-3)} \int_{-\infty}^\infty \dd y\,\eE^{-y/(2b-3)} \bigg[\sum_{n=1}^{b-2} \sin\Big(\frac{\pi n}{2(2b-3)}\Big)\log \Asf^n + \tfrac12 \Asf^{b}\bigg],
\label{eq:log.b.int.A22.dual}
\end{alignat}
where we used
\be
\tilde K_n\big(z+(2b-3)\log N\big) \simeq \frac{2 \sin\big(\frac{\pi n}{2(2b-3)}\big)\eE^{-z/(2b-3)}}{\pi N(2b-3)}.
\ee

To apply the dilogarithm technique, we define the integral
\be
\mathcal J = \int_{-\infty}^\infty \dd y \bigg[\sum_{n=1}^{b}(\log \asf^n)' \log \Asf^n  -\log \asf^n (\log \Asf^n)'\bigg].
\ee
The integral $\mathcal J$ is evaluated in two ways. The first consists of replacing $\log \asf^n$ and its derivative by its expression \eqref{eq:scalingNLIEs.A22.dual}. Many terms cancel out because of the symmetries $K_{nm}(z) = K_{nm}(-z) = K_{mn}(z)$ of the kernel functions \eqref{eq:Kij.A22.dual}. The only surviving contributions come from the driving terms, and the result reads
\be
\mathcal J = \frac8{2b-3} \int_{-\infty}^\infty \dd y \, \eE^{-y/(2b-3)} \bigg[\sum_{n=1}^{b-2} \sin\Big(\frac {\pi n}{2(2b-3)}\Big) \log \Asf^n + \tfrac12 \Asf^{b}\bigg].
\ee
Up to an overall prefactor, this is precisely the integral we wish to compute in \eqref{eq:log.b.int.A22.dual}. The second way of computing the integral is to apply the derivatives explicitly, which yields
\be
\mathcal J = \int_{-\infty}^\infty \dd y \,  \bigg[\sum_{n=1}^{b-2}\frac{\dd \asf^n}{\dd y}\bigg(\frac{\log \Asf^n}{\asf^n} -\frac{\log \asf^n}{\Asf^n}\bigg) + \frac{\dd \asf^{b-1}}{\dd y}\bigg(\frac{\log \Asf^{b-1}}{\asf^{b-1}}  - \frac{\dd \Asf^{b-1}}{\dd \asf^{b-1}}\frac{\log \asf^{b-1}}{\Asf^{b-1}}\bigg) + \frac{\dd \asf^b}{\dd y} \bigg(\frac{\log \Asf^b}{\asf^b} - \frac{\log \asf^b}{\Asf^b}\bigg)\bigg].
\ee
Dividing the integral in three parts and changing the integration variables from $y$ to $\asf^n$, we find
\begin{alignat}{2}
\mathcal J &= \sum_{n=1}^{b-2}\int_{\asf^n_{-\infty}}^{\asf^n_\infty} \dd \asf^n \bigg(\frac{\log \Asf^n}{\asf^n}  -\frac{\log \asf^n}{\Asf^n}\bigg) + \int_{\asf^{b-1}_{-\infty}}^{\asf^{b-1}_\infty} \dd \asf^{b-1} \bigg(\frac{\log \Asf^{b-1}}{\asf^{b-1}} - \frac{\dd \Asf^{b-1}}{\dd \asf^{b-1}}\frac{\log \asf^{b-1}}{\Asf^{b-1}}\bigg) 
\nonumber\\[0.15cm]
&+ \int_{\asf^b_{-\infty}}^{\asf^b_\infty} \dd \asf^b  \bigg(\frac{\log \Asf^b}{\asf^b} - \frac{\log\asf^b}{\Asf^b}\bigg)
\label{eq:J1.A22.dual}
\end{alignat}
where
\begin{subequations}
\begin{alignat}{2}
\Asf^n &= 1 + \asf^n, \qquad n = 1, 2,\dots, b-2, b,
\\[0.15cm] 
\Asf^{b-1} &= (1 + \omega^b \asf^{b-1})(1 + \asf^{b-1})(1 + \omega^{-b} \asf^{b-1}).
\end{alignat}
\end{subequations}
The resulting expression for $\mathcal J$ is therefore a combination of regular integrals. Setting $\omega = \eE^{\ir \gamma}$, the integral evaluates to
\be
\label{eq:J2.A22.dual}
\mathcal J = \frac{\pi^2}3\bigg(\frac32- \frac{3\gamma^2 b}{\pi^2}\bigg) = \frac{\pi^2}3\bigg(\frac32-\frac{3\gamma^2}{2\pi(\pi-\lambda)}\bigg), 
\qquad \gamma \in \big(0,\tfrac{2\pi}{b+1}\big).
\ee
The proof of this result is given in \cref{app:A22.dual}. The final result is
\be
\label{eq:result.A22.dual}
\log \bmf(z) \simeq \frac{\pi \cosh \frac{z}{2b-3}}{6 N} \bigg(\frac32- \frac{3\gamma^2 b}{\pi^2}\bigg), \qquad 
\log T_{\rm f}(u) \simeq \frac{\pi \sin \frac{\pi(u-\pi)}{3\lambda-2\pi}}{6 N} \bigg(\frac32- \frac{3\gamma^2 b}{\pi^2}\bigg),
\ee
where the constant $\phi_0$ was found to equal zero using $T_{\rm f}(u=\pi)=1$.
This result is precisely \eqref{eq:fsc.general} with $c-24 \Delta$ and $\vartheta(u)$ given in \eqref{eq:fsc.A22}. 

We note that the factor $(u-\pi)$ in the argument of the sine function is different compared to the results for the other models, which have only a factor of $u$. This is due to the fact that for $\lambda = \frac{\pi(2b-1)}{2b}$, the value $u = \frac{3\lambda}2$ whose neighborhood we are studying is larger than $\pi$, and $u=\pi$ is the nearest value where $\log T_{\rm f}(u)$ vanishes identically. Alternatively, if one wishes to study the finite-size correction in the neighborhood of $u = \frac{3\lambda}2 - \pi$, then by the periodicity $T(u) = T(u+\pi)$, the result is identical to \eqref{eq:result.A22.dual} but with $u-\pi$ changed for $u$ in the argument of the sine function.

%
\section{Finite-size corrections for the $\boldsymbol{A_2^{(1)}}$ models}\label{sec:A21}
%

\subsection[Definition of the $A_2^{(1)}$ models]{Definition of the $\boldsymbol{A_2^{(1)}}$ models}

The loop and vertex models in the $A_2^{(1)}$ family are the $A_2^{(1)}$ (or fully packed) loop model and the 15-vertex model. The $A_2^{(1)}$ loop model is a face model on the square lattice, where each face takes on one of seven possible local configurations. The elementary face operator is defined by the linear combination
\be
\begin{pspicture}[shift=-.40](0,0)(1,1)
\facegrid{(0,0)}{(1,1)}
\psarc[linewidth=0.025]{-}(0,0){0.16}{0}{90}
\rput(.5,.5){$u$}
\end{pspicture}
\ = s_1(-u)
\bigg(\ 
\begin{pspicture}[shift=-.40](0,0)(1,1)
\facegrid{(0,0)}{(1,1)}
\rput[bl](0,0){\loopa}
\end{pspicture}
\ + \ 
\begin{pspicture}[shift=-.40](0,0)(1,1)
\facegrid{(0,0)}{(1,1)}
\rput[bl](0,0){\looph}
\end{pspicture}
\ \bigg)
+ \ 
\begin{pspicture}[shift=-.40](0,0)(1,1)
\facegrid{(0,0)}{(1,1)}
\rput[bl](0,0){\loopb}
\end{pspicture}
\ + \ 
\begin{pspicture}[shift=-.40](0,0)(1,1)
\facegrid{(0,0)}{(1,1)}
\rput[bl](0,0){\loopc}
\end{pspicture}
\ + s_0(u) \bigg(\ 
\begin{pspicture}[shift=-.40](0,0)(1,1)
\facegrid{(0,0)}{(1,1)}
\rput[bl](0,0){\loopf}
\end{pspicture} 
\ + \ 
\begin{pspicture}[shift=-.40](0,0)(1,1)
\facegrid{(0,0)}{(1,1)}
\rput[bl](0,0){\loopg}
\end{pspicture}
\ + \ 
\begin{pspicture}[shift=-.40](0,0)(1,1)
\facegrid{(0,0)}{(1,1)}
\rput[bl](0,0){\loopi}
\end{pspicture}
\ \bigg)
\label{eq:faceop.A21}
\ee
where $s_k(u)$ is defined in \eqref{eq:faceop.A11}.  The fugacities of the contractible and non-contractible loops are
\be
\beta = 2 \cos \lambda, \qquad \alpha = \omega + \omega^{-1} = 2 \cos \gamma,
\ee
where $\omega = \eE^{\ir \gamma}$ is a free parameter. The $\check R$-matrix of the 15-vertex model is
\be
\check R(u) = \left(
\begin{array}{ccccccccc}
 s_1(-u) & 0 & 0 & 0 & 0 & 0 & 0 & 0 & 0 \\
 0 & 1 & 0 & s_0(u) & 0 & 0 & 0 & 0 & 0 \\
 0 & 0 & \eE^{\ir u}  & 0 & 0 & 0 & s_0(u) & 0 & 0 \\
 0 & s_0(u) & 0 & 1 & 0 & 0 & 0 & 0 & 0 \\
 0 & 0 & 0 & 0 & s_1(-u)  & 0 & 0 & 0 & 0 \\
 0 & 0 & 0 & 0 & 0 & 1 & 0 & s_0(u) & 0 \\
 0 & 0 & s_0(u) & 0 & 0 & 0 & \eE^{-\ir u} & 0 & 0 \\
 0 & 0 & 0 & 0 & 0 & s_0(u) & 0 & 1 & 0 \\
 0 & 0 & 0 & 0 & 0 & 0 & 0 & 0 &  s_1(-u)  \\
\end{array}
\right).
\ee
The twist matrix is given in \eqref{eq:twists} with 
$\omega = \eE^{\ir \gamma}$. Both the vertex and loop
$A_2^{(1)}$ models are described by the dilute Temperley-Lieb algebra
\cite{GP93,P94}, with its parameter $\beta$ fixed to $\beta = 2 \cos
\lambda$. To be more precise, on the cylinder, the single-row transfer
matrices are elements of a subalgebra of the periodic dilute Temperley-Lieb algebra where the number of vacancies is conserved \cite{MDPR2019}. The $\check R$-matrix is then equal to $\check R(u) = \sum_{\nu = 1}^9 \tilde \rho_\nu g^{(\nu)}$, where the matrices $g^{(\nu)}$ are the representatives of the nine tiles in the vertex representation of the dilute Temperley-Lieb algebra, discussed in \cref{sec:A22def}. The functions $\tilde\rho_\nu$ are
\be
\tilde \rho_1 = \tilde\rho_8 = s_1(-u), \qquad\tilde \rho_2 = \tilde\rho_3 = 1, \qquad \tilde\rho_4 = \tilde\rho_5 = 0, \qquad \tilde\rho_6 = \tilde\rho_7 = \tilde\rho_9 = s_0(u).  
\ee

We consider the $A_2^{(1)}$ model on the domain (including Regimes I \& II)
\be
\begin{array}{rll}
0<u<\lambda,\quad&0<\lambda<\pi.
\end{array}
\ee
For the vertex model, the elementary space is spanned by the three states $\uparrow$, $0$ and $\downarrow$ and the groundstate appears in the sector with equal numbers of these three states. For the loop model, it appears in the standard module with zero defects and $\frac N3$ vacancies, $\stanW_{N,0,N/3}$. (We follow the convention used in \cite{MDPR2019} for these modules.)  
In these sectors, the spectrum of the transfer matrix $\Tb(u)$ of the $A_2^{(1)}$ models is invariant under the involution
\be
\lambda \leftrightarrow \pi-\lambda,\qquad u \leftrightarrow -u.\label{A21duality}
\ee
The roots of unity values of $\lambda$ are those for which $\frac{\lambda}{\pi}\in{\Bbb Q}$. We parameterise them in terms of two integers $p,p'$ as
\be
\lambda = \lambda_{p,p'} =  \frac{\pi(p'-p)}{p'},\qquad  \mbox{gcd}(p,p')=1.
\ee
Our calculation of the finite-size corrections below focuses on two series:
\be
\label{eq:series.A21}
\begin{array}{lll}
\textrm{Principal series:}\quad & (p,p') = (p'-1,p'),\quad & 0<u<\lambda,\\[0.15cm]
\textrm{Dual series:}\quad & (p,p') = (1,p'), & 0<u<\lambda.
\end{array}
\ee
From \eqref{A21duality}, the dual series can be alternatively specified by 
\be
\label{eq:dualseries.A21}
\begin{array}{ll}
\textrm{Dual series:}\quad & (p,p') = (p'-1,p'),\qquad \lambda-\pi<u<0.
\end{array}
\ee

In this section, we focus on the groundstate of the transfer matrix for $N \equiv 0 \textrm{ mod } 3$ for the principal series and $N \equiv 0 \textrm{ mod } 6$ in the dual series. We will focus on values of $u$ in the neighborhood of $u = \frac \lambda 2$ and will restrict to values of the twist parameter $\omega = \eE^{\ir \gamma}$ in the intervals 
\be
\gamma \in 
\left\{\begin{array}{cl}
\big(0,\frac{2\pi(p'-1)}{p'^2}\big) & \textrm{Principal series,}
\\[0.15cm]
\big(0,\frac{2\pi}{p'+1}\big) & \textrm{Dual series.}
\end{array}\right.
\ee 
We will compute the $\frac1N$ finite-size correction term for the groundstate eigenvalue $T(u)$ and will confirm the conformal prediction \eqref{eq:fsc.general} with 
\be
\label{eq:fsc.A21}
c-24 \Delta = \left\{\begin{array}{ll}
\displaystyle 2-\frac{6\gamma^2 p'}{\pi^2(p'-1)},\quad\mbox{Principal series,}\\[8pt]
\displaystyle 2-\frac{6\gamma^2 p'}{\pi^2},\qquad\quad \mbox{Dual series,}
\end{array}\right.
\qquad
\vartheta(u) = \frac {2\pi u}{3\lambda}.
\ee

\subsection{Functional relations}

The fused transfer matrices $\Tb^{m,n}(u)$ for the $A_2^{(1)}$ models are defined recursively in \cite{MDPR2019} from the fusion hierarchy relations, as functions of the two elementary transfer matrices $\Tb(u) = \Tb^{1,0}(u)$ and $\bar\Tb(u) = \Tb^{0,1}(u)$. There, it was found that the $T$-system equations involve only the transfer matrices $\Tb^{m,0}(u)$ and $\Tb^{0,n}(u)$ where one of the two indices is zero:
\be
\label{eq:Trelations.A21}
\Tb^{m,0}_0 \Tb^{m,0}_1 = f_{m}\Tb^{0,m}_0 + \Tb^{m+1,0}_0 \Tb^{m-1,0}_1, \qquad
\Tb^{0,n}_0 \Tb^{0,n}_1 = \sigma^n f_{-1}\Tb^{n,0}_1 + \Tb^{0,n+1}_0 \Tb^{0,n-1}_1.
\ee
Here, $m,n \ge 1$ and we use the compact notations
\begin{subequations}
\begin{alignat}{3}
&\Tb^{m,n}_k = \Tb^{m,n} (u+k \lambda), 
\qquad 
&&\Tb^{0,0}_k = f_{k-1}\Ib, 
\qquad 
\sigma = (-1)^{N}, 
\qquad
\\[0.15cm]
&f_k = \Big(\frac{\sin(u+k\lambda)}{\sin \lambda}\Big)^N, 
\qquad 
&&
\Tb^{-1,0}_k = \Tb^{0,-1}_k = \boldsymbol 0.
\end{alignat}
\end{subequations}

For $\lambda = \lambda_{p,p'}$, the $Y$-system is finite and is defined in terms of a set of $2p'$ functions:\footnote{For convenience, we choose here slightly different sign conventions for certain functions compared to \cite{MDPR2019}.}
\begin{subequations}
\begin{alignat}{4}
\tb^m_0 &= \frac{\Tb^{m+1,0}_0 \Tb^{m-1,0}_1}{f_{m}\Tb^{0,m}_0}, \qquad &\tbb^n_0 &= \frac{\Tb^{0,n+1}_0 \Tb^{0,n-1}_1}{\sigma^n f_{-1}\Tb^{n,0}_1}, \quad &n &= 1, \dots, p'-2, \\
\xb_0 &=  \frac{\sigma^{p}\Tb^{p'-2,0}_1}{\Tb^{0,p'-1}_0}, \qquad &\xbb_0 &= \frac{\sigma^{p+1}\Tb^{0,p'-2}_0}{\Tb^{p'-1,0}_0},
\qquad\quad &\yb_0 &= \xb_{0}\xbb_0, \qquad \zb_0 = \xb_0 \xbb_1.
\end{alignat}
\end{subequations}
The functional equations are 
\begingroup
\allowdisplaybreaks
\begin{subequations}
\label{eq:Ysys.A21.general}
\begin{alignat}{2}
\frac{\tb^n_0\tb^n_1}{\tbb^n_0} &= \frac{(\Ib+\tb^{n-1}_1)(\Ib+\tb^{n+1}_0)}{\Ib+\tbb^{n}_0}, \qquad\quad n = 1, \dots, p'-3,
\\[0.15cm]
\frac{\tb^{p'-2}_0 \tb^{p'-2}_1}{\tbb^{p'-2}_0} &= \frac{(\Ib+\tb^{p'-3}_1)(\Ib+\eE^{\ir \Lambdab_1}\xb_0)(\Ib+\eE^{\ir \Lambdab_2}\xb_0)(\Ib+\eE^{\ir \Lambdab_3}\xb_0)}{(\Ib+\tbb^{p'-2}_0)(\Ib-\yb_0)(\Ib-\zb_0)},
\\[0.15cm]
\frac{\xb_0 \xb_1}{\xbb_1} &= \frac{(\Ib + \tb^{p'-2}_1)(\Ib-\yb_1)(\Ib-\zb_0)}{(\Ib+\eE^{-\ir \Lambdab_1}\xbb_1)(\Ib+\eE^{-\ir \Lambdab_2}\xbb_1)(\Ib+\eE^{-\ir \Lambdab_3}\xbb_1)},
\\[0.15cm]
\frac{\yb_0 \yb_1}{\zb_0} &=  \frac{(\Ib + \tb^{p'-2}_1)(\Ib + \tbb^{p'-2}_0)(\Ib-\yb_0)(\Ib-\yb_1)(\Ib-\zb_0)^2}{(\Ib+\eE^{\ir \Lambdab_1}\xb_0)(\Ib+\eE^{\ir \Lambdab_2}\xb_0)(\Ib+\eE^{\ir \Lambdab_3}\xb_0)(\Ib+\eE^{-\ir \Lambdab_1}\xbb_1)(\Ib+\eE^{-\ir \Lambdab_2}\xbb_1)(\Ib+\eE^{-\ir \Lambdab_3}\xbb_1)},
\end{alignat}
and 
\begin{alignat}{2}
\frac{\tbb^n_0\tbb^n_1}{\tb^n_1} &= \frac{(\Ib+\tbb^{n-1}_1)(\Ib+\tbb^{n+1}_0)}{\Ib+\tb^{n}_1}, \qquad\quad n = 1, \dots, p'-3,
\\[0.15cm]
\frac{\tbb^{p'-2}_0 \tbb^{p'-2}_1}{\tb^{p'-2}_1} &= \frac{(\Ib+\tbb^{p'-3}_1)(\Ib+\eE^{-\ir \Lambdab_1}\xbb_1)(\Ib+\eE^{-\ir \Lambdab_2}\xbb_1)(\Ib+\eE^{-\ir \Lambdab_3}\xbb_1)}{(\Ib+\tb^{p'-2}_1)(\Ib-\yb_1)(\Ib-\zb_0)},
\\[0.15cm]
\frac{\xbb_0 \xbb_1}{\xb_0} &= \frac{(\Ib + \tbb^{p'-2}_0)(\Ib-\yb_0)(\Ib-\zb_0)}{(\Ib+\eE^{\ir \Lambdab_1}\xb_0)(\Ib+\eE^{\ir \Lambdab_2}\xb_0)(\Ib+\eE^{\ir \Lambdab_3}\xb_0)},
\\[0.15cm]
\frac{\zb_0 \zb_1}{\yb_1} &=  \frac{(\Ib + \tb^{p'-2}_1)(\Ib + \tbb^{p'-2}_1)(\Ib-\yb_1)^2(\Ib-\zb_0)(\Ib-\zb_1)}{(\Ib+\eE^{\ir \Lambdab_1}\xb_1)(\Ib+\eE^{\ir \Lambdab_2}\xb_1)(\Ib+\eE^{\ir \Lambdab_3}\xb_1)(\Ib+\eE^{-\ir \Lambdab_1}\xbb_1)(\Ib+\eE^{-\ir \Lambdab_2}\xbb_1)(\Ib+\eE^{-\ir \Lambdab_3}\xbb_1)}.
\end{alignat}
\end{subequations}
Finally, another relation that will play an important role in the derivations below is the fusion hierarchy relation for $\Tb^{p'-1,p'-1}(u)$ obtained in \cite{MDPR2019}, which can be conveniently written as
\be
\label{eq:1-z(u).A21}
\Ib-\zb_0 = \frac{f_{-1} \Tb^{p'-1,p'-1}_1}{\Tb^{p'-1,0}_1\Tb^{0,p'-1}_0}.
\ee

In the loop model, the matrices $\eE^{\ir \Lambdab_1}$, $\eE^{\ir \Lambdab_2}$
and $\eE^{\ir \Lambdab_3}$ are diagonal on the standard module $\stanW_{N,0,N/3}$ with the unique eigenvalues $\omega^{p'}$, $1$ and $\omega^{-p'}$, respectively. Likewise, for the vertex model in the zero magnetisation sector, the matrices $\eE^{\ir \Lambdab_1}$, $\eE^{\ir \Lambdab_2}$ and $\eE^{\ir \Lambdab_3}$  are diagonal with the unique eigenvalues $\omega^{p'}$, $1$ and $\omega^{-p'}$, respectively.

\subsection{The principal series}\label{sec:A21.principal}

In this subsection, we fix $p = p'-1$ with $p' \in \mathbb N_{\ge 4}$, so that $\lambda = \frac{\pi}{p'}$. We compute the $\frac1N$ term in \eqref{eq:fsc.general} explicitly for $N \equiv 0 \textrm{ mod } 3$, $\gamma \in (0,\frac{2\pi(p'-1)}{p'^2})$ and $u$ in the neighborhood of $\frac\lambda2$.

\subsubsection[Analyticity properties and symmetric $Y$-system]{Analyticity properties and symmetric $\boldsymbol Y$-system}\label{sec:analyticity.A21}

\begin{figure}
\centering
\begin{tabular}{ccccc}
\begin{tabular}{c}
$T^{1,0}(u)$ \\[0.1cm]
\includegraphics[width=.25\textwidth]{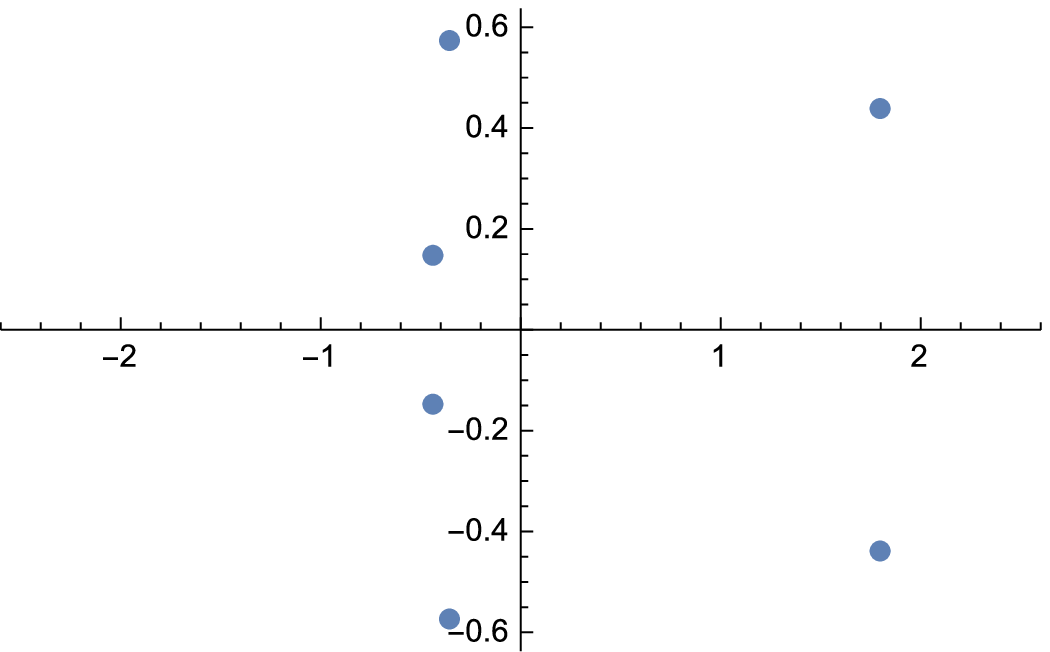} \\[0.3cm]
$T^{3,0}(u)$ \\[0.1cm]
\includegraphics[width=.25\textwidth]{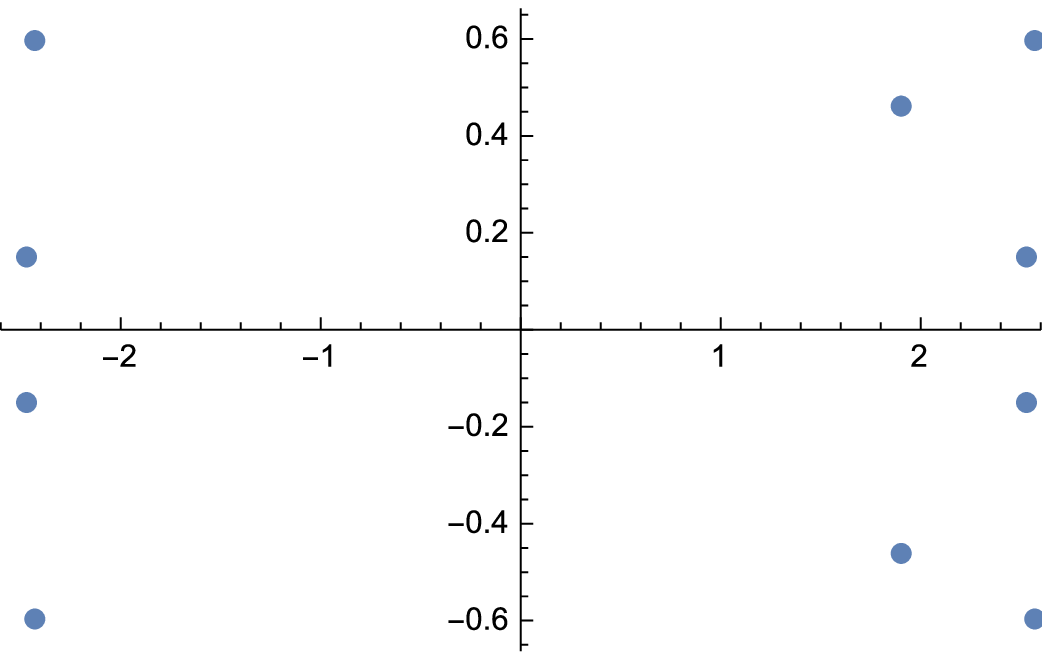}
\end{tabular}
&\quad&
\begin{tabular}{c}
$T^{2,0}(u)$ \\[0.1cm]
\includegraphics[width=.25\textwidth]{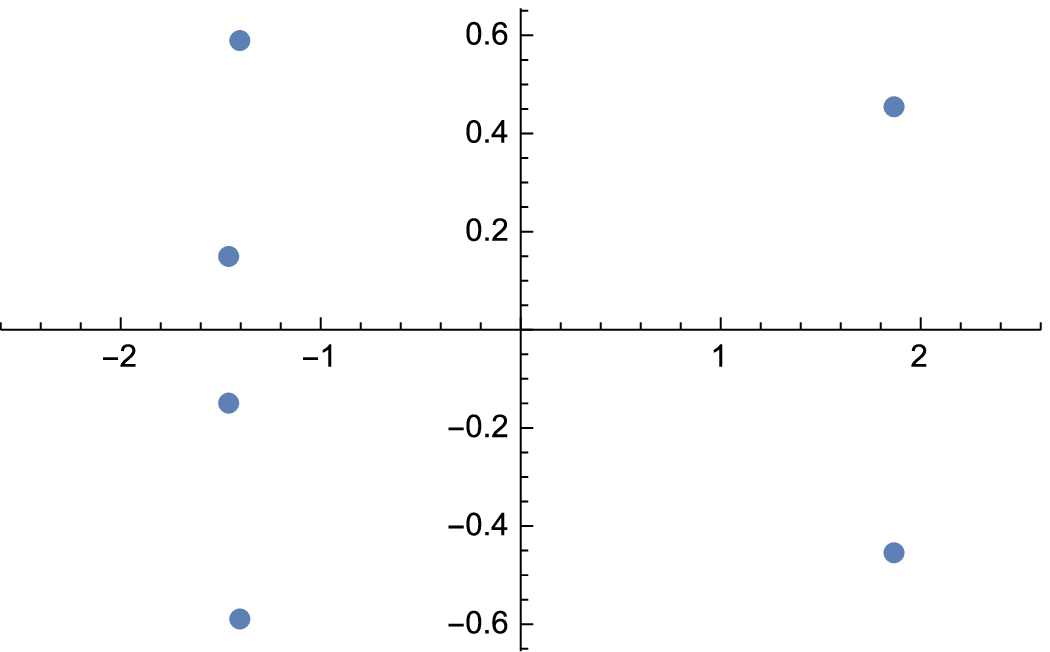}
\\[0.3cm]
$T^{4,0}(u)$ \\[0.1cm]
\includegraphics[width=.25\textwidth]{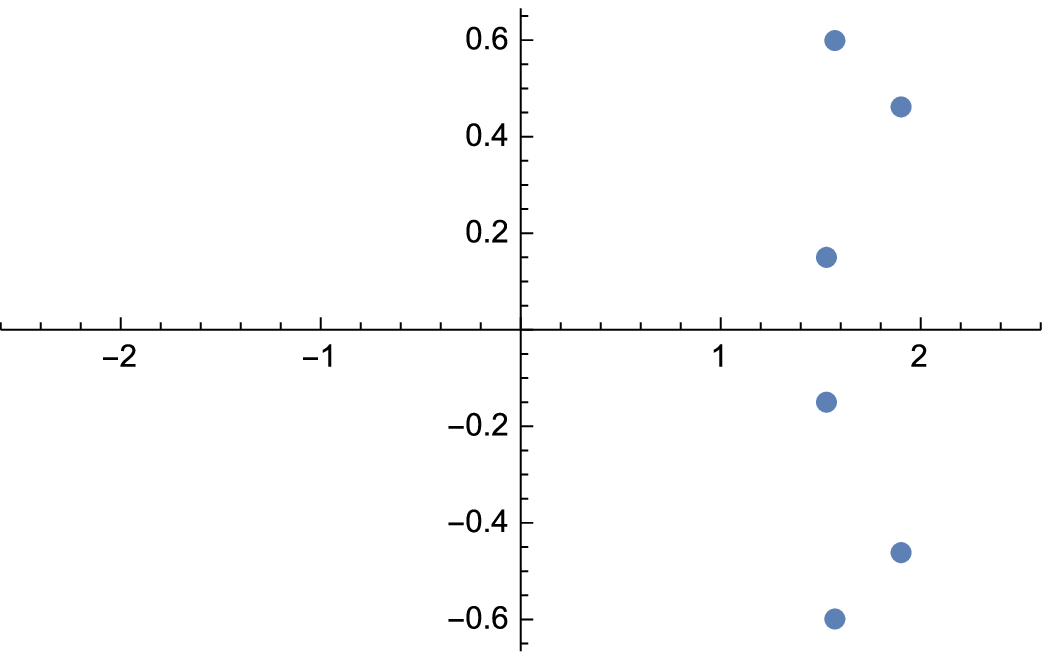}
\end{tabular}
&\quad&
\begin{tabular}{c}
$T^{4,4}(u)$ \\[0.1cm]
\includegraphics[width=.25\textwidth]{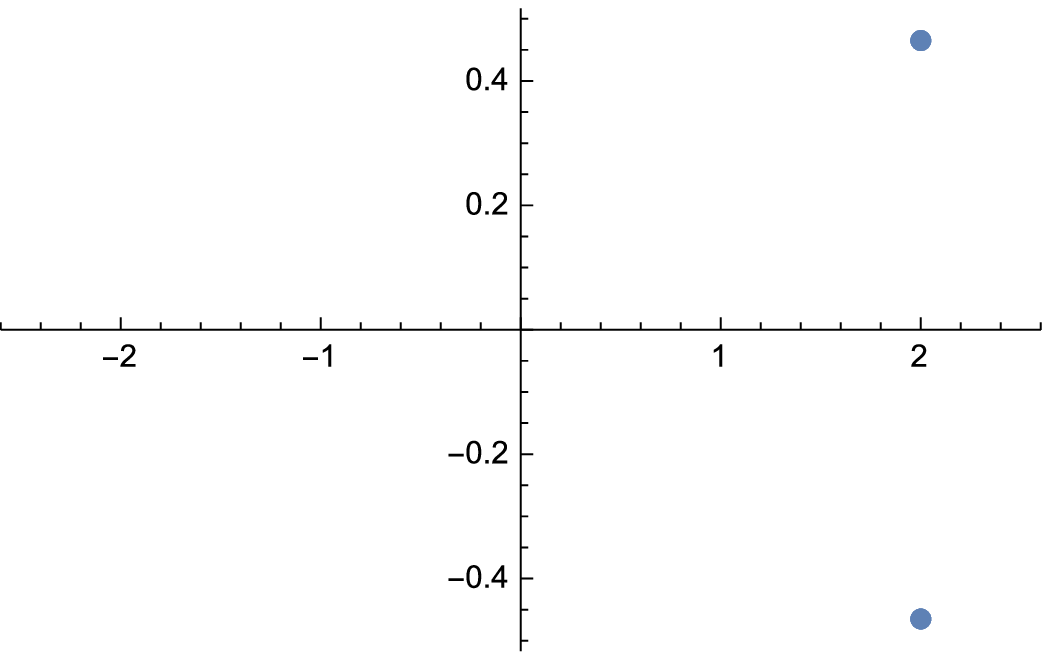} 
\\[0.3cm]
$1-y(u)$ \\[0.1cm]
\includegraphics[width=.25\textwidth]{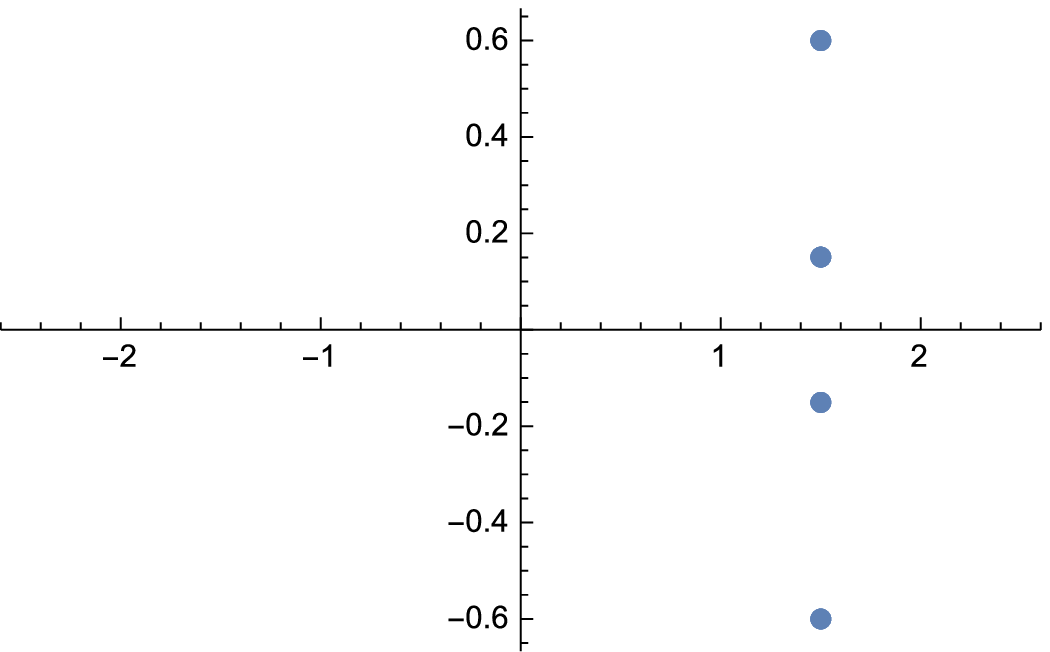}
\end{tabular}
\end{tabular}
\caption{The patterns of zeros for the groundstate of $\Tb(u)$ for $N=6$, $(p,p') = (4,5)$ and $\omega = 1$, in the complex $u$-plane. The horizontal axis is divided in units of $\lambda = \frac\pi5$. Each zero of $T^{4,4}(u)$ and $1-y(u)$ is triply degenerate. This degeneracy is lifted for $\omega\neq1$, with the zeros remaining on the same vertical lines for $\omega$ on the unit circle. The diagram for $1-y(u)$ only shows its zeros, and not its poles which can be deduced from the zeros of the other functions.}
\label{fig:patterns.A21.principal}
\end{figure}

Our computer implementation of the transfer matrices reveals that, in the complex $u$-plane, the zeros of $T^{1,0}(u)$ for the groundstate lie close to the vertical lines with Re$(u) = -\frac\lambda2, 2\lambda$. We shall therefore center its analyticity strip at $\frac{3\lambda}4$. Likewise, the zeros of $T^{2,0}(u)$ lie on the vertical lines Re$(u) = -\frac{3 \lambda}2, 2\lambda$. These patterns are repeated in each vertical strip of width $\pi$. In general, the zeros of $T^n(u)$ lie on the vertical lines Re$(u) = -(n-\frac12) \lambda, 2\lambda$, for $n = 1, \dots, p'-1$. We set the center of its analyticity strip at $(\frac52-n)\frac{\lambda}2$. The groundstate eigenvalues satisfy the crossing relations
\be
\label{eq:crossing.A21}
T^{n,0}(u) =  T^{0,n}\big((2-n)\lambda-u\big).
\ee
The positions of the zeros for $T^{0,n}(u)$ can therefore be deduced from those of $T^{n,0}(u)$ using \eqref{eq:crossing.A21}. They lie on the vertical lines Re$(u)= -n\lambda, \frac{3\lambda}2$. Furthermore, the zeros of $T^{p'-1,p'-1}(u)$ lie on the vertical line Re$(u) = 2\lambda$. To deduce the positions of the zeros and poles of all the functions of the $Y$-system, we also need to know the position of the zeros of the function $1-y(u)$. Indeed, there appears not to be a relation similar to \eqref{eq:1-z(u).A21} for this function which would otherwise give us this information. Our computer implementation reveals that its zeros lie on the vertical line Re$(u) = \frac{3\lambda}2$. To illustrate, the patterns of zeros for $N=6$, $(p,p') = (4,5)$ and $\omega=1$ are given in \cref{fig:patterns.A21.principal}.

With this information, we deduce the position of the zeros and poles of the $Y$-system functions. We find that the functions $t^1(u)$ and $\bar t^1(u)$ each have a zero of order $N$ at $u=0$, whereas they have a pole of order $N$ at $u = -\lambda$ and $u=\lambda$ respectively. Its zeros will play an important role in the following. The functions $t^n(u)$ and $\bar t^n(u)$ with $n = 2, \dots, p'-2$ have no zeros of order $N$, but have poles of order $N$ at $u = -m \lambda$ and $u=\lambda$ respectively. These poles will not play any role for this computation. The functions $x(u)$, $\bar x(u)$, $y(u)$ and $z(u)$ have no order-$N$ zeros and no order-$N$ poles. The same investigation is repeated for the patterns of zeros of the functions $1+t^n(u)$, $1+\bar t^n(u)$, $\big(1+\omega^{p'} x(u)\big)\big(1+x(u)\big)\big(1+\omega^{-p'} x(u)\big)$, $\big(1+\omega^{p'}\bar x(u)\big)\big(1+\bar x(u)\big)\big(1+\omega^{-p'} \bar x(u)\big)$, $1-y(u)$ and $1-z(u)$. Our derivation below uses certain assumptions for the analyticity strips of these functions. These are given in \cref{tab:analyticity.A21}. Crucially, except for the order-$N$ zeros of $t^1(u)$ and $\bar t^1(u)$, these analyticity strips are free of zeros and poles.
\begin{table}
\begin{center}
\begin{tabular}{c|c|c}
& width is larger than & centered at 
\\[0.1cm]\hline
&&\\[-0.3cm]
$t^n(u)$ & $ \lambda$ & $(\frac32-n)\frac\lambda2$
\\[0.1cm]
$1+t^n(u)$ & $0$ & $(\frac32-n)\frac\lambda2$
\\[0.1cm]
$x(u)$ & $\lambda$ & $(\frac52-p')\frac\lambda2$
\\[0.1cm]
$\big(1+\omega^{p'} x(u)\big)\big(1+x(u)\big)\big(1+\omega^{-p'} x(u)\big)$ & $0$ & $(\frac52-p')\frac\lambda2$
\\[0.1cm]
$y(u)$ & $\lambda$ & $(3-p')\frac\lambda2$
\\[0.1cm]
$1-y(u)$ & $\lambda$ & $(3-p')\frac\lambda2$
\\[0.1cm]
$\bar t^n(u)$ & $ \lambda$ & $(\frac12-n)\frac\lambda2$
\\[0.1cm]
$1+\bar t^n(u)$ & $0$ & $(\frac12-n)\frac\lambda2$
\\[0.1cm]
$\bar x(u)$ & $\lambda$ & $(\frac72-p')\frac\lambda2$
\\[0.1cm]
$\big(1+\omega^{p'} \bar x(u)\big)\big(1+ \bar x(u)\big)\big(1+\omega^{-p'} \bar x(u)\big)$ & $0$ & $(\frac72-p')\frac\lambda2$
\\[0.1cm]
$z(u)$ & $\lambda$ & $(2-p')\frac\lambda2$
\\[0.1cm]
$1-z(u)$ & $\lambda$ & $(2-p')\frac\lambda2$
\end{tabular}
\caption{The analyticity strips for the various functions.}
\label{tab:analyticity.A21}
\end{center}
\end{table}

We make a change of variables for the $Y$-system functions in such a way that the central lines of the analyticity strips coincide with the real axis:
\begin{subequations}
\begin{alignat}{2}
&t^n(u) = \amf^n\Big(\!-\!\tfrac {\ir \pi}{\lambda}\big(u+(n-\tfrac32) \tfrac\lambda2\big)\Big),
\qquad 
\hspace{0.25cm}\Amf^n(z) = 1+\amf^n(z),
\qquad n = 1, \dots, p'-2,
\\[0.15cm]
&x(u) = \amf^{p'-1}\Big(\!-\!\tfrac {\ir \pi}{\lambda}\big(u+(p'-\tfrac52)\tfrac{\lambda}2\big)\Big), 
\\[0.15cm]
&\Amf^{p'-1}(z) = \big(1 + \omega^{p'} \amf^{p'-1}(z)\big)\big(1 + \amf^{p'-1}(z)\big)\big(1 + \omega^{-p'} \amf^{p'-1}(z)\big),
\\[0.15cm]\label{eq:y(u).A21}
&y(u) = \amf^{p'}\Big(\!-\!\tfrac {\ir \pi}{\lambda}\big(u+(p'-3)\tfrac\lambda2\big)\Big), 
\hspace{1.08cm} \Amf^{p'}(z) = 1-\amf^{p'}(z),
\end{alignat}
and
\begin{alignat}{2}
&\bar t^n(u) = \bar\amf^n\Big(\!-\!\tfrac {\ir \pi}{\lambda}\big(u+(n-\tfrac12) \tfrac\lambda2\big)\Big),
\qquad 
\hspace{0.25cm}\bar\Amf^n(z) = 1+\bar\amf^n(z),
\qquad n = 1, \dots, p'-2,
\\[0.15cm]
&\bar x(u) =\bar\amf^{p'-1}\Big(\!-\!\tfrac {\ir \pi}{\lambda}\big(u+(p'-\tfrac72)\tfrac{\lambda}2\big)\Big), 
\\[0.15cm]
&\bar\Amf^{p'-1}(z) = \big(1 + \omega^{p'} \bar\amf^{p'-1}(z)\big)\big(1 + \bar\amf^{p'-1}(z)\big)\big(1 + \omega^{-p'} \bar\amf^{p'-1}(z)\big),
\\[0.15cm]
&z(u) = \bar\amf^{p'}\Big(\!-\!\tfrac {\ir \pi}{\lambda}\big(u+(p'-2)\tfrac\lambda2\big)\Big), 
\hspace{1.08cm} \bar\Amf^{p'}(z) = 1-\bar\amf^{p'}(z).
\end{alignat}
\end{subequations}
In terms of these functions, the $Y$-system takes a more symmetric form:
\begin{subequations}
\label{eq:sym.Y.A21}
\begin{alignat}{2}
\frac{\amf^{n}(z - \frac{\ir \pi}2) \amf^{n}(z + \frac{\ir \pi}2)}{\bar\amf^{n}(z)} &= \frac{\Amf^{n-1}(z)\Amf^{n+1}(z)}{\bar\Amf^{n}(z)}, \qquad n = 1, \dots, p'-3,
\label{eq:sym.Ya.A21}\\[0.15cm]
\frac{\amf^{p'-2}(z - \frac{\ir \pi}2) \amf^{p'-2}(z + \frac{\ir \pi}2)}{\bar\amf^{p'-2}(z)} &= \frac{\Amf^{p'-3}(z)\Amf^{p'-1}(z)}{\bar\Amf^{p'-2}(z)\bar\Amf^{p'\!}(z - \frac{\ir \pi}4)\Amf^{p'\!}(z + \frac{\ir \pi}4)}, 
\label{eq:sym.Yb.A21}\\[0.15cm]
\frac{\amf^{p'-1}(z - \frac{\ir \pi}2) \amf^{p'-1}(z + \frac{\ir \pi}2)}{\bar\amf^{p'-1}(z)} &= \frac{\Amf^{p'-2}(z)\Amf^{p'\!}(z - \frac{\ir \pi}4)\bar\Amf^{p'\!}(z + \frac{\ir \pi}4)}{\bar\Amf^{p'-1}(z)}, 
\label{eq:sym.Yc.A21}\\[0.15cm]
\frac{\amf^{p'\!}(z - \frac{\ir \pi}2) \amf^{p'\!}(z + \frac{\ir \pi}2)}{\bar\amf^{p'\!}(z)} &= \frac{\Amf^{p'-2}(z-\frac{\ir \pi}4)\bar\Amf^{p'-2}(z+\frac{\ir \pi}4)\Amf^{p'\!}(z - \frac{\ir \pi}2)\Amf^{p'\!}(z + \frac{\ir \pi}2)\big(\bar\Amf^{p'\!}(z)\big)^2}{\bar\Amf^{p'-1}(z-\frac{\ir \pi}4)\Amf^{p'-1}(z+\frac{\ir \pi}4)}.\label{eq:sym.Yd.A21}
\end{alignat}
and
\begin{alignat}{2}
\frac{\bar\amf^{n}(z - \frac{\ir \pi}2) \bar\amf^{n}(z + \frac{\ir \pi}2)}{\amf^{n}(z)} &= \frac{\bar\Amf^{n-1}(z)\bar\Amf^{n+1}(z)}{\Amf^{n}(z)},\qquad n = 1, \dots, p'-3,
\label{eq:sym.Ye.A21}\\[0.15cm]
\frac{\bar\amf^{p'-2}(z - \frac{\ir \pi}2) \bar\amf^{p'-2}(z + \frac{\ir \pi}2)}{\amf^{p'-2}(z)} &= \frac{\bar\Amf^{p'-3}(z)\bar\Amf^{p'-1}(z)}{\Amf^{p'-2}(z)\Amf^{p'\!}(z - \frac{\ir \pi}4)\bar\Amf^{p'\!}(z + \frac{\ir \pi}4)}, 
\label{eq:sym.Yf.A21}\\[0.15cm]
\frac{\bar\amf^{p'-1}(z - \frac{\ir \pi}2) \bar\amf^{p'-1}(z + \frac{\ir \pi}2)}{\amf^{p'-1}(z)} &= \frac{\bar\Amf^{p'-2}(z)\bar\Amf^{p'\!}(z - \frac{\ir \pi}4)\Amf^{p'\!}(z + \frac{\ir \pi}4)}{\Amf^{p'-1}(z)}, 
\label{eq:sym.Yg.A21}\\[0.15cm]
\frac{\bar\amf^{p'\!}(z - \frac{\ir \pi}2) \bar\amf^{p'\!}(z + \frac{\ir \pi}2)}{\amf^{p'\!}(z)} &= \frac{\bar\Amf^{p'-2}(z-\frac{\ir \pi}4)\Amf^{p'-2}(z+\frac{\ir \pi}4)\bar\Amf^{p'\!}(z - \frac{\ir \pi}2)\bar\Amf^{p'\!}(z + \frac{\ir \pi}2)\big(\Amf^{p'\!}(z)\big)^2}{\Amf^{p'-1}(z-\frac{\ir \pi}4)\bar\Amf^{p'-1}(z+\frac{\ir \pi}4)}.\label{eq:sym.Yh.A21}
\end{alignat}
\end{subequations}

In terms of the variable $z$, the analyticity strips are horizontal and centered on the real line. For $\amf^n(z)$ and $\bar\amf^n(z)$, the width of the strips is $\ir \pi$. Our computer implementation also reveals that, in the $z$-plane, the zeros of all the eigenvalues are symmetrically distributed between the right and left half-planes, but not between the upper and lower half-planes. This implies that 
\be
\label{eq:z-z.ids.A21}
\amf^n(z + \ir \xi) = \amf^n(-z + \ir \xi)^*,\qquad \bar\amf^n(z + \ir \xi) = \bar\amf^n(-z + \ir \xi)^*,\qquad z,\xi \in \mathbb R.
\ee
This is true for the groundstate eigenvalues, but not for arbitrary eigenstates of the transfer matrix.
\endgroup

\subsubsection{Bulk and finite contributions}

The eigenvalues of the two elementary transfer matrices are related to the $Y$-system functions by
\be
\frac{T^{1,0}_0T^{1,0}_1}{T^{0,1}_0} = f_1 (1 + t^{1}_0), 
\qquad
\frac{T^{0,1}_0 T^{0,1}_1}{T^{1,0}_1} = \sigma f_{-1} (1+\bar t^1_0).
\ee
The eigenvalues $T(u) = T^{1,0}(u)$ and $\bar T(u) = T^{1,0}(u)$ are written as the product of their bulk and finite contributions: 
\be
T(u) = \kappa(u)^N T_{\rm f}(u), \qquad \bar T(u) = \bar \kappa(u)^N \bar T_{\rm f}(u).
\ee
These satisfy the functional equations
\begin{subequations}
\begin{alignat}{2}
\label{eq:func.rel.A21a}
\frac{\kappa(u)\kappa(u+\lambda)}{\bar\kappa(u)} &= \frac{\sin(\lambda +u)}{\sin \lambda},
\qquad
\frac{T_{\rm f}(u)T_{\rm f}(u+\lambda)}{\bar T_{\rm f}(u)} = \big(1+t^1(u)\big),
\\[0.2cm]
\label{eq:func.rel.A21b}
\frac{\bar\kappa(u)\bar\kappa(u+\lambda)}{\kappa(u+\lambda)} &= \frac{\sin(\lambda - u)}{\sin \lambda},
\qquad
\frac{\bar T_{\rm f}(u)\bar T_{\rm f}(u+\lambda)}{T_{\rm f}(u+\lambda)} = \big(1+\bar t^1(u)\big).
\end{alignat}
\end{subequations}
The solution for the bulk contribution can be computed using Baxter's technique \cite{B82}:
\be
 \log\kappa(u)=\int_{-\infty}^\infty \frac{\dd t}t \frac{\sinh\lambda t\, \sinh u t}{\sinh\pi t\,\sinh3\lambda t}\big(\sinh(\pi \!-\! u)t+\sinh(4 \lambda \!-\! \pi \!-\! u)t+\sinh(\pi \!-\! 2\lambda \!-\! u)t\big).
\ee
This result in fact holds for $0 < \lambda < \pi$. The expression for $\log \bar\kappa(u)$ is obtained from the crossing relation $\bar\kappa(u) = \kappa(\lambda - u)$. For the finite term, we define
\be
T_{\rm f}(u) = \bmf\big(\!-\!\tfrac {\ir \pi}{\lambda}(u-\tfrac {3\lambda} 4)\big), \qquad \bar T_{\rm f}(u) = \bar\bmf\big(\!-\!\tfrac {\ir \pi}{\lambda}(u-\tfrac {\lambda} 4)\big),
\ee
and rewrite the rightmost relations in \eqref{eq:func.rel.A21a} and \eqref{eq:func.rel.A21b} as
\be
\label{eq:bbb.A21}
\frac{\bmf(z - \frac{\ir \pi}2) \bmf(z + \frac{\ir \pi}2)}{\bar\bmf(z)} = \Amf^1(z), 
\qquad
\frac{\bar\bmf(z - \frac{\ir \pi}2) \bar\bmf(z + \frac{\ir \pi}2)}{\bmf(z)} = \bar\Amf^1(z).
\ee

\subsubsection{Non-linear integral equations}

The functions $\amf^1(z)$ and $\bar\amf^1(z)$ have order-$N$ zeros on the imaginary axis inside their analyticity strips. We define new functions $\ell^n(z)$ and $\bar\ell^n(z)$ as the products and ratios of $\amf^n(z)$ and $\bar\amf^n(z)$, with the order-$N$ zeros removed:
\begin{subequations}
\begin{alignat}{3}
\ell^1(z) &= \frac{\asf^1(z)\bar\asf^1(z)}{\eta(z-\frac{\ir \pi}4)\eta(z+\frac{\ir \pi}4)}, \qquad &\ell^n(z) &= \asf^n(z)\,\bar\asf^n(z), \qquad &n = 2, \dots, p',
\\
\bar\ell^1(z) &= \frac{\asf^1(z)}{\bar\asf^1(z)}\frac{\nu(z-\frac{\ir \pi}4)}{\nu(z+\frac{\ir \pi}4)}, \qquad &\bar\ell^n(z) &= \frac{\asf^n(z)}{\bar\asf^n(z)}, \qquad &n = 2, \dots, p'.
\end{alignat}
\end{subequations}
The functions $\eta(z)$ and $\nu(z)$ are
\begin{subequations}
\label{eq:eta.nu.A21}
\begin{alignat}{3}
\eta(z) &= \tanh\tfrac z 3, \qquad && \eta(z-\tfrac{3 \ir \pi}4)\,\eta(z+\tfrac{3 \ir \pi}4)=1,
\\[0.15cm]
\nu(z) &= \tanh\tfrac {2z} 3, \qquad && \frac{\nu(z+\tfrac{3 \ir \pi}4)}{\nu(z-\tfrac{3 \ir \pi}4)} = 1.
\end{alignat}
\end{subequations}
The $Y$-system equations for the functions $\ell^n(z)$ and $\bar \ell^n(z)$
are obtained by taking products and ratios of the first and second set of relations in \eqref{eq:sym.Y.A21}. It in fact splits in two separate sets of equations, the first depending on the functions $\ell^n(z)$ and $\Amf^n(z)\bar\Amf^n(z)$ and the second on the functions $\bar\ell^n(z)$ and $\Amf^n(z)/\bar\Amf^n(z)$. Moreover, these equations do not depend explicitly on $\eta(z)$ and $\nu(z)$ due to the convenient definitions \eqref{eq:eta.nu.A21} for these functions. 

The functions $\ell^n(z)$ and $\bar\ell^n(z)$ are analytic and non-zero inside their respective analyticity strips. As discussed in \cref{sec:braid.and.bulk.A21}, for generic values of $\omega$, the functions $\asf^n(z)$ and $\bar\asf^n(z)$ have asymptotic values for $z \to \pm \infty$ that are finite and nonzero. The same then holds for the functions $\ell^n(z)$, $\bar\ell^n(z)$, $\Amf^n(z)$ and $\bar\Amf^n(z)$, allowing us to define the Fourier transforms
\begin{subequations}
\begin{alignat}{2}
&L^n(k) = \frac1{2\pi} \int_{-\infty}^\infty \dd z\, \eE^{-\ir k z}\big[\log \ell^n(z)\big]', \qquad 
&&A^n(k) = \frac1{2\pi} \int_{-\infty}^\infty \dd z\, \eE^{-\ir k z}\big[\log \big(\Amf^n(z)\bar\Amf^n(z)\big)\big]',
\\[0.15cm]
&\bar L^n(k) = \frac1{2\pi} \int_{-\infty}^\infty \dd z\, \eE^{-\ir k z}\big[\log \bar \ell^n(z)\big]', \qquad 
&&\bar A^n(k) = \frac1{2\pi} \int_{-\infty}^\infty \dd z\, \eE^{-\ir k z}\bigg[\log \bigg(\frac{\Amf^n(z)}{\bar\Amf^n(z)}\bigg)\bigg]',
\end{alignat}
\end{subequations}
The non-linear integral equations for the eigenvalues are obtained by first
taking the Fourier transform of the logarithmic derivative of the $Y$-system
equations yielding
\begin{subequations}
\label{eq:LKA}
\be
\label{eq:LKAa}
\left(\begin{smallmatrix}
L^1 \\[0.05cm] L^2 \\[0.05cm] \svdots \\[0.05cm] L^{p'-2} \\[0.05cm] L^{p'-1} \\[0.05cm] L^{p'}
\end{smallmatrix}\right)
= \frac1{2\cosh\!\frac {\pi k}2-1} 
 \left(\begin{smallmatrix}
-1 &&& 1 &&& 0 &&&  &&&  \\[0.1cm]
1 &&& -1 &&& 1 &&& \sddotss &  \\[0.1cm]
0 &&& 1 &&& -1 &&& 1 &&& 0 \\[0.1cm]
 &&& \sddotss &&& 1 &&& -1 &&& 1 &&& -2 \cosh\!\frac{\pi k}4 \\[0.1cm]
 &&&  &&& 0 &&& 1 &&&-1 &&& 2 \cosh\!\frac{\pi k}4 \\[0.1cm]
 &&&  &&&  &&& 2 \cosh\!\frac{\pi k}4 &&& -2 \cosh\!\frac{\pi k}4 &&& 2 \cosh\!\frac{\pi k}2\,+\,2
\end{smallmatrix}\right)
\left(\begin{smallmatrix}
A^1 \\[0.05cm] A^2 \\[0.05cm] \svdots \\[0.05cm] A^{p'-2} \\[0.05cm] A^{p'-1}\\[0.05cm] A^{p'}
\end{smallmatrix}\right)
\ee
and
\be
\label{eq:LKAb}
\left(\begin{smallmatrix}
\bar L^1 \\[0.05cm] \bar L^2 \\[0.05cm] \svdots \\[0.05cm] \bar L^{p'-2} \\[0.05cm] \bar L^{p'-1} \\[0.05cm] \bar L^{p'}
\end{smallmatrix}\right)
= \frac1{2\cosh\!\frac {\pi k}2+1} 
 \left(\begin{smallmatrix}
1 &&& 1 &&& 0 &&&  &&&  \\[0.1cm]
1 &&& 1 &&& 1 &&& \sddotss &  \\[0.1cm]
0 &&& 1 &&& 1 &&& 1 &&& 0 \\[0.1cm]
 &&& \sddotss &&& 1 &&& 1 &&& 1 &&& -2 \sinh\!\frac{\pi k}4 \\[0.1cm]
 &&&  &&& 0 &&& 1 &&&1 &&& -2 \sinh\!\frac{\pi k}4 \\[0.1cm]
 &&&  &&&  &&& -2 \sinh\!\frac{\pi k}4 &&& -2 \sinh\!\frac{\pi k}4 &&& 2 \cosh\!\frac{\pi k}2\,-\,2
\end{smallmatrix}\right)
\left(\begin{smallmatrix}
\bar A^1 \\[0.05cm] \bar A^2 \\[0.05cm] \svdots \\[0.05cm] \bar A^{p'-2} \\[0.05cm] \bar A^{p'-1}\\[0.05cm] \bar A^{p'}
\end{smallmatrix}\right).
\ee
\end{subequations}
We note that, up to a rescaling of $k$, the $Y$-system relations \eqref{eq:LKAa} are identical to the similar relations \eqref{eq:scalingNLIEs.A22} for the $A_2^{(2)}$ model. Writing the result in matrix form as
\be
\label{eq:LKA21}
\vec L = \hat K^{(1)} \cdot \vec A, \qquad \vec {\bar L} = \hat K^{(2)} \cdot \vec {\bar A},
\ee
we note that the matrices $\hat K^{(1)}$ and $\hat K^{(2)}$ have the symmetries
\be
\label{eq:K12.symm.A21}
\big(\sigma\hat K^{(j)}\big)^\intercal = \sigma\hat K^{(j)}\big|_{k \to -k}, \qquad j = 1, 2, \qquad \sigma = \textrm{diag}(1,1, \dots, 1, -1).
\ee
We apply the inverse transform, integrate with respect to $z$, and find
\begin{subequations}
\label{eq:NLIEs.A21}
\begin{alignat}{2}
\log \amf^{n}(z) - \phi_n &= \tfrac12\big(\fmf^n(z)+\gmf^n(z)\big) + \tfrac12 \sum_{m=1}^{p'} \big(K^{(1)}_{nm} + K^{(2)}_{nm}\big) * \log \Amf^{m} + \big(K^{(1)}_{nm} - K^{(2)}_{nm}\big) * \log \bar\Amf^{m},
\\
\log \bar \amf^{n}(z) - \bar\phi_n &= \tfrac12\big(\fmf^n(z)-\gmf^n(z)\big) + \tfrac12 \sum_{m=1}^{p'} \big(K^{(1)}_{nm} - K^{(2)}_{nm}\big) * \log \Amf^{m} + \big(K^{(1)}_{nm} + K^{(2)}_{nm}\big) * \log \bar\Amf^{m},
\end{alignat}
\end{subequations}
where $n = 1, \dots, p',$ and $\phi_n$ and $\bar \phi_n$ are integration constants. The driving terms are
\begin{subequations}
\begin{alignat}{2}
\fmf^n(z) &= \left\{
\begin{array}{cl}
N \log \Big[\eta\big(z-\frac{\ir\pi}4\big)\eta\big(z+\frac{\ir\pi}4\big)\Big]\quad& n = 1,\\[0.3cm]
0 & n = 2,\dots, p',
\end{array}\right.
\\[0.15cm]
\gmf^n(z) &= \left\{
\begin{array}{cl}
\displaystyle N \log \bigg[\frac{\nu\big(z+\frac{\ir\pi}4\big)}{\nu\big(z-\frac{\ir\pi}4\big)}\bigg]\quad& n = 1,\\[0.4cm]
0 & n = 2,\dots, p'.
\end{array}\right.
\end{alignat}
\end{subequations}
The kernel functions are given by
\be
\label{eq:Kij.A21.principal}
K^{(j)}_{nm}(z) = \frac1{2\pi}\int_{-\infty}^\infty \dd k\, \eE^{\ir k z}\hat K^{(j)}_{nm}, \qquad j = 1,2, \qquad (n,m) \neq (p',p').
\ee
Extra care has to be taken for the kernel functions $K^{(j)}_{p',p'}$, as in this case the corresponding functions in $k$-space tend to $1$ for $k \to \pm \infty$. This results in extra algebraic terms in the non-linear integral equations (similarly to \eqref{eq:scalingNLIEs.A22e}), which can be encoded in the kernel functions as
\be
K^{(j)}_{p',p'}(z) = \delta_{z,0}+\frac1{2\pi}\int_{-\infty}^\infty \dd k\, \eE^{\ir k z}(\hat K^{(j)}_{p',p'}-1), \qquad j = 1,2.
\ee
These extra terms are diagonal and do not play a role in the computation of the dilog integral in \cref{sec:dilog.A21}.

\subsubsection{Scaling functions and scaling non-linear integral equations}

In \eqref{eq:NLIEs.A21}, the dependence on $N$ appears only in the driving terms. For $z$ of order $\pm \frac32\log N$ with $N$ large, the functions $\fsf^n(z)$ behave as exponentials:
\be
\fsf^n(z) = \lim_{N\to \infty} \fmf^n\big(\!\pm\!(z +\tfrac32 \log N)\big) = 
 \left\{\begin{array}{cl}
-2 \sqrt 3\, \eE^{-2z/3}& n = 1,\\[0.2cm]
0& n = 2, \dots, p'.
\end{array}\right.
\ee
In contrast, in the same regime, the functions $\gmf^n(z)$ vanish:
\be
\lim_{N\to \infty} \gmf^n\big(\!\pm\!(z +\tfrac32 \log N)\big) = 0.
\ee
To compute the finite-size correction at order $\frac 1N$, we assume that the unknown functions appearing in \eqref{eq:NLIEs.A21} are well-defined in this limit. For the groundstate, the patterns of zeros are all symmetric with respect to a reflection along the imaginary $z$-axis, and we have the equalities \eqref{eq:z-z.ids.A21}. As a result, the scaling functions behave identically in both limits:
\be
\asf^n(z) = \lim_{N\to \infty} \amf^n\big(\!\pm\!(z + \tfrac32\log N)\big), \qquad \Asf^n(z) = \lim_{N\to \infty} \Amf^n\big(\!\pm\!(z +  \tfrac32\log N)\big),
\ee
\be
\bar \asf ^n(z) = \lim_{N\to \infty} \bar\amf^n\big(\!\pm\!(z + \tfrac32\log N)\big), \qquad \bar \Asf^n(z) = \lim_{N\to \infty} \bar\Amf^n\big(\!\pm\!(z +  \tfrac32\log N)\big),
\ee
where $n = 1, \dots, p'$ and $z$ is taken to be real. These functions satisfy the following set of integral equations:
\begin{subequations}
\label{eq:scalingNLIEs.A21}
\begin{alignat}{2}
\log \asf^n(z) - \phi_n &= \tfrac12\fsf^n(z) + \tfrac12 \sum_{m=1}^{p'} \big(K^{(1)}_{nm} + K^{(2)}_{nm}\big) * \log \Asf^{m} + \big(K^{(1)}_{nm} - K^{(2)}_{nm}\big) * \log \bar\Asf^{m}, 
\\
\log \bar \asf^{n}(z) - \bar\phi_n &= \tfrac12\fsf^n(z) + \tfrac12 \sum_{m=1}^{p'} \big(K^{(1)}_{nm} - K^{(2)}_{nm}\big) * \log \Asf^{m} + \big(K^{(1)}_{nm} + K^{(2)}_{nm}\big) * \log \bar\Asf^{m},
\end{alignat}
\end{subequations}
where $n = 1, \dots, p'$.
Multiplying the equations for $\log \asf^{p'\!}(z)$ and $\log
\bar\asf^{p'\!}(z)$ by an overall minus sign, we obtain a set of equations
where the kernel terms are symmetric, which follows from \eqref{eq:K12.symm.A21}.

\subsubsection{Braid and bulk behavior}\label{sec:braid.and.bulk.A21}

The scaling functions have finite asymptotics for $z \to \pm\infty$. For $z \to \infty$, these are obtained directly from the braid limits of the transfer matrix eigenvalues:
\begin{subequations}
\begin{alignat}{2}
\asf^n_\infty &=\bar \asf^n_\infty = \frac{(\omega^{n/2}-\omega^{-n/2})(\omega^{(n+3)/2}-\omega^{-(n+3)/2})}{(\omega^{1/2}-\omega^{-1/2})(\omega-\omega^{-1})}, \qquad n = 1, \dots, p'-2,
\\[0.15cm]
\asf^{p'-1}_\infty &= \bar \asf^{p'-1}_\infty =\frac{\omega^{(p'-1)/2}-\omega^{-(p'-1)/2}}{\omega^{(p'+1)/2}-\omega^{-(p'+1)/2}}, 
\qquad \asf^{p'}_\infty = (\asf^{p'-1}_\infty)^2.
\end{alignat}
\end{subequations}
For $\gamma \in (0, \frac {2\pi}{p'+1})$, these constants are all positive and finite.
These values allow us to compute the constants $\phi_n$ by studying the $z\to \infty$ asymptotics of \eqref{eq:scalingNLIEs.A21}. After a straightforward computation, we find that the constants vanish:
\be
\phi_n = \bar \phi_n =  0, \qquad n = 1, \dots, p', \qquad \gamma \in (0,\tfrac {2\pi}{p'+1}).
\ee

For the bulk behavior at $z \to -\infty$, we recall that the functions $\amf^1(z)$ and $\bar\amf^1(z)$ each have a zero of order $N$ near the origin, so that $\asf^1_{-\infty} = \bar \asf^1_{-\infty}= 0$. The asymptotic values $\asf^n_{-\infty}$ and $\bar\asf^n_{-\infty}$ are constant solutions to the $Y$-system \eqref{eq:sym.Y.A21}. We select the only solution for which the functions $\asf^n_{\infty}$ and $\bar \asf^n_{\infty}$, with $n=2, \dots, p'$, are positive in the neighborhood of $\omega = 1$:
\begin{subequations}
\begin{alignat}{2}
\asf^n_{-\infty} = \bar \asf^n_{-\infty} &= \frac{(\bar\omega^{(n-1)/2}-\bar\omega^{-(n-1)/2})(\bar\omega^{(n+2)/2}-\bar\omega^{-(n+2)/2})}{(\bar\omega^{1/2}-\bar\omega^{-1/2})(\bar\omega-\bar\omega^{-1})}, 
\\[0.15cm]
\asf^{p'-1}_{-\infty} = \bar \asf^{p'-1}_{-\infty} &=\frac{\bar\omega^{(p'-2)/2}-\bar\omega^{-(p'-2)/2}}{\bar\omega^{p'/2}-\bar\omega^{-p'/2}}, 
\qquad \asf^{p'}_{-\infty} = \bar \asf^{p'}_{-\infty} = (\asf^{p'-1}_{-\infty})^2,
\qquad \bar\omega = \omega^{p'/(p'-1)}.
\end{alignat}
\end{subequations}
These asymptotic values are positive and finite on the range $\gamma \in (0,\frac{2\pi(p'-1)}{p'^2})$.

\subsubsection{Finite-size correction and the dilogarithm technique}\label{sec:dilog.A21}

Applying the Fourier transform and subsequently the inverse transform of the logarithmic derivative of \eqref{eq:bbb.A21}, we find
\begin{subequations}
\label{eq:bbarb.A21}
\begin{alignat}{2}
\log \bmf(z) - \phi_0 &= \tilde K^{(1)} * \log \Amf^1(z) + \tilde K^{(2)} * \log \bar\Amf^1(z),
\\[0.15cm]
\log \bar\bmf(z) - \bar \phi_0 &= \tilde K^{(2)} * \log \Amf^1(z) + \tilde K^{(1)} * \log \bar\Amf^1(z),
\end{alignat}
\end{subequations}
where $\phi_0$ and $\bar\phi_0$ are integration constants. The kernels are
\begin{subequations}
\begin{alignat}{2}
\tilde K^{(1)} &= \frac1{2\pi} \int_{-\infty}^\infty \dd k\, \eE^{\ir k z}\frac{2 \cosh \frac{\pi k}2}{2 \cosh\pi k + 1} = \frac{1}{\pi \sqrt 3} \frac{\cosh\frac z3}{\cosh z},
\\[0.15cm]
\tilde K^{(2)} &= \frac1{2\pi} \int_{-\infty}^\infty \dd k\, \eE^{\ir k z}\frac{1}{2 \cosh\pi k + 1} = \frac{1}{\pi \sqrt 3} \frac{\sinh\frac z3}{\sinh z}.
\end{alignat}
\end{subequations}
We express the large-$N$ asymptotics of $\log \bmf(z)$ in terms of integrals involving the scaling functions $\Asf^1(z)$ and $\bar\Asf^1(z)$: 
\begin{alignat}{2}
\log \bmf(z) - \phi_0 &= \int_{-\frac32\log N}^\infty \dd y \Big(K^{(1)}(y+\tfrac32\log N - z) \log \Amf^1(y+\tfrac32\log N) \nonumber
\\ & \hspace{2.5cm}+ K^{(2)}(y+\tfrac32\log N - z) \log \bar\Amf^1(y+\tfrac32\log N) \nonumber
\\[0.2cm] & \hspace{2.5cm}+ K^{(1)}(-y-\tfrac32\log N - z) \log \Amf^1(-y-\tfrac32\log N)\nonumber
\\[0.1cm] & \hspace{2.5cm}+ K^{(2)}(-y-\tfrac32\log N - z) \log \bar\Amf^1(-y-\tfrac32\log N)\Big)
\nonumber\\[0.15cm]
&\simeq \frac{1}{\pi \sqrt 3 N}(\eE^{2z/3}+\eE^{-2z/3})\int_{-\infty}^\infty \dd y\, \eE^{-2y/3} \big(\log \Asf^1(y) + \log \bar\Asf^1(y)\big),
\label{eq:log.b.int.A21}
\end{alignat}
where we used
\be
K^{(1)}(z + \tfrac32\log N) \simeq K^{(2)}(z + \tfrac32\log N) \simeq \frac{\eE^{-2z/3}}{\pi \sqrt 3 N }.
\ee
Here, $\simeq$ indicates that higher-order terms in $\frac1N$ are omitted.

To apply the dilogarithm technique, we define the integral
\begin{alignat}{2}
\mathcal J &= \int_{-\infty}^\infty \dd y &\bigg[\sum_{n=1}^{p'-1}\Big((\log \asf^n)' \log \Asf^n  -\log \asf^n (\log \Asf^n)'\Big) - \Big(\big(\log \asf^{p'}\big)' \log \Asf^{p'} -\log \asf^{p'}\, (\log \Asf^{p'})' \Big) \bigg]\nonumber\\
&+ \int_{-\infty}^\infty \dd y &\bigg[\sum_{n=1}^{p'-1}\Big((\log \bar\asf^n)' \log \bar\Asf^n  -\log \bar\asf^n (\log \bar\Asf^n)'\Big) - \Big(\big(\log \bar\asf^{p'}\big)' \log \bar\Asf^{p'} -\log \bar\asf^{p'}\, (\log \bar\Asf^{p'})' \Big) \bigg].
\end{alignat}
This integral is evaluated in two ways. The first consists of replacing $\log \asf^n$ and $\log \bar\asf^n$ by their expressions \eqref{eq:scalingNLIEs.A21}. Many terms cancel out because of the symmetries of the kernels. The only surviving contributions come from the driving terms, and the result reads
\be
\mathcal J = \frac 4{\sqrt 3} \int_{-\infty}^\infty \dd y \, \eE^{-2y/3} \big(\log \Asf^1(y) + \log \bar\Asf^1(y)\big).
\ee
Up to an overall prefactor, this is precisely the integral we wish to compute in \eqref{eq:log.b.int.A21}. The second way of computing the integral is to apply the derivatives explicitly, which yields
\begin{alignat}{2}
\mathcal J &= \int_{-\infty}^\infty \dd y \,  \bigg[\sum_{n=1}^{p'-2}\frac{\dd \asf^n}{\dd y}\bigg(\frac{\log \Asf^n}{\asf^n} -\frac{\log \asf^n}{\Asf^n}\bigg) + \frac{\dd \asf^{p'-1}}{\dd y}\bigg(\frac{\log \Asf^{p'-1}}{\asf^{p'-1}}  - \frac{\dd \Asf^{p'-1}}{\dd \asf^{p'-1}}\frac{\log \asf^{p'-1}}{\Asf^{p'-1}}\bigg) 
\nonumber\\&\hspace{11cm}
- \frac{\dd \asf^{p'}}{\dd y} \bigg(\frac{\log \Asf^{p'}}{\asf^{p'}} + \frac{\log \asf^{p'}}{\Asf^{p'}}\bigg)\bigg]\nonumber\\
&+ \int_{-\infty}^\infty \dd y \,  \bigg[\sum_{n=1}^{p'-2}\frac{\dd \bar\asf^n}{\dd y}\bigg(\frac{\log \bar\Asf^n}{\bar\asf^n} -\frac{\log \bar\asf^n}{\bar\Asf^n}\bigg) + \frac{\dd \bar\asf^{p'-1}}{\dd y}\bigg(\frac{\log \bar\Asf^{p'-1}}{\bar\asf^{p'-1}}  - \frac{\dd \bar\Asf^{p'-1}}{\dd \bar\asf^{p'-1}}\frac{\log \bar\asf^{p'-1}}{\bar\Asf^{p'-1}}\bigg) 
\\&\hspace{11cm}
- \frac{\dd \bar\asf^{p'}}{\dd y} \bigg(\frac{\log \bar\Asf^{p'}}{\bar\asf^{p'}} + \frac{\log \bar\asf^{p'}}{\bar\Asf^{p'}}\bigg)\bigg].\nonumber
\end{alignat}
We split the integrals in three parts and change the integration variables from $y$ to $\asf^n$ or to $\bar\asf^n$. Because the functions $\asf^n(z)$ and $\bar\asf^n(z)$ have identical asymptotics, we find two copies of the same integrals:
\begin{alignat}{2}
\mathcal J = 2 \Bigg[&\sum_{n=1}^{p'-2}\int_{\asf^n_{-\infty}}^{\asf^n_\infty} \dd \asf^n \bigg(\frac{\log \Asf^n}{\asf^n}  -\frac{\log \asf^n}{\Asf^n}\bigg) + \int_{\asf^{p'-1}_{-\infty}}^{\asf^{p'-1}_\infty} \dd \asf^{p'-1} \bigg(\frac{\log \Asf^{p'-1}}{\asf^{p'-1}} - \frac{\dd \Asf^{p'-1}}{\dd \asf^{p'-1}}\frac{\log \asf^{p'-1}}{\Asf^{p'-1}}\bigg) 
\nonumber\\[0.15cm]
&- \int_{\asf^{p'}_{-\infty}}^{\asf^{p'}_\infty} \dd \asf^{p'}  \bigg(\frac{\log \Asf^{p'}}{\asf^{p'}} + \frac{\log \asf^{p'}}{\Asf^{p'}}\bigg)\Bigg],
\end{alignat}
where
\begin{subequations}
\begin{alignat}{2}
\Asf^n &= 1 + \asf^n, \qquad n = 1, \dots, p'-2,
\\[0.15cm] 
\Asf^{p'-1} &= (1 + \omega^{p'} \asf^{p'-1})(1 + \asf^{p'-1})(1 + \omega^{-p'} \asf^{p'-1}),
\\[0.15cm]  
\Asf^{p'} &= (1 - \asf^{p'}).
\end{alignat}
\end{subequations}
The resulting expression for $\mathcal J$ is therefore a combination of regular integrals. This integral is exactly twice the quantity in \eqref{eq:J1.A22} and \eqref{eq:J2.A22}, with $p'$ replacing $b$. Setting $\omega = \eE^{\ir \gamma}$, $\mathcal J$ evaluates to 
\be
\mathcal J = \frac{\pi^2}3\bigg(2- \frac{6\gamma^2 p'}{\pi^2 (p'-1)}\bigg) = \frac{\pi^2}3\bigg(2-\frac{6\gamma^2}{\pi(\pi-\lambda)}\bigg), \qquad \gamma \in \big(0,\tfrac{2\pi(p'-1)}{p'^2}\big).
\ee
The proof of this result is given in \cref{app:A21.principal}. The final result is
\be
\log \bmf(z) \simeq \frac{\pi \cosh (\frac{2z}3)}{6 N} \bigg(2- \frac{6\gamma^2 p'}{\pi^2 (p'-1)}\bigg),
\ee
and therefore
\be
\log T_{\rm f}(u) \simeq \frac{\pi \sin\!\frac{2\pi u}{3\lambda}}{6 N} \bigg(2- \frac{6\gamma^2 p'}{\pi^2 (p'-1)}\bigg),
\qquad
\log \bar T_{\rm f}(u) \simeq \frac{\pi \sin\!\frac{2\pi (\lambda - u)}{3\lambda}}{6 N} \bigg(2- \frac{6\gamma^2 p'}{\pi^2 (p'-1)}\bigg),
\ee
where we recall the crossing relation $\bar T_{\rm f}(u) = T_{\rm f}(\lambda - u)$. The constant $\phi_0$ was found to equal zero using $T_{\rm f}(u=0)=1$.
This result is precisely \eqref{eq:fsc.general} with $c-24 \Delta$ and $\vartheta(u)$ given in \eqref{eq:fsc.A21}.

\subsection{The dual series}\label{sec:A21.dual}

In this subsection, we fix $p =1$, so that $\lambda = \frac{\pi(p'-1)}{p'}$, and consider $p' \in \mathbb N_{\ge 4}$. We also define $\bar\lambda = \pi - \lambda = \frac{\pi}{p'}$. We compute the $\frac1N$ term in \eqref{eq:fsc.general} explicitly for $N \equiv 0 \textrm{ mod } 6$, $\gamma \in \big(0,\frac{2\pi}{p'+1}\big)$ and $u$ in the neighborhood of $\frac \lambda 2$.

\subsubsection[Analyticity properties and symmetric $Y$-system]{Analyticity properties and symmetric $\boldsymbol Y$-system}\label{sec:analyticity.A21.dual}

\begin{figure}
\centering
\begin{tabular}{ccccc}
\begin{tabular}{c}
$T^{1,0}(u)$ \\[0.1cm]
\includegraphics[width=.25\textwidth]{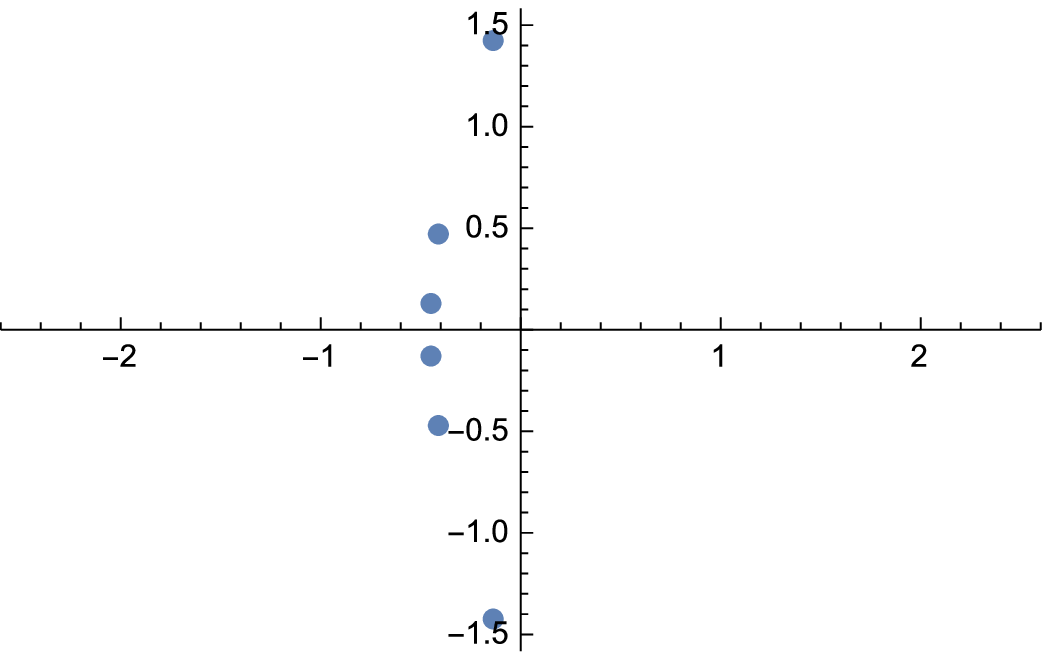} \\[0.3cm]
$T^{3,0}(u)$ \\[0.1cm]
\includegraphics[width=.25\textwidth]{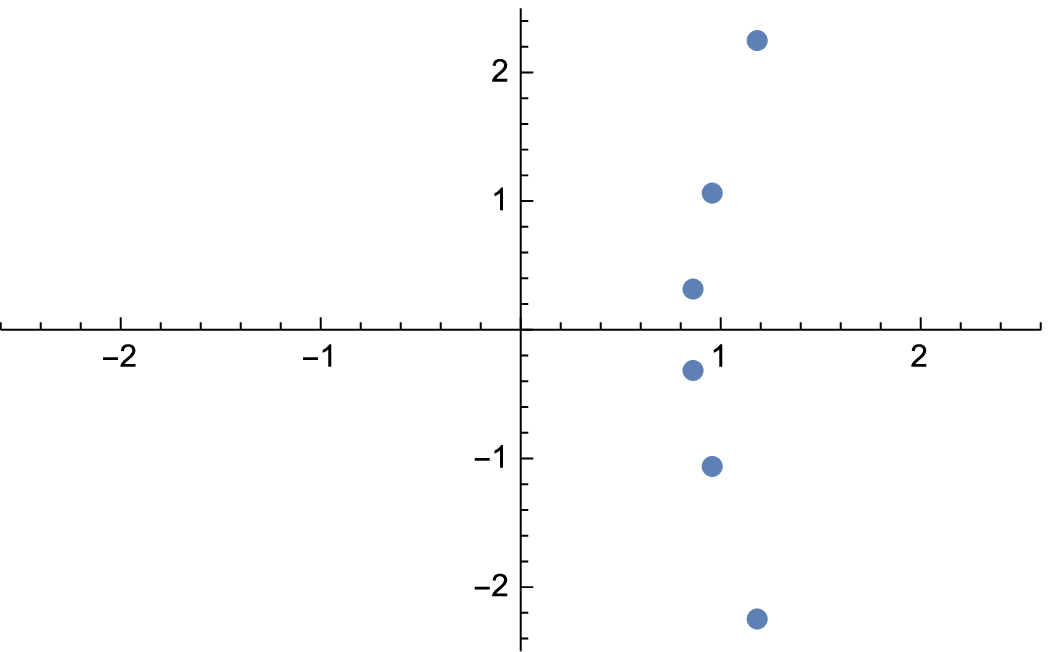}
\end{tabular}
&\quad&
\begin{tabular}{c}
$T^{2,0}(u)$ \\[0.1cm]
\includegraphics[width=.25\textwidth]{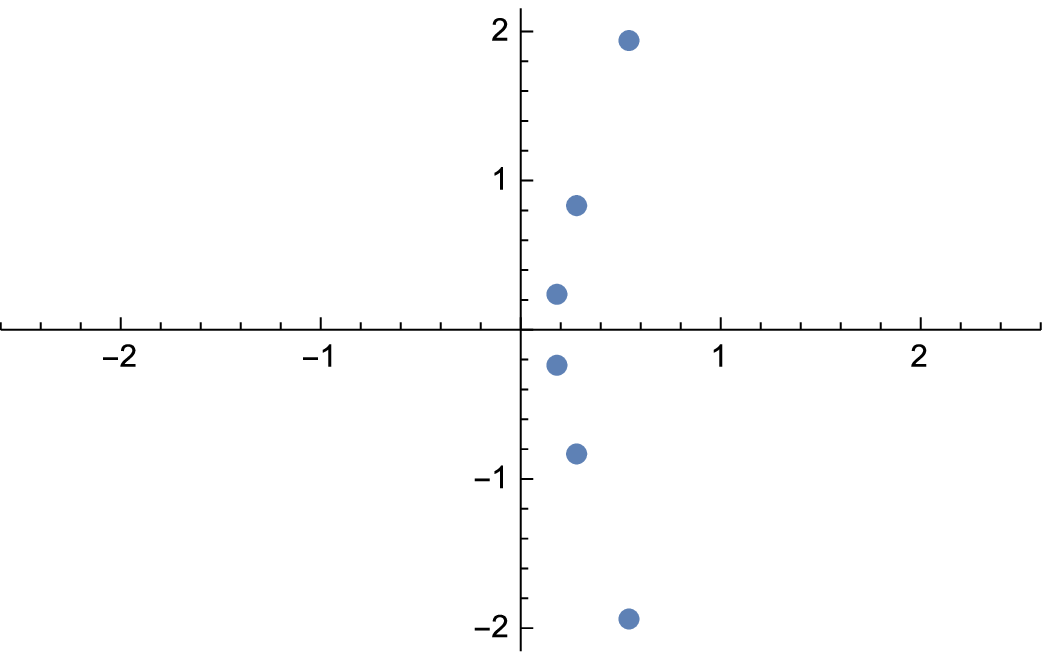}
\\[0.3cm]
$T^{4,0}(u)$ \\[0.1cm]
\includegraphics[width=.25\textwidth]{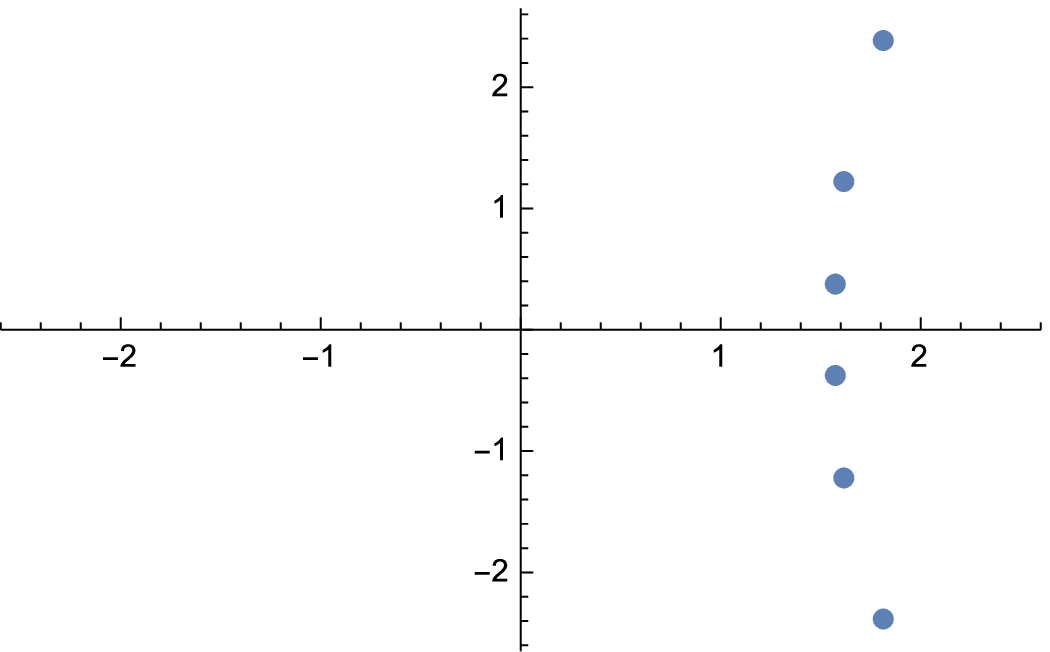}
\end{tabular}
&\quad&
\begin{tabular}{c}
$T^{4,4}(u)$ \\[0.1cm]
\includegraphics[width=.25\textwidth]{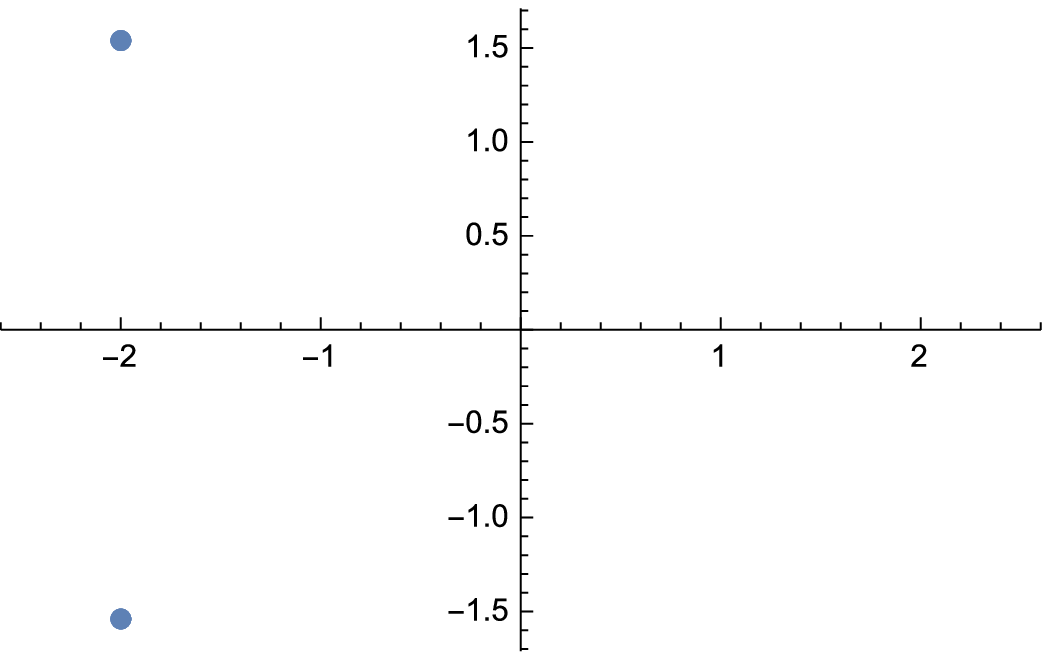} 
\\[0.3cm]
$1-y(u)$ \\[0.1cm]
\includegraphics[width=.25\textwidth]{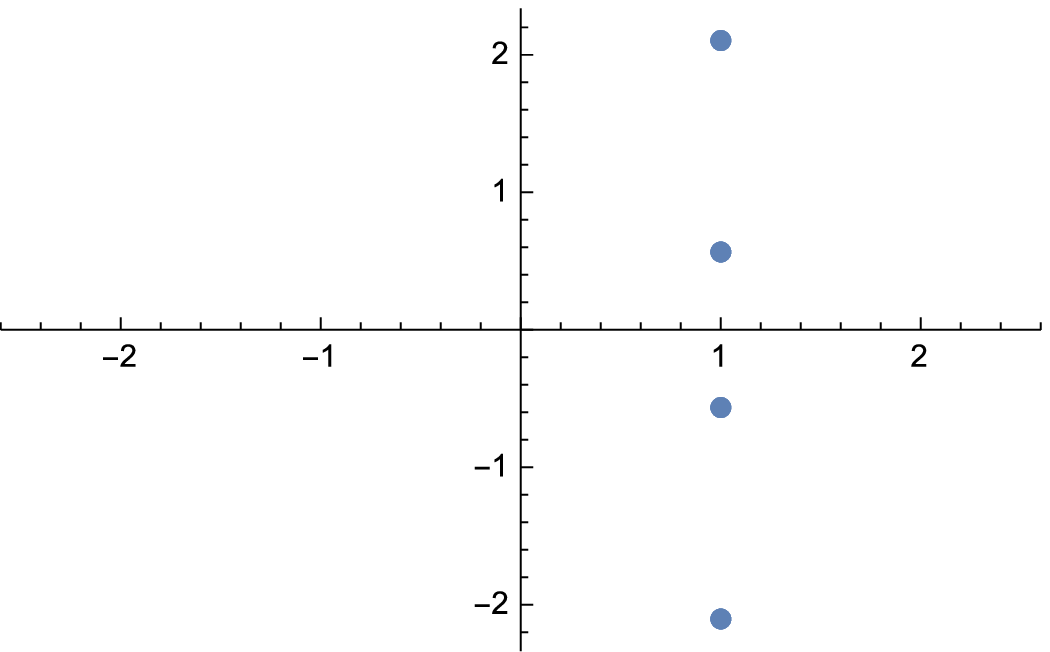}
\end{tabular}
\end{tabular}
\caption{The patterns of zeros for the groundstate of $\Tb(u)$ for $N=6$, $(p,p') = (1,5)$ and $\omega = 1$, in the complex $u$-plane. The horizontal axis is divided in units of $\bar \lambda = \frac\pi5$. Each zero of $T^{4,4}(u)$ and $1-y(u)$ is triply degenerate. This degeneracy is lifted for $\omega\neq1$, with the zeros remaining on the same vertical lines for $\omega$ on the unit circle. The diagram for $1-y(u)$ only shows its zeros, and not its poles which can be deduced from the zeros of the others functions.}
\label{fig:patterns.A21.dual}
\end{figure}

Our computer implementation outputs the patterns of zeros for the fused transfer matrices, for the groundstate. For this case, the zeros do not appear precisely on vertical lines, in the complex $u$-plane, in constrast with the other cases. We find that the zeros of $T^{1,0}(u)$ for the groundstate lie in the region Re$(u) \in (- \frac{\bar\lambda}2,\frac{\bar\lambda}2)$. We choose the center of its analyticity strip to be on the vertical line Re$(u) = \frac\lambda2 - \frac{\bar\lambda}4$. We observe that there is a strip of width larger than $\bar \lambda$ centered at this position where there are no zeros. Likewise, our computer implementation reveals that $T^{2,0}(u)$ has an analyticity strip of width larger than $\bar \lambda$ centered at Re$(u) = \frac\lambda2 + \frac{\bar\lambda}4$. In general, we observe that the function $T^{n,0}(u)$ has its zeros in the region Re$(u) \in \big(\frac{(n-2)\bar\lambda}2,\frac{n\bar\lambda}2\big)$, and has an analyticity strip of width larger than $\bar \lambda$ centered at Re$(u) = \frac\lambda2 + (n-\frac32)\frac{\bar\lambda}2$. The positions of the zeros and of the analyticity strips for $T^{0,n}(u)$ are obtained from those of $T^{n,0}(u)$ and the crossing relation \eqref{eq:crossing.A21}. As a result, $T^{0,n}(u)$ has an analyticity strip of width larger than $\bar \lambda$ centered at Re$(u) = \frac\lambda2 + (n-\frac12)\frac{\bar\lambda}2$. Furthermore, $T^{p'-1,p'-1}(y)$ has its zeros on the vertical line Re$(u) = -2 \bar \lambda$. To deduce the positions of the zeros and poles of all the functions of the $Y$-system, we also need to know the position of the zeros of the function $1-y(u)$, as we lack a formula similar to \eqref{eq:1-z(u).A21} for this function. Our computer implementation reveals that its zeros lie on the vertical line Re$(u) = \frac{(p'-3)\bar\lambda}2$. The patterns of zeros for $N=6$, $(p,p') = (1,5)$ and $\omega=1$ are given in \cref{fig:patterns.A21.dual}.

This information allows us to read off the positions of the zeros and poles of the functions of the $Y$-system. Let us describe the order-$N$ poles and zeros. The function $t^1(u)$ has a zero of order $N$ at $u=0$ and a pole of order $N$ at $u = \bar \lambda$. The function $t^n(u)$ with $n = 2, \dots, p'-2$ has no zeros of order~$N$, but has a pole of order $N$ at $u = n\bar\lambda$. Similarly, the function $\bar t^1(u)$ has a zero at $u = 0$ and a pole at $u = (p'-1)\lambda$. The function $\bar t^n(u)$ has no zero of order $N$, but has a pole at $u = (p'-1)\lambda$. The functions $x(u)$, $\bar x(u)$, $y(u)$ and $z(u)$ have neither zeros nor poles of order $N$. The same applies to the functions $\big(1+\omega^{p'} x(u)\big)\big(1+x(u)\big)\big(1+\omega^{-p'} x(u)\big)$, $\big(1+\omega^{p'}\bar x(u)\big)\big(1+\bar x(u)\big)\big(1+\omega^{-p'} \bar x(u)\big)$ and $1-y(u)$, which are free of order-$N$ zeros and poles. Finally, the function $1-z(u)$ satisfies the identity \eqref{eq:1-z(u).A21} and therefore has an order-$N$ zero at $u = (p'-1)\bar \lambda$. 

The poles of these functions will play an important role in what follows, as some of them turn out to lie inside the analyticity strips. We choose to work with the functions $t^n(u)^{-1}$, $\bar t^n(u)^{-1}$, $x(u)^{-1}$, ${\bar x}(u)$ and
\be
w(u) = \frac{1-z(u)}{z(u)}, \qquad \bar w(u) = \frac{y(u)}{1-y(u)}.
\ee
Our asymmetrical choice for $x(u)^{-1}$ and ${\bar x}(u)$ might be surprising, but leads to the correct calculation of the finite-size corrections below. The $Y$-system equations \eqref{eq:Ysys.A21.general} are then rewritten in terms of these new functions. Our derivation below uses certain assumptions for the analyticity strips of these functions. These are given in \cref{tab:analyticity.A21.dual}. 
Crucially, these analyticity strips are free of order-$N$ zeros and poles, with the exception of the order-$N$ zeros of $t^{p'-2}(u)^{-1}$ and $w(u)$. The analyticity strips are free of simple zeros, except for the functions $w(u)$, $\big(1+\omega^{p'} x(u)^{-1}\big)\big(1+x(u)^{-1}\big)\big(1+\omega^{-{p'}} x(u)^{-1}\big)$ and $\big(1+\omega^{p'}{\bar x}(u)\big)\big(1+{\bar x}(u)\big)\big(1+\omega^{-p'} \bar x(u)\big)$. These functions all have simple zeros on the central line of the analyticity strip. Moreover, the number of such zeros grows linearly with $N$.

As we shall see, the derivation below works despite the presence of the simple zeros in the central line of the analyticity strips for these three functions. A key remark is that in the $Y$-system relations that we will solve below, these three functions always appear at most once, instead of multiple times with shifted arguments like the other functions. In the calculation below, the presence of their zeros inside the analyticity strips will be addressed by shifting the arguments of the functions in the complex plane, in such a way that the integration path does not cross these zeros.

\begin{table}
\begin{center}
\begin{tabular}{c|c|c}
& width is larger than & centered at 
\\[0.1cm]\hline
&&\\[-0.3cm]
$t^n(u)^{-1}, \quad n \le p'-3$ & $ \bar\lambda$ & $\frac\lambda2+(n-\frac12)\frac{\bar\lambda}2$
\\[0.1cm]
$t^{p'-2}(u)^{-1}$ & $ \frac{3\bar\lambda}2$ & $\frac\lambda2+(p'-\frac52)\frac{\bar\lambda}2$
\\[0.1cm]
$1+t^n(u)^{-1}, \quad n \le p'-4$ & $0$ & $\frac\lambda2+(n-\frac12)\frac{\bar\lambda}2$
\\[0.1cm]
$1+t^{p'-3}(u)^{-1}$ & $\frac{\bar\lambda}2$ & $\frac\lambda2+(p'-\frac72)\frac{\bar\lambda}2$
\\[0.1cm]
$1+t^{p'-2}(u)^{-1}$ & $\frac{\bar\lambda}2$ & $\frac\lambda2+(p'-\frac52)\frac{\bar\lambda}2$
\\[0.1cm]
$x(u)^{-1}$ & $2\bar\lambda$ & $\frac\lambda2+(p'-1)\frac{\bar\lambda}2$
\\[0.1cm]
$\big(1+\omega^{p'} x(u)^{-1}\big)\big(1+x(u)^{-1}\big)\big(1+\omega^{-{p'}} x(u)^{-1}\big)$ & $0$ & $\frac\lambda2+(p'-1)\frac{\bar\lambda}2$
\\[0.1cm]
$w(u)$ & $0$ & $\frac\lambda2+(p'-1)\frac{\bar\lambda}2$
\\[0.1cm]
$1+w(u)$ & $0$ & $\frac\lambda2+(p'-1)\frac{\bar\lambda}2$
\\[0.1cm]
${\bar t}^n(u)^{-1}, \quad n \le p'-3$ & $ \bar\lambda$ & $\frac\lambda2+(n+\frac12)\frac{\bar\lambda}2$
\\[0.1cm]
${\bar t}^{p'-2}(u)^{-1}$ & $ \frac{3\bar\lambda}2$ & $\frac\lambda2+(p'-\frac32)\frac{\bar\lambda}2$
\\[0.1cm]
$1+{\bar t}^n(u)^{-1}, \quad n \le p'-4$ & $0$ & $\frac\lambda2+(n+\frac12)\frac{\bar\lambda}2$
\\[0.1cm]
$1+{\bar t}^{p'-3}(u)^{-1}$ & $\frac{\bar\lambda}2$ & $\frac\lambda2+(p'-\frac52)\frac{\bar\lambda}2$
\\[0.1cm]
$1+{\bar t}^{p'-2}(u)^{-1}$ & $\frac{\bar\lambda}2$ & $\frac\lambda2+(p'-\frac32)\frac{\bar\lambda}2$
\\[0.1cm]
${\bar x}(u)$ & $2\bar\lambda$ & $\frac\lambda2+(p'-2)\frac{\bar\lambda}2$
\\[0.1cm]
$\big(1+\omega^{p'}{\bar x}(u)\big)\big(1+{\bar x}(u)\big)\big(1+\omega^{-p'}{\bar x}(u)\big)$ & $0$ & $\frac\lambda2+(p'-2)\frac{\bar\lambda}2$
\\[0.1cm]
$\bar w(u)$ & $\bar \lambda$ & $\frac\lambda2+(p'-2)\frac{\bar\lambda}2$
\\[0.1cm]
$1+\bar w(u)$ & $0$ & $\frac\lambda2+(p'-2)\frac{\bar\lambda}2$
\end{tabular}
\caption{The analyticity strips for the various functions.}
\label{tab:analyticity.A21.dual}
\end{center}
\end{table}

We make a change of variables for the $Y$-system functions in such a way that the central lines of the analyticity strips coincide with the real axis:
\begin{subequations}
\begin{alignat}{2}
&t^n(u)^{-1} = \amf^n\Big(\!-\!\tfrac {\ir \pi}{\lambdabr}\big(u-\tfrac \lambda 2 -(n-\tfrac12) \tfrac{\bar \lambda}2\big)\Big),
\qquad 
\hspace{0.25cm}\Amf^n(z) = 1+\amf^n(z),
\qquad n = 1, \dots, p'-2,
\\[0.15cm]
&x(u)^{-1} = \amf^{p'-1}\Big(\!-\!\tfrac {\ir \pi}{\lambdabr}\big(u-\tfrac \lambda 2 - (p'-1)\tfrac{\bar\lambda}2\big)\Big), 
\\[0.15cm]
&\Amf^{p'-1}(z) = \big(1 + \omega^{p'} \amf^{p'-1}(z)\big)\big(1 + \amf^{p'-1}(z)\big)\big(1 + \omega^{-p'} \amf^{p'-1}(z)\big),\label{eq:y(u).A21.dualc}
\\[0.15cm]\label{eq:y(u).A21.dual}
&w(u) = \amf^{p'}\Big(\!-\!\tfrac {\ir \pi}{\lambdabr}\big(u-\tfrac \lambda 2-(p'-1)\tfrac{\bar\lambda}2\big)\Big), 
\hspace{1.0cm} \Amf^{p'}(z) = 1+\amf^{p'}(z),
\end{alignat}
and
\begin{alignat}{2}
&\bar t^n(u)^{-1} = \bar\amf^n\Big(\!-\!\tfrac {\ir \pi}{\lambdabr}\big(u-\tfrac \lambda 2-(n+\tfrac12) \tfrac{\bar\lambda}2\big)\Big),
\qquad 
\hspace{0.25cm}\bar\Amf^n(z) = 1+\bar\amf^n(z),
\qquad n = 1, \dots, p'-2,
\\[0.15cm]
&\bar x(u) =\bar\amf^{p'-1}\Big(\!-\!\tfrac {\ir \pi}{\lambdabr}\big(u-\tfrac \lambda 2-(p'-3)\tfrac{\bar\lambda}2\big)\Big), 
\\[0.15cm]
&\bar\Amf^{p'-1}(z) = \big(1 + \omega^{p'} \bar\amf^{p'-1}(z)\big)\big(1 + \bar\amf^{p'-1}(z)\big)\big(1 + \omega^{-p'} \bar\amf^{p'-1}(z)\big),
\\[0.15cm]
&\bar w(u) = \bar\amf^{p'}\Big(\!-\!\tfrac {\ir \pi}{\lambdabr}\big(u-\tfrac \lambda 2-(p'-2)\tfrac{\bar\lambda}2\big)\Big), 
\hspace{1.0cm} \bar\Amf^{p'}(z) = 1+\bar\amf^{p'}(z).
\end{alignat}
\end{subequations}
In terms of these functions, the $Y$-system reads
\begin{subequations}
\label{eq:sym.Y.A21.dual}
\begin{alignat}{2}
\frac{\amf^{n-1}(z)\amf^{n+1}(z)}{\amf^{n}(z - \frac{\ir \pi}2) \amf^{n}(z + \frac{\ir \pi}2)} &= \frac{\Amf^{n-1}(z)\Amf^{n+1}(z)}{\bar\Amf^{n}(z)},
\label{eq:sym.Ya.A21.dual}\\[0.15cm]
\frac{\amf^{p'-3}(z-\frac{\ir \pi}4)\amf^{p'-1}(z)\bar\amf^{p'-1}(z - \ir \pi) \bar\amf^{p'-1}(z)\amf^{p'\!}(z)}{\amf^{p'-2}(z - \frac{3\ir \pi}4) \amf^{p'-2}(z + \frac{\ir \pi}4)\bar\amf^{p'\!}(z - \frac{\ir \pi}2)} &= \frac{\Amf^{p'-3}(z-\frac{\ir \pi}4)\Amf^{p'-1}(z)}{\bar\Amf^{p'-2}(z-\frac{\ir \pi}4)}, 
\label{eq:sym.Yb.A21.dual}\\[0.15cm]
\frac{\amf^{p'-2}(z+\frac{\ir \pi}4)\bar\amf^{p'\!}(z + \frac{\ir \pi}2)}{\big(\bar\amf^{p'-1}(z)\big)^3\amf^{p'\!}(z)} &= \frac{\Amf^{p'-2}(z + \frac{\ir \pi}4)}{\bar\Amf^{p'-1}(z)}, 
\label{eq:sym.Yc.A21.dual}\\[0.15cm]
\frac{\amf^{p'-1}(z)}{\bar\amf^{p'-1}(z)} &= \Amf^{p'\!}(z),\label{eq:sym.Yd.A21.dual}
\end{alignat}
and
\begin{alignat}{2}
\frac{\bar\amf^{n-1}(z)\bar\amf^{n+1}(z)}{\bar\amf^{n}(z - \frac{\ir \pi}2) \bar\amf^{n}(z + \frac{\ir \pi}2)} &= \frac{\bar\Amf^{n-1}(z)\bar\Amf^{n+1}(z)}{\Amf^{n}(z)},
\label{eq:sym.Ya.A21.dual.2}\\[0.15cm]
\frac{\bar\amf^{p'-3}(z + \frac{\ir \pi}4)\big(\bar\amf^{p'-1}(z)\big)^2\amf^{p'\!}(z)}{\bar\amf^{p'-2}(z - \frac{\ir \pi}4) \bar\amf^{p'-2}(z + \frac{3\ir \pi}4)\amf^{p'-1}(z)\amf^{p'-1}(z + \ir \pi)\bar\amf^{p'\!}(z + \frac{\ir \pi}2)} &= \frac{\bar\Amf^{p'-3}(z + \frac{\ir \pi}4)\bar\Amf^{p'-1}(z)}{\Amf^{p'-2}(z + \frac{\ir \pi}4)}, 
\label{eq:sym.Yb.A21.dual.2}\\[0.15cm]
\frac{\bar\amf^{p'-2}(z - \frac{\ir \pi}4)\bar\amf^{p'\!}(z - \frac{\ir \pi}2)}{\amf^{p'\!}(z)} &= \frac{\bar\Amf^{p'-2}(z - \frac{\ir \pi}4)}{\Amf^{p'-1}(z)}, 
\label{eq:sym.Yc.A21.dual.2}\\[0.15cm]
\frac{\amf^{p'-1}(z + \tfrac{\ir \pi}2)\bar\amf^{p'}(z)}{\bar\amf^{p'-1}(z - \tfrac{\ir \pi}2)} &= \bar\Amf^{p'\!}(z),\label{eq:sym.Yd.A21.dual.2}
\end{alignat}
\end{subequations}
for $n = 1, \dots, p'-2$.

In terms of the variable $z$, the analyticity strips are horizontal and centered on the real line. For $\amf^n(z)$ and $\bar\amf^n(z)$ with $n = 1, \dots, p'-3$, the width of the strips is $\ir \pi$. Our computer implementation also reveals that, in the $z$-plane, the zeros of all the functions are symmetrically distributed between the right and left half-planes, but not between the upper and lower half-planes. This implies that
\be
\label{eq:z-z.ids.A21.dual}
\amf^n(z + \ir \xi) = \amf^n(-z + \ir \xi)^*,\qquad \bar\amf^n(z + \ir \xi) = \bar\amf^n(-z + \ir \xi)^*,\qquad z,\xi \in \mathbb R.
\ee
This is true for the groundstate eigenvalues, but not for arbitrary eigenstates of the transfer matrix.

\subsubsection{Bulk and finite contributions}

The eigenvalues of the transfer matrices are related to the $Y$-system functions by
\begin{subequations}
\begin{alignat}{2}
\frac{T^{n,0}_0 T^{n,0}_1}{T^{n+1,0}_0 T^{n-1,0}_1} &= 1 + (t^{n}_0)^{-1},\qquad n = 1, \dots, p'-2,
\\
\frac{T^{0,p'-2}_0 T^{0,p'-2}_1 w_0}{f_{-1}T^{p'-2,0}_1\bar w_0} &= (1+\omega^{p'} x_0^{-1})(1+ x_0^{-1})(1+\omega^{-p'} x_0^{-1}),
\\
 \frac{T^{0,p'-1}_0 T^{p'-1,0}_1}{T^{p'-2,0}_1T^{0,p'-2}_1} & = 1+ w_0,
\end{alignat}
and
\begin{alignat}{2}
\frac{T^{0,n}_0 T^{0,n}_1}{T^{0,n+1}_0 T^{0,n-1}_1} &= 1 + ({\bar t}^{n}_0)^{-1},\qquad n = 1, \dots, p'-2,
\\
\frac{T^{p'-2,0}_1 T^{p'-2,0}_2 (T^{0,p'-2}_1)^2 w_0 }{f_{-1}(T^{p'-1,0}_1)^3\bar w_1} &= (1+\omega^{p'} {\bar x}_1)(1+{\bar x}_1)(1+\omega^{-p'} {\bar x}_1),
\\
\frac{T^{0,p'-1}_0 T^{p'-1,0}_0\bar w_0}{T^{p'-2,0}_1T^{0,p'-2}_0} &= 1+\bar w_0.
\end{alignat}
\end{subequations}
The functions $T^{n,0}(u)$, $T^{0,n}(u)$, $w(u)$ and $\bar w(u)$ are written as the product of their bulk and finite contributions: 
\begin{subequations}
\begin{alignat}{2}
T^{n,0}(u) &= \big(\kappa^{n}(u)\big)^N T^{n}_{\rm f}(u), \qquad w(u) = \big(\kappa_w(u)\big)^N w_{\rm f}(u),
\\[0.15cm]
T^{0,n}(u) &= \big(\bar\kappa^{n}(u)\big)^N \bar T^{n}_{\rm f}(u), \qquad \bar w(u) = \big(\kappa_{\bar w}(u)\big)^N \bar w_{\rm f}(u).
\end{alignat}
\end{subequations}
The bulk terms satisfy the functional equations
\begin{subequations}
\begin{alignat}{2}
\frac{\kappa^{n}(u) \kappa^{n}(u-\bar\lambda)}{\kappa^{n+1}(u) \kappa^{n-1}(u-\bar\lambda)} &= 1,\qquad n = 1, \dots, p'-2,
\\[0.15cm]
\frac{\bar\kappa^{p'-2}(u)\bar\kappa^{p'-2}(u-\bar\lambda)\kappa_w(u)}{\kappa^{p'-2}(u-\bar\lambda)\kappa_{\bar w}(u)} &= 
\frac{\sin (u-\lambda)}{\sin \lambda},
\\[0.15cm]
\frac{\kappa^{p'-1}(u-\bar \lambda)\bar\kappa^{p'-1}(u)}{\kappa^{p'-2}(u-\bar\lambda)\bar\kappa^{p'-2}(u-\bar \lambda)} &= 
1,
\end{alignat}
and
\begin{alignat}{2}
\frac{\bar\kappa^{n}(u)\bar\kappa^{n}(u-\bar\lambda)}{\bar\kappa^{n+1}(u) \bar\kappa^{n-1}(u-\bar\lambda)} &= 1,\qquad n = 1, \dots, p'-2,
\\[0.15cm]
\frac{\kappa^{p'-2}(u-\bar\lambda)\kappa^{p'-2}(u-2\bar\lambda)\big(\bar\kappa^{p'-2}(u-\bar\lambda)\big)^2\kappa_{w}(u)}{\big(\kappa^{p'-1}(u-\bar \lambda)\big)^3\kappa_{\bar w}(u-\bar \lambda)} &= 
\frac{\sin (u-\lambda)}{\sin \lambda},
\\[0.15cm]
\frac{\kappa^{p'-1}(u)\bar\kappa^{p'-1}(u)\kappa_{\bar w}(u)}{\kappa^{p'-2}(u-\bar\lambda)\bar\kappa^{p'-2}(u)} &= 1. 
\end{alignat}
\end{subequations}
Likewise, the finite terms satisfy the functional relations
\begin{subequations}
\label{eq:func.rel.A21.dual}
\begin{alignat}{2}
\frac{T_{\rm f}^{n}(u) T_{\rm f}^{n}(u-\bar\lambda)}{T_{\rm f}^{n+1}(u) T_{\rm f}^{n-1}(u-\bar\lambda)} &= 1 + t^{n}(u)^{-1},\qquad n = 1, \dots, p'-2,
\\[0.15cm]
\frac{\bar T^{p'-2}_{\rm f}(u) \bar T^{p'-2}_{\rm f}(u-\bar \lambda) w_{\rm f}(u)}{T^{p'-2}_{\rm f}(u-\bar \lambda) \bar w_{\rm f}(u)} &= \big(1+\omega^{p'} x(u)^{-1}\big)\big(1+x(u)^{-1}\big)\big(1+\omega^{-p'} x(u)^{-1}\big),
\\[0.15cm]
\frac{\bar T_{\rm f}^{p'-1}(u) T_{\rm f}^{p'-1}(u-\bar \lambda)}{T_{\rm f}^{p'-2}(u-\bar \lambda)\bar T_{\rm f}^{p'-2}(u-\bar \lambda)} & = 1+ w(u),
\end{alignat}
and
\begin{alignat}{2}
\frac{\bar T^{n}_{\rm f}(u) \bar T^{n}_{\rm f}(u-\bar \lambda)}{\bar T^{n+1}_{\rm f}(u) \bar T^{n-1}_{\rm f}(u-\bar \lambda)} &= 1 + {\bar t}^{n}(u)^{-1},\qquad n = 1, \dots, p'-2,
\\
\frac{T^{p'-2}_{\rm f}(u-\bar \lambda) T^{p'-2}_{\rm f}(u-2\bar \lambda) \big(\bar T^{p'-2}_{\rm f}(u-\bar \lambda)\big)^2 w_{\rm f}(u) }{\big(T^{p'-1}_{\rm f}(u-\bar \lambda)\big)^3\bar w_{\rm f}(u-\bar \lambda)} &= (1+\omega^{p'} {\bar x}(u-\bar\lambda))(1+ {\bar x}(u-\bar\lambda))(1+\omega^{-p'} {\bar x}(u-\bar\lambda)),
\\
\frac{\bar T^{p'-1}_{\rm f}(u) T^{p'-1}_{\rm f}(u)\bar w_{\rm f}(u)}{T^{p'-2}_{\rm f}(u-\bar \lambda)\bar T^{p'-2}_{\rm f}(u)} &= 1 + \bar w(u).
\end{alignat}
\end{subequations}
The initial conditions are $\kappa^{0}(u) = \bar\kappa^{0}(u) = \frac{\sin(u-\lambda)}{\sin \lambda}$ and $T^{0}_{\rm f}(u) = \bar T^{0}_{\rm f}(u) = 1$.
For the finite terms, we define
\begin{subequations}
\begin{alignat}{3}
T^{n}_{\rm f}(u) &= \bmf^n\Big(\!-\!\tfrac {\ir \pi}{\textrm{\raisebox{-0.06cm}{$\bar \lambda$}}}\big(u-\tfrac\lambda 2- (n-\tfrac32) \tfrac{\bar\lambda}2\big)\Big), \qquad
&&w_{\rm f}(u) = \bmf^{p'\!}\Big(\!-\!\tfrac {\ir \pi}{\textrm{\raisebox{-0.06cm}{$\bar \lambda$}}}\big(u-\tfrac\lambda 2- (p'-1) \tfrac{\bar\lambda}2\big)\Big),\\[0.15cm]
\bar T^{n}_{\rm f}(u) &= \bar \bmf^n\Big(\!-\!\tfrac {\ir \pi}{\textrm{\raisebox{-0.06cm}{$\bar \lambda$}}}\big(u-\tfrac\lambda 2- (n-\tfrac12) \tfrac{\bar\lambda}2\big)\Big),\qquad
&&\bar w_{\rm f}(u) = \bar \bmf^{p'\!}\Big(\!-\!\tfrac {\ir \pi}{\textrm{\raisebox{-0.06cm}{$\bar \lambda$}}}\big(u-\tfrac\lambda 2- (p'-2) \tfrac{\bar\lambda}2\big)\Big),
\end{alignat}
\end{subequations}
where $n = 1, \dots, p'-1$. We rewrite the relations \eqref{eq:func.rel.A21.dual} as
\begin{subequations}
\label{eq:bbb.A21.dual}
\begin{alignat}{2}
\frac{\bmf^n(z - \frac{\ir \pi}2) \bmf^n(z + \frac{\ir \pi}2)}{\bmf^{n-1}(z)\bmf^{n+1}(z)} &= \Amf^n(z), \qquad n = 1, \dots, p'-2,
\\[0.15cm]
\frac{\bar\bmf^{p'-2}(z - \frac{3\ir \pi}4)\bar\bmf^{p'-2}(z + \frac{\ir \pi}4)\bmf^{p'\!}(z)}{\bmf^{p'-2}(z - \frac{\ir \pi}4)\bar\bmf^{p'\!}(z - \frac{\ir \pi}2)} &= \Amf^{p'-1}(z),
\\[0.1cm]\
\frac{\bmf^{p'-1}(z+\frac{\ir\pi}4 )\bar\bmf^{p'-1}(z-\frac{\ir\pi}4)}{\bmf^{p'-2}(z-\frac{\ir \pi}4)\bar\bmf^{p'-2}(z+\frac{\ir \pi}4)} &= \Amf^{p'}(z),
\end{alignat}
and
\begin{alignat}{2}
\frac{\bar\bmf^n(z - \frac{\ir \pi}2) \bar\bmf^n(z + \frac{\ir \pi}2)}{\bar\bmf^{n-1}(z)\bar\bmf^{n+1}(z)} &= \bar\Amf^n(z), \qquad n = 1, \dots, p'-2,
\\[0.15cm]
\frac{\bmf^{p'-2}(z + \frac{3\ir \pi}4)\bmf^{p'-2}(z - \frac{\ir \pi}4)\big(\bar\bmf^{p'-2}(z + \frac {\ir \pi }4)\big)^2\bmf^{p'\!}(z)}{\big(\bmf^{p'-1}(z + \frac{\ir \pi}4)\big)^3\bar\bmf^{p'\!}(z + \frac{\ir \pi}2)} &= \bar\Amf^{p'-1}(z),
\\[0.1cm]
\frac{\bmf^{p'-1}(z-\frac{\ir \pi}4)\bar\bmf^{p'-1}(z+\frac{\ir \pi}4)\bar\bmf^{p'}(z)}{\bmf^{p'-2}(z+\frac{\ir \pi}4)\bar\bmf^{p'-2}(z-\frac{\ir \pi}4)} &= \bar\Amf^{p'}(z).
\end{alignat}
\end{subequations}

\subsubsection{Non-linear integral equations}

The functions $\amf^n(z)$ and $\bar\amf^n(z)$ have order-$N$ zeros on the imaginary axis. We define new functions $\ell^n(z)$ and $\bar\ell^n(z)$ where these order-$N$ zeros are removed:
\begin{subequations}
\label{eq:ell.a.A21.dual}
\begin{alignat}{2}
\ell^n(z) &= \frac{\amf^n(z)}{\Big[\eta\big(z-\frac{\ir \pi}2(p'-n-\tfrac32)\big)\Big]^{N}},\qquad
\ell^{p'-1}(z) = \amf^{p'-1}(z), \qquad \ell^{p'}(z) = \frac{\amf^{p'}(z)}{\big[\eta(z)\big]^{N}},
\\[0.15cm]
\bar\ell^n(z) &= \frac{\bar\amf^n(z)}{\Big[\eta\big(z+\frac{\ir \pi}2(p'-n-\tfrac32)\big)\Big]^{N}},
\qquad
\bar\ell^{p'-1}(z) = \bar\amf^{p'-1}(z), \qquad \bar\ell^{p'}(z) = \bar\amf^{p'}(z),
\end{alignat}
\end{subequations}
where $n = 1, \dots, p'-2$ and
\be
\eta(z) = \tanh\tfrac z{2p'-3}.
\ee
In terms of the these functions, we have the $Y$-system relations
\begin{subequations}
\begin{alignat}{2}
\frac{\ell^{2}(z)}{\ell^{1}(z - \frac{\ir \pi}2) \ell^{1}(z + \frac{\ir \pi}2)} &= \frac{\Amf^{2}(z)}{\bar\Amf^{1}(z)} \times \big[\eta\big(z-\tfrac{\ir \pi}2(p'-\tfrac32)\big)\big]^N, \\[0.15cm]
\frac{\bar\ell^{2}(z)}{\bar\ell^{1}(z - \frac{\ir \pi}2) \bar\ell^{1}(z + \frac{\ir \pi}2)} &= \frac{\bar\Amf^{2}(z)}{\Amf^{1}(z)} \times \big[\eta\big(z+\tfrac{\ir \pi}2(p'-\tfrac32)\big)\big]^N.
\end{alignat}
\end{subequations}
The other $Y$-system relations are obtained from the relations \eqref{eq:sym.Y.A21.dual} for $n>1$, by replacing the functions $\amf^n(z)$ and $\bar \amf^n(z)$ on the left sides by the corresponding functions $\ell^n$ and $\bar\ell^n$. As discussed in \cref{sec:braid.and.bulk.A21.dual}, for generic values of $\omega$, these functions have asymptotic values for $z \to \pm \infty$ that are finite and nonzero. We define the Fourier transform of their logarithmic derivative:
\begin{subequations}
\begin{alignat}{2}
&L^n(k) = \frac1{2\pi} \int \dd z\, \eE^{-\ir k z}\big[\log \ell^n(z)\big]', \qquad 
&&A^n(k) = \frac1{2\pi} \int \dd z\, \eE^{-\ir k z}\big[\log \Amf^n(z)\big]',
\\[0.15cm]
&\bar L^n(k) = \frac1{2\pi} \int \dd z\, \eE^{-\ir k z}\big[\log \bar \ell^n(z)\big]', \qquad 
&&\bar A^n(k) = \frac1{2\pi} \int \dd z\, \eE^{-\ir k z}\big[\log \bar\Amf^n(z)\big]',
\end{alignat}
\end{subequations}
where $n = 1, \dots, p'$. Here the integrals are performed from $-\infty + \ir \epsilon^1_n$ to $\infty + \ir \epsilon^1_n$ for the functions $L^n$ and $A^n$, and from $-\infty + \ir \epsilon^2_n$ to $\infty + \ir \epsilon^2_n$ for the functions $\bar L^n$ and $\bar A^n$, for certain small real parameters $\epsilon^1_n,\epsilon^2_n$. This allows the integration paths to avoid the zeros of the functions $\amf^{p'}(z)$, $\Amf^{p'-1}(z)$ and $\bar\Amf^{p'-1}(z)$ that lie on the real line. (The other functions are all analytic and non-zero inside their respective analyticity strips.) We shall see later that this peculiar form of the Fourier transforms is useful, with parameters $\epsilon^1_n$ and $\epsilon^2_n$ that can be chosen separately for different values of $n$. It will moreover turn out to be useful for the calculation to choose $\epsilon^1_n$ and $\epsilon^2_n$ to be negative numbers for all values of $n$. The inverse transforms are
\begin{subequations}
\begin{alignat}{2}
\big(\log \ell^n(z+\ir \epsilon^1_n)\big)' &=  \int_{-\infty}^\infty \dd k\, \eE^{\ir k (z+ \ir \epsilon_n^{1})} L^n(k), \qquad
\big(\log \Amf^n(z+\ir \epsilon^1_n)\big)' &=  \int_{-\infty}^\infty \dd k\, \eE^{\ir k (z+ \ir \epsilon_n^{1})} A^n(k),
\\[0.15cm]
\big(\log \bar\ell^n(z+\ir \epsilon^2_n)\big)' &=  \int_{-\infty}^\infty \dd k\, \eE^{\ir k (z+ \ir \epsilon_n^{2})} \bar L^n(k), \qquad
\big(\log \bar\Amf^n(z+\ir \epsilon^2_n)\big)' &=  \int_{-\infty}^\infty \dd k\, \eE^{\ir k (z+ \ir \epsilon_n^{2})} \bar A^n(k).
\end{alignat}
\end{subequations}
We also compute the Fourier transforms
\begin{subequations}
\begin{alignat}{2}
C(k) &= \frac N{2\pi} \int_{-\infty}^{\infty} \dd z\, \eE^{-\ir k z}\big[\log \eta\big(z-\tfrac{\ir \pi}{2}(p'-\tfrac32)\big)\big]' =  \frac{\ir N}{2 \cosh\big(\tfrac{\pi k}4(2p'-3)\big)},\\[0.15cm]
\bar C(k) &= \frac N{2\pi} \int_{-\infty}^{\infty} \dd z\, \eE^{-\ir k z}\big[\log \eta\big(z+\tfrac{\ir \pi}{2}(p'-\tfrac32)\big)\big]'=  -\frac{\ir N}{2 \cosh\big(\tfrac{\pi k}4(2p'-3)\big)} = - C(k).
\end{alignat}
\end{subequations}

The non-linear integral equations for the eigenvalues are obtained by first taking the Fourier transform of the
logarithmic derivative of the $Y$-system equations yielding
\begin{alignat}{2}
\label{eq:big.matrices.A21.dual}
&\hspace{-0.7cm}\begin{pspicture}(0,0)(0,0)
\psline[linewidth=0.5pt]{-}(0.4,0.20)(15.9,0.20)
\psline[linewidth=0.5pt]{-}(8.,-2.5)(8.,2.7)
\end{pspicture}
\left(\begin{smallmatrix}
-2 \cosh \!\frac{\pi k}2 & 1 & & & & & 0 & 0 & & &\\
1 & \hspace{0.3cm}\sddots\hspace{0.3cm} & 1 & & & & 0 &  \hspace{0.3cm}\sddots\hspace{0.3cm} & 0 & & &\\
& 1 & -2 \cosh \!\frac{\pi k}2 & 1 &  & & & 0 & 0 & 0 & &\\
& & \eE^{\frac{\pi k}4} & -\eE^{\frac{3\pi k}4}-\eE^{-\frac{\pi k}4} & 1 &1 & & &  0 & 0 & 1+\eE^{\pi k} & -\eE^{\frac{\pi k}2}\\
& &  & \eE^{-\frac{\pi k}4} & 0 & -1 & & & & 0 & -3 & \eE^{-\frac{\pi k}2} \\[0.1cm]
& & &  & 1 & 0 & & &  &  & -1 & 0
\\[0.2cm]
0 & 0 & & & & & -2 \cosh \!\frac{\pi k}2 & 1 & & & &\\[0.1cm]
0 &  \hspace{0.3cm}\sddots\hspace{0.3cm} & 0 & & & & 1 & \hspace{0.3cm}\sddots\hspace{0.3cm} & 1 & & &\\[0.1cm]
& 0 & 0 & 0 & & & & 1 & -2 \cosh \!\frac{\pi k}2 & 1 &  &\\
& & 0 & 0 & -1-\eE^{-\pi k} & 1 & & & \eE^{-\frac{\pi k}4} & -\eE^{-\frac{3\pi k}4}-\eE^{\frac{\pi k}4} & 2 & -\eE^{-\frac{\pi k}2}
\\
& &  & 0 & 0 & -1 & & &  & \eE^{\frac{\pi k}4} & 0 & \eE^{\frac{\pi k}2}\\
& &  &  & \eE^{-\frac{\pi k}2} & 0 & & &  &  & -\eE^{\frac{\pi k}2} & 1
\end{smallmatrix}\right)
\left(\begin{smallmatrix}
L^1 \\[0.1cm] L^2 \\[0.1cm] \svdots \\ L^{p'-2} \\ L^{p'-1} \\ L^{p'} \\[0.1cm]\hline\\[0.05cm] \bar L^1 \\[0.1cm] \bar L^2 \\[0.1cm]  \svdots \\ \bar L^{p'-2} \\ \bar L^{p'-1} \\ \bar L^{p'}
\end{smallmatrix}\right)
\nonumber\\[0.3cm]
 &\hspace{4cm}= 
 \begin{pspicture}(0,0)(0,0)
\psline[linewidth=0.5pt]{-}(0.4,0.1)(6.8,0.1)
\psline[linewidth=0.5pt]{-}(3.65,-2.2)(3.65,2.4)
\end{pspicture}
\left(\begin{smallmatrix}
0 & 1 &  &  &  &  & -1 &  &  &  &  &  \\[0.1cm]
1 & \sddotss & 1 &  &  &  &  & \hspace{0.1cm}\sddotss\hspace{-0.1cm} &  &  &  &  \\[0.1cm]
 & 1 & 0 & 1 &  &  &  &  & -1 &  &  &  \\
 &  & \eE^{\frac{\pi k}4} & 0 & 1 &  &  &  &  & -\eE^{\frac{\pi k}4} &  &  \\
 &  &  & \eE^{-\frac{\pi k}4} & 0 & 0 &  &  &  &  & -1 &  \\[0.1cm]
 &  &  &  & 0 & 1 &  &  &  &  &  & 0 \\[0.2cm]
-1 &  &  &  &  &  &  & 1 &  &  &  &  \\[0.1cm]
 & \hspace{0.1cm}\sddotss\hspace{-0.1cm} &  &  &  &  & 1 & \sddotss & 1 &  &  &  \\[0.1cm]
 &  & -1 &  &  &  &  & 1 & 0 & 1 &  &  \\
 &  &  & -\eE^{-\frac{\pi k}4} &  &  &  &  & \eE^{-\frac{\pi k}4}  & 0 & 1 &  \\
 &  &  &  & -1 &  &  &  &  & \eE^{\frac{\pi k}4}  & 0 & 0 \\[0.1cm]
 &  &  &  &  & 0 &  &  &  &  & 0 & 1 
\end{smallmatrix}\right)
\left(\begin{smallmatrix}
A^1 \\[0.1cm] 
A^2 \\[0.1cm] \svdots \\ 
A^{p'-2} \\ 
A^{p'-1} \\ 
A^{p'} \\[0.1cm]
\hline\\[0.05cm] 
{\bar A}^1 \\[0.1cm] 
{\bar A}^2 \\[0.1cm] \svdots \\ 
{\bar A}^{p'-2} \\ 
{\bar A}^{p'-1} \\ 
{\bar A}^{p'}
\end{smallmatrix}\right) \ + \ \left(\begin{smallmatrix}
C \\[0.12cm] 0 \\[0.12cm] \svdots \\[0.12cm] 0 \\[0.12cm] 0 \\[0.12cm] 0 \\[0.12cm]\hline\\[0.05cm] \bar C \\[0.12cm] 0 \\[0.12cm] \svdots \\[0.12cm] 0 \\[0.12cm] 0 \\[0.12cm] 0
\end{smallmatrix}\right).\end{alignat}
We note that the $Y$-system relations \eqref{eq:sym.Y.A21.dual} are written in such a way that the functions $\asf^{p'}(z)$, $\Asf^{p'-1}(z)$ and $\bar\Asf^{p'-1}(z)$ appear without shifts in their arguments. The calculation leading to \eqref{eq:big.matrices.A21.dual} thus does not require one to shift the path of integration for the Fourier transform of these functions, which would be problematic because of the real zeros of these functions.

We denote by $\hat M$ and $\hat N$ the matrices on the left and right sides of \eqref{eq:big.matrices.A21.dual}. We invert $\hat M$ and apply it to both sides of \eqref{eq:big.matrices.A21.dual} to find
\be
\begin{pmatrix} \vec L \\ \vec {\bar L} \end{pmatrix} = \hat K \cdot \begin{pmatrix} \vec A \\ \vec {\bar A}\end{pmatrix} +\hat M^{-1} \cdot \begin{pmatrix} \vec C \\ \vec {\bar C}\end{pmatrix}
\ee
where $\hat K = \hat M^{-1}\hat N$. The matrix elements of $\hat K$ and $\hat M^{-1}$ can be computed explicitly. In particular, we find that the matrix $\hat K$ satisfies the symmetry
\be
\label{sigma.and.K}
(\sigma\hat K)^\intercal = \sigma\hat K \big |_{k\to - k}, \qquad \sigma = \rm{diag}(1,1, \dots, 1,1, -1,-1).
\ee
Let us write $\hat K$ as 
\be
\label{eq:hatK.symm.A21.dual}
\hat K = \begin{pmatrix}
\hat K^{(11)} & \hat K^{(12)} \\
\hat K^{(21)} & \hat K^{(22)}
\end{pmatrix}
\ee
where the matrices $\hat K^{(ij)}$ have size $p' \times p'$. Their matrix entries can be computed explicitly. We find that
\be
\lim_{k\to \infty}\hat K^{(11)}_{p'-1,p'} = \lim_{k\to \infty}\hat K^{(12)}_{p',p'-1} = 1,
\qquad
\lim_{k\to -\infty}\hat K^{(11)}_{p',p'-1} = - \lim_{k\to- \infty}\hat K^{(21)}_{p'-1,p'} = 1.
\ee
The other entries of $\hat K^{(ij)}$ vanish for $k \to \pm \infty$. We apply the inverse transform and find that the kernel functions are 
\be
\label{eq:Kij.A21.dual}
K^{(ij)}_{nm}(z) = \frac1{2\pi}\int_{-\infty}^{\infty} \dd k\, \eE^{\ir k (z + \ir(\epsilon^i_n-\epsilon^j_m) )}\hat K^{(ij)}_{nm}, \qquad i = 1,2, \quad j = 1,2.
\ee
We choose the parameters $\epsilon_n^i$ such that 
\be
\epsilon_{p'-1}^1 > \epsilon_{p'}^1  > \epsilon_{p'-1}^2.
\ee 
With this choice, the integrals \eqref{eq:Kij.A21.dual} for $K^{(11)}_{p'-1,p'}$, $K^{(11)}_{p',p'-1}$, $K^{(12)}_{p',p'-1}$ and $K^{(21)}_{p'-1,p'}$ are well-defined, as their integrands vanish at both terminals. To apply the inverse transform of the terms involving $C$ and $\bar C$, we compute the difference between the rows $1$ and $p'+1$ of $\hat M^{-1}$. Let us define 
\be
\Delta^n(k) = (\hat M^{-1})_{n,1} - (\hat M^{-1})_{n,p'+1}, 
\qquad
\bar\Delta^n(k) = (\hat M^{-1})_{n+p',1} - (\hat M^{-1})_{n+p',p'+1}, \qquad
n = 1, \dots, p'.
\ee 
These functions have the explicit expressions
\begin{subequations}
\be
\Delta^n(k) =
\frac{\sinh\frac{\pi k}2(p'-1)}{\sinh\frac{3\pi k}2(p'-1)} \times
\left\{\begin{array}{cl}
-\,\eE^{\frac{\pi k(2p'-2-n)}2}-\eE^{-\frac{\pi k(2p'-2-n)}2}-\eE^{\frac{\pi k(n-1)}2}-\eE^{-\frac{\pi kn}2} + \eE^{-\frac{\pi k(n+1)}2}
& n \le  p'-2,\\[0.15cm]
-\,\eE^{\frac{\pi k(2p'-3)}4} - \eE^{-\frac{\pi k(2p'-3)}4} & n = p'-1,\\[0.15cm]
2\,\eE^{\frac{\pi k(2p'-3)}4} + \eE^{-\frac{\pi k(2p'-3)}4} - \eE^{\frac{\pi k(2p'-1)}4} + \eE^{-\frac{\pi k(2p'-1)}4} 
& n = p',\\[0.15cm]
\end{array}\right.
\ee
and
\be
\bar\Delta^n(k) =
\frac{\sinh\frac{\pi k}2(p'-1)}{\sinh\frac{3\pi k}2(p'-1)} \times
\left\{\begin{array}{cl}
\eE^{\frac{\pi k(2p'-2-n)}2}+\eE^{-\frac{\pi k(2p'-2-n)}2}+\eE^{-\frac{\pi k(n-1)}2}+\eE^{\frac{\pi kn}2} - \eE^{\frac{\pi k(n+1)}2}
& n \le  p'-2,\\[0.15cm]
-\,\eE^{\frac{\pi k(2p'-3)}4} - \eE^{-\frac{\pi k(2p'-3)}4} & n = p'-1,\\[0.15cm]
-\,\eE^{\frac{\pi k(2p'-1)}4} + \eE^{-\frac{\pi k(2p'-1)}4} + \eE^{\frac{\pi k(2p'-5)}4} - \eE^{-\frac{\pi k(2p'-5)}4}  
& n = p'.\\[0.15cm]
\end{array}\right.
\ee
\end{subequations}
Using this data, we compute the inverse Fourier transforms
\be
\Cmf^n(z) = \int_{-\infty}^{\infty} \dd k\, \eE^{\ir k z}C(k)\Delta^n(k).\ee
These can be computed explicitly using the integral
\begin{alignat}{2}
\mathcal I(z,\alpha) &= \int_{-\infty}^\infty \dd k\, \frac{\eE^{\ir k z}\eE^{\alpha k}\sinh\frac{\pi k}2(p'-1)}{\sinh\frac{3\pi k}2(p'-1) \cosh\frac{\pi k}2(p'-\frac32)} \nonumber\\[0.15cm]
&= \frac{2 \pi \ir}{1+\eE^{-4(z-\ir \alpha)}} \sum_j {\rm Res}\bigg(\frac{\eE^{\ir k (z-\ir \alpha)}\sinh\frac{\pi k}2(p'-1)}{\sinh\frac{3\pi k}2(p'-1) \cosh\frac{\pi k}2(p'-\frac32)},k_j\bigg)
\label{eq:I4}
\end{alignat}
where the sum runs over the poles $k_j$ of the integrand on the imaginary axis between $k = 0$ and $k = 4\ir$.
The resulting non-linear integral equations are
\begin{subequations}
\label{eq:NLIEs.A21.dual}
\begin{alignat}{2}
\big(\log \amf^{n}(z+\ir \epsilon_n^1)\big)' &= \fmf^n(z+\ir \epsilon_n^1)'+\Cmf^n(z+\ir \epsilon_n^1) \nonumber\\
&+ \sum_{m=1}^{p'} K^{(11)}_{nm} * (\log \Amf^{m})'(z+\ir \epsilon_m^1) + K^{(12)}_{nm} * (\log \bar\Amf^{m})'(z+\ir \epsilon_m^2),
\\
\big(\log \bar \amf^{n}(z+\ir \epsilon_n^2)\big)' &= \bar\fmf^n(z+\ir \epsilon_n^2)'+\bar\Cmf^n(z+\ir \epsilon_n^2) \nonumber\\
&+ \sum_{m=1}^{p'} K^{(21)}_{nm}  * (\log \Amf^{m})'(z+\ir \epsilon_m^1) + K^{(22)}_{nm}  * (\log \bar\Amf^{m})'(z+\ir \epsilon_m^2),
\end{alignat}
\end{subequations}
where $n = 1, \dots, p'$, and
\begin{subequations}
\begin{alignat}{2}
\fmf^n(z) &= \left\{
\begin{array}{cl}
N \log \eta\big(z-\frac{\ir\pi}2(p'-n-\tfrac32)\big)\quad& n = 1,\dots, p'-2,\\[0.3cm]
0 & n = p'-1,\\[0.2cm]
N \log \eta(z) & n = p',
\end{array}\right.
\\[0.15cm]
\bar\fmf^n(z) &= \left\{
\begin{array}{cl}
N \log \eta\big(z+\frac{\ir\pi}2(p'-n-\tfrac32)\big)\quad& n = 1,\dots, p'-2,\\[0.3cm]
0 & n = p'-1,p'.
\end{array}\right.
\end{alignat}
\end{subequations}
The convolution terms in \eqref{eq:NLIEs.A21.dual} are symmetric, namely
\be
\label{eq:K.syms.A21.dual}
(\sigma K)^\intercal(z) = (\sigma K)(-z), \qquad 
K(z) = \begin{pmatrix}
K^{(11)}(z) & K^{(12)}(z) \\
K^{(21)}(z) & K^{(22)}(z)
\end{pmatrix}.
\ee

\subsubsection{Scaling functions and scaling non-linear integral equations}

In \eqref{eq:NLIEs.A21.dual}, the dependence on $N$ appears only in the driving terms. Crucially, the scaling behavior of the functions $\fsf^n,\bar\fsf^n$ and $\Csf^n,\bar\Csf^n$ is different. The former has a non-trivial exponential scaling behavior for $z$ of order $\pm (p'-\frac32)\log N$. The latter also has an exponential behavior, but for $z$ of order $\pm \frac32(p'-1) \log N$. This is the dominant scaling behavior, namely 
\be
\lim_{N\to \infty} \fmf^n\big(\!\pm\!(z + \tfrac32(p'-1)\log N)+ \ir \epsilon^1_n\big) = 
\lim_{N\to \infty} \bar\fmf^n\big(\!\pm\!(z + \tfrac32(p'-1)\log N)+ \ir \epsilon^2_n\big) = 0.
\ee
The functions $\Cmf^n(z)$ have different scaling behaviors in the two limits. We therefore define
\begin{subequations}
\begin{alignat}{2}
\Csf^{n,\pm}(z \pm \ir \epsilon^1_n) &= \lim_{N\to \infty}\Cmf^n\big(\!\pm\!(z + \tfrac32(p'-1)\log N)+ \ir \epsilon^1_n\big), \\
\bar\Csf^{n,\pm}(z \pm \ir \epsilon^2_n) &= \lim_{N\to \infty} \bar\Cmf^n\big(\!\pm\!(z + \tfrac32(p'-1)\log N)+ \ir \epsilon^2_n\big).
\end{alignat}
\end{subequations}
These can be computed directly from the asymptotic behavior of $\mathcal I(z,\alpha)$ at $\pm \infty$, which are respectively dictated by the first and last pole in \eqref{eq:I4}:
\be
\lim_{N \to \infty}\mathcal I\big(\!\pm\!(z+\tfrac32(p'-1)\log N),\alpha\big) = \frac{2\,\eE^{-2(z\mp\ir \alpha)/(3(p'-1))}}{\sqrt 3(p'-1)\sin(\frac\pi 6\frac{p'}{p'-1})}.
\ee
After simplifiation, we find
\begin{subequations}
\be
\Csf^{n,\rho}(z+\ir \rho\, \epsilon^1_n) = \rho\,\frac {2 \,\eE^{-2(z+\ir \rho\, \epsilon^1_n)/(3(p'-1))}}{\sqrt 3(p'-1)}
\left\{\begin{array}{cl}
2\,\eE^{-\rho\, \ir \pi(2p'-1)/(6(p'-1))}\sin\big(\frac{\pi n}{3(p'-1)}\big)& n = 1, \dots, p'-2,\\[0.25cm]
-\rho\, \ir & n = p'-1,\\[0.25cm]
\sqrt 3\, \eE^{\rho\, \ir \pi/3}& n = p'.
\end{array}\right.
\ee
and
\be
\bar\Csf^{n,\rho}(z+ \ir \rho\,\epsilon^2_n) = \rho\, \frac {2\,\eE^{-2(z+\ir \rho\, \epsilon^2_n)/(3(p'-1))}}{\sqrt 3(p'-1)}
\left\{\begin{array}{cl}
2\,\eE^{\rho\, \ir \pi(2p'-1)/(6(p'-1))}\sin\big(\frac{\pi n}{3(p'-1)}\big)& n = 1, \dots, p'-2,\\[0.25cm]
-\rho\,\ir & n = p'-1,\\[0.25cm]
2 \sin\big(\frac{\pi}{3(p'-1)}\big)& n = p',
\end{array}\right.
\ee
\end{subequations}
where $\rho \in \{+,-\}$.

To compute the finite-size correction at order $\frac 1N$, we assume that the unknown functions appearing in \eqref{eq:NLIEs.A21.dual} are well-defined in these limits. We define both limits separately:
\begin{subequations}
\begin{alignat}{2}
\asf^{n,\pm}(z\pm\ir \epsilon^1_n) &= \lim_{N\to \infty} \amf^n\big(\!\pm\!(z +\tfrac32(p'-1) \log N)+\ir \epsilon^1_n\big), 
\\[0.2cm]
\Asf^{n,\pm}(z\pm\ir \epsilon^1_n) &= \lim_{N\to \infty} \Amf^n\big(\!\pm\!(z +\tfrac32(p'-1) \log N)+\ir \epsilon^1_n\big),
\\[0.2cm]
\bar \asf^{n,\pm}(z\pm\ir \epsilon^2_n) &= \lim_{N\to \infty} \bar\amf^n\big(\!\pm\!(z +\tfrac32(p'-1) \log N)+\ir \epsilon^2_n\big), 
\\[0.2cm]
\bar\Asf^{n,\pm}(z\pm\ir \epsilon^2_n) &= \lim_{N\to \infty} \bar\Amf^n\big(\!\pm\!(z +\tfrac32(p'-1) \log N)+\ir \epsilon^2_n\big),
\end{alignat}
\end{subequations}
where $n = 1, \dots, p'$. In the language of conformal field theory, the $+$ and $-$ limits correspond to {\it left-movers} and {\it right-movers}, respectively. These satisfy the integral equations
\begin{subequations}
\label{eq:scalingNLIEs.A21.dual}
\begin{alignat}{2}
\log \asf^{n,\pm}(z\pm\ir \epsilon_n^1) - \phi_n^{\pm} &= \gsf^{n,\pm}(z\pm\ir \epsilon_n^1) +  \sum_{m=1}^{p'} \big(K^{(11)}_{nm} * \log \Asf^{m,\pm}\big)(z\pm\ir \epsilon_m^1) + \big(K^{(12)}_{nm} * \log \bar\Asf^{m,\pm}\big)(z\pm\ir \epsilon_m^2), 
\\
\log \bar \asf^{n,\pm}(z\pm\ir \epsilon_n^2) - \bar\phi_n^{\pm} &= \bar\gsf^{n,\pm}(z\pm\ir \epsilon_n^2) +  \sum_{m=1}^{p'} \big(K^{(21)}_{nm} * \log \Asf^{m,\pm}\big)(z\pm\ir \epsilon_m^1) + \big(K^{(22)}_{nm} * \log \bar\Asf^{m,\pm}\big)(z\pm\ir \epsilon_m^2),
\end{alignat}
\end{subequations}
where $\phi_n^{\pm},\bar\phi_n^{\pm}$ are integration constants and 
\be
\gsf^{n,\rho}(z) = -\rho\,\frac{3(p'-1)}{2}\,\Csf^{n,\rho}(z), \qquad \bar\gsf^{n,\pm}(z) = -\rho\,\frac{3(p'-1)}{2}\,\bar\Csf^{n,\rho}(z).
\ee
We note that the overall sign $\rho$ comes from the terms with derivatives in \eqref{eq:NLIEs.A21.dual}, which pick up an extra minus sign for the $\rho = -$ case.

\subsubsection{Braid and bulk behavior}\label{sec:braid.and.bulk.A21.dual}

The scaling functions have finite asymptotics for $z \to \pm\infty$. For $z \to \infty$, these are obtained directly from the braid limits of the transfer matrix eigenvalues:
\begin{subequations}
\begin{alignat}{2}
\asf^{n,\rho}_\infty &= \bar \asf^{n,\rho}_\infty = \frac{(\omega^{1/2}-\omega^{-1/2})(\omega-\omega^{-1})}{(\omega^{n/2}-\omega^{-n/2})(\omega^{(n+3)/2}-\omega^{-(n+3)/2})}, \qquad n = 1, \dots, p'-2, 
\\[0.15cm]
\asf^{p'-1,\rho}_\infty &= (\bar\asf^{p'-1,\rho}_\infty)^{-1} =\frac{\omega^{(p'+1)/2}-\omega^{-(p'+1)/2}}{\omega^{(p'-1)/2}-\omega^{-(p'-1)/2}}, \\[0.15cm]
\qquad \asf^{p'\!,\rho}_\infty &= (\bar\asf^{p'\!,\rho}_\infty)^{-1} = \frac{(\omega^{p'}-\omega^{-p'})(\omega-\omega^{-1})}{(\omega^{(p'-1)/2}-\omega^{-(p'-1)/2})^2}.
\end{alignat}
\label{eq:braid.an.A12}
\end{subequations}
For $\gamma \in \big(0, \frac {2\pi}{p'+1}\big)$, these numbers are strictly positive and finite. These values allow us to compute the constants $\phi^\rho_n$ and $\bar \phi^\rho_n$ by studying the $z\to \infty$ asymptotics of \eqref{eq:scalingNLIEs.A21.dual}. We find that the constants all vanish:
\be
\phi^{\rho}_n = \bar\phi^{\rho}_n = 0.
\ee
Indeed, we have
\begin{alignat}{2}
\phi_n^\rho &= \lim_{z\to \infty} \bigg[\log \asf^{n,\rho}(z+\ir \rho\,\epsilon^1_n) -  \sum_{m=1}^{p'} \Big((K^{(11)}_{mn} * \log \Asf^{m,\rho})(z+\ir\rho\, \epsilon^1_m) + (K^{(12)}_{mn} * \log \bar\Asf^{m,\rho})(z+\ir \rho\,\epsilon^2_m)\Big)\bigg]
\nonumber\\
&=  \log \asf^{n,\rho}_\infty - \sum_{m=1}^{p'} \big(\hat K^{(11)}_{mn}(0)  \log \Asf^{m,\rho}_\infty + \hat K^{(12)}_{mn}(0) * \log \bar\Asf^{m,\rho}_\infty\big) = 0.
\end{alignat}
At the last step, we used the $Y$-system relations for the braid values, which is indeed described by the matrices $\hat K^{(ij)}(0)$. The same arguments apply to show that $\bar\phi^{\rho}_n$ is zero.

The behavior of the functions $\asf^{n,\rho}(z)$ and $\bar\asf^{n,\rho}(z)$ for $z \to - \infty$ is dictated by the driving terms in the non-linear integral equations \eqref{eq:scalingNLIEs.A21.dual}:
\begin{subequations}
\begin{alignat}{3}
\label{eq:bulk.an.A21.dual}
\asf^{n,\rho}_{-\infty} &= \bar \asf^{n,\rho}_{-\infty} = 0, \qquad &&n = 1, \dots, p'.
\end{alignat}
\end{subequations}
We note that the bulk values for $n=p'-1$ vanish because we chose $\epsilon^1_{p'-1}$ and $\epsilon^2_{p'-1}$ to be negative numbers.

\subsubsection{Finite-size correction and the dilogarithm technique}\label{sec:dilog.A21.dual}

We define the Fourier transform of the logarithmic derivative of the functions $\bmf^n(z)$ and $\bar\bmf^n(z)$: 
\be
B^n(k) = \frac1{2\pi} \int \dd z\, \eE^{-\ir k z}\big[\log  \bmf^n(z)\big]', \qquad
\bar B^n(k) = \frac1{2\pi} \int \dd z\, \eE^{-\ir k z}\big[\log\bar\bmf^n(z)\big]'.
\ee 
These integrals are performed from $-\infty$ to $\infty$, with imaginary shifts in the complex plane. The integration paths then avoid the poles of the $\bmf^{p'}(z)$ on the real line. For convenience, we choose that these contours are not shifted for the eigenvalues of the elementary transfer matrix.
Applying the Fourier transform to the relations \eqref{eq:bbb.A21.dual}, we find
\be
M \cdot\begin{pmatrix} \vec B \\ \vec {\bar B} \end{pmatrix} =  \begin{pmatrix} \vec A \\ \vec {\bar A} \end{pmatrix}
\ee
with
\be
\hspace{-0.1cm}M =
\begin{pspicture}(0,0)(0,0)
\psline[linewidth=0.5pt]{-}(0.5,0.15)(13.7,0.15)
\psline[linewidth=0.5pt]{-}(6.97,-2.65)(6.97,2.78)
\end{pspicture}
\left(\begin{smallmatrix}
2 \cosh \!\frac{\pi k}2 & -1 & & & && & 0 & 0 & & &\\
-1 & \hspace{0.3cm}\sddots\hspace{0.3cm} & -1 & && & & 0 &  \hspace{0.3cm}\sddots\hspace{0.3cm} & 0 & & &\\
& -1 & 2 \cosh \!\frac{\pi k}2 & -1 &  & && & 0 & 0 & 0 & &\\
& & -1 & 2\cosh \!\frac{\pi k}2 & -1 &0 & && &  0 & 0 & 0 & 0\\
& &  &- \eE^{\frac{\pi k}4} & 0 & 1 & && & & \eE^{-\frac{\pi k}4} + \eE^{\frac{3\pi k}4} & 0 & -\eE^{\frac{\pi k}2} \\[0.1cm]
& & & -\eE^{\frac{\pi k}4}& \eE^{-\frac{\pi k}4} & 0 && & &  & -\eE^{-\frac{\pi k}4} & \eE^{\frac{\pi k}4} & 0
\\[0.16cm]
0 & 0 & & & & && 2 \cosh \!\frac{\pi k}2 & -1 & & & &\\[0.1cm]
0 &  \hspace{0.3cm}\sddots\hspace{0.3cm} & 0 & & && & -1 & \hspace{0.3cm}\sddots\hspace{0.3cm} & -1 & & &\\[0.1cm]
& 0 & 0 & 0 & & && & -1 & 2 \cosh \!\frac{\pi k}2 & -1 &  &\\
& & 0 & 0 & 0 & 0 && & & -1 & 2 \cosh \!\frac{\pi k}2 & -1 &0\\
& &  & \eE^{\frac{\pi k}4}+ \eE^{-\frac{3\pi k}4} & -3\,\eE^{-\frac{\pi k}4} & 1 & && &  & 2\,\eE^{-\frac{\pi k}4} & 0 & -\eE^{-\frac{\pi k}2}\\
& &  & -\eE^{-\frac{\pi k}4} & \eE^{\frac{\pi k}4} & 0 & & &&  & -\eE^{\frac{\pi k}4} & \eE^{-\frac{\pi k}4} & 1
\end{smallmatrix}\right).
\ee
The matrix $M^{-1}$ can be computed explicitly. The entries of its first row are
\be
\label{eq:M.kernels.dual.A21}
(M^{-1})_{1,n} =  \frac{\sinh\frac{\pi k}2(p'-1)}{\sinh\frac{3\pi k}2(p'-1)}
\left\{\begin{array}{cl}
\eE^{\frac{\pi k(2p'-2-n)}2} + \eE^{\frac{\pi k n}2} + \eE^{-\frac{\pi k(2p'-2-n)}2}\quad& n = 1, \dots, p'-2,\\[0.2cm]
\eE^{\frac{\pi k(2p'-3)}4}\quad& n = p'-1,\\[0.2cm]
\eE^{-\frac{\pi k(2p'-1)}4} - \eE^{\frac{\pi k(2p'-3)}4}\quad & n = p',\\[0.2cm]
\eE^{-\frac{\pi k(n-p'+1)}2}-\eE^{-\frac{\pi k(n-p'-1)}2}\quad& n = p'+1, \dots, 2p'-2,\\[0.2cm]
-\,\eE^{\frac{\pi k(2p'-3)}4}\quad& n = 2p'-1,\\[0.2cm]
\eE^{\frac{\pi k(2p'-1)}4} - \eE^{\frac{\pi k(2p'-5)}4}\quad & n = 2p'.
\end{array}\right.
\ee
As a result, we find
\be
\label{eq:logb'}
\log \bmf^1(z) - \phi_0  = \sum_{n=1}^{p'} \big(\tilde K_n * \log \Amf^n\big)(z + \ir \epsilon_n^1) + \big(\tilde {\bar K}_{n} * \log \bar \Amf^n\big)(z + \ir \epsilon_n^2)
\ee
where $\phi_0$ is a constant and
\be
\tilde K_n (z) = \frac1{2\pi} \int_{-\infty}^{\infty} \dd k\, \eE^{\ir  k (z - \ir\epsilon^1_n)} M^{-1}_{1,n}\,,
\qquad
\tilde {\bar K}_n (z) = \frac1{2\pi} \int_{-\infty}^{\infty} \dd k\, \eE^{\ir  k (z - \ir\epsilon^2_n)} M^{-1}_{1,n+p'}\,, 
\ee
with $n = 1, \dots, p'$. These can be computed using the integral
\be
\frac1{2\pi} \int_{-\infty}^\infty \dd k \,\eE^{\ir k z}\eE^{\alpha k}\frac{\sinh\frac{\pi k}2(p'-1)}{\sinh\frac{3\pi k}2(p'-1)} = \frac{1}{ \pi\sqrt 3 (p'-1)}\frac{\sinh \frac{z-\ir \alpha}{3(p'-1)}}{\sinh \frac{z-\ir \alpha}{p'-1}}.
\ee
Writing the convolution integrals explicitly, we find
\begingroup
\allowdisplaybreaks
\begin{alignat}{2}
\log \bmf^1(z) - \phi_0  &= \int_{-\frac32(p'-1)\log N}^\infty \dd y \bigg[\sum_{n=1}^{p'} \tilde K_n\big(z-y-\tfrac32(p'-1)\log N\big) \log \Amf^n\big(y+\tfrac32(p'-1)\log N+\ir \epsilon_n^1\big) 
\nonumber\\[0.15cm]
&\hspace{2.2cm} +  \tilde {\bar K}_{n}\big(z-y-\tfrac32(p'-1)\log N\big) \log \bar\Amf^n\big(y+\tfrac32(p'-1)\log N+\ir \epsilon_n^2\big) 
\nonumber\\[0.15cm]
&\hspace{2.2cm}+  \tilde K_n\big(z+y + \tfrac32(p'-1)\log N\big) \log \Amf^n\big(\!-y-\tfrac32(p'-1)\log N+\ir \epsilon_n^1\big) 
\nonumber\\[0.15cm]
&\hspace{2.2cm}+ \tilde {\bar K}_{n}\big(z+y+\tfrac32(p'-1)\log N\big) \log \bar\Amf^n\big(\!-y-\tfrac32(p'-1)\log N+\ir \epsilon_n^2\big)\bigg]
\nonumber\\
& \simeq \frac1N\int_{-\infty}^\infty \dd y \bigg[\sum_{n=1}^{p'} \tilde K^-_n(-z+y) \log \Asf^{n,+}(y+\ir \epsilon_n^1) + \tilde {\bar K}^-_n(-z+y) \log \bar\Asf^{n,+}(y+\ir \epsilon_n^2) 
\nonumber\\[0.15cm]
&\hspace{2.2cm}+ \tilde K^+_n(z+y) \log \Asf^{n,-}(y-\ir \epsilon_n^1) + \tilde {\bar K}^+_n(z+y) \log \bar\Asf^{n,-}(y-\ir \epsilon_n^2)\bigg]
\label{eq:log.b.int.A21.dual}
\end{alignat}
\endgroup
where we defined
\be
\tilde K_n^\pm(z) = \lim_{N\to \infty} N \tilde K_n\big(\pm(z+\tfrac32(p'-1)\log N)\big), \qquad
\tilde {\bar K}^\pm_n(z) = \lim_{N\to \infty} N \tilde {\bar K}_n\big(\pm(z+\tfrac32(p'-1)\log N)\big).
\ee 
These can be computed explicitly:
\begin{subequations}
\begin{alignat}{2}
\tilde K^\rho_n(z) &= \frac {\rho\,\ir \eE^{-2(z- \ir \rho\, \epsilon_n^1)/(3(p'-1))}}{\sqrt 3\pi(p'-1)}
\left\{\begin{array}{cl}
2\,\eE^{-\rho\, \ir \pi/3}\sin\big(\frac{\pi n}{3(p'-1)}\big)& n = 1, \dots, p'-2,\\[0.25cm]
\eE^{-\rho\, \ir \pi/(6(p'-1))} & n = p'-1,\\[0.25cm]
-\sqrt 3\, \eE^{-\rho\, \ir \pi/(6(p'-1))}& n = p',
\end{array}\right.
\\[0.15cm]
\tilde {\bar K}^\rho_n(z) &= -\frac {\rho\,\ir \eE^{-2(z - \ir \rho\, \epsilon_n^2)/(3(p'-1))}}{\sqrt 3\pi(p'-1)}
\left\{\begin{array}{cl}
2\,\eE^{-\rho\, \ir \pi/3}\sin\big(\frac{\pi n}{3(p'-1)}\big)& n = 1, \dots, p'-2,\\[0.25cm]
\eE^{-\rho\, \ir \pi/(6(p'-1))} & n = p'-1,\\[0.25cm]
-2\,\eE^{\rho\, \ir \pi(2p'-3)/(6(p'-1))}& n = p',
\end{array}\right.
\end{alignat}
\end{subequations}
with $\rho \in \{+,-\}$.

To apply the dilogarithm technique, we define the integrals
\begin{alignat}{2}
\mathcal J^\rho = \int_{-\infty}^\infty \dd y \bigg[\sum_{n=1}^{p'}&\,\sigma^n\Big((\log \asf^{n,\rho})' \log \Asf^{n,\rho} -\log \asf^{n,\rho} (\log \Asf^{n,\rho})'\Big) 
\nonumber\\[0.15cm]&
+ \bar\sigma^{n}\Big((\log \bar\asf^{n,\rho})' \log \bar\Asf^{n,\rho}  -\log \bar\asf^{n,\rho} (\log \bar\Asf^{n,\rho})'\Big)\bigg].
\label{eq:Jrho}
\end{alignat}
where
\be
\sigma^n = 1, \qquad \bar\sigma^n = \left\{\begin{array}{cl}
1 & n = 1, \dots, p'-2,\\[0.15cm]
-1 &  n=p'-1,p',
\end{array}\right.
\ee
are the diagonal entries of the matrix $\sigma$ defined in \eqref{sigma.and.K}. In \eqref{eq:Jrho}, the arguments of the functions $\asf^{n,\rho}, \Asf^{n,\rho}$ and $\bar\asf^{n,\rho}, \bar\Asf^{n,\rho}$  are omitted and understood to be $y + \ir \rho\,\epsilon^{1}_n$ and $y + \ir \rho\,\epsilon^{2}_n$, respectively. 

We evaluate $\mathcal J^\rho$ in two ways. The first consists of replacing $\log \asf^{n,\rho}$ and $\log \bar\asf^{n,\rho}$ by their expressions~\eqref{eq:scalingNLIEs.A21.dual}. Many terms cancel out because of the symmetries \eqref{eq:K.syms.A21.dual} of the kernel. The only surviving contributions come from the driving terms, and the result reads
\be
\mathcal J^\rho = 2 \int_{-\infty}^\infty \dd y \, \bigg[\sum_{n=1}^{p'} \sigma^n \big(\gsf^{n,\rho}\big)'\log \Asf^{n,\rho} + \bar \sigma^n\big(\bar\gsf^{n,\rho}\big)'\log \bar\Asf^{n,\rho} \bigg],
\ee
where the functions $\gsf^{n,\rho}$ and $\bar \gsf^{n,\rho}$ have the arguments $y+\rho\ir \, \epsilon_n^1$ and $y+ \ir\rho\, \epsilon_n^2$ respectively.
Up to overall prefactors, these are precisely the integrals we wish to compute in \eqref{eq:log.b.int.A21.dual}. This is due to the relation
\be
\begin{pmatrix}\tilde K^\rho_n(z)\\[0.1cm]\tilde {\bar K}^\rho_n(z)\end{pmatrix}
 = \frac{\eE^{-\rho\,\ir \pi p'/(6(p'-1))}}{2\pi}
\begin{pmatrix}
\sigma^n\big(\gsf^{n,-\rho}(z)\big)'\\[0.1cm]
\bar\sigma^n\big(\bar \gsf^{n,-\rho}(z)\big)'
\end{pmatrix}, 
\qquad \rho = +,-.
\ee
This yields
\be
\log \bmf^1(z) - \phi_0 = \frac1{4 \pi N} \Big(\eE^{(4z+\ir \pi p')/(6(p'-1))}\mathcal J^+ + \eE^{-(4z+\ir \pi p')/(6(p'-1))}\mathcal J^-\Big).
\ee
The second way of computing the integral is to apply the derivatives explicitly, which yields
\begin{alignat}{2}
\mathcal J^\rho = \int_{-\infty}^\infty \dd y \,  \bigg[&
\sum_{n=1}^{p'-2}\frac{\dd \asf^{n,\rho}}{\dd y}\bigg(\frac{\log \Asf^{n,\rho}}{\asf^{n,\rho}} -\frac{\log \asf^{n,\rho}}{\Asf^{n,\rho}}\bigg) 
+ \sum_{n=1}^{p'-2}\frac{\dd \bar\asf^{n,\rho}}{\dd y}\bigg(\frac{\log \bar\Asf^{n,\rho}}{\bar\asf^{n,\rho}} -\frac{\log \bar\asf^{n,\rho}}{\bar\Asf^{n,\rho}}\bigg) 
 \nonumber\\&
+ \frac{\dd \asf^{p'-1,\rho}}{\dd y}\bigg(\frac{\log \Asf^{p'-1,\rho}}{\asf^{p'-1,\rho}} - \frac{\dd \Asf^{p'-1,\rho}}{\dd \asf^{p'-1,\rho}}\frac{\log \asf^{p'-1,\rho}}{\Asf^{p'-1,\rho}}\bigg)
+ \frac{\dd \asf^{p'\!,\rho}}{\dd y} \bigg(\frac{\log \Asf^{p'\!,\rho}}{\asf^{p'\!,\rho}} - \frac{\log \asf^{p'\!,\rho}}{\Asf^{p'\!,\rho}}\bigg) 
\label{eq:step1.A21.dual}\\&
 - \frac{\dd \bar\asf^{p'-1,\rho}}{\dd y}\bigg(\frac{\log \bar\Asf^{p'-1,\rho}}{\bar\asf^{p'-1,\rho}}  - \frac{\dd \bar\Asf^{p'-1,\rho}}{\dd \bar\asf^{p'-1,\rho}}\frac{\log\bar\asf^{p'-1,\rho}}{\bar\Asf^{p'-1,\rho}}\bigg) 
 - \frac{\dd \bar\asf^{p'\!,\rho}}{\dd y} \bigg(\frac{\log \bar\Asf^{p'\!,\rho}}{\bar\asf^{p'\!,\rho}} - \frac{\log \bar\asf^{p'\!,\rho}}{\bar\Asf^{p'\!,\rho}}\bigg)\bigg]. \nonumber
\end{alignat}
Dividing the integral in six parts and changing the integration variables from $y$ to $\asf^{n,\rho}$ and $\bar\asf^{n,\rho}$, we find 
\begin{alignat}{2}
\mathcal J^\rho &= \sum_{n=1}^{p'-2}\int_{\asf^{n,\rho}_{-\infty}}^{\asf^{n,\rho}_\infty} \dd \asf^{n,\rho} \bigg(\frac{\log \Asf^{n,\rho}}{\asf^{n,\rho}} -\frac{\log \asf^{n,\rho}}{\Asf^{n,\rho}}\bigg)
+ \sum_{n=1}^{p'-2}\int_{\bar\asf^{n,\rho}_{-\infty}}^{\bar\asf^{n,\rho}_\infty} \dd \bar\asf^{n,\rho} \bigg(\frac{\log \bar\Asf^{n,\rho}}{\bar\asf^{n,\rho}} -\frac{\log \bar\asf^{n,\rho}}{\bar\Asf^{n,\rho}}\bigg)
\label{eq:factor.of.2.A21.dual}\\[0.15cm]&
+ \int_{\asf^{p'-1,\rho}_{-\infty}}^{\asf^{p'-1,\rho}_\infty} \dd \asf^{p'-1,\rho} \bigg(\frac{\log \Asf^{p'-1,\rho}}{\asf^{p'-1,\rho}} - \frac{\dd \Asf^{p'-1,\rho}}{\dd \asf^{p'-1,\rho}}\frac{\log \asf^{p'-1,\rho}}{\Asf^{p'-1,\rho}}\bigg) 
+ \int_{\asf^{p',\rho}_{-\infty}}^{\asf^{p',\rho}_\infty} \dd \asf^{p',\rho}  \bigg(\frac{\log \Asf^{p',\rho}}{\asf^{p',\rho}} - \frac{\log\asf^{p',\rho}}{\Asf^{p',\rho}}\bigg) 
\nonumber\\[0.15cm]&
- \int_{\bar\asf^{p'-1,\rho}_{-\infty}}^{\bar\asf^{p'-1,\rho}_\infty} \dd \bar\asf^{p'-1,\rho} \bigg(\frac{\log \bar\Asf^{p'-1,\rho}}{\bar\asf^{p'-1,\rho}} - \frac{\dd \bar\Asf^{p'-1,\rho}}{\dd \bar\asf^{p'-1,\rho}}\frac{\log \bar\asf^{p'-1,\rho}}{\bar\Asf^{p'-1,\rho}}\bigg) 
- \int_{\bar\asf^{p',\rho}_{-\infty}}^{\bar\asf^{p',\rho}_\infty} \dd \bar\asf^{p',\rho}  \bigg(\frac{\log \bar\Asf^{p',\rho}}{\bar\asf^{p',\rho}} - \frac{\log\bar\asf^{p',\rho}}{\bar\Asf^{p',\rho}}\bigg) 
\nonumber
\end{alignat}
where
\begin{subequations}
\begin{alignat}{2}
\Asf^{n,\rho} &= 1 + \asf^{n,\rho}, \qquad \bar\Asf^{n,\rho} = 1 + \bar\asf^{n,\rho}, \qquad n = 1, 2,\dots, p'-2, p',
\\[0.15cm] 
\Asf^{p'-1,\rho} &= (1 + \omega^{p'} \asf^{p'-1,\rho})(1 + \asf^{p'-1,\rho})(1 + \omega^{-p'} \asf^{p'-1,\rho}),
\\[0.1cm] 
\bar\Asf^{p'-1,\rho} &= (1 + \omega^{p'} \bar\asf^{p'-1,\rho})(1 + \bar\asf^{p'-1,\rho})(1 + \omega^{-p'} \bar\asf^{p'-1,\rho}).
\end{alignat}
\end{subequations}
We note that in order to go from \eqref{eq:step1.A21.dual} to \eqref{eq:factor.of.2.A21.dual}, which only involves integrals with real terminals, one must consider how the functions $\asf^n(y+\ir \epsilon^1_n)$ and $\bar\asf^n(y+\ir \epsilon^2_n)$ wind around the points $0$ and $-1$, where the integrand has poles. Here we observe using our computer implementation that the only functions with a non-trivial winding are the functions $\asf^{p'-1}$ and $\bar\asf^{p'-1}$. For $\epsilon^1_{p'-1},\epsilon^2_{p'-1}<0$, they do not wind around the point $-1$, but they do wind around the origin. They in fact both wind around this point the same number of times and in the same direction. Because their contributions in \eqref{eq:step1.A21.dual} come with opposite signs, the resulting residues cancel out and the result \eqref{eq:factor.of.2.A21.dual} is indeed correct.

The resulting expressions for $\mathcal J^\rho$ are therefore a combination of regular integrals. They are in fact equal for both values of $\rho$. Setting $\omega = \eE^{\ir \gamma}$, they evaluate to
\be
\mathcal J^\rho =  \frac{\pi^2}3\bigg(2-\frac{6\gamma^2 p'}{\pi^2}\bigg) =  \frac{\pi^2}3\bigg(2-\frac{6\gamma^2}{\pi(\pi-\lambda)}\bigg), 
\qquad \gamma \in \big(0,\tfrac{2\pi}{p'+1}\big).
\ee
The proof of this result is given in \cref{app:A21.dual}. The final result is
\be
\log \bmf(z) - \phi_0 \simeq \frac{\pi \cosh \frac{4z+\ir \pi p'}{6(p'-1)}}{6 N} \bigg(2- \frac{6\gamma^2 p'}{\pi^2}\bigg),
\ee
and therefore
\be
\log T_{\rm f}(u) \simeq \frac{\pi \sin \frac{2 \pi u}{3\lambda}}{6 N} \bigg(2- \frac{6\gamma^2 p'}{\pi^2}\bigg), \qquad
\log \bar T_{\rm f}(u) \simeq \frac{\pi \sin \frac{2 \pi (\lambda - u)}{3\lambda}}{6 N} \bigg(2- \frac{6\gamma^2 p'}{\pi^2}\bigg),
\ee
where we used the crossing relation $\bar T_{\rm f}(u) = T_{\rm f}(\lambda - u)$.
The constant $\phi_0$ was found to equal zero using $T_{\rm f}(u=0)=1$.
This result is precisely \eqref{eq:fsc.general} with $c-24 \Delta$ and $\vartheta(u)$ given in \eqref{eq:fsc.A21}. 

\section{Conclusion}\label{sec:conclusion}

In this paper, we computed the groundstate finite-size corrections for the twisted $A_1^{(1)}$, $A_2^{(1)}$, $A_2^{(2)}$ vertex and loop models at roots of unity. This was achieved by converting the recently obtained $Y$-systems into nonlinear integral equations in the form of TBA equations. Using various dilogarithm identities, stated and proved in \cref{app:dilogs}, these can be solved analytically for the conformal data $c-24 \Delta$ with the methods of Kl\"umper and Pearce. Here $c$ is the central charge and $\Delta$ is the conformal weight associated with the groundstate. The resulting expressions for the principal and dual series are in agreement with previous analytic results for the central charges and conformal weights obtained by different methods. 

Strictly speaking, one cannot obtain separately the central charge $c$ and the conformal weights $\Delta$ from the finite-size correction, as only the difference $c-24\Delta$ can be measured. Without further information, there are two possible consistent interpretations of this conformal data. The first interpretation is that the continuum scaling limit of a twisted model is described by a unitary CFT. In this interpretation, this CFT describes a lattice model labeled by a pair $(\lambda, \gamma)$. The minimal conformal weight associated with the groundstate is $\Delta=\Delta_\text{min} = 0$. The quantity $c-24 \Delta$ appearing in the finite-size correction on the right sides of \eqref{eq:central.charges.2} is then simply the central charge $c(\lambda, \gamma)$ of the $(\lambda, \gamma)$ model. This is the usual interpretation for vertex models with generic parameters.

A second interpretation is possible, where the conformal field theories, instead of being labeled by pairs $(\lambda, \gamma)$, are labeled by the single parameter $\lambda$. The groundstate of the theory then corresponds to a special choice $\gamma = \gamma^*$ of the twist parameter, and is assumed to have the conformal weight $\Delta = 0$.  This second interpretation is for instance appropriate for the loop models whose continuum scaling limits are described by logarithmic CFTs. In such cases, the twist $\gamma=\gamma^*$ is usually tuned to ensure the equality of the contractible and non-contractible fugacities $\alpha=\beta$. For the loop models considered here, the required specialisations for the twist parameter $\gamma$ are given in \eqref{TwistChoices}. The central charges $c(\lambda)$ of the corresponding logarithmic CFTs are then obtained from the differences $c-24 \Delta$ in \eqref{eq:central.charges.2} by specialising to $\gamma = \gamma^*$:
\begin{subequations}
\begin{alignat}{4}
&A_1^{(1)}:\qquad&
\displaystyle c(\lambda)  &= 
\displaystyle 1-\frac{6\lambda^2}{\pi(\pi-\lambda)},\hspace{1.1cm}\mbox{$0<\lambda<\pi$,}\\[2pt]
&A_2^{(1)}: \qquad&
\displaystyle c(\lambda) &= 
\displaystyle 2-\frac{6\lambda^2}{\pi(\pi-\lambda)}, \hspace{1.1cm}\mbox{$0<\lambda<\pi$,}
\\
&A_2^{(2)}:\qquad&
\displaystyle c(\lambda) &= \left\{\begin{array}{ll}
\displaystyle 1-\frac{3(\pi-4\lambda)^2}{\pi(\pi-2\lambda)},&\mbox{$0<\lambda<\frac{\pi}{2}$,}\\[8pt]
\displaystyle \frac32-\frac{3(3\pi-4\lambda)^2}{2\pi(\pi-\lambda)},&\mbox{$\frac{2\pi}{3}<\lambda<\pi$.}
\end{array}\right.
\end{alignat}
\end{subequations}
They depend only on $\lambda$. The other values of $\gamma$ are then described in the CFT in terms of the presence of fields in the theory. On the cylinder, setting the fugacity $\alpha \neq \beta$ to non-contractible loops corresponds to inserting a field at infinity that has a conformal dimension $\Delta(\lambda,\gamma)$. The differences appearing in \eqref{eq:central.charges.2} are then equal to $c(\lambda) - 24 \Delta(\lambda,\gamma)$, allowing us to solve for $\Delta(\lambda,\gamma)$.

This second interpretation also produces a consistent framework for the vertex models. In this case, the special value of the twist parameter is $\gamma^* = 0$ and corresponds to the untwisted model. The corresponding central charges $c(\lambda)$ of the vertex model are obtained by specialising \eqref{eq:central.charges.2} to this value,
\begin{subequations}
\label{eq:central.charges.3}
\begin{alignat}{4}
&A_1^{(1)}:\qquad&
\displaystyle c(\lambda) &= 
\displaystyle 1,\hspace{1.0cm}\mbox{$0<\lambda<\pi$,}
\\[0.3cm]
&A_2^{(1)}: \qquad&
\displaystyle c(\lambda) &= 
2, \hspace{1.0cm} 0<\lambda<\pi,
\\[0.3cm]
&A_2^{(2)}:\qquad&
\displaystyle c(\lambda) &= \left\{\begin{array}{cc}
\displaystyle 1,& 0<\lambda<\frac{\pi}{2},\\[0.15cm]
\displaystyle \tfrac32,&\frac{2\pi}{3}<\lambda<\pi.
\end{array}\right.
\end{alignat}
\end{subequations}
They are independent of $\lambda$.
On the cylinder, a non-zero value for the twist $\gamma$ is then accounted for in the CFT by the insertion of a field at infinity with a conformal dimension $\Delta(\lambda, \gamma)$. The differences appearing in \eqref{eq:central.charges.2} are again equal to $c(\lambda) - 24 \Delta(\lambda,\gamma)$, allowing us to solve for $\Delta(\lambda,\gamma)$.

By analysing just the groundstate, we have found results consistent with the known finite-size corrections of the twisted $A_1^{(1)}$, $A_2^{(1)}$, $A_2^{(2)}$ vertex and loop models. The advantage of the current methods, however, is that they can be applied very generally to calculate the conformal energies of all finite excitations and hence conformal partition functions. This was achieved in the case of critical bond percolation on the square lattice ${\cal LM}(2,3)$~\cite{MDKP2017}. In principle, this analysis can also be applied to critical site percolation on the triangular lattice. This is the limiting case ${\cal DLM}(2,3)$~\cite{MDP19} of the $A_2^{(2)}$~loop model with $\lambda\to\frac{\pi}{3}$, where the limit is taken after the face operator is renormalised to remove the factors $\sin 3\lambda$ in the denominators. This mapping of the usual lattice model of site percolation to the loop model is described in \cite{FeherNien2015}. The analysis of finite excitations and conformal partition functions of site percolation would provide a highly nontrivial test of the meaning of {\em universality} in the setting of these logarithmic CFTs. We hope to return to this problem at a later time.

Although we focussed on the principal and dual series, it is expected that the methods of this paper will generalize to all non-principal series described by fractional values of $\frac\lambda\pi$.
In the context of the $A_1^{(1)}$ models, this entails solving the so-called ``snake" $Y$-systems discussed in \cite{Tateo,KSS}. For the $A_2^{(1)}$ and $A_2^{(2)}$ models, the analog of the ``snake" $Y$-systems are not yet known. It will be interesting to derive these systems of functional equations and to study prototypical examples. Lastly, it would also be of interest to apply these techniques to the $A_2^{(2)}$ models in the non-compact regime ($\frac\pi 2 < \lambda < \frac{2\pi}3$), where many eigenvalues share the same $\frac1N$ finite-size correction and the degeneracy is only lifted at order $\frac1{N(B+\log N)^2}$, for some constant $B$ \cite{VJS2014}.

\goodbreak
\section*{Acknowledgments} 

AMD was an FNRS Postdoctoral Researcher under the project CR28075116. AK~is grateful to DFG (Deutsche Forschungsgemeinschaft) for financial support in the framework of the research unit FOR 2316. All three authors acknowledge support from the EOS-contract O013018F and hospitality from the Universit\'e catholique de Louvain where early stages of this project were done. AMD and PAP acknowledge the hospitality and support of the Bergische Universit\"at Wuppertal where later stages of this work were done. Part of this work was also carried out while PAP was visiting the APCTP as an ICTP Visiting Scholar. We also thank an anonymous referee for drawing our attention to an inconsistency for the dual $A_2^{(1)}$ models in an earlier version.

\appendix
\section{Proofs of the dilogarithm identities}\label{app:dilogs}

\subsection{Solution for the general case}

\subsubsection{Standardised non-linear integral equations}
\label{sec:standardization}

For studying the finite-size corrections of the eigenvalues, we obtained non-linear integral equations for the functions
$\amf^n(z)$. These take the general form
\be
\label{NLIE.A1}
\log \amf^n(z) = \dmf^n(z) + \sum_m K_{n,m} \ast \log \Amf^{m} (z),
\ee
where $K$ is a kernel matrix with sufficiently well-behaved entries
$K_{n,m}$. The argument $z$ of the convolution, indicated by $\ast$, must not
be misunderstood as
the argument of the second function of the convolution. The objects
$\dmf^n(z)$ are explicit driving terms depending
on the argument $z$.

For most values of $n$, the functions $\Amf^{n}$ are related to $\amf^n$ as
\be
\label{NLIE.A2}
\Amf^n(z)=1+\amf^n(z).
\ee
In certain special cases, the function $\Amf^n(z)$ is defined as $\Amf^n(z)=1-\amf^n(z)$. 
Finally, in each of the six derivations performed in \cref{sec:A11,sec:A22,sec:A21}, one of the functions 
$\Amf^n(z)$ is defined as a quadratic or cubic polynomial in $\amf^n(z)$:
\be
\label{eq:Amf.special.case}
\Amf^{\rho-1}(z) = \prod_j \big(1 + \omega^{\rho j} \amf(z)\big), \qquad 
\rho = \left\{\begin{array}{ll}
p', & A_1^{(1)}, A_2^{(1)},
\\[0.1cm]
b, & A_2^{(2)}.
\end{array}\right.
\ee

The first step of our calculation will be to recast these non-linear integral equations in a standardised form
where the functions $\Amf^n(z)$ are defined as $\Amf^n(z)=F^n\big(\amf^n(z)\big)$ with $F^n(\amf)$ either
equal to $1+\amf$ or $1-\amf$ for all $n$. This will be achieved by defining new functions $\amf^{(\rho-1,j)}(z)$ and $\Amf^{(\rho-1,j)}(z)$ in such a way that the
function $\Amf^{\rho-1}(z)$ in \eqref{eq:Amf.special.case} takes the form
\be
\Amf^{\rho-1}(z) = \prod_j \Amf^{(\rho-1,j)}(z), \qquad  \Amf^{(\rho-1,j)}(z)= 1 + \amf^{(\rho-1,j)}(z), \qquad \amf^{(\rho-1,j)}(z) = \omega^{\rho j} \amf(z).
\ee
The non-linear integral equations are then easily rewritten in terms of these new functions. They are in fact precisely of the form \eqref{NLIE.A1}, where the indices $n$ and $m$ in this equation are allowed to take the values $(\rho-1,j)$ in addition to the usual integer values (with the value $\rho-1$ now removed). The resulting non-linear integral equation for $\amf^{(\rho-1,j)}(z)$ has a term proportional to $j\rho \log \omega$ on the right-hand side, which for convenience we incorporate in the driving term $\dmf(z)$ in \eqref{NLIE.A1}.
 
We will further use the fact that the kernel matrix $K$ is symmetric:
$K^\intercal(-z)=K(z)$. In three of the cases to be treated, $K$ is not symmetric, but
there exists a set of numbers $\sigma^n \in \{+1,-1\}$ such that the product of
the diagonal matrix $\sigma=\textrm{diag}(\sigma^1, \sigma^2, \dots)$ with $K$ is a
symmetric kernel matrix:
\be
\label{NLIE.A3}
(\sigma K)^\intercal(-z) = \sigma K(z).
\ee
In standard cases, the matrix $\sigma$ is simply the identity matrix. From the
derivations of \cref{sec:A11,sec:A22,sec:A21}, we observe that in the cases with
$\sigma^n = -1$, the
corresponding function $F^n(\amf)$ may
equal $1-\asf$ or $1+\asf$. We therefore write generally 
\be
\label{eq:Fn}
F^n(\amf) = 1 + \varepsilon^n\amf
\ee
where $\varepsilon^n \in \{+1,-1\}$.

For the study of the finite-size corrections to the eigenvalues, we make use of
the observation that a suitable scaling limit of the functions $\amf^n(z)$
exists. To this end, we look at the scaling limit of the driving terms
$\dmf^n(z)$ upon replacing $z$ by $z+r\log N$, for a
suitable choice of the constant~$r$, and take the limit $N\to\infty$. We have 
\be
\label{NLIE.A4}
\dsf^n(z)=\lim_{N\to\infty}\dmf^n(z+r\log N)=-\delta^n\eE^{-z/r}+\phi^n(\omega)
\ee
where $\delta^n$ has a non-negative real part. The extra terms $\phi^n(\omega)$ are independent of $z$ and are present only when $n$ takes the values $(\rho-1,j)$ with $j \neq 0$.
We now assume that the scaling limit of the functions $\amf^n$ also exists:
\be
\label{NLIE.A5}
\asf^n(z)=\lim_{N\to\infty}\amf^n(z+r\log N).
\ee
These functions satisfy the integral equations
\be
\label{NLIE.A1sc}
\log \asf^n(z) = \dsf^n(z) + \sum_m K_{n,m} \ast \log \Asf^{m}(z).
\ee
Of course, these are still non-linear because of the same relations
between $\asf^n$ and $\Asf^n$ as between $\amf^n$ and $\Amf^n$. We are not able 
to solve these equations in a general manner. Yet, we will evaluate exactly a certain integral
involving the auxiliary functions $\log \Asf^{n}$. The necessary manipulations
use among other properties the symmetry \eqref{NLIE.A3} of the matrix $\sigma K(z)$.

\subsubsection{First use of the kernel's symmetry: derivation of the integral identity}\label{sec:first.use}

We treat the most general case where $\sigma$ is the diagonal matrix $\textrm{diag}(\sigma^1, \sigma^2, \dots)$. 
We take the derivative of \eqref{NLIE.A1sc} with respect to the argument $z$, which is indicated by a
prime $'$,  and multiply with $\sigma$. To show the structure of the
equation most clearly, we suppress in the formula the explicit dependence on $z$:
\be
\label{NLIE.A1scd}
\sigma^n\left(\log \asf^n\right)' = \sigma^n(\dsf^n)' + \sum_m(\sigma K)_{n,m} \ast (\log \Asf^{m})'.
\ee
Next we multiply by the function $\log \Asf^{n}$, sum over $n$ and integrate over
the argument $z$ from $-\infty$ to $\infty$. We indicate this last operation by the short-hand notation $\int$, omitting the integration terminals and the infinitesimal factor $\dd z$. This yields
\be
\label{NLIE.A6}
\int \sum_n\sigma^n\left(\log \Asf^{n}\right)\left(\log \asf^n\right)' = \int \sum_n\sigma^n\left(\log \Asf^{n}\right)(\dsf^n)' + \int \sum_{n,m}\left(\log \Asf^{n}\right)(\sigma K)_{n,m} \ast (\log \Asf^{m})'.
\ee
Next we take \eqref{NLIE.A1sc} as it is and multiply it with $\sigma$. Then we
multiply this by the derivative of the function $\log \Asf^{n}$ with respect to
$z$, sum over $n$ and integrate:
\be
\label{NLIE.A7}
\int\sum_n\sigma^n\left(\log \Asf^{n}\right)'\left(\log \asf^n\right) = \int\sum_n\sigma^n\left(\log \Asf^{n}\right)'(\dsf^n) + \int\sum_{n,m}\left(\log \Asf^{n}\right)'(\sigma K)_{n,m} \ast (\log \Asf^{m}).
\ee
A most important observation is that the two double integrals in
\eqref{NLIE.A6} and \eqref{NLIE.A7} are identical: due to the symmetry
\eqref{NLIE.A3}, the integrals turn into each other upon simultaneously exchanging the
discrete summation variables $n,m$ and the two (not shown) variables of
integration. Hence the difference of \eqref{NLIE.A6} and \eqref{NLIE.A7}
yields an identity of single integrals
\be
\label{NLIE.A8}
\int\sum_n\sigma^n\left[\left(\log \Asf^{n}\right)\left(\log \asf^n\right)' - \left(\log \Asf^{n}\right)'\left(\log \asf^n\right)\right] = \int\sum_n\sigma^n\left[\left(\log \Asf^{n}\right)(\dsf^n)' -\left(\log \Asf^{n}\right)'(\dsf^n)\right].
\ee
The merits of this equation are two-fold: (i) the left-hand side can be evaluated as it is an
integral over a perfect differential, and (ii) the right-hand side is closely
related to the amplitude of the $1/N$ finite-size correction term of the eigenvalue. Let us look at
the integrals on the left-hand side and show the dependence on $z$ explicitly:
\be
\label{NLIE.A9}
\hbox{lhs of}\ \eqref{NLIE.A8}=\sum_n\sigma^n\int \dd z\left[\big(\log \Asf^{n}(z)\big)\big(\log \asf^n(z)\big)' - \big(\log \Asf^{n}(z)\big)'\big(\log \asf^n(z)\big)\right].
\ee
Here we perform a change of the variable of integration from $z$ to the
functions $\asf^n$ respectively $\Asf^n$. With $(\asf^n)' \dd z = \dd \asf^n$
and $(\Asf^n)' \dd z = \dd \Asf^n$, we obtain
\be
\hbox{lhs of}\ \eqref{NLIE.A8}=
\sum_n\sigma^n\int \left[\dd \asf^n \frac{\log \Asf^{n}}{\asf^n} -\dd \Asf^{n}\frac{\log \asf^n}{\Asf^n}\right].
\ee
The variables $\asf^n$ and $\Asf^n$ are related by the simple function $F^n$ in \eqref{eq:Fn}. Here we find for the left-hand side of \eqref{NLIE.A8}
\be
\label{NLIE.A11}
\hbox{lhs of}\ \eqref{NLIE.A8}=
\sum_n\sigma^n\int_{\asf^n(-\infty)}^{\asf^n(\infty)} \dd \asf\left[\frac{\log
    (1+\varepsilon^n \asf)}{\asf} - \varepsilon^n\frac{\log \asf}{1+\varepsilon^n \asf}\right],
\ee
where we specified the terminals of the integrals and replaced the variables
of integration $\asf^n$ by a uniformized variable $\asf$. We will evaluate these integrals explicitly as we have explicit expressions
for the terminals $\asf^n(\pm\infty)$. However a suitable mathematical reasoning
will replace the cumbersome calculations.

Before treating the expression \eqref{NLIE.A11}, we
want to slightly simplify the right-hand side of \eqref{NLIE.A8}. An
integration by parts yields
\be
\label{NLIE.A12}
\hbox{rhs of}\ \eqref{NLIE.A8}=
2\int\sum_n\sigma^n\left(\log \Asf^{n}(z)\right)(\dsf^n(z))'\dd z -\sum_n\sigma^n\left(\log \Asf^{n}(z)\right)(\dsf^n(z))\Big|_{-\infty}^\infty.
\ee
The surface term may look delicate as certain factors $\dsf^n(z)$
diverge to $-\infty$ for $z\to-\infty$. However, in such cases, the corresponding function $\asf^n(z)$ tends to zero in this same limit, 
the factor $\log \Asf^{n}(z)$ vanishes much faster than $\dsf^n(z)$ diverges, and the product $\big(\log \Asf^{n}(z)\big)\dsf^n(z)$ converges to zero. 
It appears that the exponentials in \eqref{NLIE.A4} drop out in the surface term, and the
constants $\phi^n(\omega)$ naturally drop out in the integral term resulting in
\be
\hbox{rhs of}\ \eqref{NLIE.A8}=
\frac 2r\int \dd z\, \eE^{-z/r}\sum_n\sigma^n\delta^n\log \Asf^{n}(z)-\sum_n\sigma^n\phi^n(\omega)\log \Asf^{n}(z)\Big|_{-\infty}^\infty.
\ee
Let us state the intermediate result
\begin{alignat}{2}
&\frac 2r\int \dd z\, \eE^{-z/r}\sum_n\sigma^n\delta^n\log \Asf^{n}(z)-\sum_n\sigma^n\phi^n(\omega)\log \Asf^{n}(z)\Big|_{-\infty}^\infty
\nonumber\\[0.15cm]
&=\sum_n\sigma^n\int_{\asf^n(-\infty)}^{\asf^n(\infty)} \dd \asf\left[\frac{\log
    (1+\varepsilon^n \asf)}{\asf} - \varepsilon^n\frac{\log \asf}{1+\varepsilon^n \asf}\right]
=: I(\omega).
\label{NLIE.A13}
\end{alignat}
In the next subsection, we will show how to evaluate this sum of
  dilogarithmic integrals. Here we like to mention, without using it for the
  evaluation, a compact way of writing the integrals in terms of the Rogers dilogarithm~\cite{Lewin,Kirillov,Zagier}
  \be
  L(z)=Li_2(z)+\frac12\log z\,\log(1-z)=-\frac12\int_0^z da \left[\frac{\log (1-a)}{a}+\frac{\log
    a}{1-a}\right]\,,\ z\in\mathbb C\setminus (-\infty,0]\cup [1,\infty).
  \ee
  This is most obvious for the terms in \eqref{NLIE.A13} with $\varepsilon^n=-1$:
\be
\int_{\asf^n(-\infty)}^{\asf^n(\infty)} \dd \asf\left[\frac{\log
    (1- \asf)}{\asf} +\frac{\log \asf}{1- \asf}\right]
  =-2\,L(z)\bigg|_{\asf^n(-\infty)}^{\asf^n(\infty)}\,.
  \ee
  For terms in \eqref{NLIE.A13} with $\varepsilon^n=+1$ we have:
  \be
\int_{\asf^n(-\infty)}^{\asf^n(\infty)} \dd \asf\left[\frac{\log
      (1+ \asf)}{\asf} - \frac{\log \asf}{1+ \asf}\right]
  =2\,L\left(\frac z{1+z}\right)\bigg|_{\asf^n(-\infty)}^{\asf^n(\infty)}\,.
  \ee
  This is proven by taking the derivative of $2L\big(\frac z{1+z}\big)$ and
  finding the integrand of the left-hand side.
In total, we obtain
\be
\frac12  I(\omega)=
\sum_{n \mathrm{\, with} \atop \varepsilon^n=+1}\sigma^n L\left(\frac z{1+z}\right)\bigg|_{\asf^n(-\infty)}^{\asf^n(\infty)}
-\sum_{n \mathrm{\, with}\atop \varepsilon^n=-1}
\sigma^nL\left(z\right)\bigg|_{\asf^n(-\infty)}^{\asf^n(\infty)}\,.
\label{IintermsofRog}
\ee  

\subsubsection{Second use of the kernel's symmetry: evaluation of the sum of dilogarithms}

Now let us evaluate $I(\omega)$. We will see that in all of the cases of
interest to us this quantity is independent of $\omega$. Furthermore we will
see that an evaluation at $\omega=\infty$ is done most economically.
We take the derivative of $I(\omega)$ and note that just the integration
terminals depend on $\omega$, hence
\begin{alignat}{2}
\ddo I(\omega)&=\sum_n\sigma^n\left[\left(\ddo\log \asf^n(+\infty)\right)\log \Asf^n(+\infty)
  -\big(\log \asf^n(+\infty)\big)\ddo\log \Asf^n({+\infty})\right]
\nonumber\\[0.15cm]
&-\sum_n\sigma^n\left[\left(\ddo\log \asf^n(-\infty)\right)\log \Asf^n(-\infty)
  -\big(\log \asf^n(-\infty)\big)\ddo\log \Asf^n({-\infty})\right].
\label{NLIE.A14} 
\end{alignat}
This expression can be simplified by a second use of the fact that the
integral kernel $K$ in \eqref{NLIE.A1sc} is symmetric after multiplication
by $\sigma$, as in \eqref{NLIE.A3}. We take the limit $z\to
+\infty$ in \eqref{NLIE.A1sc}:
\be
\label{NLIE.B1sc}
\log \asf^n(+\infty) = \phi^n(\omega) + \sum_m \kappa_{n,m} \cdot \log \Asf^{m}(+\infty),
\ee
where the convolution turned into a simple multiplication with the number
$\kappa_{n,m}=\int_{-\infty}^{\infty} \dd z \, K_{n,m}(z)$. We note that the matrix
$\sigma\kappa$ is symmetric. We take the derivative of \eqref{NLIE.B1sc} with
respect to $\omega$, multiply with $\sigma^n$ and find
\be
\label{NLIE.B1scd}
\sigma^n\ddo \log \asf^n(+\infty) = \sigma^n\ddo\phi^n(\omega) + \sum_m(\sigma
\kappa)_{n,m} \ddo \log \Asf^{m}(+\infty).
\ee
Next we multiply with $\log \Asf^{n}(+\infty)$ and sum over $n$:
\be
\label{NLIE.B6}
\sum_n\sigma^n\log \Asf^{n}(\infty)\ddo\log \asf^n(\infty) =
\sum_n\sigma^n\log \Asf^{n}(\infty)\ddo\phi^n(\omega)+
\sum_{n,m}\log \Asf^{n}(\infty)(\sigma \kappa)_{n,m} \ddo \log \Asf^{m}(\infty).
\ee
Then we take \eqref{NLIE.B1sc}, multiply with  $\sigma^n\ddo\log \Asf^{n}(+\infty)$ and sum over $n$:
\be
\label{NLIE.B7}
\sum_n\sigma^n\ddo\big(\log \Asf^{n}(\infty)\big) \log \asf^n(\infty) = \sum_n\sigma^n\ddo\big(\log \Asf^{n}(\infty)\big)\phi^n(\omega) + \sum_{n,m}\ddo\big(\log \Asf^{n}(\infty)\big)(\sigma \kappa)_{n,m} \log \Asf^{m}(\infty).
\ee
Similarly to above, we see that the double sums on the right-hand sides of
\eqref{NLIE.B6} and \eqref{NLIE.B7} are identical. Hence the difference of
both equations yields
\begin{alignat}{2}
&\sum_n\sigma^n\left[\log \Asf^{n}(\infty)\ddo\log \asf^n(\infty)-\ddo(\log
  \Asf^{n}(\infty)) \log \asf^n(\infty)\right] \nonumber\\[0.15cm]
&=
\sum_n\sigma^n\left[\log \Asf^{n}(\infty)\ddo\phi^n(\omega)-\ddo(\log \Asf^{n}(\infty))\phi^n(\omega)\right].
\label{NLIE.B8}
\end{alignat}
A similar equation holds in the limit $z\to -\infty$:
\begin{alignat}{2}
&\sum_n\sigma^n\left[\log \Asf^{n}(-\infty)\ddo\log \asf^n(-\infty)-\ddo(\log
  \Asf^{n}(-\infty)) \log \asf^n(-\infty)\right] \nonumber\\[0.15cm]
&=
\sum_n\sigma^n\left[\log \Asf^{n}(-\infty)\ddo\phi^n(\omega)-\ddo(\log \Asf^{n}(-\infty))\phi^n(\omega)\right].
\label{NLIE.B9}
\end{alignat}
In deriving this equation, driving terms
$\dsf^n(z)=-\delta^n\eE^{-z/r}+\phi^n(\omega)$ appear which differ (especially in
the limit $z\to -\infty$) from $\phi^n(\omega)$. This happens for instance in 
the counterpart of \eqref{NLIE.B1sc} for the cases
with ${\rm Re}(\delta^n)>0$ (the case ${\rm Re}(\delta^n)<0$ does not occur). For these cases, $\log
\Asf^n(-\infty)$ is zero and its product with the exponential driving terms (and other terms) consistently
vanishes in the limit $z \to - \infty$, ensuring that the possibly problematic extra terms drop out.

For \eqref{NLIE.A14}, we find with \eqref{NLIE.B8} and \eqref{NLIE.B9}
\be
\label{NLIE.B10}
\ddo I(\omega)=\sum_n\sigma^n\left[\log{\Asf^{n}(z)}\ddo\phi^n(\omega)-\ddo\left(\log{\Asf^{n}(z)}\right)\phi^n(\omega)\right]\Bigg|_{-\infty}^\infty.
\ee
Let us focus first on the $A_1^{(1)}$ and $A_2^{(2)}$ models. In our concrete calculations below, we will find that for only two values of
the index $n$, say
$p$ and $m$, the term $\phi^n(\omega)$ is non-zero. These will in fact correspond to some of 
the indices $(\rho-1,j)$ that appear in our standardisation procedure of the non-linear integral equations, as described in \cref{sec:standardization}. In these two
cases, we will find that $\phi^n(\omega)$ for $n=p$ and $m$ are equal up to a sign:
$\phi^m(\omega)=-\phi^p(\omega)$. Furthermore we have $\sigma^p=\sigma^m=:\sigma^{p,m}$ in all cases.
Hence
\be
\label{NLIE.B11}
\ddo I(\omega)=\sigma^{p,m}\left[\log\frac{\Asf^{p}(z)}{\Asf^{m}(z)}\ddo\phi^p(\omega)-\ddo\left(\log\frac{\Asf^{p}(z)}{\Asf^{m}(z)}\right)\phi^p(\omega)\right]\Bigg|_{-\infty}^\infty.
\ee
Here the sum over copies is trivial for the $A_1^{(1)}$ and $A_2^{(2)}$ models, and covers the unbarred and barred functions for the $A_2^{(1)}$ model.
Concrete calculations reveal that 
$\log\big(\Asf^{p}(\pm\infty)/\Asf^{m}(\pm\infty)\big)$ and $\phi^p(\omega)$ are of the form
$\mu_\pm\log\omega$ and  $\nu\log\omega$ for certain constants $\mu_\pm$ and
$\nu$. Therefore the right-hand side of \eqref{NLIE.B11} is clearly zero.
We rewrite \eqref{NLIE.A13} as
\be
\frac 2r\int \dd z\, \eE^{-z/r}\sum_n\sigma^n\delta^n\log \Asf^{n}(z)
=I(\omega)+\sigma^{p,m}\nu(\mu_+-\mu_-)(\log\omega)^2
=I(\infty)-\sigma^{p,m}\nu(\mu_+-\mu_-)\gamma^2.
\label{NLIE.B12}
\ee
where finally we have inserted the physical value
$\omega=\eE^{\ir\gamma}$ except in $I(\omega)$, which is independent of $\omega$,
and where $\omega=\infty$ will lead to a simpler computation scheme. 

For the $A_2^{(1)}$ model, the same arguments apply, except that there are four functions with non-zero values of $\phi^n(\omega)$ and $\bar \phi^n(\omega)$, namely $\asf^{(p'-1,j)}$ and $\bar\asf^{(p'-1,j)}$ with $j=+1,-1$. In this case, we have two functions $I(w)$ and $\bar I(w)$, with the latter defined similarly to \eqref{NLIE.A13} but with $\sigma^n,\varepsilon^n$ replaced by $\bar\sigma^n,\bar\varepsilon^n$ and the integration terminals set to $\bar \asf^n(\pm\infty)$. The above proof generalises to show that the sum $I(w)+\bar I(w)$ is independent of~$\omega$. Moreover, we have 
\begin{subequations}
\begin{alignat}{2}
\log\big(\Asf^{p}(\pm\infty)/\Asf^{m}(\pm\infty)\big) &= \mu_\pm \log \omega,\qquad
\phi^{p/m}(\omega) = \pm\nu \log \omega,\qquad
\sigma^p=\sigma^m=:\sigma^{p,m},
\\[0.15cm]
\log\big(\bar\Asf^{p}(\pm\infty)/\bar\Asf^{m}(\pm\infty)\big) &= \bar\mu_\pm \log \omega,\qquad 
\bar\phi^{p/m}(\omega) = \pm\bar\nu \log \omega,\qquad
\bar\sigma^p=\bar\sigma^m=:\bar\sigma^{p,m},
\end{alignat}
\end{subequations}
which yields
\begin{alignat}{2}
\frac 2r\int \dd z\, \eE^{-z/r}&\bigg(\sum_n\sigma^n\delta^n\log \Asf^{n}(z)+
\sum_n\bar\sigma^n\bar\delta^n\log \bar\Asf^{n}(z)\bigg)
\nonumber\\[0.15cm]
&=I(\omega)+\bar I(\omega)+\sigma^{p,m}\nu(\mu_+-\mu_-)(\log\omega)^2 + \bar\sigma^{p,m}\bar\nu(\bar\mu_+-\bar\mu_-)(\log\omega)^2
\nonumber\\[0.15cm]
&=I(\infty)+\bar I(\infty)-\sigma^{p,m}\nu(\mu_+-\mu_-)\gamma^2-\bar\sigma^{p,m}\bar\nu(\bar\mu_+-\bar\mu_-)\gamma^2.
\label{NLIE.B12.A21}
\end{alignat}

Returning to the $A_1^{(1)}$ and $A_2^{(2)}$ cases, we note that we can make a statement stronger than $\frac{\dd}{\dd z} I(\omega) = 0$.
For the two contributions to
  $I(\omega)$ from the upper and the lower terminals separately, namely
\begin{alignat}{3}
I_\pm(\omega)&:=\sum_n\sigma^n\int_{0}^{\asf^n(\pm\infty)} \dd \asf\left[\frac{\log
    (1+\varepsilon^n \asf)}{\asf} - \varepsilon^n\frac{\log \asf}{1+\varepsilon^n
    \asf}\right]\nonumber\\
&=2\sum_{n \mathrm{\, with} \atop \varepsilon^n=+1}\sigma^nL\left(\frac {\asf^n(\pm\infty)}{1+\asf^n(\pm\infty)}\right)
-
2\sum_{n \mathrm{\, with}\atop \varepsilon^n=-1}\sigma^n
L\left(\asf^n(\pm\infty)\right),\label{IomegaPMdef}\\
I(\omega)&=I_+(\omega)-I_-(\omega),
\label{IintermsofRogPM}
\end{alignat}
the above reasoning applies. The two contributions on the right hand side of
\eqref{NLIE.A14} were treated separately down to \eqref{NLIE.B10} and
\eqref{NLIE.B11}. Also these last identities may be formulated separately for
$I_\pm(\omega)$
from which we find
\be
I_\pm(\omega)=\mathrm{constant}\,.\label{IomegaPMconst}
\ee
This separate independence of $\omega$ is not of relevance for the computation
of the conformal data. It leads however to stronger identities for dilogarithms
that we derive and state in the course of our calculations.

\subsection{Results for specific cases}

\subsubsection[Case $A_1^{(1)}$: Principal series]{Case $\boldsymbol{A_1^{(1)}}$: Principal series}\label{app:A11.principal}

We rewrite \eqref{eq:NLIEs.A11} by the above explained standard. We avoid the
definition of \eqref{eq:A11b} for $\Asf^{p'-1}$ by defining
\be
\asf^{(p'-1,\pm 1)}=\omega^{\pm p'}\asf^{p'-1}, \qquad \Asf^{(p'-1,\pm 1)} = 1 + \asf^{(p'-1,\pm 1)}.
\ee
The resulting set of functions 
now has the standardized form discussed in \cref{sec:first.use}. The number of functions
$\asf^n$ is now $p'$, as it is for $\Asf^n$. Note that we doubled the index $p'-1$ to
$(p'-1,-1)$, and $(p'-1,+1)$. From \cref{sec:A11.principal}, we read off the values
\be
r = 1, \qquad \sigma^n=\varepsilon^n=+1, \qquad
\delta^n = \left\{\begin{array}{cl}
2 & n = 1,\\[0.1cm]
0 & \textrm{otherwise.}
\end{array}
\right.
\ee
For the indices $m=(p'-1,-1)$ and $p=(p'-1,+1)$, explicit calculations yield
\be
\frac{\Asf^{p}(\infty)}{\Asf^{m}(\infty)}=\omega^{2p'-2},
\qquad
\frac{\Asf^{p}(-\infty)}{\Asf^{m}(-\infty)}=\omega^{2p'(p'-2)/(p'-1)},
\qquad 
\phi^{p/m}=\pm p'\log\omega,
\ee
and hence 
\be
\mu_+=2p'-2, \qquad \mu_-=\frac{2p'(p'-2)}{p'-1}, \qquad \nu=p'.
\ee

For the calculation of $I(\infty)$, we see that the lower and upper terminals of integration
appearing in \eqref{NLIE.A13} are identical and equal to $+\infty$, except for the
index $n=1$ where
the lower terminal is 0 and the upper terminal is~$+\infty$. As
\be
\int_{0}^{\infty} \dd \asf\left[\frac{\log
    (1+\asf)}{\asf} - \frac{\log \asf}{1+\asf}\right]=\frac{\pi^2}3,
\ee
we find $I(\infty)=\pi^2/3$ and from \eqref{NLIE.B12}
\be
4\int_{-\infty}^\infty \dd z\,\eE^{-z}\log
\Asf^1(z)=\frac{\pi^2}3\left(1-\frac 6{\pi^2}\frac{p'}{p'-1}\gamma^2\right).
\ee
The underlying identities for dilogarithms
\eqref{IomegaPMdef} and \eqref{IomegaPMconst} read in this case
  $I_+(\omega)=(p'-1)\pi^2/3$ and $I_-(\omega)=(p'-2)\pi^2/3$. We state the identity for $I_+(\omega)$ explicitly
by using the terminals $\asf^j_{+\infty} (j=1,\dots,p'-1)$ as given in \eqref{eq:braid.an.A11}
and writing $N$ for $p'$ and $\eE^{\ir\phi}$ for $\omega$:
\begin{alignat}{1}
\sum_{n =1}^{N-2}L\left(\frac{\sin{(n+2)}\phi\,\sin{n}\phi}{\sin^2{(n+1)}\phi}\right)+
L\left(\eE^{\ir\phi}\frac{\sin{(N-1)}\phi}{\sin{N}\phi}\right)
+L\left(\eE^{-\ir\phi}\frac{\sin{(N-1)}\phi}{\sin{N}\phi}\right)
=(N-1)\frac{\pi^2}6.\label{A11dilogIdP}
\end{alignat}
This identity holds for general continuous
$\phi$ from a domain containing the imaginary axis and for general integer
$N\ge2$. Note that the identity may be analytically continued in $\phi$ to the entire
complex plane, however avoiding the singularities of the dilogarithmic
function and choosing the appropriate branch. It then also holds on a real
interval in a neighborhood of $\phi = 0$, relevant to our calculation
involving $\gamma$.
The identity following from $I_-(\omega)=(p'-2)\pi^2/3$ is the same as
\eqref{A11dilogIdP} with the parameters $N=p'-1$ and
$\eE^{\ir\phi}=\bar\omega$, see \eqref{eq:lowlim.A11}.

\subsubsection[Case $A_1^{(1)}$: Dual series]{Case $\boldsymbol{A_1^{(1)}}$: Dual series}\label{app:A11.dual}

We rewrite \eqref{eq:NLIEs.A11.dual} by the above explained standard. We avoid the
definition of \eqref{eq:A11b.dual} for $\Asf^{p'-1}$ by defining
\be
\asf^{(p'-1,\pm 1)}=\omega^{\pm p'}\asf^{p'-1}, \qquad \Asf^{(p'-1,\pm 1)} = 1 + \asf^{(p'-1,\pm 1)}.
\ee
The number of functions
$\asf^n$ is now $p'$, as it is for $\Asf^n$. Note that we doubled the index $p'-1$ to
$(p'-1,-1)$, and $(p'-1,+1)$. From \cref{sec:A11.dual}, we read off the values
\be
r = p'-1, \qquad
\sigma^n=\varepsilon^n=+1, \qquad
\delta^n = \left\{\begin{array}{cl}
4\sin(\frac{\pi n}{2p'-2}) & n = 1,\dots, p'-2,\\[0.1cm]
2 & n=(p'-1,-1), (p'-1,+1).
\end{array}\right.
\ee
For the indices $m=(p'-1,-1)$ and $p=(p'-1,+1)$, explicit calculations yield
\be
\frac{\Asf^{p}(\infty)}{\Asf^{m}(\infty)}=\omega^{2},
\qquad
\frac{\Asf^{p}(-\infty)}{\Asf^{m}(-\infty)}=1,
\qquad
\phi^{p/m}=\pm p'\log\omega,
\ee
and hence 
\be
\mu_+=2, \qquad \mu_-=0, \qquad \nu=p'.
\ee

For the calculation of $I(\infty)$, we see that the lower and upper terminals of integration
appearing in \eqref{NLIE.A13} are identical except for the index $n=(p'-1,+1)$
where
the lower terminal is 0 and the upper terminal is $\infty$. As
\be
\int_{0}^{\infty} \dd \asf\left[\frac{\log
    (1+\asf)}{\asf} - \frac{\log \asf}{1+\asf}\right]=\frac{\pi^2}3,
\ee
we find $I(\infty)=\pi^2/3$ and from \eqref{NLIE.B12}
\begin{alignat}{2}
&\frac8{p'-1} \int_{-\infty}^\infty \dd z \, \eE^{-z/(p'-1)}
\bigg[\sum_{n=1}^{p'-2} \sin\frac {\pi n}{2(p'\!-\!1)}\, \log \Asf^n(z) +
  \tfrac12 \log \Asf^{(p'-1,-1)}(z)+
  \tfrac12 \log \Asf^{(p'-1,+1)}(z)\bigg]\nonumber\\[0.15cm]
&=\frac{\pi^2}3\left(1-\frac 6{\pi^2}
p'\gamma^2\right).
\end{alignat}
The underlying identities for dilogarithms
\eqref{IomegaPMdef} and \eqref{IomegaPMconst}
read in this case $I_+(\omega)=I(\omega)=\pi^2/3$ and (trivially) $I_-(\omega)=0$.
We state the identity for $I_+(\omega)$ explicitly by using the terminals $\asf^j_{+\infty}
(j=1,\dots,p'-1)$ as given in
\eqref{eq:braid.an.A11.dual}
and writing $N$ for $p'$ and $\eE^{\ir\phi}$ for $\omega$:
\be
\sum_{n =1}^{N-2}L\left(\frac{\sin^2\phi}{\sin^2{(n+1)}\phi}\right)+
L\left(\eE^{\ir(N-1)\phi}\,\frac{\sin\phi}{\sin{N}\phi}\right)
+L\left(\eE^{-\ir(N-1)\phi}\,\frac{\sin\phi}{\sin{N}\phi}\right)
=\frac{\pi^2}6\,.
\ee
This identity holds for general continuous
$\phi$ from a domain containing the imaginary axis and for general integer
$N\ge2$. Note that the identity may be analytically continued in $\phi$ to the entire
complex plane, however avoiding the singularities of the dilogarithmic
function and choosing the appropriate branch. It then also holds on a real
interval in a neighborhood of $\phi = 0$, relevant to our calculation
involving $\gamma$.

\subsubsection[Case $A_2^{(2)}$: Principal series]{Case $\boldsymbol{A_2^{(2)}}$: Principal series}\label{app:A22.principal}

We rewrite \eqref{eq:NLIEs.A22} by the above explained standard. We avoid the
definition of \eqref{eq:A22c} for $\Asf^{b-1}$ by defining
\begin{subequations}
\begin{alignat}{3}
\asf^{(b-1,0)}&=\asf^{b-1},\qquad &\Asf^{(b-1,0)}&= 1+\asf^{(b-1,0)},\\[0.2cm]
\asf^{(b-1,\pm 1)} &=\omega^{\pm b}\asf^{b-1}, \qquad &\Asf^{(b-1,\pm 1)} &= 1+\asf^{(b-1,\pm 1)}.
\end{alignat}
\end{subequations}
The number of functions $\asf^n$ is now $b+2$, as it is for $\Asf^n$. Note that we tripled the index $b-1$ to
$(b-1,-1)$, $(b-1,0)$ and $(b-1,+1)$.
From \cref{sec:A22.principal}, we read off the values
\be
r=1, 
\qquad
\sigma^n=\varepsilon^n= \left\{\begin{array}{cl}
-1 & n = b,\\[0.1cm]
1 &\textrm{otherwise},
\end{array}\right.
\qquad
\delta^n = 
\left\{\begin{array}{cl}
2\sqrt{3} & n = 1,\\[0.1cm]
0 &\textrm{otherwise}.
\end{array}\right.
\ee
For the indices $m=(b-1,-1)$ and $p=(b-1,+1)$, explicit calculations yield 
\be
\frac{\Asf^{p}(\infty)}{\Asf^{m}(\infty)}=\omega^{b-1},
\qquad
\frac{\Asf^{p}(-\infty)}{\Asf^{m}(-\infty)}=\omega^{b(b-2)/(b-1)},
\qquad
\phi^{p/m}=\pm b\log\omega,
\ee
and hence 
\be
\mu_+=b-1, \qquad
\mu_-=\frac{b(b-2)}{b-1}, \qquad
\nu=b.
\ee

For the calculation of $I(\infty)$, we see that the lower and upper terminals of integration
appearing in \eqref{NLIE.A13} are identical except for the index $n=1$ where
the lower terminal is 0 and the upper terminal is $\infty$. As
\be
\int_{0}^{\infty} \dd \asf\left[\frac{\log
    (1+\asf)}{\asf} - \frac{\log \asf}{1+\asf}\right]=\frac{\pi^2}3,
\ee
we find $I(\infty)=\pi^2/3$ and from \eqref{NLIE.B12}
\be
4\sqrt{3}\int_{-\infty}^\infty \dd z\,\eE^{-z}\log
\Asf^1(z)=\frac{\pi^2}3\left(1-\frac 3{\pi^2}\frac b{b-1}\gamma^2\right).
\ee
The underlying identities for dilogarithms
\eqref{IomegaPMdef} and \eqref{IomegaPMconst} read in this case
  $I_+(\omega)=(b-1)\pi^2/3$ and $I_-(\omega)=(b-2)\pi^2/3$. We state the identity for $I_+(\omega)$ explicitly
by using the terminals $\asf^j_{+\infty} (j=1,\dots,b-1)$ as given in \eqref{eq:braid.an.A22}
and writing $N$ for $b$ and $\eE^{\ir\phi}$ for $\omega$:
\begin{alignat}{2}
  &\sum_{n =1}^{N-2}L\left(
\frac{\sin\frac{n}{2}\phi\,\sin\frac{n+3}{2}\phi}
{\sin\frac{n+1}{2}\phi\,\sin\frac{n+2}{2}\phi}
  \right)+
L\left(\frac{\sin\frac{N-1}{2}\phi}
{2\sin\frac{N}{2}\phi\,\cos\frac{1}{2}\phi}\right)
\label{A22dilogIdP}\\
&+
L\left(\eE^{\ir\frac{N+1}2\phi}\,\frac{\sin\frac{N-1}{2}\phi}
{\sin{N}\phi}\right)
+
L\left(\eE^{-\ir\frac{N+1}2\phi}\,\frac{\sin\frac{N-1}{2}\phi}
{\sin{N}\phi}\right)
+
L\left(\frac{\sin^2\frac{N-1}{2}\phi}
{\sin^2\frac{N+1}{2}\phi}\right)
=(N-1)\frac{\pi^2}6\,.\nonumber
\end{alignat}
This identity holds for general continuous
$\phi$ from a domain containing the imaginary axis and for general integer
$N\ge2$. Note that the identity may be analytically continued in $\phi$ to the entire
complex plane, however avoiding the singularities of the dilogarithmic
function and choosing the appropriate branch. It then also holds on a real
interval in a neighborhood of $\phi = 0$, relevant to our calculation
involving $\gamma$.
The identity following from $I_-(\omega)=(b-2)\pi^2/3$ is the same as
\eqref{A22dilogIdP} with the parameters $N=b-1$ and $\eE^{\ir\phi}=\bar\omega$, see
\eqref{eq:lowlims.A22}.

\subsubsection[Case $A_2^{(2)}$: Dual series]{Case $\boldsymbol{A_2^{(2)}}$: Dual series}\label{app:A22.dual}

We rewrite \eqref{eq:NLIEs.A22.dual} by the above explained standard. We avoid the
definition of \eqref{eq:A22c.dual} for $\Asf^{b-1}$ by defining
\begin{subequations}
\begin{alignat}{3}
\asf^{(b-1,0)}&=\asf^{b-1},\qquad &\Asf^{(b-1,0)}&= 1+\asf^{(b-1,0)},\\[0.2cm]
\asf^{(b-1,\pm 1)} &=\omega^{\pm b}\asf^{b-1}, \qquad &\Asf^{(b-1,\pm 1)} &= 1+\asf^{(b-1,\pm 1)}.
\end{alignat}
\end{subequations}
The number of functions $\asf^n$ is now $b+2$, as it is for $\Asf^n$. Note that we tripled the index $b-1$ to
$(b-1,-1)$, $(b-1,0)$ and $(b-1,+1)$. From \cref{sec:A22.dual}, we read off the values
\be
r=2b-3, \qquad
\sigma^n=\varepsilon^n=+1,
\qquad
\delta^n = 
\left\{\begin{array}{cl}
4\sin\big(\frac{\pi n}{2b-3}\big) & n = 1, \dots, b-2, \\[0.1cm]
0 & n = (b-1,-1), (b-1,0), (b-1,+1), \\[0.1cm]
2 & n = b.
\end{array}\right.
\ee 
For the indices $m=(b-1,-1)$ and $p=(b-1,+1)$, explicit calculations yield
\be
\frac{\Asf^{p}(\infty)}{\Asf^{m}(\infty)}=\omega^{b+1},
\qquad
\frac{\Asf^{p}(-\infty)}{\Asf^{m}(-\infty)}=\omega^{b},
\qquad
\phi^{p/m}=\pm b\log\omega,
\ee
and hence 
\be
\mu_+=b+1, \qquad \mu_-=b, \qquad \nu = b.
\ee

For the calculation of $I(\infty)$, we see that the lower and upper terminals of integration
appearing in \eqref{NLIE.A13} are identical except for the indices $n=(b-1,0)$
and $n=b$ where the lower terminals are 1 and 0 and the upper terminals are $\infty$. As
\be
\left(\int_{1}^{\infty}+\int_{0}^{\infty}\right) \dd \asf\left[\frac{\log
    (1+\asf)}{\asf} - \frac{\log \asf}{1+\asf}\right]=\frac{\pi^2}2,
\ee
we find $I(\infty)=\pi^2/2$ and from \eqref{NLIE.B12}
\be
\frac8{2b-3} \int_{-\infty}^\infty \dd z \, \eE^{-z/(2b-3)}
\bigg[\sum_{n=1}^{b-2} \sin\frac {\pi n}{2b\!-\!3}\, \log \Asf^n(z) +
  \tfrac12 \log \Asf^{b}(z)\bigg]=\frac{\pi^2}3\left(\frac 32-\frac 3{\pi^2}
b\gamma^2\right).
\ee
The underlying identities for dilogarithms
\eqref{IomegaPMdef} and \eqref{IomegaPMconst} read in this case
  $I_+(\omega)=\pi^2$ and $I_-(\omega)=\pi^2/2$. We state the identity for $I_+(\omega)$ explicitly
by using the terminals $\asf^j_{+\infty} (j=1,\dots,b-1)$ as given in \eqref{eq:uplims.A22}
and writing $N$ for $b$ and $\eE^{\ir\phi}$ for $\omega$:
\begin{alignat}{2}
  &\sum_{n =1}^{N-2}L\left(
\frac{\sin\frac{1}{2}\phi\,\sin\phi}
{\sin\frac{n+1}{2}\phi\,\sin\frac{n+2}{2}\phi}
  \right)+
L\left(\frac{\sin\frac{N+1}{2}\phi}
{2\sin\frac{N}{2}\phi\,\cos\frac{1}{2}\phi}\right)
\label{A22dilogIdPdual}\\
&+
L\left(\eE^{\ir\frac{N-1}{2}\phi}\,\frac{\sin\frac{N+1}{2}\phi}
{\sin{N}\phi}\right)
+
L\left(\eE^{-\ir\frac{N-1}{2}\phi}\,\frac{\sin\frac{N+1}{2}\phi}
{\sin{N}\phi}\right)
+
L\left(\frac{\sin N\phi\,\sin\phi}
{\sin^2\frac{N+1}{2}\phi}\right)
=\frac{\pi^2}2\,.\nonumber
\end{alignat}
This identity holds for general continuous
$\phi$ from a domain containing the imaginary axis and for general integer
$N\ge2$. Note that the identity may be analytically continued in $\phi$ to the entire
complex plane, however avoiding the singularities of the dilogarithmic
function and choosing the appropriate branch. It then also holds on a real
interval in a neighborhood of $\phi = 0$, relevant to our calculation
involving $\gamma$.
The identity following from $I_-(\omega)=\pi^2/2$ is
\be
L\left(\frac12\right)+L\left(\frac{\eE^{\ir\frac N2\phi}}{2\cos\frac
  N2\phi}\right)+L\left(\frac{\eE^{-\ir\frac N2\phi}}{2\cos\frac
  N2\phi}\right)
=\frac{\pi^2}4\,.
\ee

\subsubsection[Case $A_2^{(1)}$: Principal series]{Case $\boldsymbol{A_2^{(1)}}$: Principal series}\label{app:A21.principal}

The non-linear integral equations \eqref{eq:NLIEs.A21} may be understood as a doubling of those for
the $A_2^{(2)}$ case, see \eqref{eq:NLIEs.A22}, with $p'$ playing the
role of $b$ in the case of $A_2^{(2)}$. Here we are dealing with two copies
of the functions $\asf^n$, namely $\asf^n$ and $\bar \asf^n$, which satisfy coupled
non-linear integral equations. We use the definitions of the $A_2^{(2)}$ dual case for the
objects appearing in the dilog identity, see \cref{app:A22.principal}, with the number of functions 
doubled and with $p'$ playing the role of $b$. 
The total number of functions is now $2(p'+2)$. The data needed to apply \eqref{NLIE.B12.A21} is obtained from \cref{sec:A21.principal}:
\be
r=\frac32, \qquad
\sigma^n = \bar \sigma^n = \varepsilon^n=\bar\varepsilon^n=
\left\{\begin{array}{cl}
-1 & n = p',\\[0.1cm]
1 & \textrm{otherwise,}
\end{array}\right.
\qquad
\delta^n = \bar \delta^n = 
\left\{\begin{array}{cl}
\sqrt 3 & n = 1,\\[0.1cm]
0 & \textrm{otherwise,}
\end{array}\right.
\ee
and
\be
\mu_+=\bar\mu_+=p'-1, \qquad
\mu_-=\bar\mu_-=\frac{p'(p'-2)}{p'-1}, \qquad
\nu=\bar\nu=p'.
\ee
We note that to obtain the correct values for $\delta^n$, one must not forget the factors of $\frac12$ appearing in front of $\fsf(z)$ in \eqref{eq:scalingNLIEs.A21}.
We obtain
\be
\frac 4{\sqrt{3}}\int_{-\infty}^\infty \dd z\,\eE^{-2z/3}\left(\log
\Asf^1(z)+\log
\bar \Asf^1(z)\right)=\frac{\pi^2}3\left(2-\frac 6{\pi^2}\frac {p'}{p'-1}\gamma^2\right).
\ee

\subsubsection[Case $A_2^{(1)}$: Dual series]{Case $\boldsymbol{A_2^{(1)}}$: Dual series}\label{app:A21.dual}

We proceed as in the
case $A_2^{(1)}$ principal series and deal with two sets of functions, namely
$\asf^n$ and $\bar \asf^n$. These satisfy coupled non-linear integral equations. 
The data needed to apply \eqref{NLIE.B12.A21} is obtained from
\cref{sec:A21.dual}. For the sake of a transparent presentation we focus on
one case of the two scaling limits corresponding to left- and right-movers
  and choose $\rho=+$. For convenience, we omit the extra upper label ``$+$'' of the
  functions.

We rewrite \eqref{eq:scalingNLIEs.A21.dual} by the above explained
standard. We avoid the definition of \eqref{eq:y(u).A21.dualc} for $\Asf^{p'-1}$ by
defining
\begin{subequations}
\begin{alignat}{3}
\asf^{(p'-1,0)}&=\asf^{p'-1},\qquad &\Asf^{(p'-1,0)}&= 1+\asf^{(p'-1,0)},\\[0.2cm]
\asf^{(p'-1,\pm 1)} &=\omega^{\pm p'}\asf^{p'-1}, \qquad &\Asf^{(p'-1,\pm 1)} &= 1+\asf^{(p'-1,\pm 1)}.
\end{alignat}
\end{subequations}
The number of functions $\asf^n$ is now $p'+2$, as it is for $\Asf^n$. Note that we tripled the index $p'-1$ to
$(p'-1,-1)$, $(p'-1,0)$ and $(p'-1,+1)$. From \cref{sec:A21.dual}, we read off the values
\begin{subequations}
\begin{align}
r&=\frac32(p'-1),\qquad
\quad
{\sigma^n=\varepsilon^n=+1}, \\
\delta^{n} &= -\sqrt 3 \,\eE^{-\ir \epsilon^1_n/r}
\left\{\begin{array}{cl}
2\,\eE^{-\ir \pi(2p'-1)/(6(p'-1))}\sin\big(\frac{\pi n}{3(p'-1)}\big)& n = 1, \dots, p'-2,\\[0.25cm]
- \ir & (p'-1,-1), (p'-1,0), (p'-1,+1),\\[0.25cm]
\sqrt 3\, \eE^{\ir \pi/3}& n = p'.
\end{array}\right.
\end{align}
\end{subequations}
For the indices $m=(p'-1,-1)$ and $p=(p'-1,+1)$, explicit calculations yield
\be
\frac{\Asf^{p}(\infty)}{\Asf^{m}(\infty)}=\omega^{p'+1},
\qquad
\frac{\Asf^{p}(-\infty)}{\Asf^{m}(-\infty)}=1,
\qquad
\phi^{p/m}=\pm p'\log\omega,
\ee
and hence 
\be
\sigma^{p,m}=1, \qquad \mu_+=p'+1, \qquad \mu_-=0, \qquad \nu = p'\,,
\label{paramnonbar}
\ee
where $\sigma^{p,m}$ is the sign $\sigma^{p'-1}=1$.

Analogously we deal with the second set of functions.
The number of functions ${\bar \asf}^n$ is now $p'+2$, as it is for ${\bar \Asf}^n$. Note that we tripled the index $p'-1$ to
$(p'-1,-1)$, $(p'-1,0)$ and $(p'-1,+1)$:
\begin{subequations}
\begin{alignat}{3}
{\bar \asf}^{(p'-1,0)}&={\bar \asf}^{p'-1},\qquad &{\bar \Asf}^{(p'-1,0)}&= 1+{\bar \asf}^{(p'-1,0)},\\[0.2cm]
{\bar \asf}^{(p'-1,\pm 1)} &=\omega^{\pm p'}{\bar \asf}^{p'-1}, \qquad &{\bar \Asf}^{(p'-1,\pm 1)} &= 1+{\bar \asf}^{(p'-1,\pm 1)}.
\end{alignat}
\end{subequations}
From \cref{sec:A21.dual}, we read off the values
\begin{subequations}
\begin{align}
r&=\frac32(p'-1),\qquad
\bar\sigma^n = \left\{\begin{array}{cl}
1& n = 1, \dots, p'-2,\\[0.25cm]
-1 & (p'-1,-1), (p'-1,0), (p'-1,+1), p',
\end{array}\right.
\quad
\bar\varepsilon^n=+1,
\\[0.1cm]
\bar\delta^{n} &= -\sqrt 3 \,\eE^{-\ir \epsilon^2_n/r}
\left\{\begin{array}{cl}
2\,\eE^{\ir \pi(2p'-1)/(6(p'-1))}\sin\big(\frac{\pi n}{3(p'-1)}\big)& n = 1, \dots, p'-2,\\[0.25cm]
- \ir & (p'-1,-1), (p'-1,0), (p'-1,+1),\\[0.25cm]
2\,\sin\big(\frac{\pi}{3(p'-1)}\big) & n = p'.
\end{array}\right.
\end{align}
\end{subequations}
For the indices $m=(p'-1,-1)$ and $p=(p'-1,+1)$, explicit calculations yield
\be
\frac{{\bar \Asf}^{p}(\infty)}{{\bar \Asf}^{m}(\infty)}=\omega^{p'-1},
\qquad
\frac{{\bar \Asf}^{p}(-\infty)}{{\bar \Asf}^{m}(-\infty)}=1,
\qquad
\bar\phi^{p/m}=\pm p'\log\omega,
\ee
and hence 
\be
\bar\sigma^{p,m}=-1, \qquad \bar\mu_+=p'-1, \qquad \bar\mu_-=0, \qquad \bar\nu = p'\,,
\label{parambar}
\ee
where $\bar\sigma^{p,m}$ is the sign $\bar\sigma^{p'-1}=-1$.

The left side of \eqref{NLIE.B12.A21} thus has $2(p'+2)$
non-zero contributions (these include the tripled contributions for the functions $\asf^{(p'-1,j)}$ and $\bar\asf^{(p'-1,j)}$). This yields
\begin{subequations}
\begin{alignat}{2}
&\frac 2r\int \dd z\, \eE^{-z/r}\bigg(\sum_n\sigma^n\delta^n\log \Asf^{n}(z)+
\sum_n\bar\sigma^n\bar\delta^n\log \bar\Asf^{n}(z)\bigg)\nonumber\\
&=I(\infty)+\bar I(\infty)-\sigma^{p,m}\nu(\mu_+-\mu_-)\gamma^2
-\bar\sigma^{p,m}\bar\nu(\bar\mu_+-\bar\mu_-)\gamma^2,\nonumber\\
&=I(\infty)+\bar I(\infty) -2 p' \gamma^2\,,
\label{NLIE.B12mod}
\end{alignat}
\end{subequations}
where for the last expression we have inserted the parameters
\eqref{paramnonbar} and \eqref{parambar}. Now the sum
$I(\infty) + \bar I(\infty)$ has to be evaluated. This is easily done by
noting that all lower terminals of integration and most of the
upper terminals of integration are zero, the latter in the limit
$\omega\to\infty$, except for the cases
\begin{subequations}
\begin{alignat}{2}
\lim_{\omega\to\infty}\asf^{(p'-1,0)}(\infty)&=\infty,\quad
\lim_{\omega\to\infty}\asf^{(p'-1,1)}(\infty)=\infty,\quad
\lim_{\omega\to\infty}\asf^{p'}(\infty)=\infty,\label{possigma}\\
\lim_{\omega\to\infty}\bar\asf^{(p'-1,1)}(\infty)&=\infty.\label{negsigma}
\end{alignat}
\end{subequations}
Note that for the functions in \eqref{possigma}, the signs
$\sigma^n=+1$ apply, and for \eqref{negsigma} it is
$\bar\sigma^{p}=-1$. Therefore
\be
I(\infty)+\bar I(\infty)=2\,L(1)\cdot(3-1)=\frac 23\pi^2.
\ee
Finally we have
\be
\textrm{rhs\ of\ \eqref{NLIE.B12mod}} =
\frac{\pi^2}3\left(2-6p'\frac{\gamma^2}{\pi^2}\right).
\ee
Lastly we state the identity for $I_+(\omega)$ explicitly ($I_-(\omega)$ is
trivially 0)
by using the terminals $\asf^j_{+\infty},\bar\asf^j_{+\infty}$ with $j=1,\dots,p'$ as given in
\eqref{eq:braid.an.A12}
and writing $N$ for $p'$ and $\eE^{\ir\phi}$ for $\omega$:
\begin{alignat}{2}
  &2\sum_{n =1}^{N-2}L\left(
\frac{\sin\frac{1}{2}\phi\,\sin\phi}
{\sin\frac{n+1}{2}\phi\,\sin\frac{n+2}{2}\phi}
\right)\nonumber\\
&+
L\left(\frac{\sin\frac{N+1}{2}\phi}
{2\sin\frac{N}{2}\phi\,\cos\frac{1}{2}\phi}\right)
+
L\left(\eE^{\ir\frac{N-1}{2}\phi}\,\frac{\sin\frac{N+1}{2}\phi}
{\sin{N}\phi}\right)
+
L\left(\eE^{-\ir\frac{N-1}{2}\phi}\,\frac{\sin\frac{N+1}{2}\phi}
{\sin{N}\phi}\right)
+
L\left(\frac{\sin N\phi\,\sin\phi}
{\sin^2\frac{N+1}{2}\phi}\right)\nonumber\\
&-
L\left(\frac{\sin\frac{N-1}{2}\phi}
{2\sin\frac{N}{2}\phi\,\cos\frac{1}{2}\phi}\right)
-
L\left(\eE^{\ir\frac{N+1}{2}\phi}\,\frac{\sin\frac{N-1}{2}\phi}
{\sin{N}\phi}\right)
-
L\left(\eE^{-\ir\frac{N+1}{2}\phi}\,\frac{\sin\frac{N-1}{2}\phi}
{\sin{N}\phi}\right)
-
L\left(\frac{\sin^2\frac{N-1}{2}\phi}
{\sin^2\frac{N+1}{2}\phi}\right)\nonumber\\
&=\frac{\pi^2}3\,.\label{A21dilogIdPdual}
\end{alignat}


\end{document}